%% file: TMVAUsersGuide.tex
\pdfoutput=1
\documentclass[11pt,twoside]{myArticle}       % CUSTOMIZED (!) article

\usepackage{amsmath,amssymb,amsfonts} % Typical maths resource packages
\usepackage{graphicx}
\usepackage{epsfig}
\usepackage{color}                    % For creating coloured text and background
\usepackage{hyperref}                 % For creating hyperlinks in cross references
\usepackage{myFancyhdr}               % CUSTOMIZED (!) fancy headers and footer
\usepackage{makeidx}                  % Package for indexing
\usepackage{helvet}                   % For sans-serif mode
\usepackage{setspace}                 % Allows to customize line distances
\usepackage{fancybox}                 % For code example and option boxes
\usepackage{float}                    % Floating objects
\usepackage{xspace}                   % ...
\usepackage{wasysym}                  % permille sign
\usepackage[numbers,sort&compress]{natbib}
\usepackage{hypernat}
\usepackage{subfigure}
\usepackage{rotating}                 % for sideways tables/figures
\newcommand\TMVAVersion{4.0.1\xspace}
\definecolor{darkblue1}{rgb}{0,0,.7}
\definecolor{darkblue}{rgb}{0,0,.3}
\definecolor{darkred}{rgb}{0.5,0,0}
\pagecolor{white} % Background color
\color{black}     % Text color
\hypersetup{breaklinks=true, 
            colorlinks=true, 
            linkcolor=darkblue1, 
            menucolor=darkblue1, 
            urlcolor=darkblue1,
            citecolor=darkblue1,
            pdftitle={TMVA - Users Guide},
            pdfauthor={TMVA},
            pdfsubject={TMVA - Users Guide},
            pdfkeywords={},
            pdfproducer={TMVA}
}
\bibstyle{plain}
\input TMVAdef
\makeindex
\begin{document}
% ============================================================================

%
% ------------ title
%
\input Title
%
% ------------ contents, figures, tables
%
\pagenumbering{roman}

\setcounter{tocdepth}{3}
\addtolength{\parskip}{-0.40\baselineskip}
{\footnotesize\sl
\twocolumn
\tableofcontents
\onecolumn
}
\addtolength{\parskip}{0.40\baselineskip} 
% \listoffigures
% \listoftables

\newpage

\pagenumbering{arabic}
%
% ------------ introduction
%
\input Introduction
%
% ------------ technicalities
%
\input UsingTMVAQuickStart

\input UsingTMVA

\vfill\pagebreak
\input DataPreprocessing

\input CommonTools

\vfill\pagebreak
%
% ------------ MVA methods
%
\input MethodsIntro

%
\input Cuts

\input Likelihood

\input PDERS

\input PDEFoam
\input KNN
\input HMatrix

\input Fisher
\input LD
\input FDA
\input MLPs
\input SVM
\input BDTs

\input RuleFit
\input Combining
%
% ------------ conclusions
% 
\input Conclusions
%
% ------------ references
% 
\newpage
\input Appendix
%
% ------------ references
% 
\newpage

\input Bibliography
%
% ------------ print and include index
%
\addcontentsline{toc}{section}{Index}
\printindex

% ============================================================================
\end{document}

%% file: TMVAdef.tex
\renewcommand{\headrulewidth}{0.5pt}    %Underlines the header. (Set to 0pt if not required).
\renewcommand{\footrulewidth}{0.5pt}    %Underlines the footer. (Set to 0pt if not required).
\newcommand{\HRule}{\rule{\linewidth}{1mm}}

\newcommand\allFontSize{\small}

\usepackage[bf]{caption}
\renewcommand{\captionfont}{\allFontSize}

\newcommand\detailsSize{\allFontSize}
\newenvironment{details}%
{\begin{myquote}\detailsSize}{\end{myquote}}

\newcommand\codeexampleCaptionSize{\allFontSize}
\newfloat{codeexample}{H}{loc}
\floatname{codeexample}{\sf\allFontSize Code Example}

\newlength{\tmvaboxwidth}
\definecolor{DarkGray}{rgb}{0.4,0.42,0.45}
\definecolor{LightGray}{rgb}{0.97,0.98,0.98}

\newenvironment{tmvacode}%
{\VerbatimEnvironment\normalsize
\setlength{\tmvaboxwidth}{\textwidth}\addtolength{\tmvaboxwidth}{-2.3\fboxsep}%
\begin{Sbox}\begin{minipage}[t]{\tmvaboxwidth}\begin{Verbatim}}%
{\end{Verbatim}\end{minipage}\end{Sbox}\vspace{1ex} \fcolorbox{DarkGray}{LightGray}{\TheSbox}}

\newcommand\optionCaptionSize{\allFontSize}
\newfloat{option}{H}{loc}
\floatname{option}{\sf\allFontSize Option Table}

\def\sze{\footnotesize}
\usepackage{tabularx}
\newenvironment{optiontableAuto}%
{\setlength{\tmvaboxwidth}{\textwidth}\addtolength{\tmvaboxwidth}{-2.3\fboxsep}%
\begin{Sbox}\begin{tabular*}{\tmvaboxwidth}{>{\rule[-0.5ex]{.0ex}{3.0ex}\tt\sze}p{3cm}>{\sze}p{0.7cm}>{\tt\sze}p{2.0cm}>{\tt\sze}p{2.7cm}>{\sze}p{5.4cm}}
&&\\[-0.8cm]
\rm\sze Option  & \rm\sze Array    & \rm\sze Default    & \rm\sze Predefined Values    & \rm\sze Description\\[0.1cm]\hline
&&\\[-0.50cm]}%
{\\[-0.1cm]\end{tabular*}\end{Sbox}\vspace{1ex} \fbox{\TheSbox}}

{\setlength{\tmvaboxwidth}{\textwidth}\addtolength{\tmvaboxwidth}{-2.3\fboxsep}%
\begin{Sbox}\begin{tabular*}{\tmvaboxwidth}{>{\rule[-0.5ex]{.0ex}{3.0ex}\tt\sze}p{3cm}>{\tt\sze}p{4.5cm}>{\sze}p{5.7cm}}
&&\\[-0.7cm]
\rm\sze Option  & \rm\sze  Values   & \rm\smszeall Description\\[0.1cm]\hline
&&\\[-0.45cm]}%
{\\[-0.1cm]\end{tabular*}\end{Sbox}\vspace{1ex} \fbox{\TheSbox}}

\newenvironment{optiontableDescr}%
{\setlength{\tmvaboxwidth}{\textwidth}\addtolength{\tmvaboxwidth}{-2.3\fboxsep}%
\begin{Sbox}\begin{tabular*}{\tmvaboxwidth}{>{\rule[-0.5ex]{.0ex}{3.0ex}\tt\sze}p{3cm}>{\tt\sze}p{3.5cm}>{\sze}p{6.7cm}}
&&\\[-0.7cm]
\rm\sze Option  & \rm\sze Values    & \rm\sze Description\\[0.1cm]\hline
&&\\[-0.45cm]}%
{\\[-0.1cm]\end{tabular*}\end{Sbox}\vspace{1ex} \fbox{\TheSbox}}

{\setlength{\tmvaboxwidth}{\textwidth}\addtolength{\tmvaboxwidth}{-2.3\fboxsep}%
\begin{Sbox}\begin{tabular*}{\tmvaboxwidth}{>{\rule[-0.5ex]{.0ex}{3.5ex}\tt\sze}p{3.0cm}>{\tt\sze}p{3.0cm}>{\sze}p{7.2cm}}
&&\\[-0.7cm]
\rm\sze Option  & \rm\sze Values    & \rm\sze Description\\[0.1cm]\hline
&&\\[-0.45cm]}%
{\\[-0.1cm]\end{tabular*}\end{Sbox}\vspace{1ex} \fbox{\TheSbox}}

{\setlength{\tmvaboxwidth}{\textwidth}\addtolength{\tmvaboxwidth}{-2.3\fboxsep}%
\begin{Sbox}\begin{tabular*}{\tmvaboxwidth}{>{\rule[-0.5ex]{.0ex}{3.0ex}\tt\sze}p{2.5cm}>{\tt\sze}p{3.5cm}>{\sze}p{7.2cm}}
&&\\[-0.7cm]
\rm\sze Option  & \rm\sze Values    & \rm\sze Description\\[0.1cm]\hline
&&\\[-0.45cm]}%
{\\[-0.1cm]\end{tabular*}\end{Sbox}\vspace{1ex} \fbox{\TheSbox}}

{\setlength{\tmvaboxwidth}{\textwidth}\addtolength{\tmvaboxwidth}{-2\fboxsep}%
\begin{Sbox}\begin{tabular*}{\tmvaboxwidth}{>{\rule[-0.5ex]{.0ex}{3.0ex}\tt\sze}p{4.5cm}>{\tt\sze}p{3.2cm}>{\sze}p{5.5cm}}
&&\\[-0.7cm]
\rm\sze Option  & \rm\sze Values    & \rm\sze Description\\[0.1cm]\hline
&&\\[-0.45cm]}%
{\\[-0.1cm]\end{tabular*}\end{Sbox}\vspace{1ex} \fbox{\TheSbox}}

\newenvironment{optiontableLong2}%
{\setlength{\tmvaboxwidth}{\textwidth}\addtolength{\tmvaboxwidth}{-2\fboxsep}%
\begin{Sbox}\begin{tabular*}{\tmvaboxwidth}{>{\rule[-0.5ex]{.0ex}{3.0ex}\tt\sze}p{4.0cm}>{\tt\sze}p{3.0cm}>{\sze}p{7.6cm}}
&&\\[-0.7cm]
\rm\sze Option  & \rm\sze Values    & \rm\sze Description\\[0.1cm]\hline
&&\\[-0.45cm]}%
{\\[-0.1cm]\end{tabular*}\end{Sbox}\vspace{1ex} \fbox{\TheSbox}}

\newcommand\programCaptionSize{\allFontSize}
\newfloat{programs}{H}{loc}
\floatname{programs}{\sf Table}

\usepackage{tabularx}
\newenvironment{programtable}%
{\setlength{\tmvaboxwidth}{\textwidth}\addtolength{\tmvaboxwidth}{-2\fboxsep}%
\begin{Sbox}\begin{tabular*}{\tmvaboxwidth}{>{\rule[-0.5ex]{.0ex}{3.0ex}\tt\small}p{4.3cm}>{\small}p{10.6cm}}
&\\[-0.7cm]
\rm\small Macro & \rm\small Description\\[0.1cm]\hline
&\\[-0.45cm]}%
{\\[-0.1cm]\end{tabular*}\end{Sbox}\vspace{1ex} \fbox{\TheSbox}}

\newcommand\email{\begingroup \urlstyle{rm}\Url}
\newcommand\emailsf{\begingroup \urlstyle{sf}\Url}
\newcommand\code{\begingroup \urlstyle{tt}\Url}
\newcommand\spacecode[1]{{\tt #1}} % use if space must be obeyed
\newcommand\T{\UrlTildeSpecial}

\newcommand\urlsm[1]{{\small\url{#1}}}
\newcommand\ie{i.e.\xspace}
\newcommand\eg{e.g.\xspace}
\newcommand\ea{{\em et al.}\xspace}
\newcommand\cf{cf.\xspace}
\newcommand\Lik{{\cal L}}
\newcommand\RLik{\ensuremath{ y_\Lik}\xspace}
\newcommand\RPDERS{\ensuremath{ y_\text{\mbox{PDE-RS}}}\xspace}
\newcommand\RPDEFoam{\ensuremath{ y_\text{\mbox{PDE-Foam}}}\xspace}
\newcommand\HMATRIX{\ensuremath{ y_H}\xspace}

\newcommand\xPDF{\ensuremath{\hat x}\xspace}
\newcommand\yPDFSB{\ensuremath{{\hat y}_{S(B)}}\xspace}
\newcommand\yPDFS{\ensuremath{{\hat y}_{S}}\xspace}
\newcommand\yPDFB{\ensuremath{{\hat y}_{B}}\xspace}
\newcommand\Rarity{{\cal R}}

\newcommand{\Signif}{\ensuremath{\cal S}\xspace}
\newcommand{\Fisher}{\ensuremath {y_\text{Fi}}\xspace}
\newcommand{\yMVA}{\ensuremath{y}\xspace}
\newcommand{\yBoost}{\ensuremath{ y_\text{Boost}\xspace}}
\newcommand{\yRF}{\ensuremath{ y_{\rm RF}}\xspace}
\newcommand{\yANN}{\ensuremath{ y_{\rm ANN}}\xspace}
\newcommand{\yANNa}{\ensuremath{ y_{{\rm ANN},a}}\xspace}
\newcommand\proba{\ensuremath{ P_{\!S}}\xspace}
\newcommand\FisherMean{\ensuremath{ \overline y_\text{Fi}\xspace}}
\newcommand\beq{\begin{equation}}
\newcommand\eeq{\end{equation}}
\newcommand\beqn{\begin{eqnarray}}
\newcommand\eeqn{\end{eqnarray}}
\newcommand\beqns{\begin{eqnarray*}}
\newcommand\eeqns{\end{eqnarray*}}
\newcommand{\intl}{\int\limits}

\newcommand{\mc}{\multicolumn}
\newcommand\rs{\raisebox{1.3ex}[-1.5ex]}
\newcommand{\e}{\varepsilon}
\newcommand{\eS}{\ensuremath{ \e_{\!S}}\xspace}
\newcommand{\eB}{\ensuremath{ \e_{\!B}}\xspace}
\newcommand{\rB}{\ensuremath{ r_{\!B}}\xspace}
\newcommand{\NS}{\ensuremath{ N_{\!S}}\xspace}
\newcommand{\NB}{\ensuremath{ N_{\!B}}\xspace}
\newcommand{\NSB}{\ensuremath{ N_{\!S(B)}}\xspace}
\newcommand{\WS}{\ensuremath{ W_{\!S}}\xspace}
\newcommand{\WB}{\ensuremath{ W_{\!B}}\xspace}
\newcommand{\WSB}{\ensuremath{ W_{\!S(B)}}\xspace}
\newcommand{\fS}{\ensuremath{ f_{\!S}}\xspace}
\newcommand{\Separation}{{\ensuremath{\langle S^2\rangle}}\xspace}
\newcommand{\Nvar}{{\ensuremath{n_\text{var}}}\xspace}
\newcommand{\Ntar}{{\ensuremath{n_\text{tar}}}\xspace}
\newcommand{\Ntrain}{{\ensuremath{N}}\xspace}
\newcommand{\Dfrac}[2]{\frac{\displaystyle#1}{\displaystyle#2}}

\newcommand\AD{0.15cm}
\newcommand\BD{-0.3cm}
\newcommand\Good{$\star\star$}
\newcommand\OK{$\star$}
\newcommand\Bad{$\circ$}
\newcommand{\YES}{$\bullet$}
\newcommand{\NO}{$\circ$}
\newcommand{\Hessian}{\ensuremath{H}\xspace}
\newcommand{\IHessian}{\ensuremath{\Hessian^{-1}}\xspace}
\newcommand{\IHessiank}{\ensuremath{\Hessian^{-1(k)}}\xspace}
\newcommand{\IHessiankmone}{\ensuremath{\Hessian^{-1(k-1)}}\xspace}

%% file: Title.tex
\vspace{-1cm}
\begin{flushright}
{\sf\em arXiv:physics/0703039 [Data Analysis, Statistics and Probability]} \\
{\sf\em CERN-OPEN-2007-007} \\
{\sf\em TMVA version \TMVAVersion} \\
{\sf\em \today} \\
\def\UrlFont{\sf\em}
\url{http://tmva.sourceforge.net} 
\end{flushright}
\def\UrlFont{\rm}

\def\miniPageOffset{0.2cm}
\def\miniPageWidth{13.5cm}
\vspace*{\stretch{20}}
\HRule
\begin{flushleft}
\hspace{\miniPageOffset}\begin{minipage}{\miniPageWidth}
{\sf\Huge\bfseries\boldmath TMVA 4} \\[0.2cm]
{\sf\Large\bfseries\boldmath Toolkit for Multivariate Data Analysis with ROOT} \\[1cm]
{\sf\Huge\bfseries\boldmath Users Guide} 
\end{minipage}
\end{flushleft}
\HRule
\vspace{2.0cm}
%\begin{flushleft}
\begin{flushright}
% \hspace{\miniPageOffset}\begin{minipage}{\miniPageWidth}
{\sf\Large  A.~Hoecker,~P.~Speckmayer,~J.~Stelzer,~J.~Therhaag,~E.~von Toerne,~H.~Voss} 
% \end{minipage}

\vspace{1.2cm}
% \hspace{\miniPageOffset}\begin{minipage}{\miniPageWidth}
{\sf\em\large Contributed to TMVA have:} \\[0.4cm]
{\sf\large 
M.~Backes,
T.~Carli,
O.~Cohen,
A.~Christov, 
D.~Dannheim,
K.~Danielowski,\\[0.1cm]
S.~Henrot-Versill\'e, 
M.~Jachowski, 
K.~Kraszewski,
A.~Krasznahorkay Jr.,  \\[0.1cm]    
M.~Kruk,
Y.~Mahalalel, 
R.~Ospanov, 
X.~Prudent, 
A.~Robert,
D.~Schouten,  \\[0.1cm]  
F.~Tegenfeldt,
A.~Voigt,
K.~Voss,
M.~Wolter, 
A.~Zemla
}

% \end{minipage}
\vspace*{\stretch{1}}
\end{flushright}
%\end{flushleft}

\vfill

\thispagestyle{empty}
\newpage

% authors

\def\UrlFont{\sf}
{\small\sf
{\sf\bfseries Abstract} --- 
In high-energy physics, with the search for ever smaller signals in
ever larger data sets, it has become essential to extract a maximum of
the available information from the data.  Multivariate classification
methods based on machine learning techniques have become a fundamental
ingredient to most analyses.  Also the multivariate classifiers
themselves have significantly evolved in recent years. Statisticians
have found new ways to tune and to combine classifiers to further gain
in performance. Integrated into the analysis framework ROOT, TMVA is
a toolkit which hosts a large variety of multivariate classification
algorithms. Training, testing, performance evaluation and application of 
all available classifiers is carried out simultaneously via user-friendly 
interfaces. With version 4, TMVA has been extended to multivariate 
regression of a real-valued target vector. Regression is invoked through 
the same user interfaces as classification. TMVA 4 also features 
more flexible data handling allowing one to arbitrarily form 
combined MVA methods. A generalised boosting method is the first 
realisation benefiting from the new framework.
}
\vspace{1.5cm}
\begin{center}
{\small\sf
{\sf\bfseries TMVA \TMVAVersion\ -- Toolkit for Multivariate Data Analysis with ROOT}  \\
Copyright\index{Copyright} 
\copyright\  2005-2009, Regents of 
CERN (Switzerland),  
DESY (Germany),
MPI-Kernphysik Heidelberg (Germany),
University of Bonn (Germany),
and University of Victoria (Canada). \\
BSD license: \url{http://tmva.sourceforge.net/LICENSE}. 

{\sf\bfseries Authors:} \\
Andreas Hoecker (CERN, Switzerland) \emailsf{<andreas.hoecker@cern.ch>}, \\
Peter Speckmayer (CERN, Switzerland) \emailsf{<peter.speckmayer@cern.ch>}, \\
J\"org Stelzer (CERN, Switzerland) \emailsf{<joerg.stelzer@cern.ch>},\\
Jan Therhaag (Universit\"at Bonn, Germany) \emailsf{<therhaag@physik.uni-bonn.de>}, \\
Eckhard von Toerne (Universit\"at Bonn, Germany) \emailsf{<evt@physik.uni-bonn.de>},\\
Helge Voss (MPI f\"ur Kernphysik Heidelberg, Germany) \emailsf{<helge.voss@cern.ch>},\\
Moritz Backes (Geneva University, Switzerland) \emailsf{moritz.backes@cern.ch},\\
Tancredi Carli (CERN, Switzerland) \emailsf{<tancredi.carli@cern.ch>}, \\
Or Cohen (CERN, Switzerland and Technion, Israel) \emailsf{<or.cohen@cern.ch>}, \\
Asen Christov (Universit\"at Freiburg, Germany) \emailsf{<christov@physik.uni-freiburg.de>}, \\
Krzysztof Danielowski (IFJ and AGH/UJ, Krakow, Poland) \emailsf{<Krzysztof.Danielowski@cern.ch>}, \\
Dominik Dannheim (CERN, Switzerland) \emailsf{<Dominik.Dannheim@cern.ch>}, \\
Sophie Henrot-Versill\'e (LAL Orsay, France) \emailsf{<versille@lal.in2p3.fr>}, \\
Matthew Jachowski (Stanford University, USA) \emailsf{<jachowski@stanford.edu>}, \\
Kamil Kraszewski (IFJ and AGH/UJ, Krakow, Poland) \emailsf{<kamil.bartlomiej.kraszewski@cern.ch>}, \\
Attila Krasznahorkay Jr. (CERN, CH, and Manchester U., UK)  \emailsf{<Attila.Krasznahorkay@cern.ch>}, \\
Maciej Kruk (IFJ and AGH/UJ, Krakow, Poland) \emailsf{<maciej.mateusz.kruk@cern.ch>}, \\
Yair Mahalalel (Tel Aviv University, Israel) \emailsf{<yair@mahalalel.com>}, \\
Rustem Ospanov (University of Texas, USA) \email{<rustem@fnal.gov>}, \\
Xavier Prudent (LAPP Annecy, France) \emailsf{<prudent@lapp.in2p3.fr>}, \\
Doug Schouten (S. Fraser U., Canada) \emailsf{<dschoute@sfu.ca>}, \\
Fredrik Tegenfeldt (Iowa University, USA) \emailsf{<fredrik.tegenfeldt@cern.ch>}, \\
Arnaud Robert (LPNHE Paris, France) \emailsf{<arobert@lpnhe.in2p3.fr>}, \\
Alexander Voigt (CERN, Switzerland) \emailsf{<alexander.voigt@cern.ch>}, \\
Kai Voss (University of Victoria, Canada) \emailsf{<kai.voss@cern.ch>}, \\
Marcin Wolter (IFJ PAN Krakow, Poland) \emailsf{<marcin.wolter@ifj.edu.pl>}, \\
Andrzej Zemla (IFJ PAN Krakow, Poland) \emailsf{<zemla@corcoran.if.uj.edu.pl>}, \\
and valuable contributions from many users, please see acknowledgements.
}
\end{center}

\thispagestyle{empty}
\newpage

%%% Local Variables: 
%%% mode: latex
%%% TeX-master: "TMVAUsersGuide"
%%% End: 

%% file: Introduction.tex
\section{Introduction}
\label{sec:introduction}

The Toolkit for Multivariate Analysis (TMVA) provides a ROOT-integrated~\cite{ROOT} 
environment for the processing, parallel evaluation and application of multivariate 
classification and -- since TMVA version 4 -- multivariate regression techniques.\footnote
{
   A classification problem corresponds in more general terms to a 
   {\em discretised regression} problem. A regression is the 
   process that estimates the parameter values of a function, which
   predicts the value of a response variable (or vector)
   in terms of the values of other variables (the {\em input} variables). 
   A typical regression problem in High-Energy Physics is for example the estimation of
   the energy of a (hadronic) calorimeter cluster from the cluster's electromagnetic 
   cell energies. The user provides a single dataset that contains the input variables 
   and one or more target variables. The interface to defining the input and target variables,
   the booking of the multivariate methods, their training and testing is very similar to 
   the syntax in classification problems. Communication between the user and TMVA proceeds
   conveniently via the Factory and Reader classes. Due to their similarity, classification 
   and regression are introduced together in this Users Guide. Where necessary, 
   differences are pointed out.
}
All multivariate methods in TMVA respond to {\em supervised} learning only, \ie,
the input information is mapped in feature space to the desired outputs. The mapping 
function can contain various degrees of approximations and may be a single global 
function, or a set of local models. 
TMVA is specifically designed for the needs of high-energy physics (HEP) applications, 
but should not be restricted to these. The package includes:
\begin{itemize}

\item Rectangular cut optimisation (binary splits, Sec.~\ref{sec:cuts}),

\item Projective likelihood estimation (Sec.~\ref{sec:likelihood}),

\item Multi-dimensional likelihood estimation (PDE range-search -- Sec.~\ref{sec:pders}, 
      PDE-Foam -- Sec.~\ref{sec:pdefoam}, and k-NN -- Sec.~\ref{sec:knn}),

\item Linear and nonlinear discriminant analysis 
      (H-Matrix -- Sec.~\ref{sec:hmatrix}, Fisher -- Sec.~\ref{sec:fisher}, 
      LD -- Sec.~\ref{sec:ld}, FDA -- Sec.~\ref{sec:fda}),

\item Artificial neural networks (three different
      multilayer perceptron implementations -- Sec.~\ref{sec:ann}),

\item Support vector machine (Sec.~\ref{sec:SVM}),

\item Boosted/bagged decision trees (Sec.~\ref{sec:bdt}),

\item Predictive learning via rule ensembles (RuleFit, Sec.~\ref{sec:rulefit}),  

\item A generic boost classifier, allowing one to boost any of the above 
      classifiers (Sec.~\ref{sec:combine}). 

\end{itemize}

The software package consists of abstract, object-oriented implementations in C++/ROOT for 
each of these multivariate analysis (MVA) techniques, as well as auxiliary tools such as 
parameter fitting and transformations. It provides training, testing and performance evaluation 
algorithms and visualisation scripts. Detailed descriptions of all the TMVA methods and 
their options for classification and (where available) regression tasks are given in 
Sec.~\ref{sec:tmvaClassifiers}. Their training and testing is 
performed with the use of user-supplied data sets in form of ROOT trees or text files, where
each event can have an individual weight. The true sample composition (for event classification) 
or target value (for regression) in these data sets must be supplied for each event. 
Preselection requirements and transformations 
can be applied to input data. TMVA supports the use of variable combinations and 
formulas with a functionality similar to the one available for the \code{Draw} command of a ROOT tree.

TMVA works in transparent factory mode to guarantee an unbiased performance comparison 
between MVA methods: they all see the same training and test data, and are 
evaluated following the same prescriptions within the same execution job. A {\em Factory} 
class organises the interaction between the user and the TMVA analysis steps. It performs 
preanalysis and preprocessing of the training data to assess basic properties of the 
discriminating variables used as inputs to the classifiers. The linear correlation 
coefficients of the input variables are calculated and displayed. For regression, also 
nonlinear correlation measures are given, such as the correlation ratio and mutual 
information between input variables and output target. A preliminary 
ranking is derived, which is later superseded by algorithm-specific variable rankings. 
For classification problems, the variables can be linearly transformed (individually 
for each MVA method) into a non-correlated variable space, projected upon their principle 
components, or transformed into a normalised Gaussian shape. Transformations can also 
be arbitrarily concatenated.

To compare the signal-efficiency and background-rejection performance of the classifiers, 
or the average variance between regression target and estimation, the analysis job prints --
among other criteria -- tabulated results for some benchmark values (see 
Sec.~\ref{sec:usingtmva:evaluation}). Moreover, a variety of graphical evaluation information
acquired during the training, testing and evaluation phases is stored in a ROOT output 
file. These results can be displayed using macros, which are conveniently executed 
via graphical user interfaces (each one for classification and regression) 
that come with the TMVA distribution (see Sec.~\ref{sec:rootmacros}).

The TMVA training job runs alternatively as a ROOT script, as a standalone executable, or 
as a python script via the PyROOT interface. Each MVA method trained in one of these 
applications writes its configuration and training results in a result (``weight'') file, 
which in the default configuration has human readable XML format.

A light-weight {\em Reader} class is provided, which reads and interprets the 
weight files (interfaced by the corresponding method), and which can 
be included in any C++ executable, ROOT macro, or python analysis job
(see Sec.~\ref{sec:usingtmva:reader}).

For standalone use of the trained MVA method, TMVA also generates lightweight C++ response 
classes (not available for all methods), which contain the encoded information from the 
weight files so that these are not required anymore. These classes do not depend on TMVA 
or ROOT, neither on any other external library (see Sec.~\ref{sec:usingtmva:standaloneClasses}).

We have put emphasis on the clarity and functionality of the Factory and Reader interfaces 
to the user applications, which will hardly exceed a few lines of code. All MVA methods
run with reasonable default configurations and should have satisfying performance for 
average applications. {\em We stress however that, to solve a concrete problem, all 
methods require at least some specific tuning to deploy their maximum classification or 
regression capabilities.} Individual optimisation and customisation of the classifiers is 
achieved via configuration strings when booking a method.

This manual introduces the TMVA Factory and Reader interfaces, and describes design and 
implementation of the MVA methods. It is not the aim here to provide a general introduction 
to MVA techniques. Other excellent reviews exist on this subject (see, \eg, 
Refs.~\cite{FriedmanBook,WebbBook,KunchevaBook}). The document begins with a quick TMVA 
start reference in Sec.~\ref{sec:quickstart}, and provides a more complete introduction 
to the TMVA design and its functionality for both, classification and regression analyses
in Sec.~\ref{sec:usingtmva}. Data preprocessing such as the transformation of input variables 
and event sorting are discussed in Sec.~\ref{sec:dataPreprocessing}. In Sec.~\ref{sec:PDF}, 
we describe the techniques used to estimate probability density functions from the training
data. Section~\ref{sec:fitting} introduces optimisation and fitting tools commonly used by 
the methods. All the TMVA methods including their configurations and tuning options are 
described in Secs.~\ref{sec:cuts}--\ref{sec:rulefit}. Guidance on which MVA method to use 
for varying problems and input conditions is given in Sec.~\ref{sec:whatMVAshouldIuse}. 
An overall summary of the implementation status of all TMVA methods is provided in 
Sec.~\ref{sec:classifierSummary}.

\subsubsection*{Copyrights and credits}
\addcontentsline{toc}{subsection}{Copyrights and credits}

\begin{details}
TMVA is an open source product. Redistribution and use of TMVA in source and binary forms, 
with or without modification, are permitted according to the terms listed in the 
BSD license\index{License}.\footnote
{
  For the BSD l
icense, see \urlsm{http://tmva.sourceforge.net/LICENSE}. 
}
Several similar combined multivariate analysis (``machine learning'') packages exist 
with rising importance in most fields of science and industry. In the HEP
community the package {\em StatPatternRecognition}~\cite{narsky,MVAreferences} 
is in use (for classification problems only). The idea of parallel training and 
evaluation of MVA-based classification in HEP has been pioneered by the {\em Cornelius} 
package, developed by the Tagging Group of the BABAR Collaboration~\cite{Cornelius}. 
See further credits and acknowledgments on page~\pageref{sec:Acknowledgments}.
\end{details}

\vfill
\pagebreak

%% file: UsingTMVAQuickStart.tex
\section{TMVA Quick Start}
\label{sec:quickstart}

To run TMVA it is not necessary to know much about its concepts or to understand 
the detailed functionality of the multivariate methods. Better, just begin 
with the quick start tutorial given below. One should note that the TMVA version
obtained from the open source software platform Sourceforge.net (where TMVA is 
hosted), and the one included in ROOT, have different directory structures for 
the example macros used for the tutorial. Wherever differences in command lines 
occur, they are given for both versions. 

Classification and regression analyses in TMVA have similar training, testing and 
evaluation phases, and will be treated in parallel in the following. 

\subsection{How to download and build TMVA}
\label{sec:download}

TMVA is developed and maintained at Sourceforge.net (\urlsm{http://tmva.sourceforge.net}). It
is built upon ROOT (\urlsm{http://root.cern.ch/}), so that for TMVA to run ROOT must be 
installed. Since ROOT version 5.11/06, TMVA
comes as integral part of ROOT and can be used from the ROOT prompt 
without further preparation. For older ROOT versions 
or {\em if the latest TMVA features are desired}, the TMVA source code can be downloaded 
from Sourceforge.net\index{Download!from Sourceforge.net}. Since we do not provide 
prebuilt libraries for any platform, the library must be built by the user (easy -- see below). 
The source code can be either 
\href{http://sourceforge.net/project/showfiles.php?group_id=152074}{downloaded}
as a gzipped tar file or via (anonymous) SVN access:\index{Download!from SVN}
\begin{codeexample}
\begin{tmvacode}
~> svn co https://tmva.svn.sourceforge.net/svnroot/tmva/tags/V04-00-01/TMVA \
          TMVA-4.0.1
\end{tmvacode}
\caption[.]{\codeexampleCaptionSize Source code download via SVN. The latest version 
         (SVN trunk) can be downloaded by typing the same command without specifying 
         a version: \spacecode{svn co http:://...tmva/trunk/TMVA}. For the latest TMVA 
         version see \url{http://tmva.sourceforge.net/}.}
\end{codeexample}
While the source code is known to compile with VisualC++ on Windows (which is
a requirement for ROOT), we do not provide project support for this platform yet.
For Unix and most Linux flavours custom Makefiles\index{Makefile} are provided 
with the TMVA distribution, so that the library can be built by typing:
\begin{codeexample}
\begin{tmvacode}
~> cd TMVA
~/TMVA> source setup.sh # for c-shell family: source setup.csh
~/TMVA> cd src
~/TMVA/src> make 
\end{tmvacode}
\caption[.]{\codeexampleCaptionSize Building the TMVA library under Linux/Unix using the provided
         Makefile. The \code{setup.[c]sh} script must be executed to 
         ensure the correct setting of symbolic links and library paths required by TMVA.}
\end{codeexample}
After compilation, the library \code{TMVA/lib/libTMVA.1.so} should be present.

\subsection{Version compatibility}
\label{sec:intro:compat}

TMVA can be run with any ROOT version equal or above v5.08\index{ROOT!compatibility}.
The few occurring conflicts due to ROOT source code evolution after v5.08 are 
intercepted in TMVA via C++ preprocessor conditions.

\subsection{Avoiding conflicts between external TMVA and ROOT's internal one}

To use a more recent version of TMVA than the one present in the local ROOT 
installation, one needs to download the desired TMVA release from Sourceforge.net, 
to compile it against the local ROOT version, and to make sure the newly built library 
\code{TMVA/lib/libTMVA.1.so} is used instead of ROOT's internal one. 
When running TMVA in a CINT macro the new library must be loaded first via:
\code{gSystem}{\tt ->}\code{Load("TMVA/lib/libTMVA.1")}. This can be done 
directly in the macro or in a file that is automatically loaded at the start 
of CINT (for an example, see the files \code{.rootrc} and \code{TMVAlogon.C} 
in the \code{TMVA/macros/} directory).  When running TMVA in an executable, 
the corresponding shared library needs to be linked.
Once this is done, ROOT's own \code{libTMVA.so} library will not be invoked anymore.

\subsection{The TMVA namespace}
\label{sec:quickstart:namespace}

All TMVA classes are embedded in the namespace \code{TMVA}. For interactive access, or 
use in macros the classes must thus be preceded by \code{TMVA::}, or one may use the 
command \code{using} \code{namespace} \code{TMVA} instead. 

\subsection{Example jobs\index{Examples}}
\label{sec:examplejob}

TMVA comes with example jobs for the training phase (this phase actually 
includes training, testing and evaluation) using the TMVA {\em Factory}\index{Factory}, 
as well as the application of the training results in a classification or regression analysis
using the TMVA {\em Reader}\index{Reader}. The first task is performed in the programs
\code{TMVAClassification}\index{TMVAClassification} or 
\code{TMVARegression}\index{TMVARegression}, respectively, and the second task in 
\code{TMVAClassificationApplication}\index{TMVAClassificationApplication} or
\code{TMVARegressionApplication}\index{TMVARegressionApplication}.

In the ROOT version of TMVA the above macros (extension \code{.C})
are located in the directory {\tt \$}\code{ROOTSYS/tmva/test/}. 

In the Sourceforge.net version these macros are located in \code{TMVA/macros/}.
At Sourceforge.net we also provide these examples in form of the C++ executables
(replace \code{.C} by \code{.cxx}), which are located in 
\code{TMVA/execs/}. To build the executables, type
\spacecode{cd \T/TMVA/execs/; make}, and then simply execute them by typing
\code{./TMVAClassification} or \code{./TMVARegression} (and similarly for the 
applications). To illustrate how TMVA can be used 
in a python script via PyROOT we also provide the script \code{TMVAClassification.py} located 
in \code{TMVA/python/}, which has the same functionality as the macro \code{TMVAClassification.C}
(the other macros are not provided as python scripts).

\subsection{Running the example}
\label{sec:qs:example}

The most straightforward way to get started with TMVA is to simply run the \code{TMVAClassification.C} 
or \code{TMVARegression.C} example macros. Both use academic toy datasets for training 
and testing, which, for classification, consists of four linearly correlated, Gaussian 
distributed discriminating input variables, with different sample means for signal and 
background, and, for regression, has two input variables with fuzzy parabolic dependence 
on the target (\code{fvalue}), and no correlations among themselves. All 
classifiers are trained, tested and evaluated using the toy datasets in the
same way the user is expected to proceed for his or her own data. It
is a valuable exercise to look at the example file in more detail. Most 
of the command lines therein should be self explaining, and one will easily 
find how they need to be customized to apply TMVA to a real use case.
A detailed description is given in Sec.~\ref{sec:usingtmva}.

The toy datasets used by the examples are included in the Sourceforge.net download.
For the ROOT distribution, the macros automatically fetch the data file from 
the web using the corresponding \code{TFile} constructor, \eg,
\code{TFile::Open("http://root.cern.ch/files/tmva_class_example.root")} for classification (\code{tmva_reg_example.root} for regression). The example %ToDo: insert this code into our example execs and macros. 
ROOT macros can be run directly in the \code{TMVA/macros/} directory (Sourceforge.net),
or in any designated test directory \code{workdir}, after adding the macro directory 
to ROOT's macro search path:
\begin{codeexample}
\begin{tmvacode}
~/workdir> echo "Unix.*.Root.MacroPath: ~/TMVA/macros" >> .rootrc
~/workdir> root -l ~/TMVA/macros/TMVAClassification.C
\end{tmvacode}
\caption[.]{\codeexampleCaptionSize Running the example \code{TMVAClassification.C} 
            using the Sourceforge.net version of TMVA (similarly for \code{TMVARegression.C}).}
\end{codeexample}
\begin{codeexample}
\begin{tmvacode}
~/workdir> echo "Unix.*.Root.MacroPath: $ROOTSYS/tmva/test" >> .rootrc
~/workdir> root -l $ROOTSYS/tmva/test/TMVAClassification.C
\end{tmvacode}
\caption[.]{\codeexampleCaptionSize Running the example \code{TMVAClassification.C}
            using the ROOT version of TMVA  (similarly for \code{TMVARegression.C}).}
\end{codeexample}
It is also possible to explicitly select the MVA methods to be processed 
(here an example given for a classification task with the Sourceforge.net version): 
\begin{codeexample}
\begin{tmvacode}
~/workdir> root -l ~/TMVA/macros/TMVAClassification.C\(\"Fisher,Likelihood\"\)
\end{tmvacode}
\caption[.]{\codeexampleCaptionSize Running the example \code{TMVAClassification.C} 
            and processing only the Fisher and likelihood classifiers. Note that
            the backslashes are mandatory. The macro \code{TMVARegression.C} can be 
            called accordingly. }
\end{codeexample}
where the names of the MVA methods are predifined in the macro. 

The training job provides formatted output logging containing analysis information
such as: linear correlation matrices for the input variables, correlation ratios 
and mutual information (see below) between input variables and regression targets,
variable ranking, summaries of the MVA configurations, goodness-of-fit evaluation
for PDFs (if requested), signal and background (or regression target) correlations 
between the various MVA methods, decision overlaps, signal efficiencies
at benchmark background rejection rates (classification) or deviations from target 
(regression), as well as other performance estimators. Comparison between the results
for training and independent test samples provides overtraining validation.
\begin{figure}[p]
  \begin{center}
	  \includegraphics[width=0.47\textwidth]{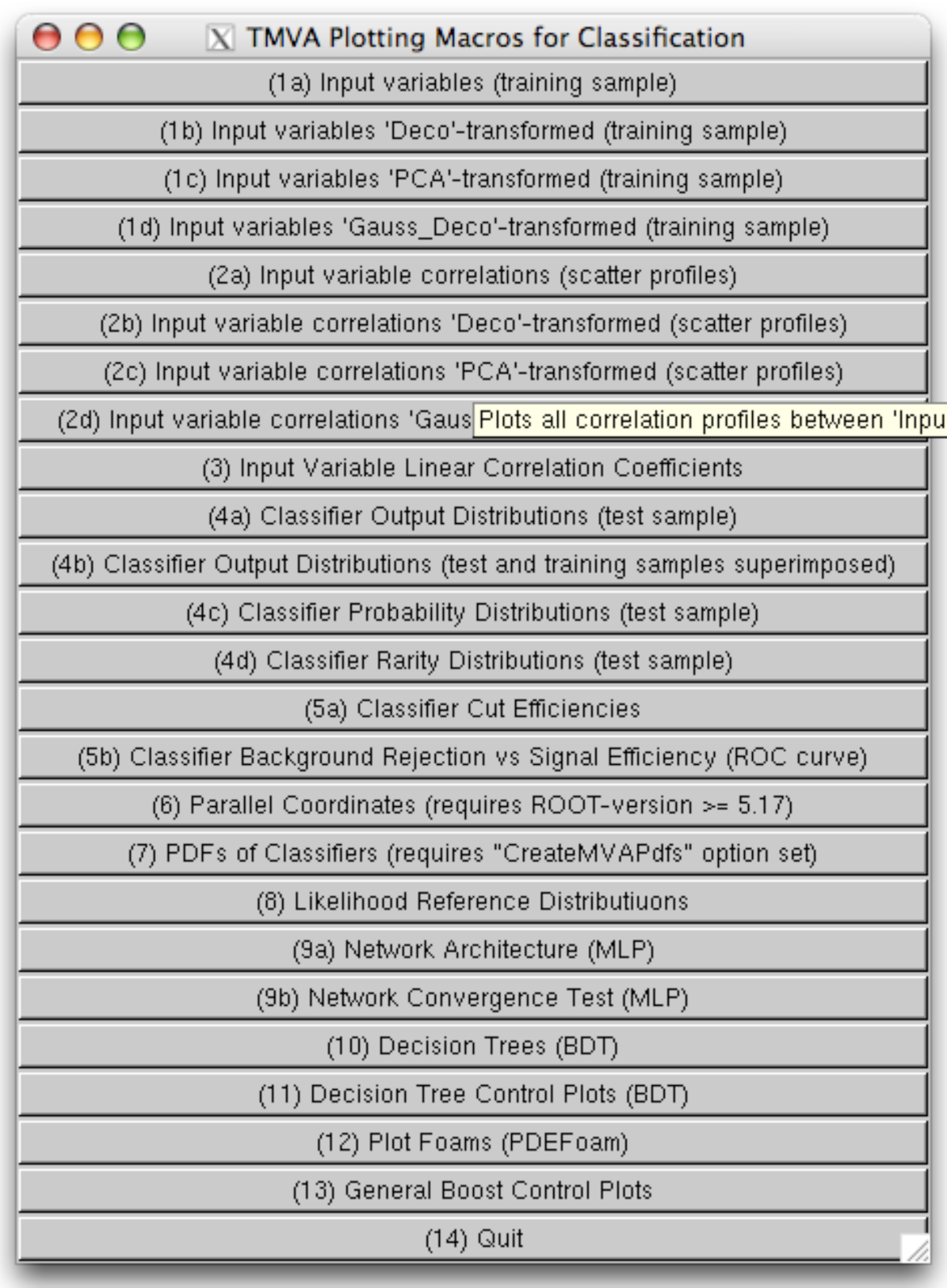}\hspace{0.3cm}
	  \includegraphics[width=0.47\textwidth]{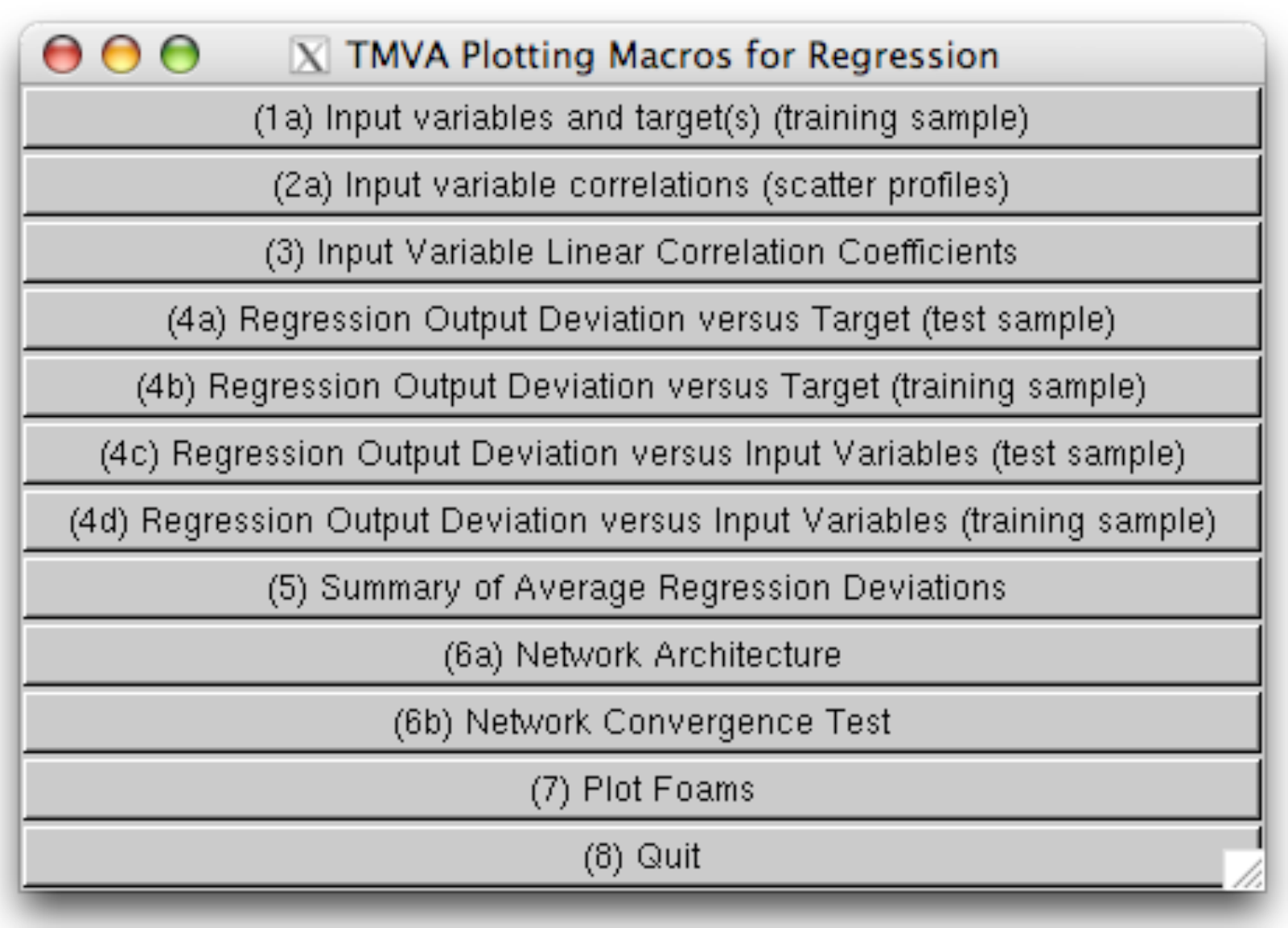}
  \end{center}
  \vspace{-0.1cm}
  \caption[.]{Graphical user interfaces (GUI) to execute  
              macros displaying training, test and evaluation results 
              (\cf\  Tables~\ref{pgr:scripttable1} and \ref{pgr:scripttable2} on page 
              \pageref{pgr:scripttable1}) for classification (left) and regression 
              problems (right).      
              The classification GUI can be launched manually by executing the scripts
              \code{TMVA/macros/TMVAGui.C}\index{TMVAGui} (Sourceforge.net version) 
              or {\tt \$}\code{ROOTSYS/tmva/test/TMVAGui.C} (ROOT version) in a ROOT 
              session. To launch the regression GUI use the macro \code{TMVARegGui.C}. \\[0.2cm]
              {\footnotesize
              \underline{Classification (left)}. The buttons behave as follows:
              (1a) plots the signal and background 
              distributions of input variables (training sample), (1b--d) the 
              same after applying the corresponding preprocessing transformation of the input 
              variables, (2a--f) scatter plots with superimposed profiles for all 
              pairs of input variables for signal and background and the applied transformations
              (training sample), (3) correlation coefficients 
              between the input variables for signal and background (training sample), 
              (4a/b) signal and background distributions for the trained classifiers (test 
              sample/test and training samples superimposed to probe overtraining), 
              (4c,d) the corresponding probability and Rarity distributions of the classifiers
              (where requested, \cf\  see Sec.~\ref{sec:otherRepresentations}), 
              (5a) signal and background efficiencies and purities versus the cut on the classifier
              output for the expected numbers of signal and background events (before applying the cut) 
              given by the user (an input dialog box pops up, where the numbers are inserted),
              (5b) background rejection versus signal efficiency obtained when cutting 
              on the classifier outputs (ROC curve, from the test sample), (6) plot of so-called
              Parallel Coordinates visualising the correlations among the input variables, and 
              among the classifier and the input variables,  (7--13) show classifier 
              specific diagnostic plots, and (14) quits the GUI. Titles greyed out indicate 
              actions that are not available because the corresponding classifier has not 
              been trained or because the transformation was not requested.\\
              \underline{Regression (right)}. The buttons behave as follows:
              (1--3) same as for classification GUI, (4a--d) show the linear deviations between 
              regression targets and estimates versus the targets or input variables for the 
              test and training samples, respectively, (5) compares the average deviations 
              between target and MVA output for the trained methods, and (6--8) are as for the classification GUI.}
}
\label{fig:tmvagui}
\end{figure}

\subsection{Displaying the results}

\begin{figure}[!t]
\begin{center}
  \includegraphics[width=0.8\textwidth]{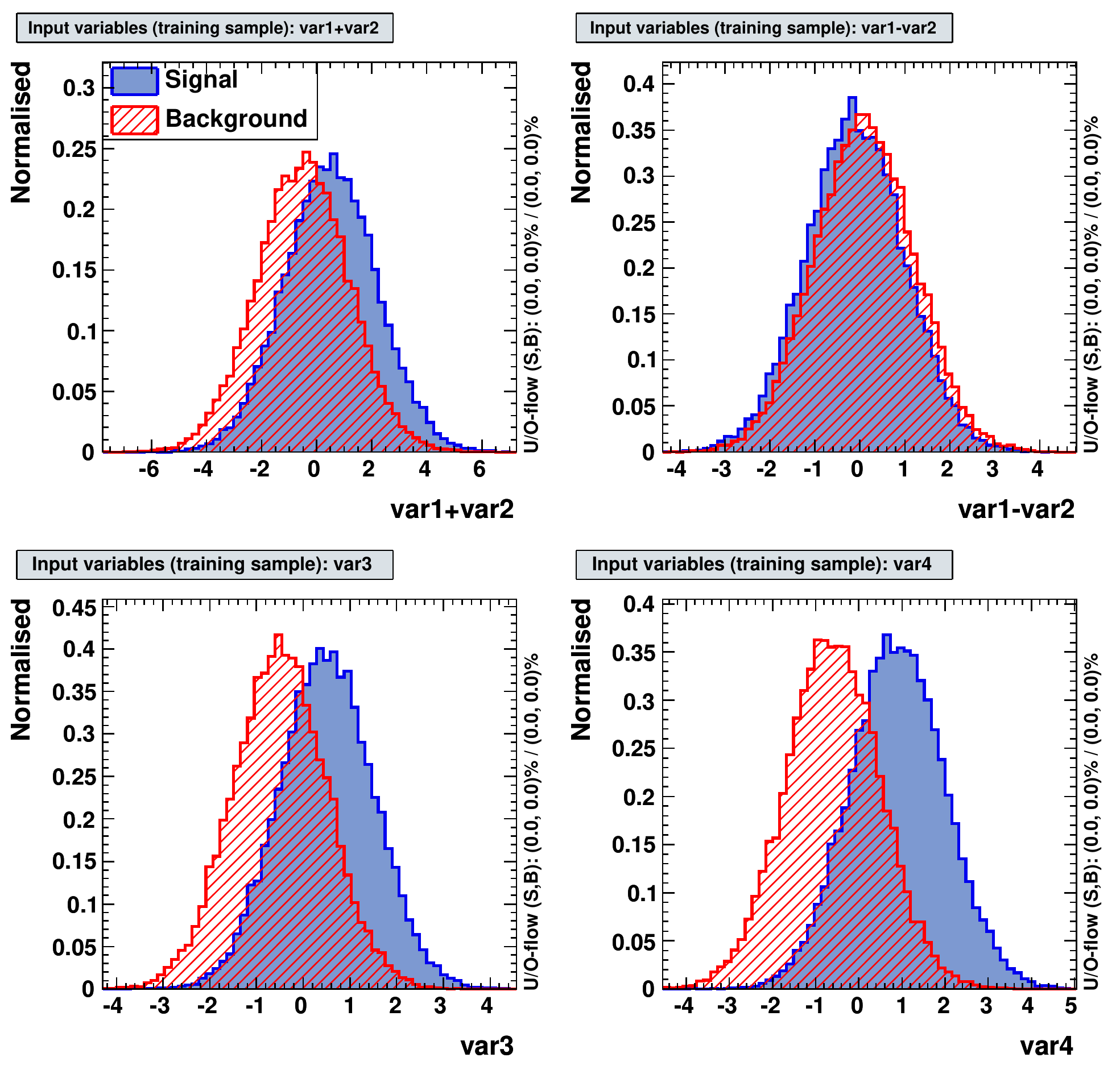}
\end{center}
  \vspace{-0.5cm}
\caption[.]{Example plots for input variable distributions. The histogram
            limits are chosen to zoom into the bulk of the distributions, which
            may lead to truncated tails. The vertical text on the 
            right-hand side of the plots indicates the under- and overflows.
            The limits in terms of multiples of the distribution's RMS can 
            be adjusted in the user script by modifying the variable
            \code{(TMVA::gConfig().GetVariablePlotting()).fTimesRMS} (\cf\  
            Code Example~\ref{ce:gconfig}). }
\label{fig:usingtmva:variables}
\end{figure}
Besides so-called ``weight'' files containing the
method-specific training results, TMVA also provides a variety of
control and performance plots that can be displayed via a set of ROOT
macros available in \code{TMVA/macros/} or
{\tt \$}\code{ROOTSYS/tmva/test/} for the Sourceforge.net and ROOT
distributions of TMVA, respectively. The macros are summarized in
Tables~\ref{pgr:scripttable1} and \ref{pgr:scripttable2} on 
page~\pageref{pgr:scripttable1}.  At the end of the example jobs a graphical 
user interface (GUI)\index{Graphical user interface (GUI)}
is displayed, which conveniently allows to run these macros (see
Fig.~\ref{fig:tmvagui}).  
 
\begin{figure}[t]
\begin{center}
  \def\thissize{0.40}
  \includegraphics[width=\thissize\textwidth]{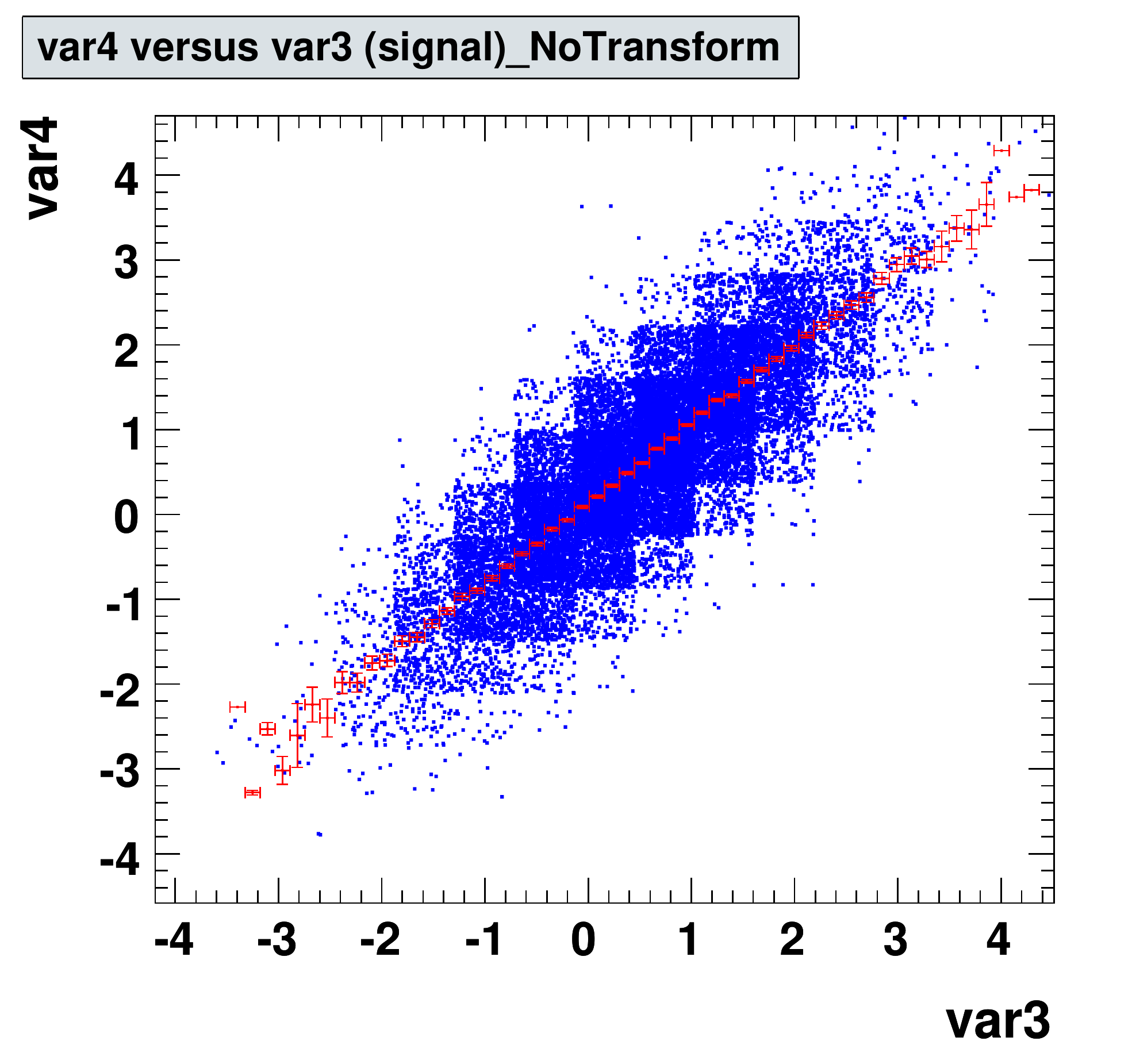}
  \includegraphics[width=\thissize\textwidth]{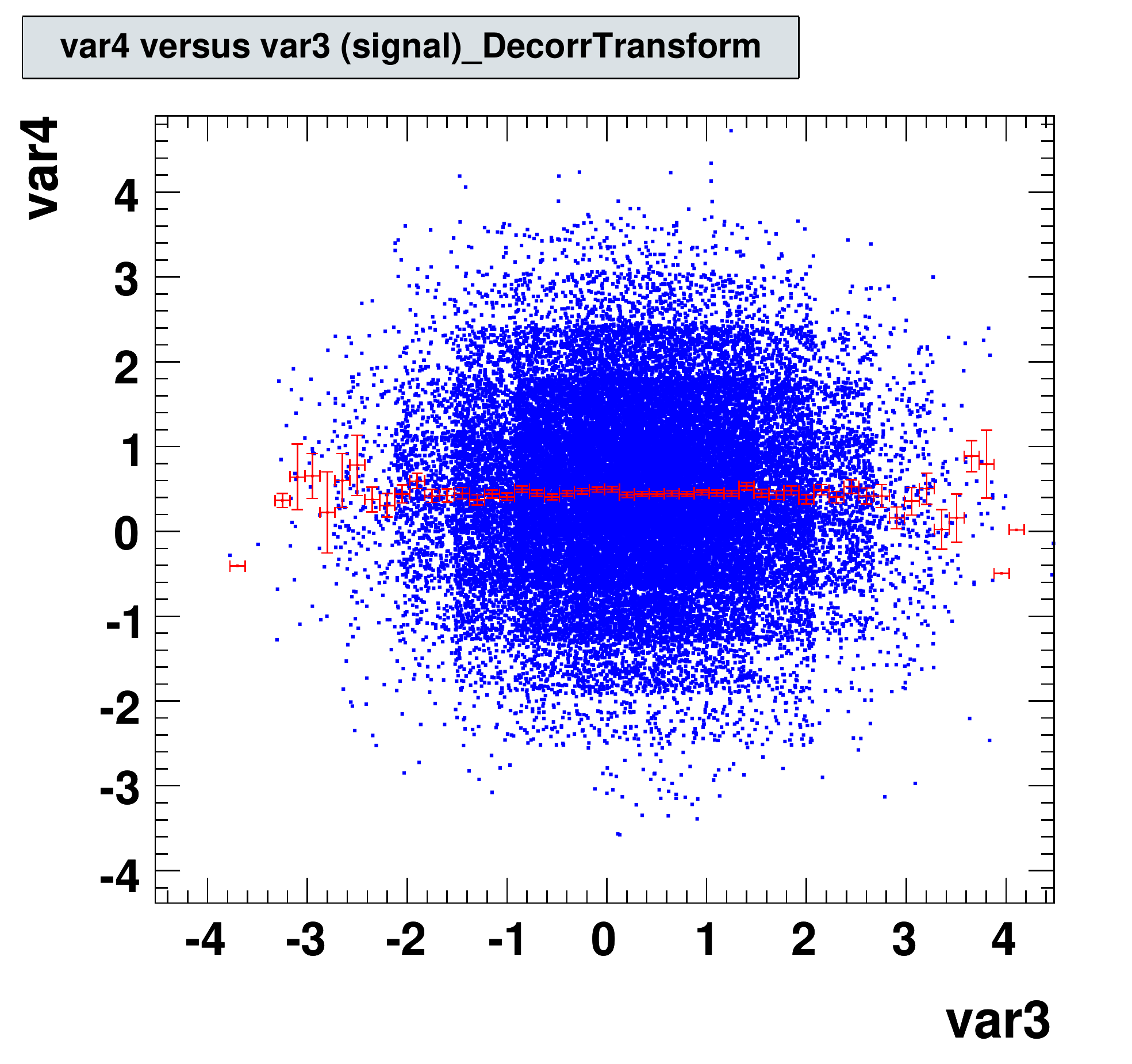}

  \vspace{0.2cm}

  \def\thissize{0.40}
  \includegraphics[width=\thissize\textwidth]{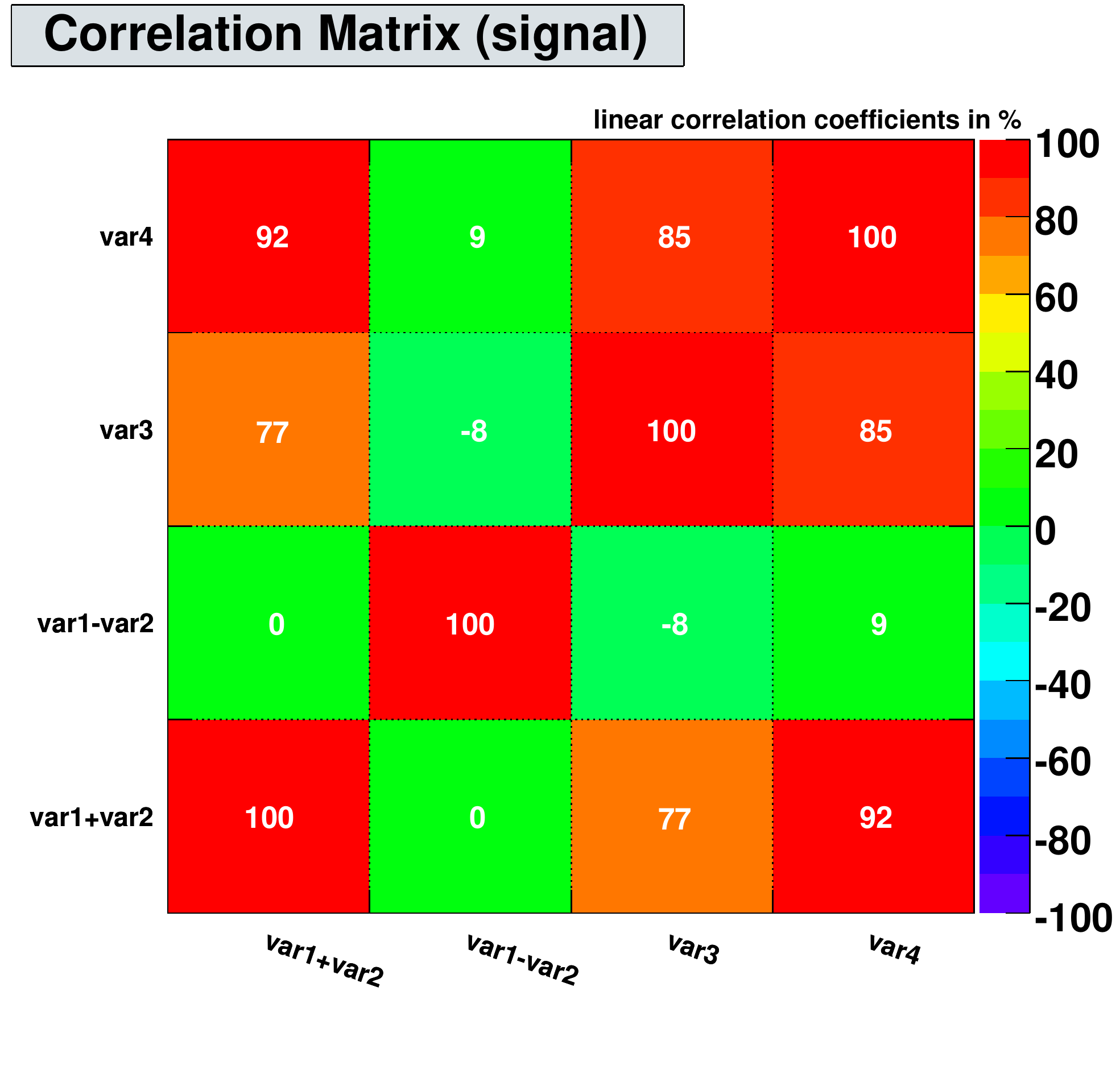}
  \includegraphics[width=\thissize\textwidth]{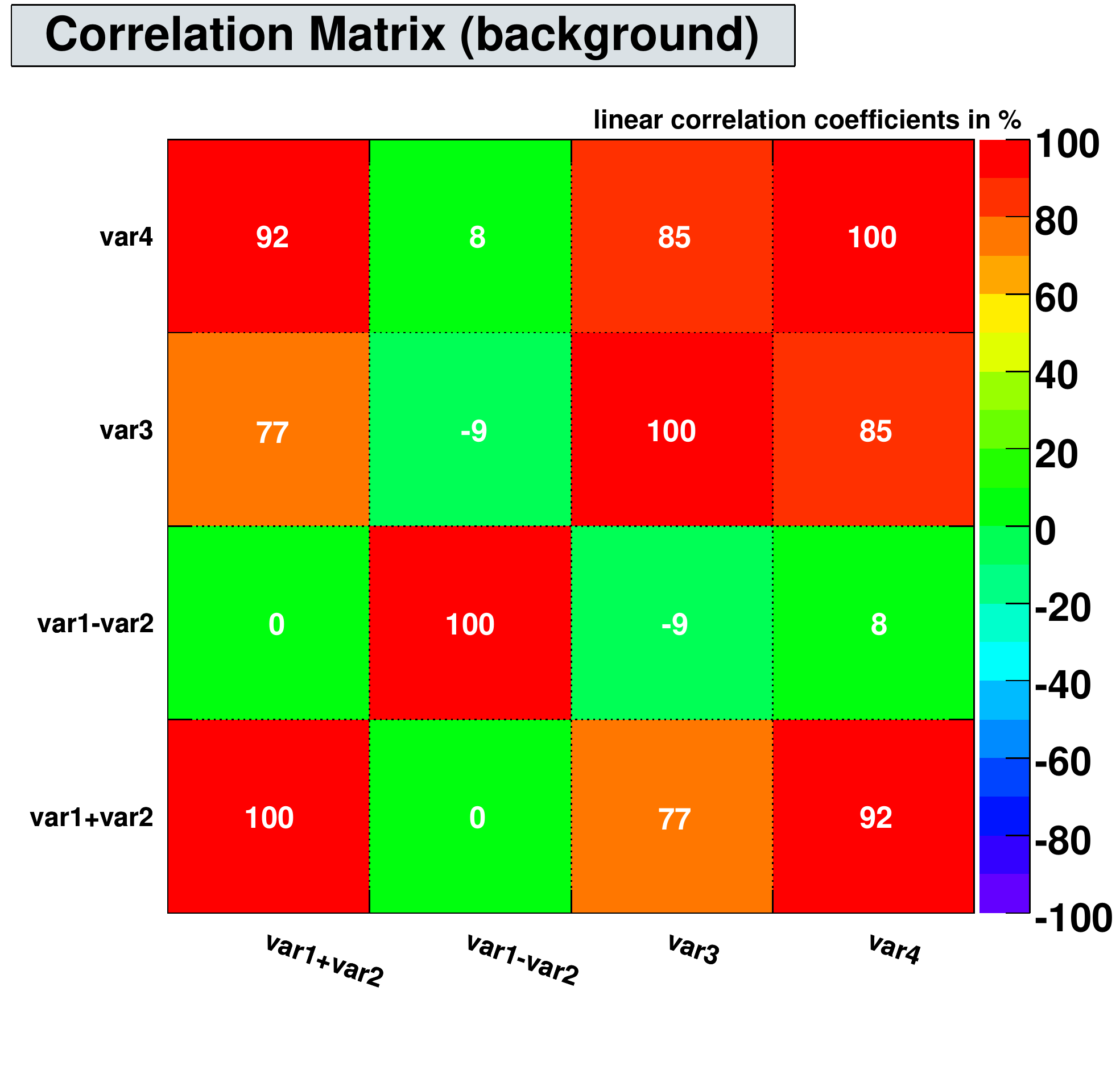}
\end{center}
\vspace{-0.7cm}
\caption[.]{Correlation between input variables. Upper left: correlations
         between var3 and var4 for the signal training sample. 
         Upper right: the same after applying a linear decorrelation transformation 
         (see Sec.~\ref{sec:decorrelation}). Lower plots: 
         linear correlation coefficients for the signal and background 
         training samples.
}
\label{fig:usingtmva:correlations}
\end{figure}

\begin{figure}[t]
\begin{center}
  \includegraphics[width=0.50\textwidth]{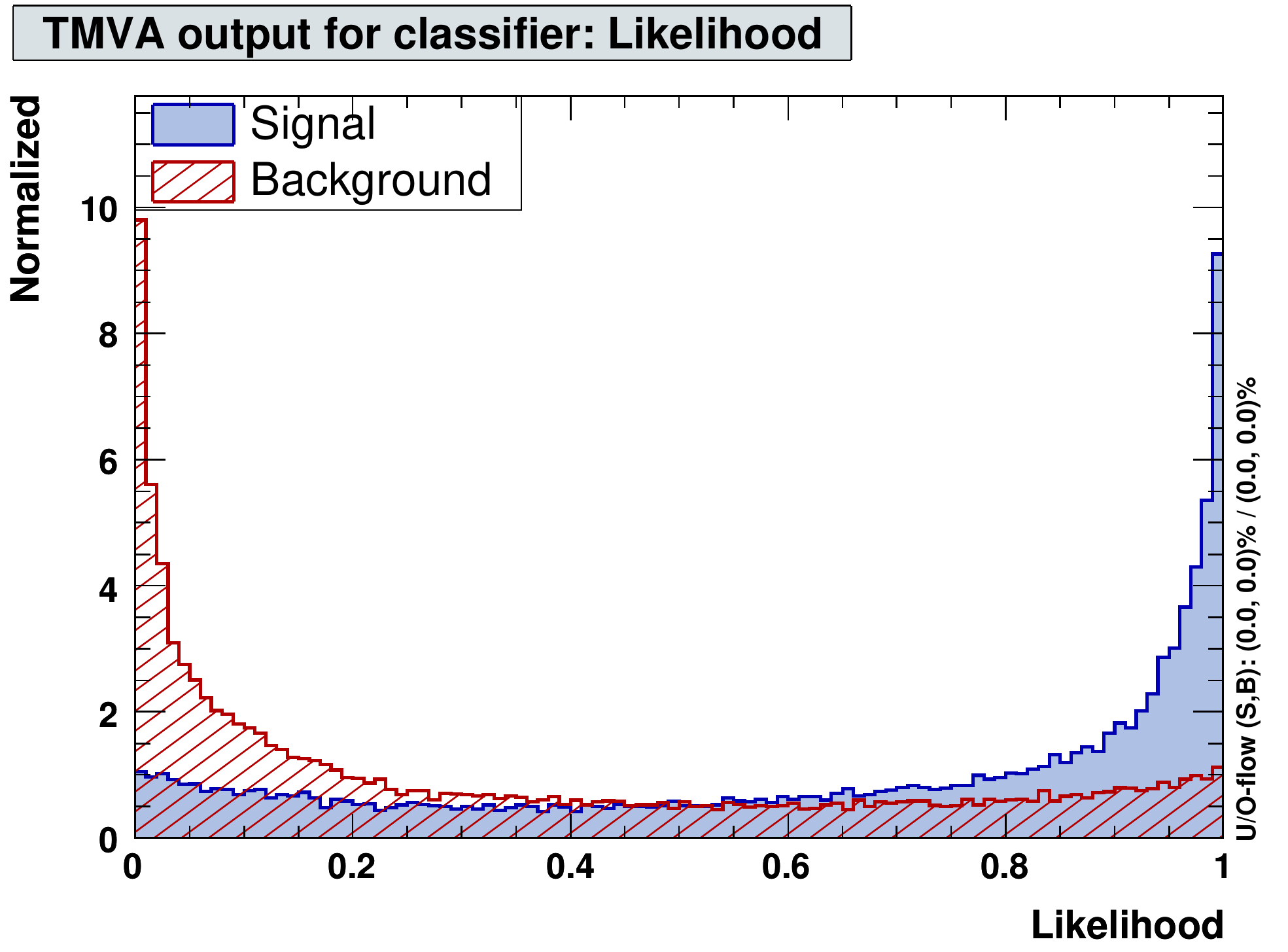}
  \hspace{-0.3cm}
  \includegraphics[width=0.50\textwidth]{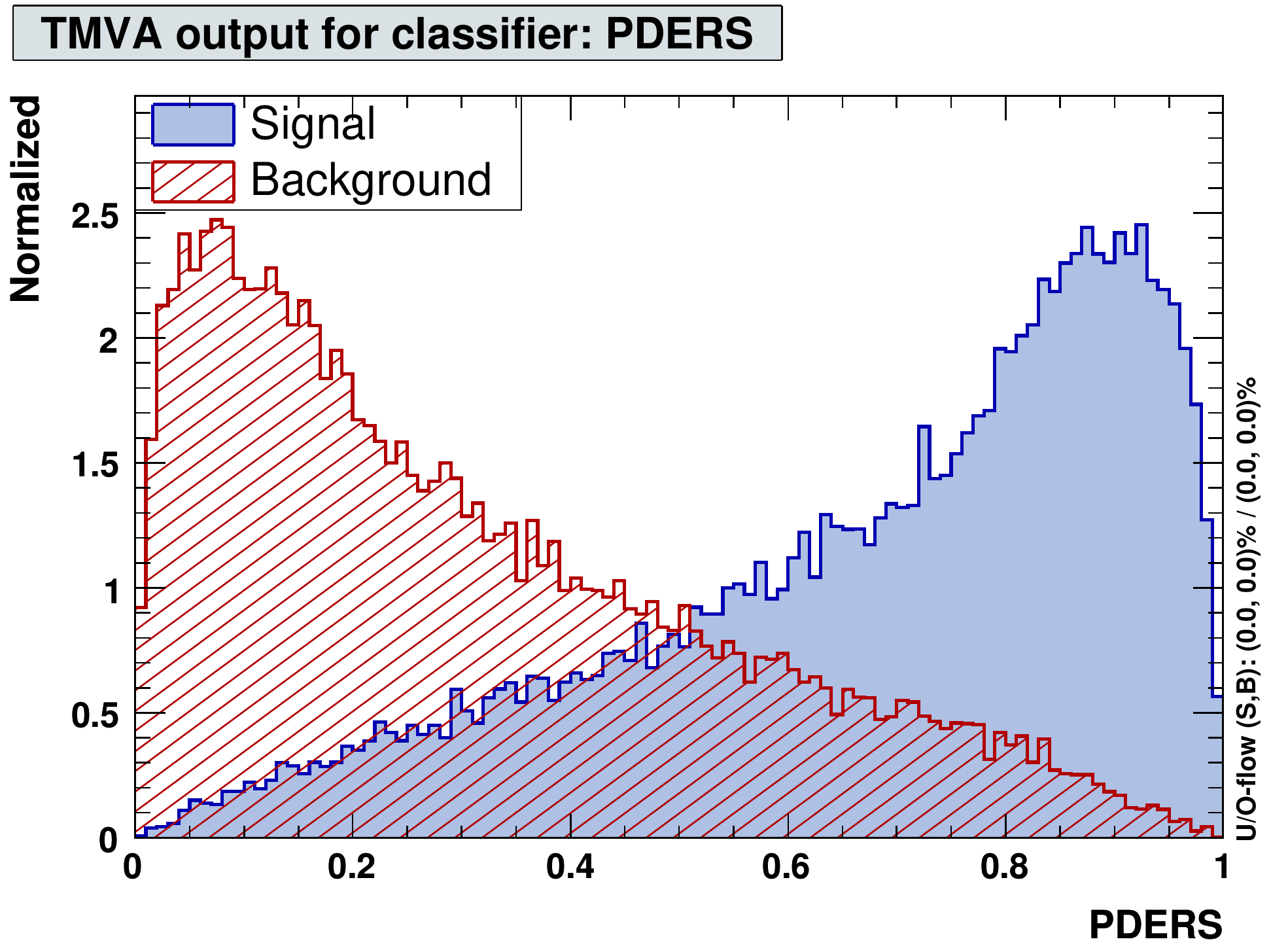}

  \vspace{0.2cm}

  \includegraphics[width=0.50\textwidth]{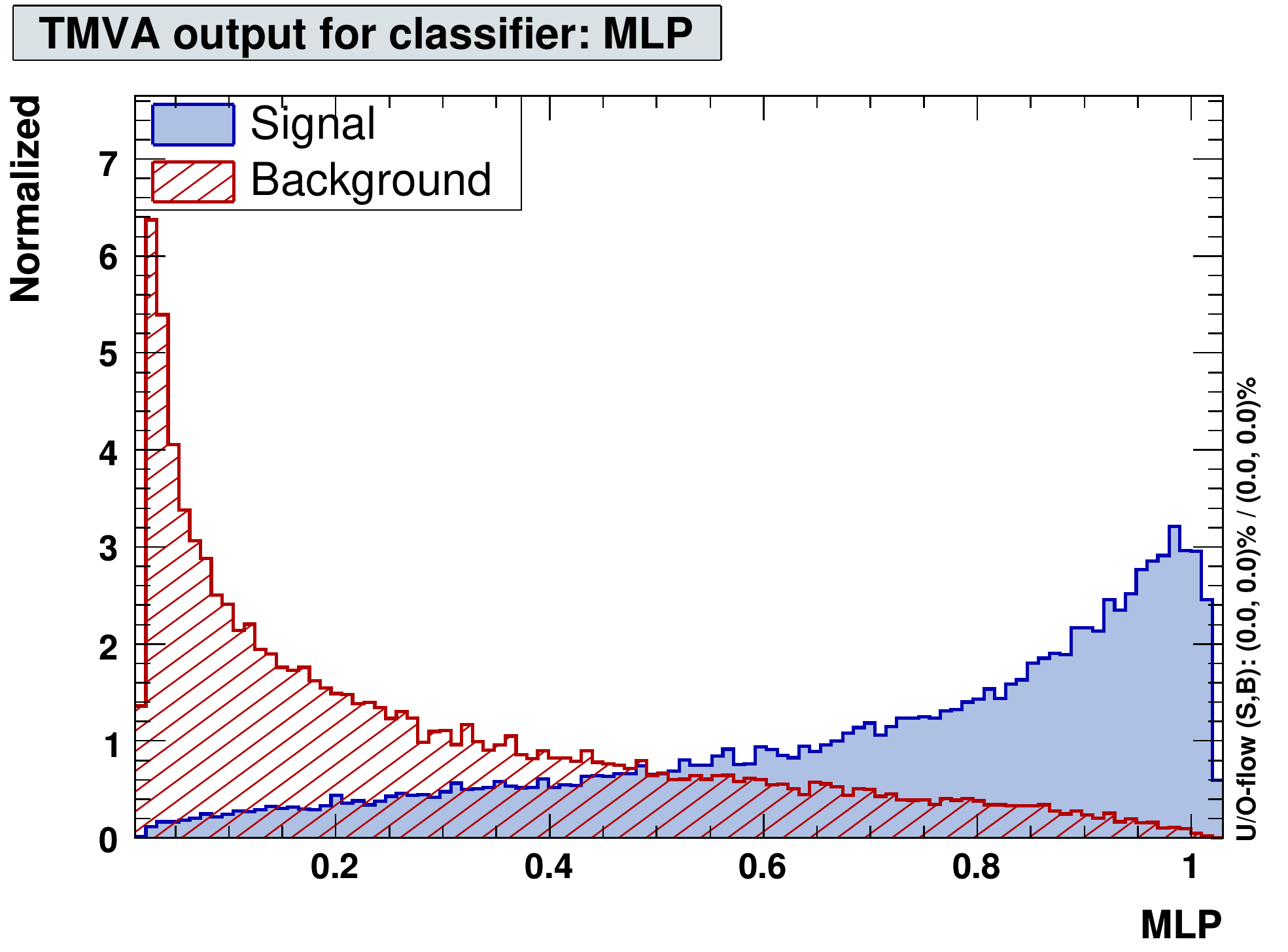}
  \hspace{-0.3cm}
  \includegraphics[width=0.50\textwidth]{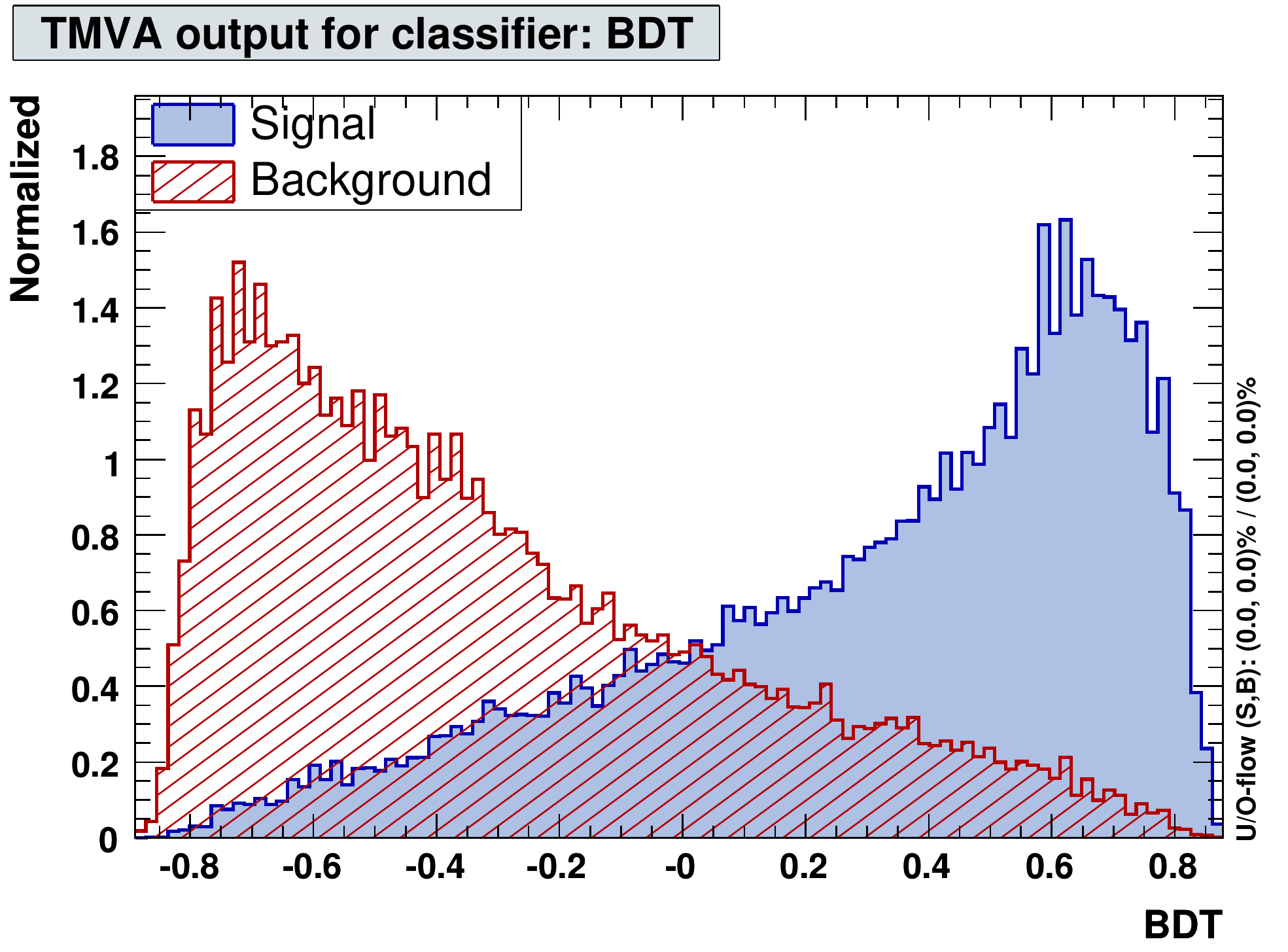}
\end{center}
\vspace{-0.5cm}
\caption[.]{Example plots for classifier output distributions for signal and 
            background events from the academic test sample. Shown are
            likelihood (upper left), PDE range search
            (upper right), Multilayer perceptron (MLP -- lower left) and boosted decision trees.}
\label{fig:usingtmva:mvas}
\end{figure}

\begin{figure}[t]
\begin{center}
  \includegraphics[width=0.65\textwidth]{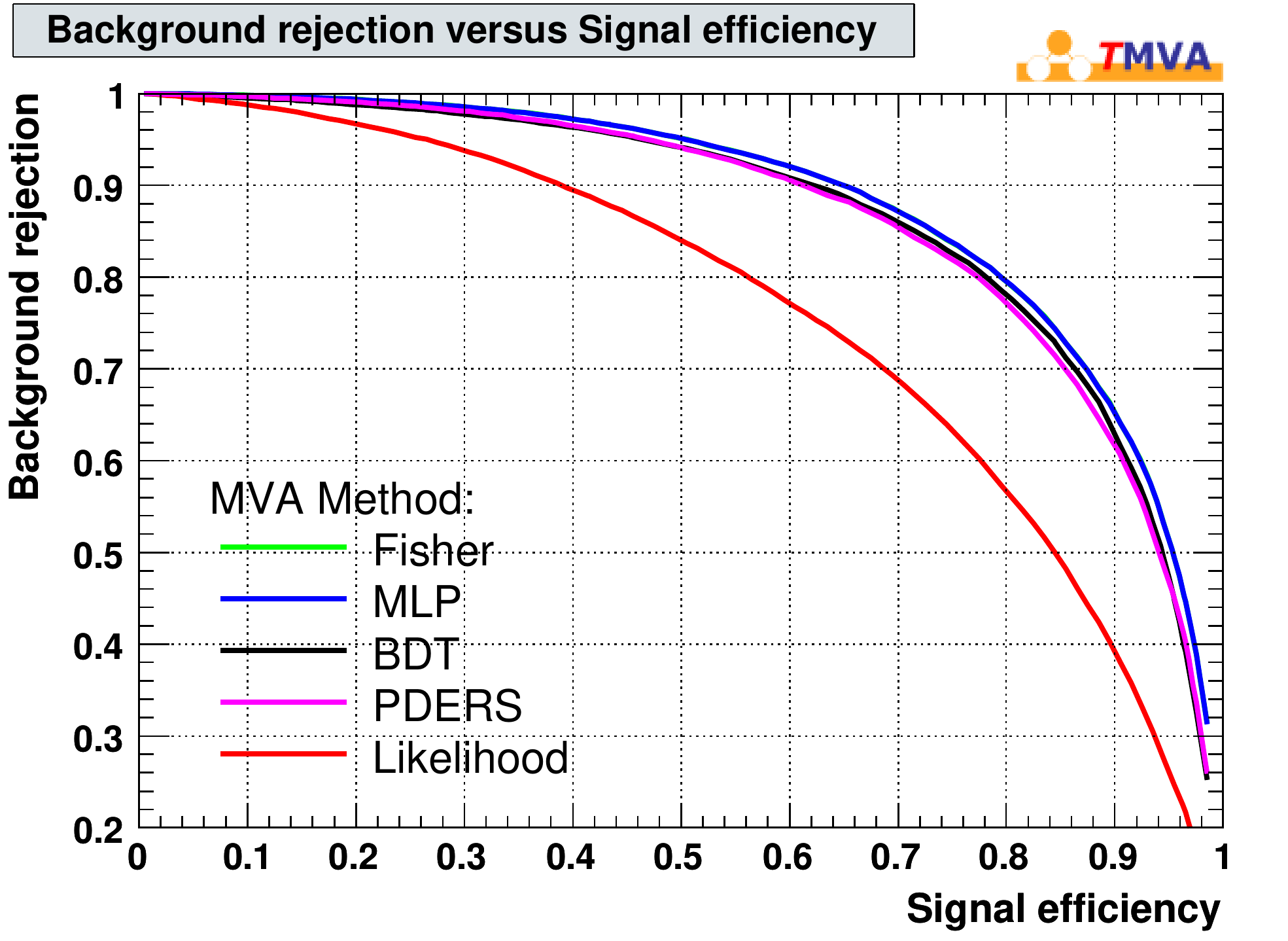}
\end{center}
\vspace{-0.5cm}
\caption[.]{Example for the background rejection versus signal efficiency obtained
            by cutting on the classifier outputs for the events of the test sample.
            \index{Performance evaluation!background rejection vs. signal efficiency} }
\label{fig:usingtmva:rejBvsS}
\end{figure}

Examples for plots produced by these macros are given in 
Figs.~\ref{fig:usingtmva:correlations}--\ref{fig:usingtmva:rejBvsS} for a 
classification problem.
The distributions of the input variables for signal and background 
according to our example job are shown in Fig.~\ref{fig:usingtmva:variables}. 
It is useful to quantify the correlations between the input variables.
These are drawn in form of a scatter plot with the superimposed profile for
two of the input variables in Fig.~\ref{fig:usingtmva:correlations} (upper left).  
As will be discussed in Sec.~\ref{sec:dataPreprocessing}, TMVA allows to perform 
a linear decorrelation transformation of the input variables prior to the MVA
training (for classification only). The result of such decorrelation is shown 
at the upper right hand plot of Fig.~\ref{fig:usingtmva:correlations}. The lower 
plots display the linear correlation coefficients between all input variables, 
for the signal and background training samples of the classification example.

Figure~\ref{fig:usingtmva:mvas} shows several classifier output 
distributions for signal and background events based on the test sample. 
By TMVA convention, signal (background) events accumulate at large 
(small) classifier output values. Hence, cutting on the output and retaining
the events with \yMVA larger than the cut requirement selects signal samples
with efficiencies and purities that respectively decrease and increase with 
the cut value. The resulting relations between background rejection versus 
signal efficiency are shown in Fig.~\ref{fig:usingtmva:rejBvsS} for all 
classifiers that were used in the example macro. This plot belongs to the 
class of {\em Receiver Operating Characteristic} (ROC)\index{Receiver 
Operating Characteristic (ROC)} diagrams,
which in its standard form shows the true positive rate versus the false 
positive rate for the different possible cutpoints of a hypothesis test.

\begin{figure}[t]
\begin{center}
  \includegraphics[width=0.50\textwidth]{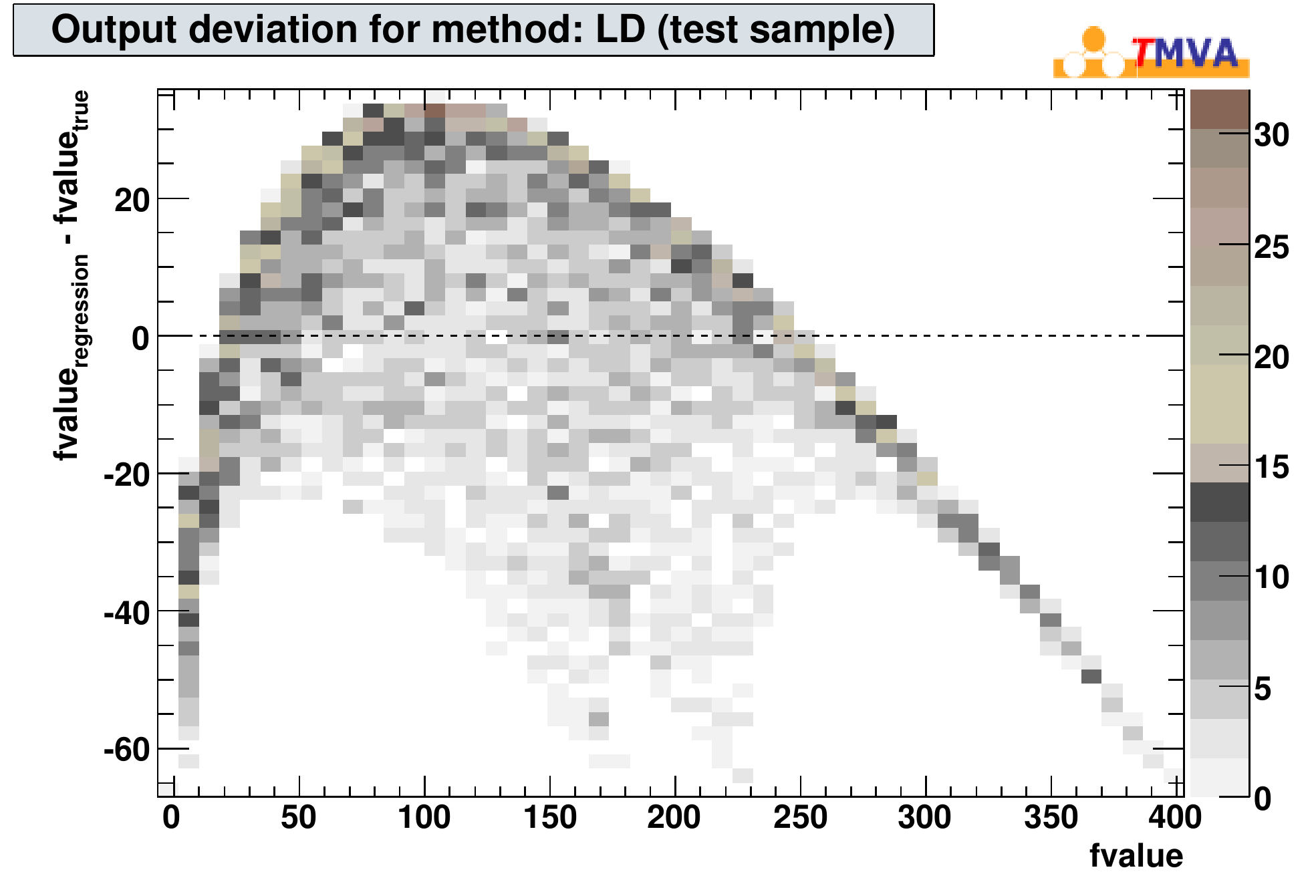}
  \hspace{-0.3cm}
  \includegraphics[width=0.50\textwidth]{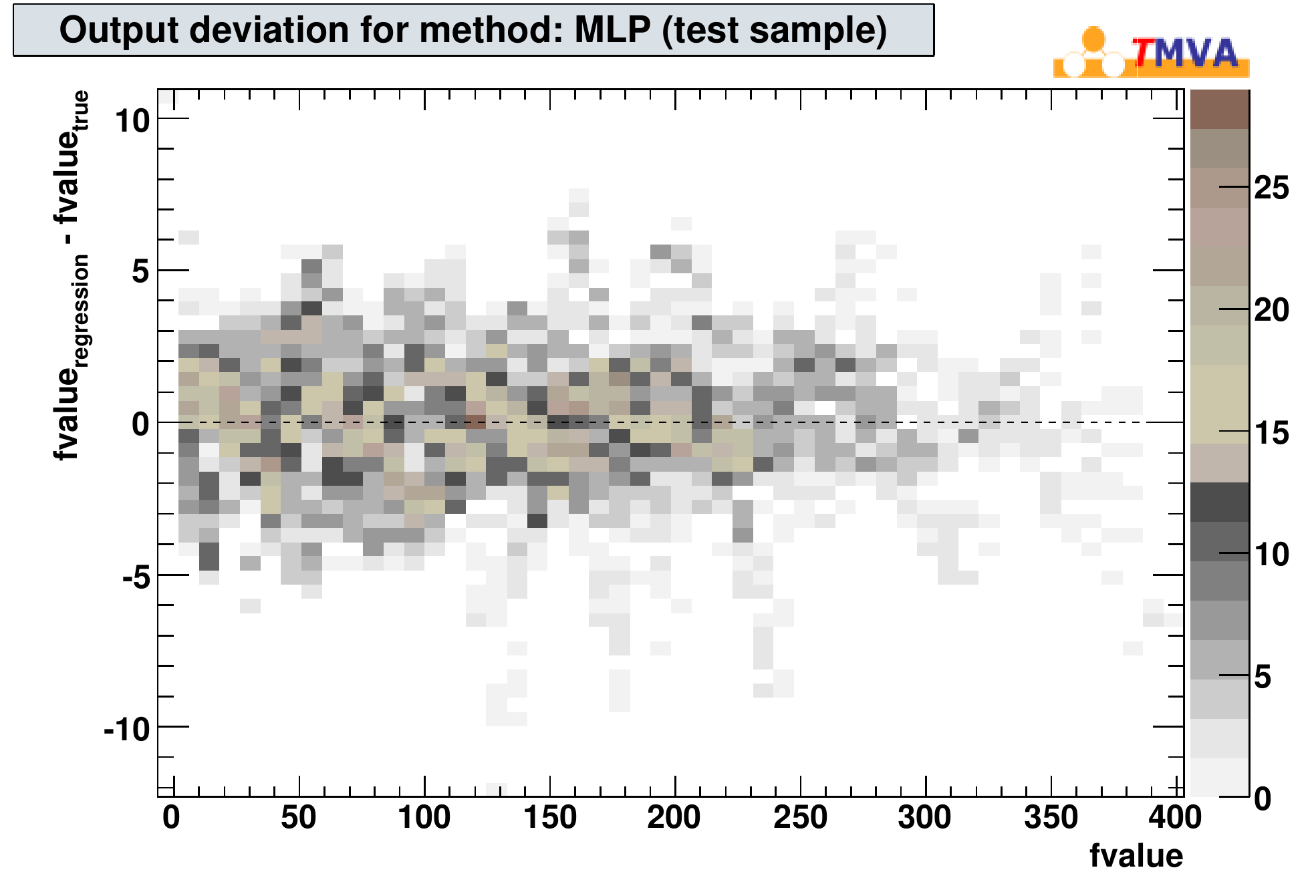}
  \vspace{0.2cm}
\end{center}
\vspace{-1.2cm}
\caption[.]{Example plots for the deviation between regression output and target values
            for a Linear Discriminant (LD -- left) and MLP (right). The dependence of the 
            input variables on the target being strongly nonlinear, LD cannot appropriately 
            solve the regression problem. }
\label{fig:usingtmva:deviation}
\end{figure}

As an example for multivariate regression, Fig.~\ref{fig:usingtmva:deviation} displays
the deviation between the regression output and target values for linear and 
nonlinear regression algorithms. 

More macros are available to validate training and response of specific 
MVA methods. For example, the macro \code{likelihoodrefs.C} compares the 
probability density functions used by the likelihood classifier to the normalised 
variable distributions of the training sample. It is also possible to visualize 
the MLP neural network architecture and to draw decision trees (see 
Table~\ref{pgr:scripttable2}).

\subsection{Getting help}

Several help sources exist for TMVA (all web address given below are also linked from the 
TMVA home page \urlsm{http://tmva.sourceforge.net}).
\begin{itemize}

\item Information on how to download and install TMVA, and the TMVA {\em Quick-start} commands 
      are also available on the web at: \urlsm{http://tmva.sourceforge.net/howto.shtml}.

\item TMVA tutorial: \urlsm{https://twiki.cern.ch/twiki/bin/view/TMVA}.

\item An up-to-date reference of all configuration options for the TMVA Factory, the fitters,
      and all the MVA methods: \urlsm{http://tmva.sourceforge.net/optionRef.html}.

\item On request, the TMVA methods provide a help message with a brief description of the 
      method, and hints for improving the performance by tuning the available configuration 
      options. The message is printed when the option "\code{H}" is added to the configuration 
      string while booking the method (switch off by setting "\code{!H}"). The very same help
      messages are also obtained by clicking the ``info'' button on the top of the reference
      tables on the options reference web page: \urlsm{http://tmva.sourceforge.net/optionRef.html}.      

\item The web address of this Users Guide:
      \urlsm{http://tmva.sourceforge.net/docu/TMVAUsersGuide.pdf}.

\item The TMVA talk collection: \urlsm{http://tmva.sourceforge.net/talks.shtml}.

\item TMVA versions in ROOT releases: 
      \urlsm{http://tmva.sourceforge.net/versionRef.html}.

\item Direct code views via ViewVC: \urlsm{http://tmva.svn.sourceforge.net/viewvc/tmva/trunk/TMVA}.

\item Class index of TMVA in ROOT: \urlsm{http://root.cern.ch/root/htmldoc/TMVA_Index.html}.

\item Please send questions and/or report problems to the {\em tmva-users} mailing list: \\
      \urlsm{http://sourceforge.net/mailarchive/forum.php?forum_name=tmva-users} (posting 
      messages requires prior subscription: 
      \urlsm{https://lists.sourceforge.net/lists/listinfo/tmva-users}).

\end{itemize}

%%% Local Variables: 
%%% mode: latex
%%% TeX-master: "TMVAUsersGuide"
%%% End: 

%% file: UsingTMVA.tex
 \section{Using TMVA}
\label{sec:usingtmva}

\begin{figure}[p]
\begin{center}
   \includegraphics[width=0.50\textwidth]{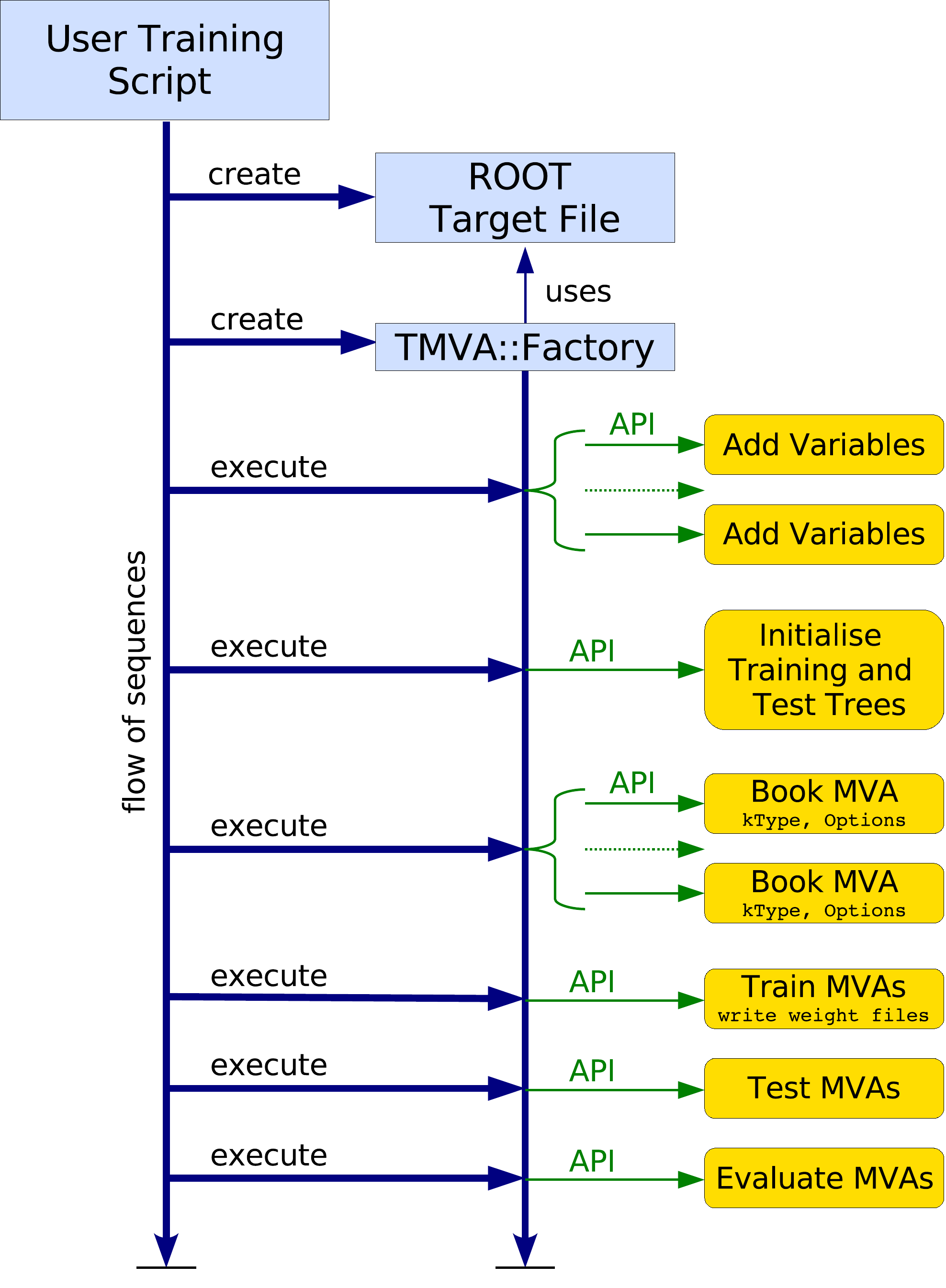}
   \hspace{-0.65cm}
   \includegraphics[width=0.516\textwidth]{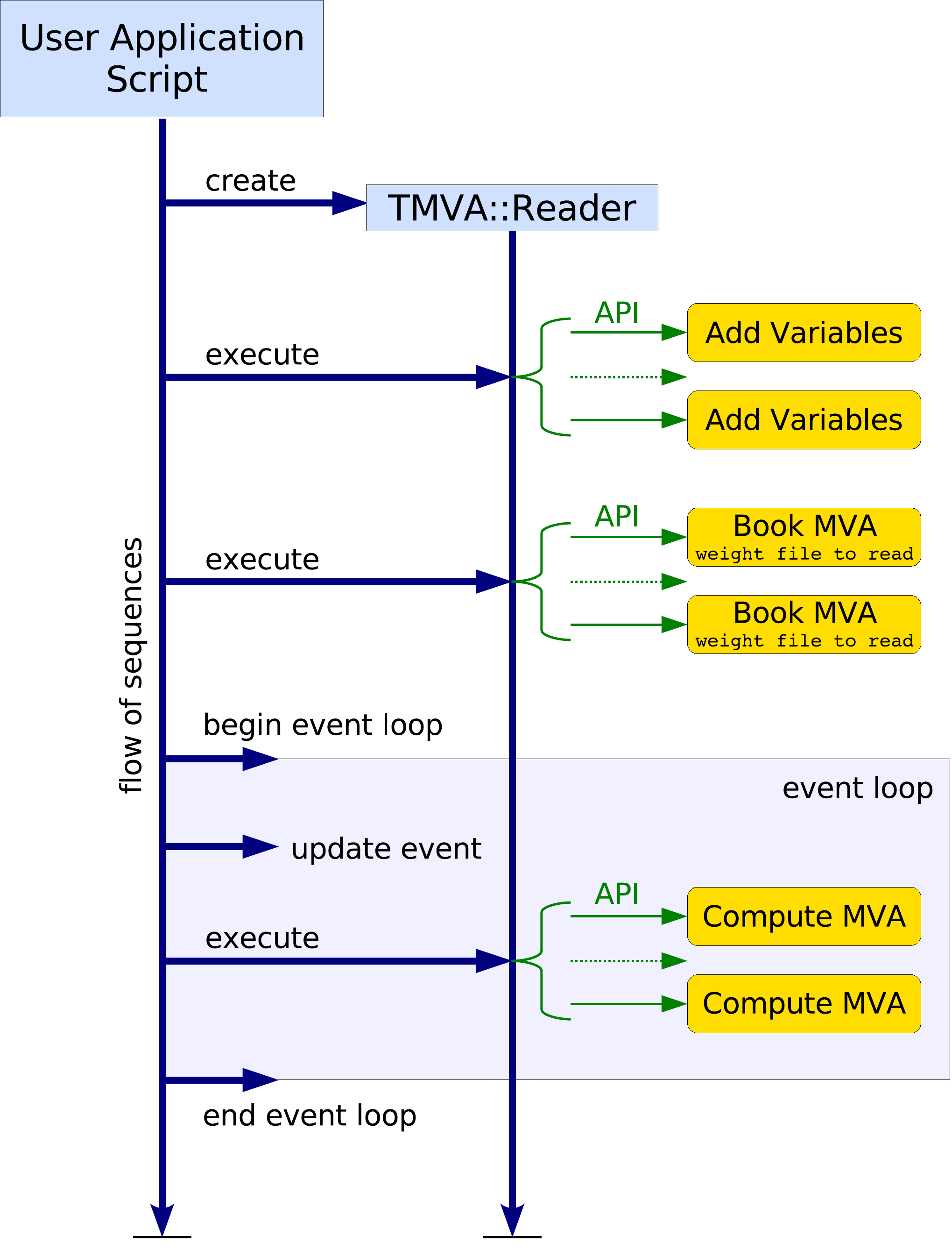}
\end{center}
\vspace{-0.5cm}
\caption[.]{\underline{Left:} Flow (top to bottom) of a typical TMVA 
         training application.
         The user script can be a ROOT macro, C++ executable, python
         script or similar. The user creates a ROOT \code{TFile}, 
         which is used by the TMVA Factory to store output histograms 
         and trees. After creation by the user, the Factory organises the 
         user's interaction with the TMVA modules. It is the only TMVA object 
         directly created and owned by the user. First the discriminating
         variables that must be \code{TFormula}-compliant functions of 
         branches in the training trees are registered. For regression also
         the target variable must be specified. Then, selected MVA methods
         are booked through a type identifier and a user-defined unique name,
         and configuration options are specified via an option string. 
         The TMVA analysis proceeds by consecutively calling the training, 
         testing and performance evaluation methods of the Factory. The training 
         results for all booked methods are written to custom weight files in 
         XML format and the evaluation histograms are stored in the output file. 
         They can be analysed with specific macros that come with TMVA (\cf\  
         Tables~\ref{pgr:scripttable1} and \ref{pgr:scripttable2}). \\
         \underline{Right:} Flow (top to bottom) of a typical 
         TMVA analysis application. The MVA methods qualified by the preceding 
         training and evaluation step are now used to classify data of unknown 
         signal and background composition or to predict a regression target. 
         First, a \code{Reader} class object is created, which 
         serves as interface to the method's response, just as was the Factory 
         for the training and performance evaluation. The discriminating variables 
         and references to locally declared memory placeholders are registered
         with the Reader. The variable names and types must be equal to those
         used for the training. The selected MVA methods are booked with their 
         weight files in the argument, which fully configures them. The user 
         then runs the event loop, where for each event the values of the input
         variables are copied to the reserved memory addresses, and the MVA
         response values (and in some cases errors) are computed.
         \index{Factory}\index{Reader}\index{TMVA analysis flow}
}
\label{fig:TMVAflow}
\end{figure}

A typical TMVA classification or regression analysis consists of two independent phases: 
the {\em training} phase, where the multivariate methods are trained, tested and 
evaluated, and an {\em application} phase, where the chosen methods are applied to 
the concrete classification or regression problem they have been trained for.
An overview of the code flow for these two phases as implemented in the 
examples \code{TMVAClassification.C}\index{TMVAClassification} and 
\code{TMVAClassificationApplication.C}\index{TMVAClassificationApplication} 
(for classification -- see Sec.~\ref{sec:examplejob}), and 
\code{TMVARegression.C}\index{TMVARegression} and 
\code{TMVARegressionApplication.C}\index{TMVARegressionApplication} (for regression)
are sketched in Fig.~\ref{fig:TMVAflow}.
 
In the training phase, the communication of the user with the data sets and the 
MVA methods is performed via a \code{Factory} object, created at the beginning of 
the program. The TMVA Factory provides member functions to specify the training 
and test data sets, to register the discriminating input (and -- in case of 
regression -- target) variables, and to 
book the multivariate methods. Subsequently the Factory calls for 
training, testing and the evaluation of the booked MVA methods. Specific 
result (``weight'') files are created after the training phase by each booked 
MVA method.

The application of training results to a data set with unknown sample composition 
(classification) / target value (regression) is governed by the \code{Reader} object. 
During initialisation, the user registers the input variables\footnote
{
  This somewhat redundant operation is required to verify the correspondence between
  the Reader analysis and the weight files used.
} 
together with their local memory addresses, and books the MVA methods that were found 
to be the most appropriate after evaluating the training results. As booking argument, the 
name of the weight file is given. The weight file provides for each of the methods full 
and consistent configuration according to the training setup and results. Within the 
event loop, the input variables are updated for each event, and the MVA response
values and, in some cases, errors are computed. 

For standalone use of the trained MVA methods, TMVA also generates lightweight 
C++ response classes, which contain the encoded information from the weight files
so that these are not required anymore (\cf\  Sec.~\ref{sec:usingtmva:standaloneClasses}).

\subsection{The TMVA Factory\index{Factory}}

The TMVA training phase begins by instantiating a \code{Factory} object
with configuration options listed in Option-Table~\ref{opt:factory}.
\begin{codeexample}
\begin{tmvacode}
TMVA::Factory* factory 
           = new TMVA::Factory( "<JobName>", outputFile, "<options>" );
\end{tmvacode}
\caption[.]{\codeexampleCaptionSize Instantiating a Factory class object. The first 
            argument is the user-defined job name that will reappear in the names of 
            the weight files containing the training results. The second argument is the
            pointer to a writable \code{TFile} output file created by the user, where 
            control and performance histograms are stored. }
\end{codeexample}

% ======= input option table ==========================================
\begin{option}[t]
\input optiontables/Factory.tex
\caption[.]{\optionCaptionSize 
     Configuration options reference for class: {\em Factory}.  
     Coloured output is switched on by default, except when running ROOT in batch
     mode (\ie, when the '\code{-b}' option of the CINT interpreter is invoked). 
     The list of transformations contains a default set of data preprocessing steps 
     for test and visualisation purposes only. The usage of preprocessing transformations
     in conjunction with MVA methods must be configured when booking the methods.
}
\label{opt:factory}
\end{option}
% =====================================================================

\subsubsection{Specifying training and test data\index{Factory!specifying input data (trees)}}

The input data sets used for training and testing of the multivariate methods
need to be handed to the Factory. TMVA supports ROOT \code{TTree} and derived 
\code{TChain} objects as well as text files. If ROOT trees are used for classification
problems, the signal and background events can be located in the same or in different 
trees. Overall weights can be specified for the signal and background training data 
(the treatment of event-by-event weights is discussed below).

Specifying {\bf classification training data} in ROOT tree format with signal 
and background events being located in different trees:
\begin{codeexample}
\begin{tmvacode}
// Get the signal and background trees from TFile source(s); 
// multiple trees can be registered with the Factory
TTree* sigTree  = (TTree*)sigSrc->Get( "<YourSignalTreeName>"   );
TTree* bkgTreeA = (TTree*)bkgSrc->Get( "<YourBackgrTreeName_A>" );
TTree* bkgTreeB = (TTree*)bkgSrc->Get( "<YourBackgrTreeName_B>" );
TTree* bkgTreeC = (TTree*)bkgSrc->Get( "<YourBackgrTreeName_C>" );

// Set the event weights per tree (these weights are applied in 
// addition to individual event weights that can be specified)
Double_t sigWeight  = 1.0; 
Double_t bkgWeightA = 1.0, bkgWeightB = 0.5, bkgWeightC = 2.0;

// Register the trees
factory->AddSignalTree    ( sigTree,  sigWeight  );
factory->AddBackgroundTree( bkgTreeA, bkgWeightA );
factory->AddBackgroundTree( bkgTreeB, bkgWeightB );
factory->AddBackgroundTree( bkgTreeC, bkgWeightC );
\end{tmvacode}
\caption[.]{\codeexampleCaptionSize Registration of signal and background ROOT trees 
            read from \code{TFile} sources. Overall signal and background weights
            per tree can also be specified.
            The \code{TTree} object may be replaced by a \code{TChain}. }
\end{codeexample}

Specifying {\bf classification training data} in ROOT tree format with signal 
and background events being located in the same tree:
\begin{codeexample}
\begin{tmvacode}
TTree* inputTree = (TTree*)source->Get( "<YourTreeName>" );

TCut signalCut = ...;  // how to identify signal events 
TCut backgrCut = ...;  // how to identify background events

factory->SetInputTrees( inputTree, signalCut, backgrCut );
\end{tmvacode}
\caption[.]{\codeexampleCaptionSize Registration of a single ROOT tree containing the 
            input data for signal {\em and} background, read from a \code{TFile} source. 
            The \code{TTree} object may be replaced by a \code{TChain}. The cuts
            identify the event species.}
\end{codeexample}

Specifying {\bf classification training data} in text format:
\begin{codeexample}
\begin{tmvacode}
// Text file format (available types: 'F' and 'I')
//   var1/F:var2/F:var3/F:var4/F
//   0.21293  -0.49200  -0.58425  -0.70591
//   ...
TString sigFile = "signal.txt";     // text file for signal
TString bkgFile = "background.txt"; // text file for background

Double_t sigWeight = 1.0; // overall weight for all signal events
Double_t bkgWeight = 1.0; // overall weight for all background events

factory->SetInputTrees( sigFile, bkgFile, sigWeight, bkgWeight );
\end{tmvacode}
\caption[.]{\codeexampleCaptionSize Registration of signal and background text files. 
            Names and types of the input variables are given in the first line, 
            followed by the values.}
\end{codeexample}

Specifying {\bf regression training data} in ROOT tree format:
\begin{codeexample}
\begin{tmvacode}
factory->AddRegressionTree( regTree, weight );  
\end{tmvacode}
\caption[.]{\codeexampleCaptionSize Registration of a ROOT tree containing the 
            input and target variables. An overall weight per tree can also be specified.
            The \code{TTree} object may be replaced by a \code{TChain}.
}
\end{codeexample}

\subsubsection{Defining input variables, targets and event weights\index{Factory!selecting input variables}}

The variables in the input trees used to train the MVA methods are registered 
with the Factory using the \code{AddVariable} method. It takes the variable name 
(string), which must have a correspondence in the input ROOT tree or input text file,
and optionally a number type (\code{'F'} (default) and \code{'I'}). The type is used 
to inform the method whether a variable takes continuous floating point or discrete 
values.\footnote
{
  For example for the projective likelihood method, a histogram out of discrete 
  values would not (and should not) be interpolated between bins. 
}
Note that \code{'F'} indicates {\em any} floating point type, \ie, \code{float} 
{\em and} \code{double}. Correspondingly, \code{'I'} stands for integer, 
{\em including} \code{int}, \code{short}, \code{char}, and the corresponding 
\code{unsigned} types. Hence, if a variable in the input tree is \code{double}, 
it should be declared \code{'F'} in the  \code{AddVariable} call. 

It is possible to specify variable expressions, just as for the \code{TTree::Draw} 
command (the expression is interpreted as a \code{TTreeFormula}, including the use 
of arrays). Expressions may be abbreviated for more concise screen output (and plotting) 
purposes by defining shorthand-notation {\em labels} via the assignment operator \code{:=}. 

In addition, two more arguments may be inserted into the \code{AddVariable} 
call, allowing the user to specify {\em titles} and {\em units} for the input variables 
for displaying purposes.

The following code example revises all possible options to declare an input variable:
\begin{codeexample}
\begin{tmvacode}
factory->AddVariable( "<YourDescreteVar>",                  'I' );
factory->AddVariable( "log(<YourFloatingVar>)",             'F' );
factory->AddVariable( "SumLabel := <YourVar1>+<YourVar2>",  'F' );
factory->AddVariable( "<YourVar3>", "Pretty Title", "Unit", 'F' );
\end{tmvacode}
\caption[.]{\codeexampleCaptionSize Declaration of variables used to train the
            MVA methods. Each variable is specified by its name in the training 
            tree (or text file), and optionally a type (\code{'F'} for 
            floating point and \code{'I'} for integer, \code{'F'} is default if 
            nothing is given). Note that \code{'F'} indicates {\em any} floating point
            type, \ie, \code{float} {\em and} \code{double}. Correspondingly, \code{'I'}
            stands for integer, {\em including} \code{int}, \code{short}, \code{char},
            and the corresponding \code{unsigned} types. Hence, even if a variable in 
            the input tree is \code{double}, it should be declared \code{'F'} here.             
            Here, \code{YourVar1} has discrete values and is thus declared 
            as an integer. Just as in the \code{TTree::Draw} command, it 
            is also possible to specify expressions of variables. The \code{:=} operator
            defines labels (third row), used for shorthand notation in screen outputs
            and plots. It is also possible to define titles and units for the variables
            (fourth row), which are used for plotting. If labels {\em and} titles are 
            defined, labels are used for abbreviated screen outputs, and titles for plotting. 
}
\label{ce:addvariable}
\end{codeexample}
It is possible to define {\em spectator variables}, which are part of the input data 
set, but which are not used in the MVA training, test nor during the evaluation. They 
are copied into the \code{TestTree}, together with the used input variables and the 
MVA response values for each event, where the spectator variables can be used for 
correlation tests or others. Spectator variables are declared as follows:
\begin{codeexample}
\begin{tmvacode}
factory->AddSpectator( "<YourSpectatorVariable>" );
factory->AddSpectator( "log(<YourSpectatorVariable>)" );
factory->AddSpectator( "<YourSpectatorVariable>", "Pretty Title", "Unit" );
\end{tmvacode}
\caption[.]{\codeexampleCaptionSize Various ways to declare a spectator variable, not 
            participating in the MVA anlaysis, but written into the final \code{TestTree}.
}
\end{codeexample}
For a regression problem, the target variable is defined similarly, without however 
specifying a number type:
\begin{codeexample}
\begin{tmvacode}
factory->AddTarget( "<YourRegressionTarget1>" );
factory->AddTarget( "log(<YourRegressionTarget2>)" );
factory->AddTarget( "<YourRegressionTarget3>", "Pretty Title", "Unit" );
\end{tmvacode}
\caption[.]{\codeexampleCaptionSize Various ways to declare the target variables used 
            to train a multivariate regression method. If the MVA method supports 
            multi-target (multidimensional) 
            regression\index{Regression!multi-target (multidimensional)}, 
            more than one regression target can be defined. 
}
\end{codeexample}
Individual events can be weighted, with the weights being a column or a function
of columns of the input data sets. To specify the weights to be used for the 
training use the command:\index{Factory!specifying event weights}
\begin{codeexample}
\begin{tmvacode}
factory->SetWeightExpression( "<YourWeightExpression>" );
\end{tmvacode}
\caption[.]{\codeexampleCaptionSize Specification of individual weights for the 
            training events. The expression must be a function of variables present in 
            the input data set.}
\end{codeexample}

\subsubsection{Negative event weights\index{Negative Event Weights}}
\label{sec:NegativeEventWeights}

In next-to-leading order Monte Carlo generators, events with
(unphysical) negative weights may occur in some phase space
regions. Such events are often troublesome to deal with, and it
depends on the concrete implementation of the MVA method, whether or
not they are treated properly. Among those methods that correctly
incorporate events with negative weights are likelihood 
and multi-dimensional probability density estimators, but also 
decision trees. A summary of this feature for all TMVA methods is given in
Table~\ref{tab:methodStatus}. In cases where a method does {\em not}
properly treat events with negative weights, it is advisable to ignore
such events for the training - but to include them in the performance
evaluation to not bias the results. This can be explicitly requested for 
each MVA method via the boolean configuration option \code{IgnoreNegWeightsInTraining} 
(\cf\  Option Table~\ref{opt:mva::methodbase} on
page~\pageref{opt:mva::methodbase}). 

\subsubsection{Preparing the training and test 
               data\index{Factory!preparing training and test data}}
\label{sec:PreparingTrainingTestData}

The input events that are handed to the Factory are internally copied
and split into one {\em training} and one {\em test} ROOT tree. This
guarantees a statistically independent evaluation of the MVA
algorithms based on the test sample.\footnote { A fully unbiased
  training and evaluation requires at least three statistically
  independent data sets. See comments in Footnote~\ref{ftn:training}
  on page~\pageref{ftn:training}.  } The numbers of events used in
both samples are specified by the user. They must not exceed the
entries of the input data sets. In case the user has provided a ROOT
tree, the event copy can (and should) be accelerated by disabling all
branches not used by the input variables.

It is possible to apply selection requirements (cuts) upon the input
events. These requirements can depend on any variable present in the
input data sets, \ie, they are not restricted to the variables used by
the methods. The full command is as follows:
\begin{codeexample}
\begin{tmvacode}
TCut preselectionCut = "<YourSelectionString>";
factory->PrepareTrainingAndTestTree( preselectionCut, "<options>" );
\end{tmvacode}
\caption[.]{\codeexampleCaptionSize Preparation of the internal TMVA
  training and test trees. The sizes (number of events) of these trees
  are specified in the configuration option string. For classification
  problems, they can be set individually for signal and
  background. Note that the preselection cuts are applied before the
  training and test samples are created, \ie, the tree sizes apply to
  numbers of {\em selected} events. It is also possible to choose
  among different methods to select the events entering the training
  and test trees from the source trees. All options are described in
  Option-Table~\ref{opt:datasetfactory}. See also the text for further
  information.}
\label{ce:treePreparation}
\end{codeexample}
For {\bf classification}, the numbers of signal and background events
used for training and testing are specified in the configuration
string by the variables \code{nTrain_Signal},
\code{nTrain_Background}, \code{nTest_Signal} and
\code{nTest_Background} (for example,
\code{"nTrain_Signal=5000:nTrain_Background=5000:nTest_Signal=4000:nTest_Background=5000"}).
The default value (zero) signifies that all available events are
taken, \eg, if \code{nTrain_Signal=5000} and \code{nTest_Signal=0},
and if the total signal sample has 15000 events, then 5000 signal
events are used for training and the remaining 10000 events are used
for testing. If \code{nTrain_Signal=0} and \code{nTest_Signal=0}, the
signal sample is split in half for training and testing.  The same
rules apply to background. Since zero is default, not specifying
anything corresponds to splitting the samples in two halves.

For {\bf regression}, only the sizes of the train and test samples are given, \eg,
\code{"nTrain_Regression=0:nTest_Regression=0"}, so that one half of the input 
sample is used for training and the other half for testing.

The option \code{SplitMode} defines how the training and test samples
are selected from the source trees. With \code{SplitMode=Random},
events are selected randomly. With \code{SplitMode=Alternate}, events
are chosen in alternating turns for the training and test samples as
they occur in the source trees until the desired numbers of training
and test events are selected. In the \code{SplitMode=Block} mode the
first \code{nTrain_Signal} and \code{nTrain_Background}
(classification), or \code{nTrain_Regression} events (regression) of
the input data set are selected for the training sample, and the next
\code{nTest_Signal} and \code{nTest_Background} or
\code{nTest_Regression} events comprise the test data. This is usually
not desired for data that contains varying conditions over the range
of the data set. For the \code{Random} selection mode, the seed of the
random generator can be set. With \code{SplitSeed=0} the generator
returns a different random number series every time.  The default seed
of 100 results in the same training and test samples each time TMVA is
run (as does any other seed apart from 0).

In some cases event weights are given by Monte Carlo generators, and
may turn out to be overall very small or large numbers. To avoid
artifacts due to this, TMVA internally renormalises the signal and
background weights so that their sums over all events equal the
respective numbers of events in the two samples. The renormalisation
is optional and can be modified with the configuration option
\code{NormMode} (\cf\ Table~\ref{opt:datasetfactory}).  Possible
settings are: \code{None}: no renormalisation is applied (the weights
are used as given), \code{NumEvents} (default): renormalisation to
sums of events as described above, \code{EqualNumEvents}: the event
weights are renormalised so that both, the sum of signal and the sum
of background weights equal the number of signal events in the sample.
% ======= input option table ==========================================
\begin{option}[t]
\input optiontables/DataSetFactory.tex
\caption[.]{\optionCaptionSize Configuration options reference in call
  \code{Factory::PrepareTrainingAndTestTree(..)}.  For regression,
  \code{nTrain_Signal} and \code{nTest_Signal} are replaced by
  \code{nTrain_Regression} and \code{nTest_Regression}, respectively,
  and \code{nTrain_Background} and \code{nTest_Background} are
  removed.  See also Code-Example~\ref{ce:treePreparation} and
  comments in the text.  }
\label{opt:datasetfactory}
\end{option}
% =====================================================================

\subsubsection{Booking MVA methods\index{Factory!booking MVA methods}}
\label{sec:usingtmva:booking}

All MVA methods are booked via the Factory by specifying the method's
type, plus a unique name chosen by the user, and a set of specific
configuration options encoded in a string qualifier.\footnote { In the
  TMVA package all MVA methods are derived from the abstract interface
  \code{IMethod} and the base class \code{MethodBase}.  }  If the same
method type is booked several times with different options (which is
useful to compare different sets of configurations for optimisation
purposes), the specified names must be different to distinguish the
instances and their weight files. A booking example for the likelihood
method is given in Code Example~\ref{codeex:factoryBooking}
below. Detailed descriptions of the configuration options are given in
the corresponding tools and MVA sections of this Users Guide, and
booking examples for most of the methods are given in
Appendix~\ref{sec:appendix:booking}.  With the MVA booking the
initialisation of the Factory is complete and no MVA-specific actions
are left to do. The Factory takes care of the subsequent training,
testing and evaluation of the MVA methods.
\begin{codeexample}
\begin{tmvacode}
factory->BookMethod( TMVA::Types::kLikelihood, "LikelihoodD", 
                     "!H:!V:!TransformOutput:PDFInterpol=Spline2:\
                      NSmoothSig[0]=20:NSmoothBkg[0]=20:NSmooth=5:\
                      NAvEvtPerBin=50:VarTransform=Decorrelate" );
\end{tmvacode} 
\caption[.]{\codeexampleCaptionSize Example booking of the likelihood
  method. The first argument is a unique type enumerator (the
  available types can be looked up in \code{src/Types.h}), the second
  is a user-defined name which must be unique among all booked MVA
  methods, and the third is a configuration option string that is
  specific to the method. For options that are not explicitly set in
  the string default values are used, which are printed to standard
  output.  The syntax of the options should be explicit from the above
  example. Individual options are separated by a ':'. Boolean
  variables can be set either explicitly as
  \code{MyBoolVar=True/False}, or just via
  \code{MyBoolVar/!MyBoolVar}.  All specific options are explained in
  the tools and MVA sections of this Users Guide. There is no
  difference in the booking of methods for classification or
  regression applications. See Appendix~\ref{sec:appendix:booking} on
  page~\pageref{sec:appendix:booking} for a complete booking list of
  all MVA methods in TMVA.}
\label{codeex:factoryBooking}
\end{codeexample}

\subsubsection{Help option for MVA booking\index{Help!method-specific help messages}
                           \index{Help!booking options}
                           \index{Help!MVA method optimisation}}
\label{sec:usingtmva:gettingHelp}

Upon request via the configuration option "\code{H}" (see code example above) the TMVA 
methods print concise help messages. These include a brief description of the 
algorithm, a performance assessment, and hints for setting the most important 
configuration options. The messages can also be evoked by the command
{\tt factory->PrintHelpMessage("<MethodName>")}.

\subsubsection{Training the MVA methods\index{Training MVA methods}}
\label{sec:usingtmva:training}

The training of the booked methods is invoked by the command:
\begin{codeexample}
\begin{tmvacode}
factory->TrainAllMethods(); 
\end{tmvacode}
\caption[.]{\codeexampleCaptionSize Executing the MVA training via the Factory.}
\end{codeexample}
The training results are stored in the weight files\index{Weight
  files} which are saved in the directory \code{weights} (which, if
not existing is created).\footnote { The default weight file directory
  name can be modified from the user script through the global
  configuration variable
  \code{(TMVA::gConfig().GetIONames()).fWeightFileDir}.  } The weight
files are named \code{Jobname_MethodName.weights.<extension>}, where
the job name has been specified at the instantiation of the Factory,
and \code{MethodName} is the unique method name specified in the
booking command. Each method writes a custom weight file in XML format
(extension is \code{xml})\index{Weight files!XML format}, where the
configuration options, controls and training results for the method
are stored.

\subsubsection{Testing the MVA methods\index{Testing multivariate methods}}

The trained MVA methods are applied to the test data set and provide
scalar outputs according to which an event can be classified as either
signal or background, or which estimate the regression
target.\footnote { In classification mode, TMVA discriminates signal
  from background in data sets with unknown composition of these two
  samples.  In frequent use cases the background (sometimes also the
  signal) consists of a variety of different populations with
  characteristic properties, which could call for classifiers with
  more than two discrimination classes. However, in practise it is
  usually possible to serialise background fighting by training
  individual classifiers for each background source, and applying
  consecutive requirements to these. Since TMVA 4, the framework
  supports multi-class classification. However, the individual MVA
  methods have not yet been prepared for it.  }  The MVA outputs are
stored in the test tree (\code{TestTree}) to which a column is added
for each booked method. The tree is eventually written to the output
file and can be directly analysed in a ROOT session. The testing of
all booked methods is invoked by the command:
\begin{codeexample}
\begin{tmvacode}
factory->TestAllMethods(); 
\end{tmvacode}
\caption[.]{\codeexampleCaptionSize Executing the validation (testing) of the MVA
            methods via the Factory.}
\end{codeexample}

\subsubsection{Evaluating the
               MVA methods\index{Evaluating MVA methods}\index{Performance evaluation}}
\label{sec:usingtmva:evaluation}

The Factory and data set classes of TMVA perform a preliminary
property assessment of the input variables used by the MVA methods,
such as computing correlation coefficients and ranking the variables
according to their separation (for classification), or according to
their correlations with the target variable(s) (for regression). The
results are printed to standard output.

The performance evaluation in terms of signal efficiency, background rejection, 
faithful estimation of a regression target, etc., of the trained and tested MVA 
methods is invoked by the command:
\begin{codeexample}
\begin{tmvacode}
factory->EvaluateAllMethods();
\end{tmvacode}
\caption[.]{\codeexampleCaptionSize Executing the performance evaluation 
            via the Factory.}
\end{codeexample}
The performance measures differ between classification and regression
problems. They are summarised below.

\subsubsection{Classification performance evaluation}

After training and testing, the linear correlation coefficients among
the classifier outputs are printed. In addition, overlap matrices are
derived (and printed) for signal and background that determine the
fractions of signal and background events that are equally classified
by each pair of classifiers. This is useful when two classifiers have
similar performance, but a significant fraction of non-overlapping
events. In such a case a combination of the classifiers (\eg, in a
{\em Committee} classifier) could improve the performance (this can be
extended to any combination of any number of classifiers).

The optimal method to be used for a specific analysis strongly depends on the
problem at hand and no general recommendations can be given. To ease the choice 
TMVA computes a number of benchmark quantities that assess the performance of the 
methods on the independent test sample. For classification these are 
\begin{itemize}

\item The {\bf signal efficiency at three representative background
  efficiencies} (the efficiency is equal to $1-{\rm rejection}$)
  obtained from a cut on the classifier output. Also given is the area
  of the background rejection versus signal efficiency function (the
  larger the area the better the performance).\index{Performance
    evaluation!background rejection vs. signal efficiency}

\item The {\bf separation}\index{Performance evaluation!separation} \Separation 
      of a classifier $y$, defined by the integral~\cite{Cornelius} 
      \beq
          \Separation = 
          \frac{1}{2} \int\frac{\left(\yPDFS(y) - \yPDFB(y)\right)^2}{\yPDFS(y) + \yPDFB(y)} dy\,, 
      \eeq
      where $\yPDFS$ and $\yPDFB$ are the signal and background PDFs of $y$, 
      respectively (\cf\  Sec.~\ref{sec:otherRepresentations}). 
      The separation is zero for identical signal and background  
      shapes, and it is one for shapes with no overlap.

\item The discrimination {\bf significance}\index{Performance
  evaluation!significance} of a classifier, defined by the difference
  between the classifier means for signal and background divided by
  the quadratic sum of their root-mean-squares.

\end{itemize}
The results of the evaluation are printed to standard output. Smooth
background rejection/efficiency versus signal efficiency curves are
written to the output ROOT file, and can be plotted using custom
macros (see Sec.~\ref{sec:rootmacros}).

\subsubsection{Regression performance evaluation}

Ranking for regression is based on the correlation strength between
the input variables or MVA method response and the regression
target. Several correlation measures are implemented in TMVA to
capture and quantify nonlinear dependencies. Their results are printed
to standard output.
\begin{itemize}

\item The {\bf Correlation}\index{Correlation} between two random variables $X$ and $Y$ is 
      usually measured with the correlation coefficient $\rho$, defined by 
      \beq
      \label{eqn:corrCoeff}
         \rho(X,Y) = \frac{{\rm cov}(X,Y)}{\sigma_X \sigma_Y}.  \eeq
         The correlation coefficient is symmetric in $X$ and $Y$, lies
         within the interval $[-1,1]$, and quantifies by definition a
         linear relationship. Thus $\rho = 0$ holds for independent
         variables, but the converse is not true in general. In
         particular, higher order functional or non-functional
         relationships may not, or only marginally, be reflected in
         the value of $\rho$ (see Fig.~\ref{fig:correlationTypes}).

\item The {\bf correlation ratio}\index{Correlation ratio} is defined by
      \beq
      \label{eqn:corrRatio}
         \eta^2(Y|X) = \frac{\sigma_{E(Y|X)}} {\sigma_Y}\,,
      \eeq
      where
      \beq
      \label{eqn:condExp}
         E(Y|X) = \int y \ P(y|x) \ dy\,, \eeq is the conditional
         expectation of $Y$ given $X$ with the associated conditional
         probability density function $P(Y|X)$. The correlation ratio
         $\eta^2$ is in general not symmetric and its value lies
         within $[0,1]$, according to how well the data points can be
         fitted with a linear or nonlinear regression curve. Thus
         non-functional correlations cannot be accounted for by the
         correlation ratio. The following relations can be derived for
         $\eta^2$ and the squared correlation coefficient
         $\rho^2$~\cite{kendall:stuart:ord:arnold:1999:2A}:
      \begin{itemize}

      \item[$\circ$] $\rho^2 = \eta^2=1$, if $X$ and $Y$ are in a
        strict linear functional relationship.

      \item[$\circ$] $\rho^2 \leq \eta^2=1$, if $X$ and $Y$ are in a
        strict nonlinear functional relationship.

      \item[$\circ$] $\rho^2 = \eta^2 < 1$, if there is no strict
        functional relationship but the regression of $X$ on $Y$ is
        exactly linear.
      
      \item[$\circ$] $\rho^2 < \eta^2 < 1$, if there is no strict
        functional relationship but some nonlinear regression curve is
        a better fit then the best linear fit.

      \end{itemize}
      Some characteristic examples and their corresponding values for $\eta^2$ are 
      shown in Fig.~\ref{fig:correlationTypes}. In the special case, where all data 
      points take the same value, $\eta$ is undefined.

\begin{figure}[t]
\begin{center}
  \includegraphics[width=6.2cm]{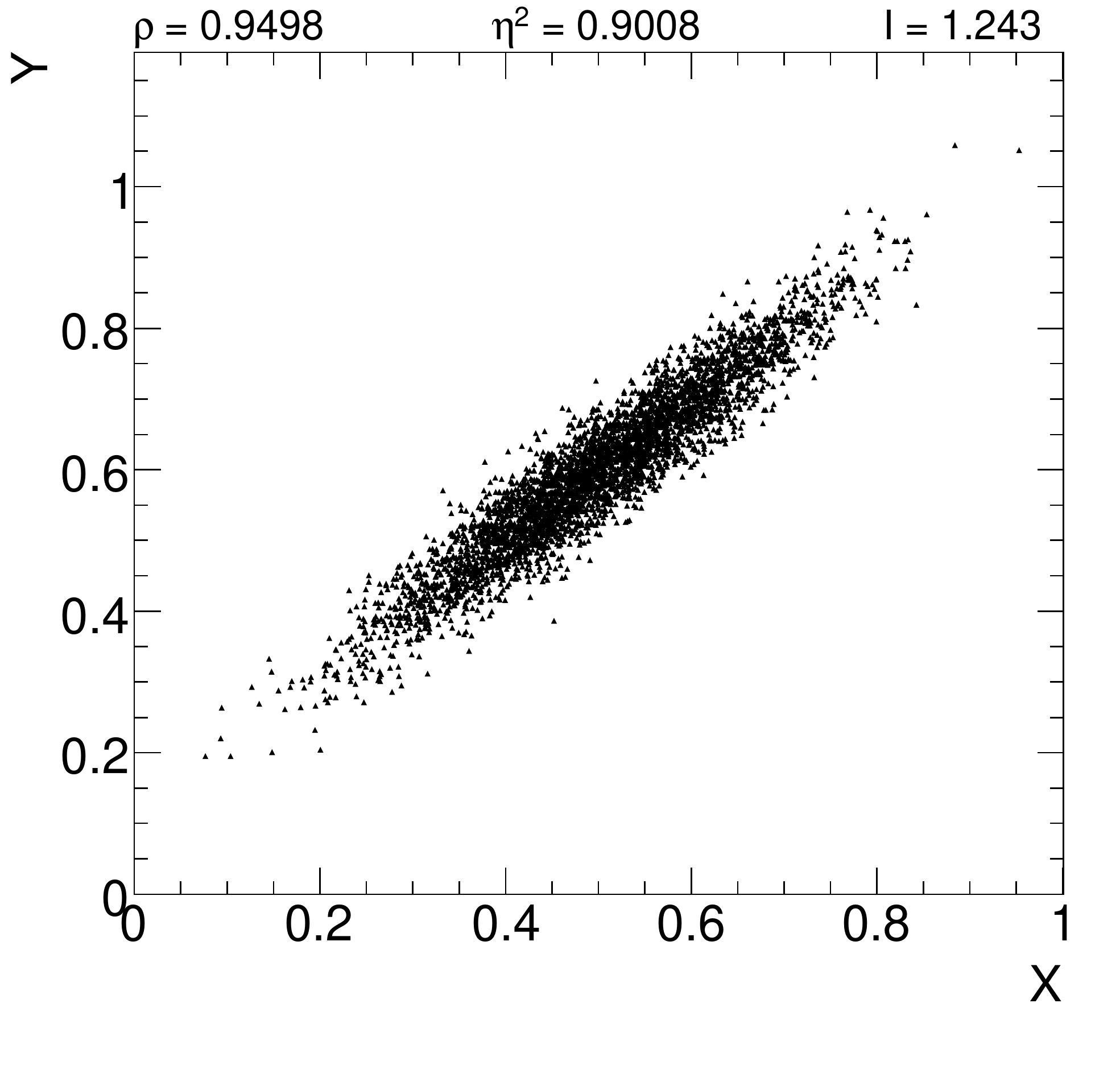} \hspace{0.3cm}
  \includegraphics[width=6.2cm]{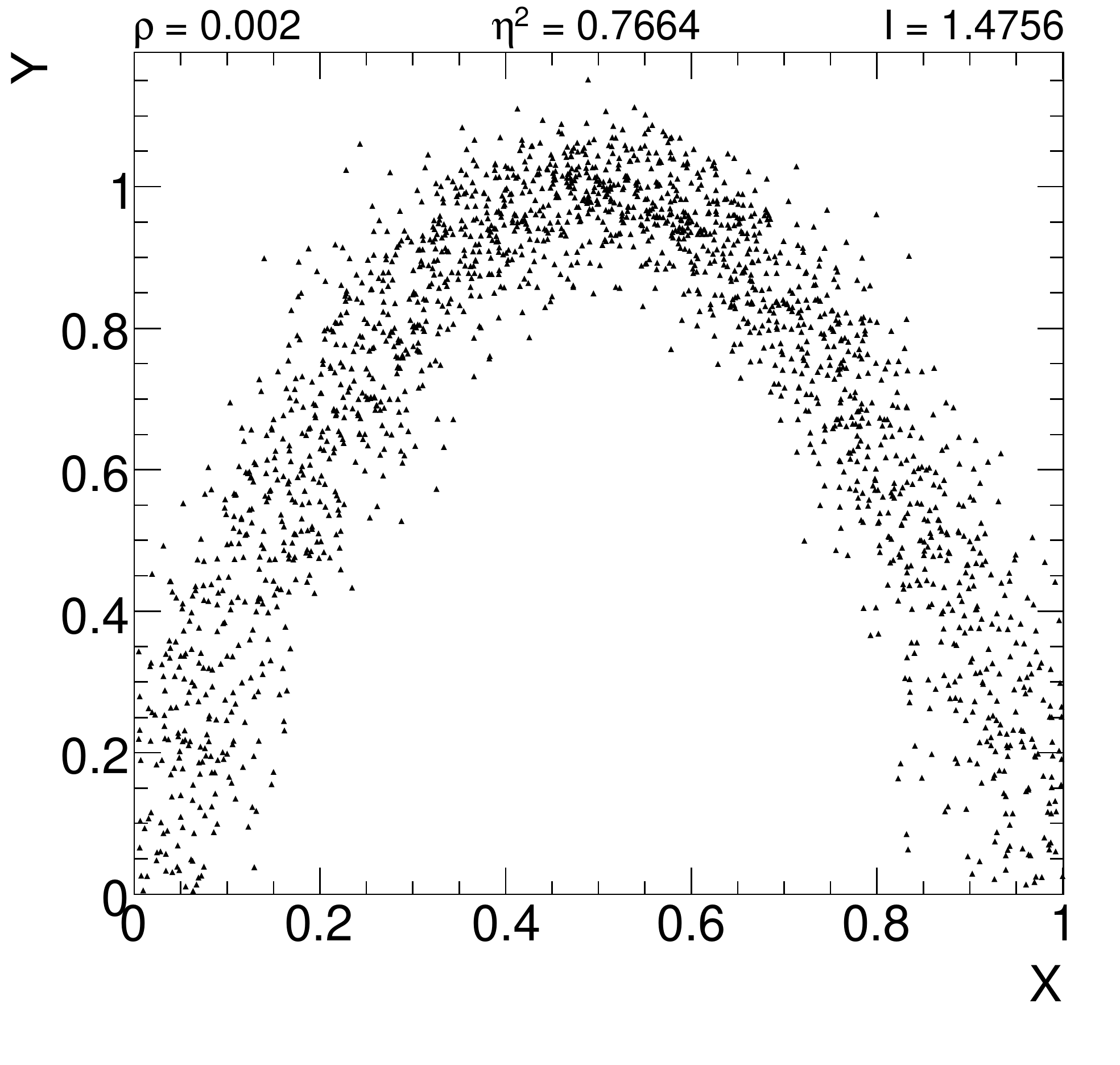} \\\vspace{+0.2cm}
  \includegraphics[width=6.2cm]{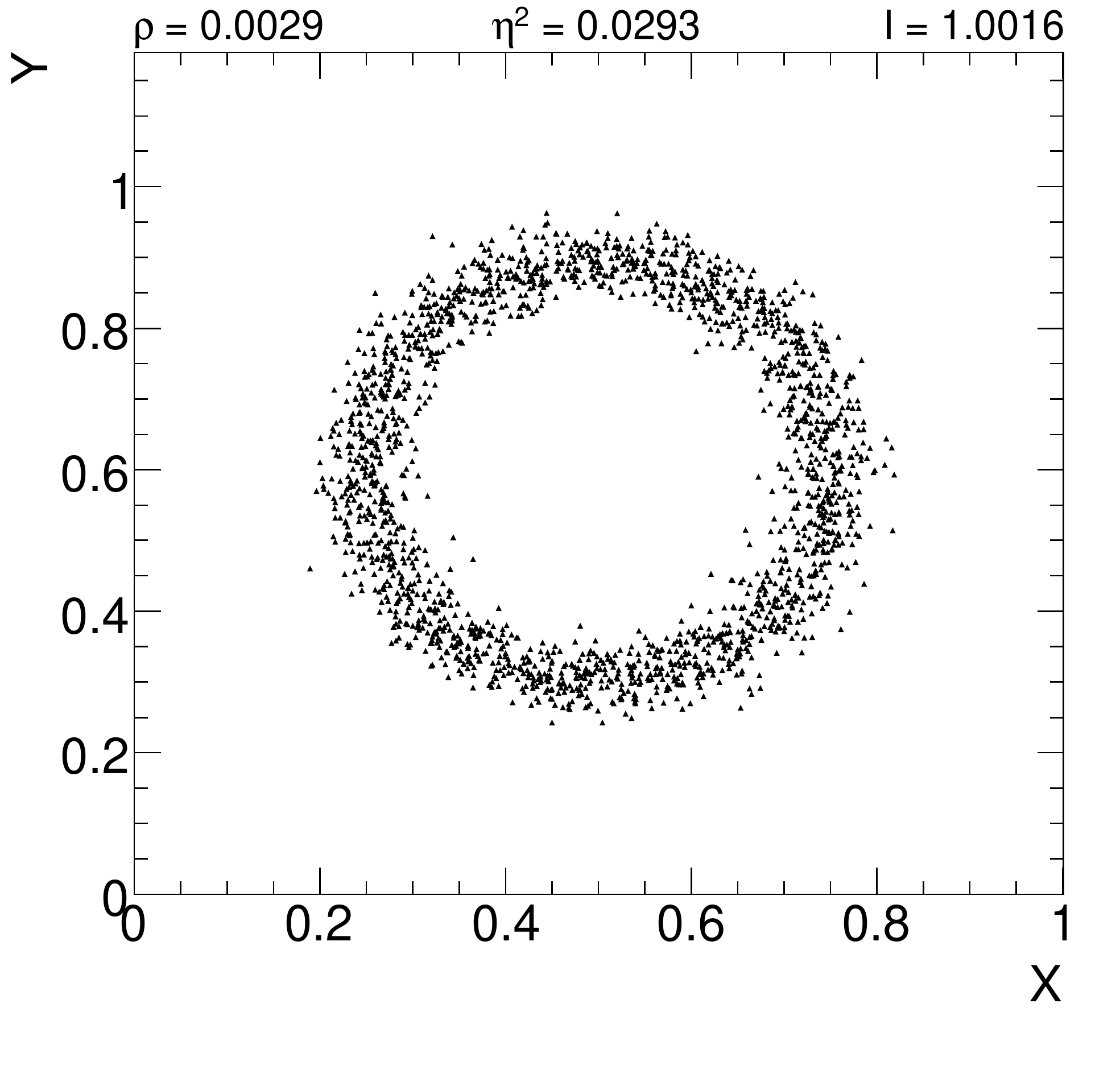} \hspace{0.3cm}
  \includegraphics[width=6.2cm]{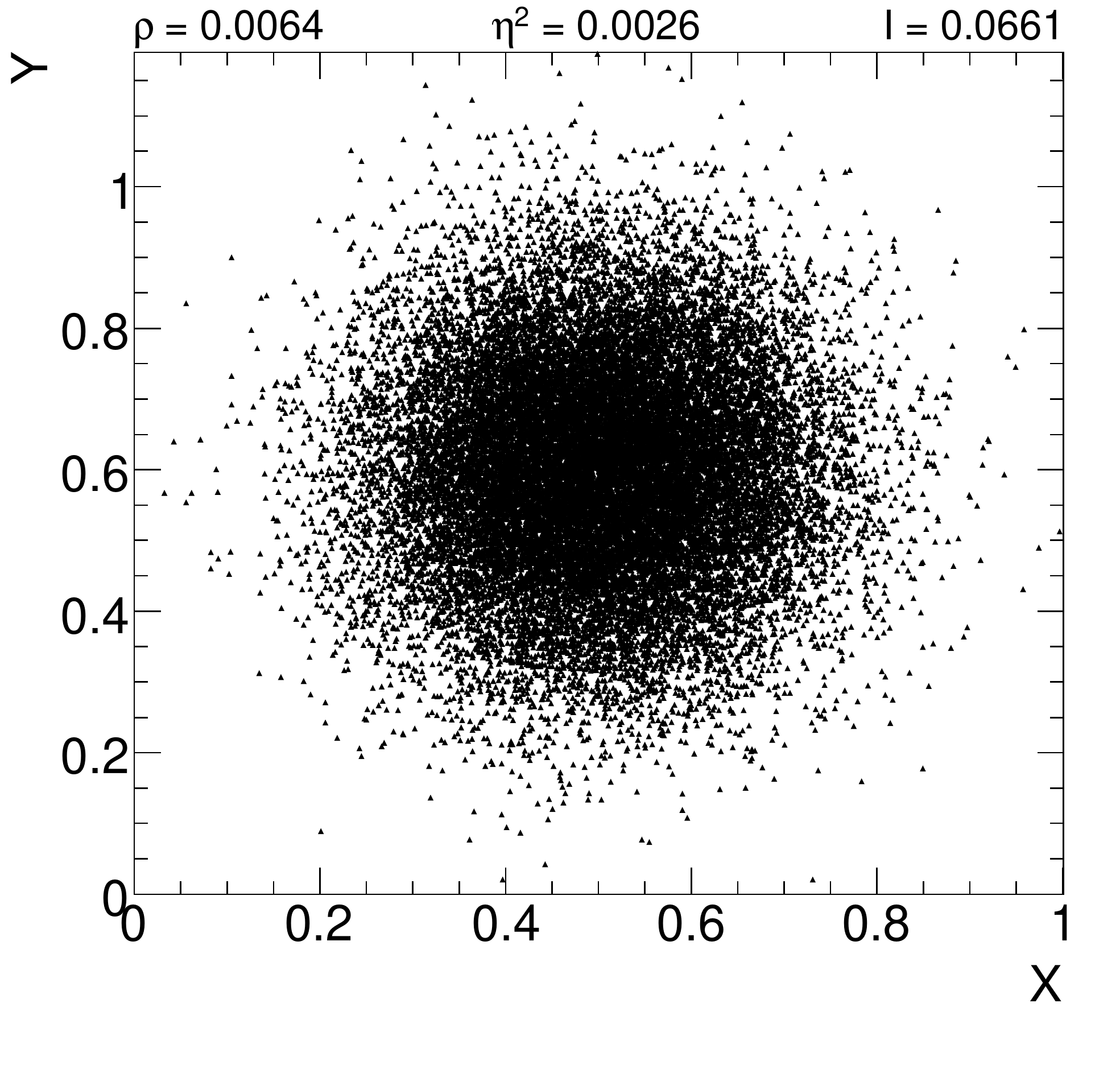} 
\end{center}
  \vspace{-0.7cm}
  \caption{Various types of correlations between two random variables
    and their corresponding values for the correlation coefficient
    $\rho$, the correlation ratio $\eta$, and mutual information
    $I$. Linear relationship (upper left), functional relationship
    (upper right), non-functional relationship (lower left), and
    independent variables (lower right).}
  \label{fig:correlationTypes}
\end{figure}
\item {\bf Mutual information} allows to detect any predictable
  relationship between two random variables, be it of functional or
  non-functional form. It is defined by~\cite{citeulike:165404}
      \beq
      \label{eqn:MI}
         I(X,Y) = \sum_{X,Y}P(X,Y) \ln \frac{P(X,Y)}{P(X) P(Y)}\,, 
      \eeq
         where $P(X,Y)$ is the joint probability density function of
         the random variables $X$ and $Y$, and $P(X)$, $P(Y)$ are the
         corresponding marginal probabilities. Mutual information
         originates from information theory and is closely related to
         entropy which is a measure of the uncertainty associated with
         a random variable. It is defined by 
      \beq
      \label{eqn:MIH}
         H(X) = - \sum_{X}P(X) \ln {P(X)}\,, 
         \eeq 
         where $X$ is the
         discrete random variable and $P(X)$ the associated
         probability density function.  The connection between the two
         quantities is given by the following transformation
      \begin{align}
         I(X,Y) &= \sum_{X,Y}P(X,Y) \ln \frac{P(X,Y)}{P(X) P(Y)}\\
           &= \sum_{X,Y}P(X,Y) \ln \frac{P(X|Y)}{P_X(X)}\\
           &= -\sum_{X,Y}P(X,Y) \ln P(X) + \sum_{X,Y}P(X,Y) \ln P(X|Y)\\
           &= -\sum_{X,Y}P(X) \ln P(X) - (-\sum_{X,Y}P(X,Y) \ln P(X|Y) ) \\
           &=H(X) - H(X|Y)\,,	   	
      \end{align}
      where $H(X|Y)$ is the conditional entropy of $X$ given $Y$. Thus
      mutual information is the reduction of the uncertainty in
      variable $X$ due to the knowledge of $Y$. Mutual information is
      symmetric and takes positive absolute values. In the case of two
      completely independent variables $I(X,Y)$ is zero.
      
      For experimental measurements the joint and marginal probability
      density functions are a priori unknown and must be approximated
      by choosing suitable binning procedures such as kernel
      estimation techniques (see, \eg,
      \cite{PhysRevE.52.2318}). Consequently, the values of $I(X,Y)$
      for a given data set will strongly depend on the statistical
      power of the sample and the chosen binning parameters.
      
      For the purpose of ranking variables from data sets of equal
      statistical power and identical binning, however, we assume that
      the evaluation from a simple two-dimensional histogram without
      further smoothing is sufficient.

\end{itemize}
A comparison of the correlation coefficient $\rho$, the correlation
ratio $\eta$, and mutual information $I$ for linearly correlated
two-dimensional Gaussian toy MC simulations is shown in
Table~\ref{tab:compLinToys}.
\begin{table}[t]
\begin{tabularx}{1.0\linewidth}{lXXXXXXXXXXX}
\hline
&&&&&&&&&&&\\[\BD]
$\rho_{\rm PDF}$ & 0.0 & 0.1 & 0.2 & 0.3 & 0.4 & 0.5 & 0.6 & 0.7 & 0.8 & 0.9 & 0.9999\\[\AD]
\hline
&&&&&&&&&&&\\[\BD]
$\rho$ & 0.006& 0.092& 0.191& 0.291& 0.391& 0.492& 0.592& 0.694& 0.795& 0.898& 1.0\\
$\eta^2$& 0.004& 0.012& 0.041& 0.089& 0.156& 0.245& 0.354& 0.484& 0.634& 0.806& 1.0\\
$I$ & 0.093& 0.099& 0.112& 0.139& 0.171& 0.222& 0.295& 0.398& 0.56& 0.861& 3.071\\[\AD]
\hline
\end{tabularx}
\caption{Comparison of the correlation coefficient $\rho$, correlation ratio $\eta$, and 
         mutual information $I$ for two-dimensional Gaussian toy Monte-Carlo distributions 
         with linear correlations as indicated ($20000~{\rm data~points}/100\times100~{\rm bins}$ .}
\label{tab:compLinToys}
\end{table}

\subsubsection{Overtraining\index{Overtraining}}
\label{sec:usingtmva:overtraining}

Overtraining occurs when a machine learning problem has too few
degrees of freedom, because too many model parameters of an algorithm
were adjusted to too few data points. The sensitivity to overtraining
therefore depends on the MVA method.  For example, a Fisher (or {\em
  linear}) discriminant can hardly ever be overtrained, whereas,
without the appropriate counter measures, boosted decision trees
usually suffer from at least partial overtraining, owing to their
large number of nodes.  Overtraining leads to a seeming increase in
the classification or regression performance over the objectively
achievable one, if measured on the training sample, and to an
effective performance decrease when measured with an independent test
sample. A convenient way to detect overtraining and to measure its
impact is therefore to compare the performance results between
training and test samples. Such a test is performed by TMVA with the
results printed to standard output.

Various method-specific solutions to counteract overtraining
exist. For example, binned likelihood reference distributions are
smoothed before interpolating their shapes, or unbinned kernel density
estimators smear each training event before computing the PDF; neural
networks steadily monitor the convergence of the error estimator
between training and test samples\footnote {
   \label{ftn:training}
   Proper training and validation requires three statistically
   independent data sets: one for the parameter optimisation, another
   one for the overtraining detection, and the last one for the
   performance validation. In TMVA, the last two samples have been
   merged to increase statistics. The (usually insignificant) bias
   introduced by this on the evaluation results does not affect the
   analysis as far as classification cut efficiencies or the
   regression resolution are independently validated with data.  }
suspending the training when the test sample has passed its minimum;
the number of nodes in boosted decision trees can be reduced by
removing insignificant ones (``tree pruning''), etc.

\subsubsection{Other representations of MVA outputs for classification: probabilities and {\em Rarity}}
\label{sec:otherRepresentations}

\begin{figure}[t]
\begin{center}
  \includegraphics[width=0.50\textwidth]{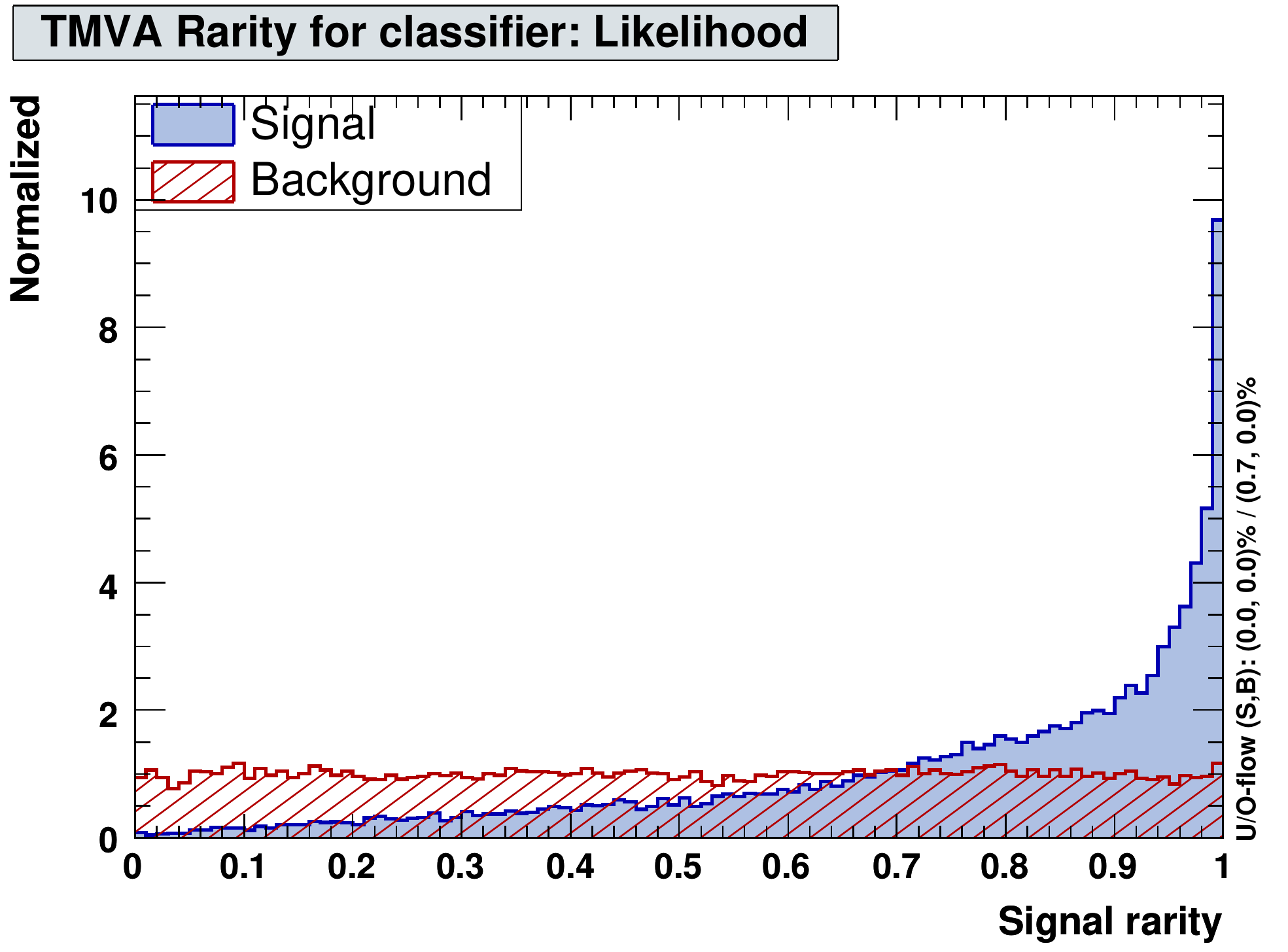}
  \hspace{-0.3cm}
  \includegraphics[width=0.50\textwidth]{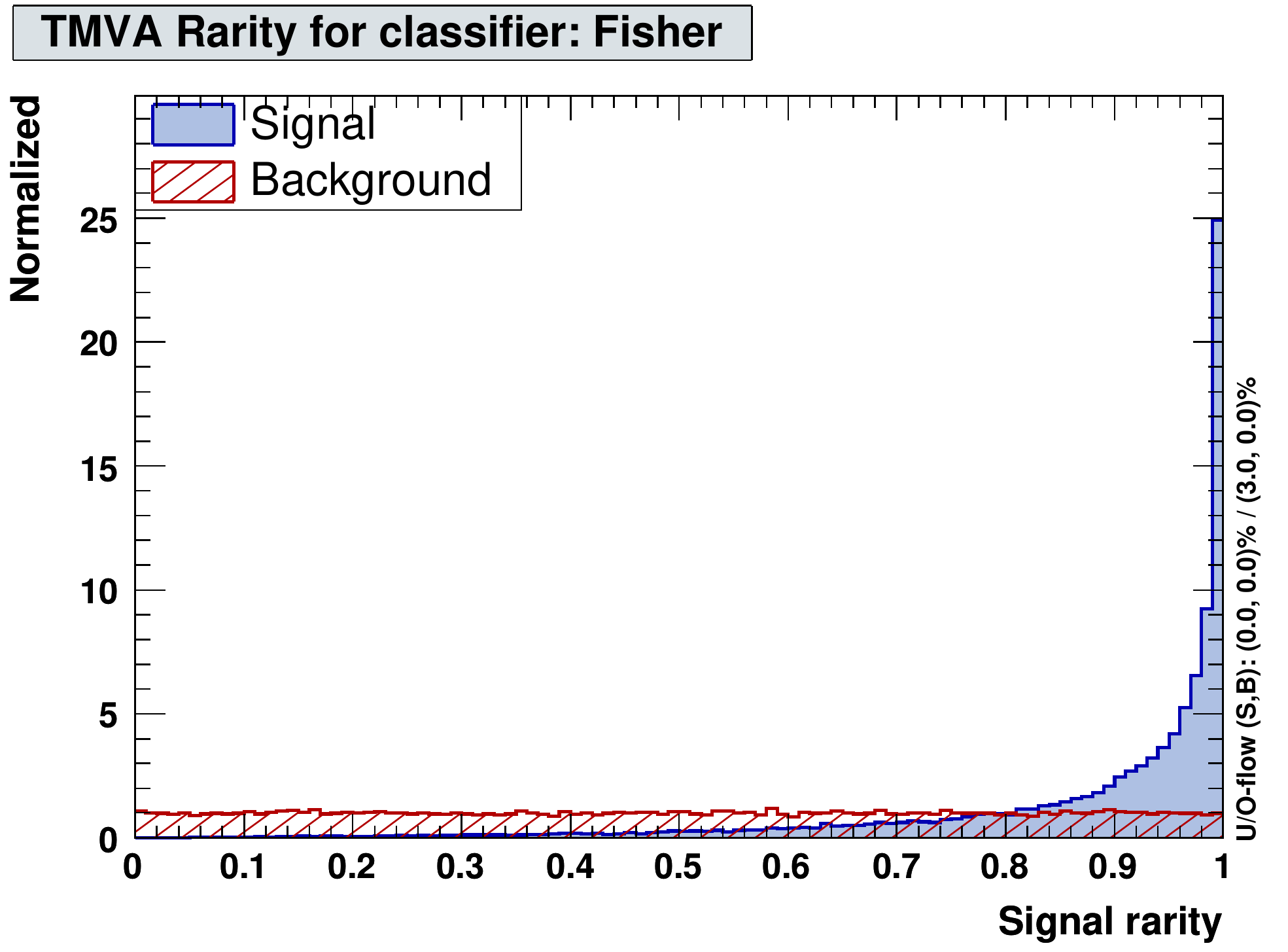}
\end{center}
\vspace{-0.5cm}
\caption[.]{Example plots for classifier Rarity distributions for signal and 
            background events from the academic test sample. Shown are
            likelihood (left) and Fisher (right).}
\label{fig:usingtmva:rarity}
\end{figure}

In addition to the MVA response value \yMVA of a classifier, which is
typically used to place a cut for the classification of an event as
either signal or background, or which could be used in a subsequent
likelihood fit, TMVA also provides the classifier's signal and
background PDFs, $\yPDFSB$. The PDFs can be used to derive
classification probabilities for individual events, or to compute any
kind of transformation of which the {\em Rarity} transformation is
implemented in TMVA.
\begin{itemize}

\item {\bf Classification probability}:\index{Classification
  probability} The techniques used to estimate the shapes of the PDFs
  are those developed for the likelihood classifier (see
  Sec.~\ref{sec:likelihood:description} for details) and can be
  customised individually for each method (the control options are
  given in Sec.~\ref{sec:tmvaClassifiers}).  The probability for event
  $i$ to be of signal type is given by\index{Signal probability}, \beq
      \label{eq:proba}
         \proba(i) = \frac{\fS \cdot\yPDFS(i)}{\fS\cdot \yPDFS(i) + (1
           - \fS)\cdot\yPDFB(i)}\,, \eeq where $\fS=\NS/(\NS+\NB)$ is
         the expected signal fraction, and $\NSB$ is the expected
         number of signal (background) events (default is
         $\fS=0.5$).\footnote { The $\proba$ distributions may exhibit
           a somewhat peculiar structure with frequent narrow
           peaks. They are generated by regions of classifier output
           values in which $\yPDFS\propto\yPDFB$ for which $\proba$
           becomes a constant.  }

\item {\bf Rarity}:\index{Rarity} 
      The Rarity $\Rarity(y)$ of a classifier $y$ is given by the integral~\cite{Rarity}
      \beq
      \label{eq:rarity}
          \Rarity(y) = \intl_{-\infty}^{y}\yPDFB(y^\prime)\,d
          y^\prime~, \eeq which is defined such that $\Rarity(y_B)$
          for background events is uniformly distributed between 0 and
          1, while signal events cluster towards 1. The signal
          distributions can thus be directly compared among the
          various classifiers.  The stronger the peak towards 1, the
          better is the discrimination. Another useful aspect of the
          Rarity is the possibility to directly visualise deviations
          of a test background (which could be physics data) from the
          training sample, by exhibition of non-uniformity.
      
      The Rarity distributions of the Likelihood and Fisher classifiers for the example 
      used in Sec.~\ref{sec:quickstart}
      are plotted in Fig.~\ref{fig:usingtmva:rarity}. Since Fisher performs better
      (\cf\  Fig.~\ref{fig:usingtmva:rejBvsS} on page~\pageref{fig:usingtmva:rejBvsS}),
      its signal distribution is stronger peaked towards 1. By construction, the 
      background distributions are uniform within statistical fluctuations.

\end{itemize}
The probability and Rarity distributions can be plotted with dedicated macros, 
invoked through corresponding GUI buttons.

\subsection{ROOT macros to plot training, testing and evaluation 
            results\index{ROOT!macros}}
\label{sec:rootmacros}

TMVA provides simple GUIs (\code{TMVAGui.C} and
\code{TMVARegGui.C}\index{Graphical user interface (GUI)}, see
Fig.~\ref{fig:tmvagui}), which interface ROOT macros that visualise
the various steps of the training analysis. The macros are
respectively located in \code{TMVA/macros/} (Sourceforge.net
distribution) and {\tt \$}\code{ROOTSYS/tmva/test/} (ROOT
distribution), and can also be executed from the command line. They
are described in Tables~\ref{pgr:scripttable1} and
\ref{pgr:scripttable2}. All plots drawn are saved as {\em png} files
(or optionally as {\em eps}, {\em gif} files) in the macro
subdirectory \code{plots} which, if not existing, is created.

The binning and histogram boundaries for some of the histograms
created during the training, testing and evaluation phases are
controlled via the global singleton class \code{TMVA::Config}. They
can be modified as follows:
\begin{codeexample}
\begin{tmvacode}
// Modify settings for the variable plotting
(TMVA::gConfig().GetVariablePlotting()).fTimesRMS = 8.0;
(TMVA::gConfig().GetVariablePlotting()).fNbins1D  = 60.0;
(TMVA::gConfig().GetVariablePlotting()).fNbins2D  = 300.0;

// Modify the binning in the ROC curve (for classification only)
(TMVA::gConfig().GetVariablePlotting()).fNbinsXOfROCCurve = 100;

// For file name settings, modify the struct TMVA::Config::IONames
(TMVA::gConfig().GetIONames()).fWeightFileDir = "myWeightFileDir";
\end{tmvacode}
\caption[.]{\codeexampleCaptionSize Modifying global parameter
  settings for the plotting of the discriminating input variables. The
  values given are the TMVA defaults. Consult the class files
  \href{http://tmva.svn.sourceforge.net/viewvc/tmva/trunk/TMVA/src/Config.h?view=markup}{Config.h}
  and
  \href{http://tmva.svn.sourceforge.net/viewvc/tmva/trunk/TMVA/src/Config.cxx?view=markup}{Config.cxx}
  for all available global configuration variables and their default
  settings, respectively.  Note that the additional parentheses are
  mandatory when used in CINT.}
\label{ce:gconfig}
\end{codeexample}
\begin{table}[h!]
\begin{programtable}
variables.C             & Plots the signal and background MVA input variables (training sample).
                          The second argument sets the directory, which determines the 
                          preprocessing type (\code{InputVariables_Id} for default identity 
                          transformation, \cf\  Sec.~\ref{sec:variableTransform}). The third
                          argument is a title, and the fourth argument is a flag whether or not 
                          the input variables served a regression analysis. \\
correlationscatter.C    & Plots superimposed scatters and profiles for all pairs of input 
                          variables used during the training phase (separate plots for 
                          signal and background in case of classification). The arguments
                          are as above.  \\ 
correlations.C          & Plots the linear correlation matrices for the input variables in the 
                          training sample (distinguishing signal and background for classification). \\
mvas.C                  & Plots the classifier response distributions of the test sample for 
                          signal and background. The second argument (\code{HistType=0,1,2,3}) allows 
                          to also plot the probability (1) and Rarity (2) distributions of 
                          the classifiers, as well as a comparison of the output distributions 
                          between test and training samples. 
                          Plotting of probability and Rarity requires 
                          the \code{CreateMVAPdfs} option for the classifier to be set to true. \\
mvaeffs.C               & Signal and background efficiencies, obtained from cutting
                          on the classifier outputs, versus the cut value. Also shown are 
                          the signal purity and the signal efficiency times signal purity
                          corresponding to the expected number of signal and background 
                          events before cutting (numbers given by user). The optimal cuts 
                          according to the best significance are printed on standard output. \\
efficiencies.C          & Background rejection (second argument \code{type=2}, default),
                          or background efficiency (\code{type=1}), versus signal efficiency 
                          for the classifiers (test sample). The efficiencies are
                          obtained by cutting on the classifier outputs. This is traditionally 
                          the best plot to assess the overall discrimination performance (ROC curve).\\   
paracoor.C              & Draws diagrams of ``Parallel coordinates''~\cite{parallelcoor} 
                          for signal and background,
                          used to visualise the correlations among the input variables, 
                          but also between the MVA output and input variables (indicating
                          the importance of the variables).                           
\end{programtable}
\caption[.]{\programCaptionSize ROOT macros for the representation of the 
         TMVA input variables and {\bf classification results}. All macros take as first 
         argument the name of the ROOT file containing the histograms (default is 
         \code{TMVA.root}). They are conveniently called via the \code{TMVAGui.C} GUI
         (the first three macros are also called from the regression GUI \code{TMVARegGui.C}).      
         Macros for the representation of regression results are given in Table~\ref{pgr:scripttable_reg}.    
         Plotting macros for MVA method specific information are listed in 
         Table~\ref{pgr:scripttable2}. 
         \index{ROOT!macros}}
\label{pgr:scripttable1}
\end{table}
\begin{table}[h!]
\begin{programtable}
deviations.C              & Plots the linear deviation between regression target value and MVA 
                            response or input variables for test and training samples. \\
regression\_averagedevs.C & Draws the average deviation between the MVA output and the 
                            regression target value for all trained methods. 
\end{programtable}
\caption[.]{\programCaptionSize ROOT macros for the representation of the 
         TMVA {\bf regression results}. All macros take as first argument the name of 
         the ROOT file containing the histograms (default is \code{TMVA.root}). They 
         are conveniently called from the \code{TMVARegGui.C} GUI. 
         \index{ROOT!macros}}
\label{pgr:scripttable_reg}
\end{table}
\begin{table}[t]
\begin{programtable}
likelihoodrefs.C  	   & Plots the reference PDFs of all 
                          input variables for the projective likelihood method and compares it 
                          to original distributions obtained from the training sample.  \\
network.C 	            & Draws the TMVA-MLP architecture including weights after training (does 
                          not work for the other ANNs).                     \\
annconvergencetest.C    & Plots the MLP error-function convergence versus the training epoch 
                          for training and test events (does not work for the other ANNs). \\
BDT.C(i)                & Draws the \code{i}th decision tree of the trained forest (default is 
                          \code{i=1}). The second argument is the weight file that contains 
                          the full architecture of the forest (default is
                          \code{weights/TMVAClassification_BDT.weights.xml}). \\
BDTControlPlots.C       & Plots distributions of boost weights throughout forest, 
                          boost weights versus decision tree, error fraction, number of nodes
                          before and after pruning and the coefficient $\alpha$.\\
mvarefs.C               & Plots the PDFs used to compute the probability response for a classifier, 
                          and compares it to the original distributions. \\
PlotFoams.C             & Draws the signal and background foams created by the method 
                          PDE-Foam.\\
rulevis.C               & Plots the relative importance of rules and linear terms.
                          The 1D plots show the accumulated importance per input variable. 
                          The 2D scatter plots show the same but correlated between the 
                          input variables. These plots help to identify regions in the 
                          parameter space that are important for the model.
\end{programtable}
\caption[.]{\programCaptionSize List of ROOT macros representing results for 
            {\bf specific MVA methods}. The macros require that these methods have been 
            included in the training. All macros take as first argument the name of the 
            ROOT file containing the histograms (default is \code{TMVA.root}).}
\label{pgr:scripttable2}
\end{table}

\subsection{The TMVA Reader\index{Reader}}
\label{sec:usingtmva:reader}

After training and evaluation, the most performing MVA methods are chosen and 
used to classify events in data samples with unknown signal and background composition,
or to predict values of a regression target. An example of how this {\em application phase} 
is carried out is given in \code{TMVA/macros/TMVAClassificationApplication.C} and
\code{TMVA/macros/TMVARegressionApplication.C} (Sourceforge.net), or 
{\tt \$}\code{ROOTSYS/tmva/test/TMVAClassificationApplication.C} and
{\tt \$}\code{ROOTSYS/tmva/test/TMVARegressionApplication.C} (ROOT).

Analogously to the Factory, the communication between the user application and 
the MVA methods is interfaced by the TMVA {\em Reader}, which is created by 
the user:
\begin{codeexample}
\begin{tmvacode}
TMVA::Reader* reader = new TMVA::Reader( "<options>" );
\end{tmvacode}
\caption[.]{\codeexampleCaptionSize Instantiating a Reader class object. The only
            options are the booleans: \code{V} for verbose, \code{Color} for coloured output, 
            and \code{Silent} to suppress all output.}
\end{codeexample}

\subsubsection{Specifying input variables\index{Reader!specifying input variables}}

The user registers the names of the input variables with the Reader. They are
required to be the same (and in the same order) as the names used for training 
(this requirement is not actually mandatory, but enforced to ensure the consistency 
between training and application). Together with the name is given the address of a 
local variable, which carries the updated input values during the event loop.
\begin{codeexample}
\begin{tmvacode}
Int_t   localDescreteVar;
Float_t localFloatingVar, locaSum, localVar3;

reader->AddVariable( "<YourDescreteVar>",                 &localDescreteVar );
reader->AddVariable( "log(<YourFloatingVar>)",            &localFloatingVar );
reader->AddVariable( "SumLabel := <YourVar1>+<YourVar2>", &locaSum          );
reader->AddVariable( "<YourVar3>",                        &localVar3        );
\end{tmvacode}
\caption[.]{\codeexampleCaptionSize Declaration of the variables and references 
            used as input to the methods (\cf Code Example~\ref{ce:addvariable}). 
            The order and naming of the variables
            must be consistent with the ones used for the training. The local 
            variables are updated during the event loop, and through the references
            their values are known to the MVA methods. The variable type must be either
            \code{float} or \code{int} (\code{double} is not supported). }
\end{codeexample}

\subsubsection{Booking MVA methods\index{Reader!booking MVA methods}}

The selected MVA methods are booked with the Reader using the weight files from the 
preceding training job:\index{Weight files!XML format}
\begin{codeexample}
\begin{tmvacode}
reader->BookMVA( "<YourMethodName>", "<path/JobName_MethodName.weights.xml>" );
\end{tmvacode}
\caption[.]{\codeexampleCaptionSize Booking a multivariate method. The first
            argument is a user defined name to distinguish between  
            methods (it does not need to be the same name as for training, 
            although this could be a useful choice). The true type of the 
            method and its full configuration are read from the weight file
            specified in the second argument. The default structure of the 
            weight file names is: \code{path/<JobName>_<MethodName>.weights.xml}.
}
\end{codeexample}

\subsubsection{Requesting the MVA response\index{Reader!requesting the MVA response}}

Within the event loop, the response value of a classifier, and -- if available -- 
its error, for a given set of input variables computed by the user, are obtained 
with the commands:
\begin{codeexample}
\begin{tmvacode}
localDescreteVar  = treeDescreteVar;       // reference could be implicit
localFloatingVar  = log(treeFloatingVar);
locaSum           = treeVar1 + treeVar2;
localVar3         = treeVar3;              // reference could be implicit

// Classifier response
Double_t mvaValue = reader->EvaluateMVA( "<YourMethodName>" );

// Error on classifier response - must be called after "EvaluateMVA"
// (not available for all methods, returns -1 in that case)
Double_t mvaErr   = reader->GetMVAError();
\end{tmvacode}
\caption[.]{\codeexampleCaptionSize Updating the local variables for an event, 
            and obtaining the corresponding classifier output and error 
            (if available -- see text). }
\end{codeexample}
The output of a classifier may then be used for example to put a cut that increases the 
signal purity of the sample (the achievable purities can be read off the evaluation
results obtained during the test phase), or it could enter a subsequent 
maximum-likelihood fit, or similar. The error reflects the uncertainty, which 
may be statistical, in the output value as obtained from the training information. 

For regression, multi-target response is already supported in TMVA, so that the 
retrieval command reads:
\begin{codeexample}
\begin{tmvacode}
// Regression response for one target
Double_t regValue = (reader->EvaluateRegression( "<YourMethodName>" ))[0];
\end{tmvacode}
\caption[.]{\codeexampleCaptionSize Obtaining the regression output (after 
            updating the local variables for an event -- see above). For mult-target 
            regression, the corresponding vector entries are filled.}
\end{codeexample}
The output of a regression method could be directly used for example as energy estimate 
for a calorimeter cluster as a function of the cell energies. 

The rectangular cut classifier is special since it returns a binary answer for a given 
set of input variables and cuts. The user must specify the desired signal efficiency
to define the working point according to which the Reader will choose the cuts:
\begin{codeexample}
\begin{tmvacode}
Bool_t passed = reader->EvaluateMVA( "Cuts", signalEfficiency );
\end{tmvacode}
\caption[.]{\codeexampleCaptionSize For the cut classifier, the second parameter 
            gives the desired signal efficiency according to which the cuts 
            are chosen. The return value is 1 for passed and 0 for retained. 
            See Footnote~\ref{ftn:cutcomp} on page~\pageref{ftn:cutcomp} for 
            information on how to determine the optimal working point for 
            known signal and background abundance.}
\end{codeexample}

Instead of the classifier response values, one may also retrieve the ratio~(\ref{eq:proba})
from the Reader, which, if properly normalised to the expected signal fraction 
in the sample, corresponds to a probability. The corresponding command
reads:\index{Reader!requesting the signal probability of a classifier}
\begin{codeexample}
\begin{tmvacode}
Double_t pSig = reader->GetProba( "<YourClassifierName>", sigFrac );
\end{tmvacode}
\caption[.]{\codeexampleCaptionSize Requesting the event's signal probability from a 
            classifier. The signal fraction is the parameter $\fS$ in Eq.~(\ref{eq:proba}).}
\end{codeexample}
Similarly, the {\em Rarity}~(\ref{eq:rarity}) of a classifier is retrieved by the 
command\index{Reader!requesting the Rarity of a classifier}
\begin{codeexample}
\begin{tmvacode}
Double_t rarity = reader->GetRarity( "<YourClassifierName>" );
\end{tmvacode}
\caption[.]{\codeexampleCaptionSize Requesting the event's Rarity from a 
            classifier. }
\end{codeexample}

\subsection{An alternative to the Reader: standalone C++ response classes
           \index{Standalone C++ response classes}}
\label{sec:usingtmva:standaloneClasses}

\begin{codeexample}[tb]
\begin{tmvacode}
// load the generated response class into macro and compile it (ROOT)
// or include it into any C++ executable
gROOT->LoadMacro( "TMVAClassification_Fisher.class.C++" ); // usage in ROOT 

// define the names of the input variables (same as for training)
std::vector<std::string> inputVars;
inputVars.push_back( "<YourVar1>" );
inputVars.push_back( "log(<YourVar2>)" );
inputVars.push_back( "<YourVar3>+<YourVar4>" );

// create a class object for the Fisher response
IClassifierReader* fisherResponse = new ReadFisher( inputVars );

// the user's event loop ...
std::vector<double> inputVec( 3 );
for (...) {
   // compute the input variables for the event
   inputVec[0] = treeVar1;
   inputVec[1] = TMath::Log(treeVar2);
   inputVec[2] = treeVar3 + treeVar4;

   // get the Fisher response
   double fiOut = fisherResponse->GetMvaValue( inputVec );
   // ... use fiOut
}
\end{tmvacode}
\caption[.]{\codeexampleCaptionSize
            Using a standalone C++ class for the classifier
            response in an application (here of the Fisher discriminant). See also the 
            example code in \code{TMVA/macros/ClassApplication.C} (Sourceforge.net).}
\label{ce:standaloneClasses}
\end{codeexample}
To simplify the portability of the trained MVA response to any application the TMVA 
methods generate after the training lightweight standalone C++ response classes 
including in the source code the content of the weight files.\footnote
{
   At present, the class making functionality has been implemented for 
   all MVA methods with the exception of cut optimisation, PDE-RS, PDE-Foam and k-NN. While 
   for the former classifier the cuts can be easily implemented into the user application, 
   and do not require an extra class, the implementation of a response class for
   PDE-RS or k-NN requires a copy of the entire analysis code, which we have not 
   attempted so far. We also point out that the use of the standalone C++ class
   for BDT is not practical due to the colossal size of the generated code. 
} 
These classes do not depend on ROOT, neither on any other non-standard library.
The names of the classes are constructed out of \code{Read+"MethodName"}, and
they inherit from the interface class \code{IClassifierReader}, which is written 
into the same C++ file. These standalone classes are {\em presently only available 
for classification}.

An example application (ROOT script here, not representative
for a C++ standalone application) for a Fisher classifier is given in 
Code-Example~\ref{ce:standaloneClasses}. The example is also available in the macro
\code{TMVA/macros/ClassApplication.C} (Sourceforge.net). These classes are C++ 
representations of the information stored in the weight files. Any change in the 
training parameters will generate a new class, which must be updated in the 
corresponding application.\footnote
{
   We are aware that requiring recompilation constitutes a significant shortcoming.
   we consider to upgrade these classes to reading the XML weight files, which
   entails significant complications if the independence of any external 
   library shall be conserved. 
} 

For a given test event, the MVA response returned by the standalone class 
is identical to the one returned by the Reader. Nevertheless, {\em we emphasise that 
the recommended approach to apply the training results is via the Reader}.

%% file: optiontables/Factory.tex
\begin{optiontableAuto}
                        V  &  \mc{1}{c}{--}  &            False  &  \mc{1}{l}{--}  &  Verbose flag \\
                    Color  &  \mc{1}{c}{--}  &             True  &  \mc{1}{l}{--}  &  Flag for coloured screen output (default: True, if in batch mode: False) \\
          Transformations  &  \mc{1}{c}{--}  &                   &  \mc{1}{l}{--}  &  List of transformations to test; formatting example: Transformations=I;D;P;G,D, for identity, decorrelation, PCA, and Gaussianisation followed by decorrelation transformations \\
                   Silent  &  \mc{1}{c}{--}  &            False  &  \mc{1}{l}{--}  &  Batch mode: boolean silent flag inhibiting any output from TMVA after the creation of the factory class object (default: False) \\
          DrawProgressBar  &  \mc{1}{c}{--}  &             True  &  \mc{1}{l}{--}  &  Draw progress bar to display training, testing and evaluation schedule (default: True) 
\end{optiontableAuto}

%% file: optiontables/DataSetFactory.tex
\begin{optiontableAuto}
                SplitMode  &  \mc{1}{c}{--}  &           Random  &  Random, Alternate, Block  &  Method of picking training and testing events (default: random) \\
                SplitSeed  &  \mc{1}{c}{--}  &              100  &  \mc{1}{l}{--}  &  Seed for random event shuffling \\
                 NormMode  &  \mc{1}{c}{--}  &        NumEvents  &  None, NumEvents, EqualNumEvents  &  Overall renormalisation of event-by-event weights (NumEvents: average weight of 1 per event, independently for signal and background; EqualNumEvents: average weight of 1 per event for signal, and sum of weights for background equal to sum of weights for signal) \\
            nTrain\_Signal  &  \mc{1}{c}{--}  &                0  &  \mc{1}{l}{--}  &  Number of training events of class Signal (default: 0 = all) \\
             nTest\_Signal  &  \mc{1}{c}{--}  &                0  &  \mc{1}{l}{--}  &  Number of test events of class Signal (default: 0 = all) \\
        nTrain\_Background  &  \mc{1}{c}{--}  &                0  &  \mc{1}{l}{--}  &  Number of training events of class Background (default: 0 = all) \\
         nTest\_Background  &  \mc{1}{c}{--}  &                0  &  \mc{1}{l}{--}  &  Number of test events of class Background (default: 0 = all) \\
                        V  &  \mc{1}{c}{--}  &            False  &  \mc{1}{l}{--}  &  Verbosity (default: true) \\
             VerboseLevel  &  \mc{1}{c}{--}  &             Info  &  Debug, Verbose, Info  &  VerboseLevel (Debug/Verbose/Info) 
\end{optiontableAuto}

%% file: DataPreprocessing.tex
\section{Data Preprocessing}
\label{sec:dataPreprocessing}

It is possible to preprocess the discriminating input variables or the training events prior 
to presenting them to a multivariate method. Preprocessing can be useful to reduce correlations 
among the variables, to transform their shapes into more appropriate forms, or to accelerate 
the response time of a method (event sorting). 

The preprocessing is completely transparent to 
the MVA methods. Any preprocessing performed for the training is automatically performed in 
the application through the Reader class. All the required information is stored in the 
weight files of the MVA method. Most preprocessing methods discussed below are only 
available for classification. An exception is the normalisation transformation, which 
exists for both classification and regression.

\subsection{Transforming input variables}
\label{sec:variableTransform}

Currently four preprocessing\index{Discriminating variables!preprocessing of}  
transformations\index{Discriminating variables!transformation of}
are implemented in TMVA:
\begin{itemize}
\item variable normalisation;
\item decorrelation via the square-root of the covariance matrix ;
\item decorrelation via a principal component decomposition;
\item transformation of the variables into Gaussian distributions (``Gaussianisation'').
\end{itemize}
With the exception of normalisation, which exists for both classification and 
regression, the other preprocessing methods are currently only available for 
classification. 

\begin{figure}[t]
  \begin{center}
	  \includegraphics[width=0.90\textwidth]{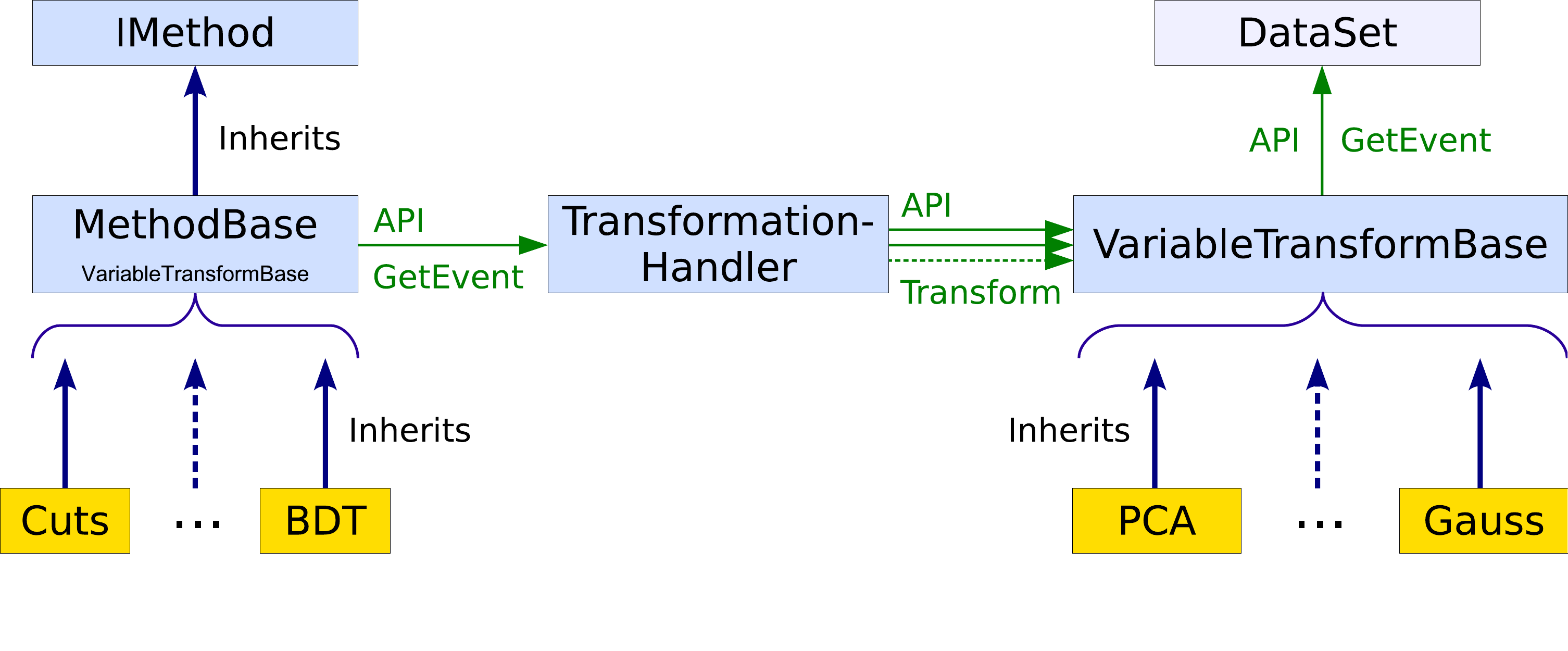}
  \end{center}
  \vspace{-1.1cm}
  \caption[.]{Schematic view of the variable transformation interface implemented in 
    TMVA. Each concrete MVA method derives from \code{MethodBase} 
    (interfaced by \code{IMethod}), which holds a protected member object 
    of type \code{TransformationHandler}. In this object a list of objects 
    derived from \code{VariableTransformBase} which are the implementations of 
    the particular variable transformations available in TMVA is stored. 
    The construction of the concrete 
    variable transformation objects proceeds in \code{MethodBase} according 
    to the transformation methods requested in the option string. The events
    used by the MVA methods for training, testing and final classification (or regression)
    analysis are read via an API of the \code{TransformationHandler} class, 
    which itself reads the events from the \code{DataSet} and applies 
    subsequently all initialised transformations. The \code{DataSet} 
    fills the current values into the reserved event 
    addresses (the event content may either stem from the training or 
    testing trees, or is set by the user application via the \code{Reader} 
    for the final classification/regression analysis). The \code{TransformationHandler}
    class ensures the proper transformation of all events seen by 
    the MVA methods.
}
\label{fig:VariableTransform}
\end{figure}

Technically, any transformation of the input variables is performed ``on the fly'' 
when the event is requested from the central \code{DataSet} class. The preprocessing 
is hence fully transparent to the MVA methods. Any preprocessing performed for the 
training is automatically also performed in the application through the Reader class. 
All the required information is stored in the weight files of the MVA method.
Each MVA method carries a variable
transformation type together with a pointer to the object of its
transformation class which is owned by the \code{DataSet}. If no
preprocessing is requested, an identity transform is applied. The
\code{DataSet} registers the requested transformations and takes care
not to recreate an identical transformation object (if requested)
during the training phase. Hence if two MVA methods wish to apply the
same transformation, a single object is shared between them. Each
method writes {\em its} transformation into its weight file once
the training has converged. For testing and application of an
MVA method, the transformation is read from the weight file and a
corresponding transformation object is created. Here each method
owns its transformation so that no sharing of potentially different
transformation objects occurs (they may have been obtained with
different training data and/or under different conditions). A
schematic view of the variable transformation interface used in TMVA
is drawn in Fig.~\ref{fig:VariableTransform}.

\subsubsection{Variable normalisation\index{Discriminating variables!normalisation of}}
\label{sec:normalisation}

Minimum and maximum values for all input variables are determined from the training 
events and used to linearly scale the input variables to lie within $[-1,1]$. 
Such a transformation is useful to allow direct comparisons between MVA weights 
applied to the variables. Large absolute weights may indicate strong separation power. 
Normalisation may also render minimisation processes, such as the adjustment of 
neural network weights, more effective. 

\subsubsection{Variable decorrelation\index{Discriminating variables!decorrelation of}}
\label{sec:decorrelation}

A drawback of, for example, the projective likelihood classifier (see 
Sec.~\ref{sec:likelihood}) is that it ignores correlations among the discriminating 
input variables. Because in most realistic use cases this is not an accurate 
conjecture it leads to performance loss. Also other classifiers, such as rectangular 
cuts or decision trees, and even multidimensional likelihood approaches underperform
in presence of variable correlations.

Linear correlations, measured in the training sample, can be taken into account 
in a straightforward manner through computing the square-root of the covariance 
matrix. The square-root of a matrix $C$ is the matrix $C^\prime$ that multiplied 
with itself yields $C$: $C=(C^\prime)^2$. TMVA computes the square-root matrix 
by means of diagonalising the (symmetric) covariance matrix
\beq
    D=S^T C S \hspace{0.5cm}\Rightarrow \hspace{0.5cm}
    C^\prime=S\sqrt{D}S^T\,,
\eeq
where $D$ is a diagonal matrix, and where the matrix $S$ is symmetric. The linear 
decorrelation of the input variables is then obtained by multiplying the 
initial variable tuple ${\bf x}$ by the inverse of the square-root matrix
\beq
      {\bf x} \mapsto (C^\prime)^{-1}{\bf x}\:.
\eeq

The transformations are performed separately for signal and background events
because their correlation patterns are usually different.\footnote
{
   Different transformations for signal and background events are only 
   useful for methods that explicitly distinguish signal and background 
   hypotheses. This is the case for the likelihood and PDE-RS classifiers.
   For all other methods the user must choose which transformation to use.
}
The decorrelation is complete only for linearly correlated and Gaussian distributed 
variables. In real-world use cases this is not often the case, so that sometimes
only little additional information can be recovered by the decorrelation
procedure. For highly nonlinear problems the performance may even become
worse with linear decorrelation. Nonlinear methods without prior variable 
decorrelation should be used in such cases.

\subsubsection{Principal component decomposition\index{Discriminating variables!PCA of}}
\label{sec:pca}

Principal component decomposition\index{Principal component decomposition, analysis}
or principal component analysis (PCA) as presently applied in TMVA is not very
different from the above linear decorrelation. In common words, PCA is a linear 
transformation that rotates a sample of data points such that the maximum 
variability is visible. It thus identifies the 
most important gradients. In the PCA-transformed coordinate system, the 
largest variance by any projection of the data comes to lie on the first 
coordinate (denoted the {\em first principal component}), the second 
largest variance on the second coordinate, and so on. PCA can thus be used
to reduce the dimensionality of a problem (initially given by the number of 
input variables) by removing dimensions with insignificant variance. This 
corresponds to keeping lower-order principal components and ignoring 
higher-order ones. This latter step however goes beyond straight variable 
transformation as performed in the preprocessing steps discussed here 
(it rather represents itself a full classification method). Hence all principal
components are retained here.

The tuples ${\bf x}_U^{\rm PC}(i)=(x_{U,1}^{\rm PC}(i),\dots,x_{U,\Nvar}^{\rm PC}(i))$ 
of principal components of a tuple of input variables ${\bf x}(i)=(x_1(i),\dots,x_{\Nvar}(i))$,
measured for the event $i$ for signal ($U=S$) and background ($U=B$),
are obtained by the transformation
\beq
\label{eq:pca}
   x_{U,k}^{\rm PC}(i) = \sum_{\ell=1}^{\Nvar}
                         \left(x_{U,\ell}(i) - \overline x_{U,\ell}\right)
                         v^{(k)}_{U,\ell}\,,
   \hspace{0.5cm}\forall k=1,\Nvar\,.
\eeq
The tuples $\overline{\bf x}_{U}$ and ${\bf v}_{U}^{(k)}$ are the sample 
means and eigenvectors, respectively. They are computed by the ROOT class 
\code{TPrincipal}. The matrix of eigenvectors 
$V_U=({\bf v}^{(1)}_{U}\!,\dots,{\bf v}^{(\Nvar)}_{U})$ 
obeys the relation
\beq
   C_U\cdot V_U = D_U\cdot V_U\,,
\eeq
where $C$ is the covariance matrix of the sample $U$, and $D_U$ is the tuple
of eigenvalues. As for the preprocessing described in 
Sec.~\ref{sec:decorrelation}, the transformation~(\ref{eq:pca}) eliminates
linear correlations for Gaussian variables.

\subsubsection{Gaussian transformation of variables 
(``Gaussianisation'')\index{Discriminating variables!Gaussianisation of}}
\label{sec:gaussianisation}

The decorrelation methods described above require linearly correlated and Gaussian 
distributed input variables. In real-life HEP applications this is however rarely the 
case. One may hence transform the variables prior to their decorrelation such that their 
distributions become Gaussian. The corresponding transformation function is conveniently
separated into two steps: first, transform a variable into a uniform distribution
using its cumulative distribution function\footnote
{
   The cumulative distribution function $F(x)$ of the variable $x$ is given by 
   the integral $F(x)=\intl_{-\infty}^{x}\!\!\xPDF(x^\prime)\,d x^\prime$, where 
   $\xPDF$ is the probability density function of $x$.
} 
obtained from the training data (this transformation is identical to the ``Rarity'' 
introduced in Sec.~\ref{sec:otherRepresentations} on page~\pageref{sec:otherRepresentations});
second, use the inverse error function to transform the uniform distribution into 
the desired Gaussian shape with zero mean and unity width. As for the
other transformations, one needs to choose which class of events is to be 
transformed and hence, for classification (Gaussianisation is not available 
for regression), it is only possible to transform signal {\em or} background into 
proper Gaussian distributions (except for classifiers testing explicitly both
hypotheses such as likelihood methods). Hence a discriminant input variable $x$ 
with probability density function  $\xPDF$ is transformed as follows 
\beq
\label{eq:gaussianisation}
   x \mapsto \sqrt{2}\cdot {\rm erf}^{-1}\!\!
             \left( 2\cdot \!\!\intl_{-\infty}^{x}\!\!\xPDF(x^\prime)\,d x^\prime -1 \right)\:.
\eeq
A subsequent decorrelation of the transformed variable tuple sees Gaussian 
distributions, but most likely non-linear correlations as a consequence of the 
transformation~(\ref{eq:gaussianisation}). The distributions obtained after the 
decorrelation may thus not be Gaussian anymore. It has been suggested that 
iterating Gaussianisation and decorrelation more than once may improve the 
performance of likelihood methods (see next section).

\subsubsection{Booking and chaining transformations}\index{Booking variable transformations}

Variable transformations to be applied prior to the MVA training (and application) 
can be defined independently for each MVA method with the booking option 
{\tt VarTransform=<type>}, where {\tt <type>} denotes the desired transformation 
(or chain of transformations). The available transformation types are normalisation, 
decorrelation, principal component analysis and Gaussianisation, which are labelled by 
\code{Norm}, \code{Deco}, \code{PCA}, \code{Gauss}, respectively, or by the short-hand 
notation \code{N}, \code{D}, \code{P}, \code{G}. 

Transformations can be chained allowing the consecutive application of all defined 
transformations to the variables for each event.
For example, the above Gaussianisation and decorrelation sequence would be programmed by 
\code{VarTransform=G,D}, or even \code{VarTransform=G,D,G,D} in case of two iterations. 

By default, the transformations are computed with the use of all training events. It is 
possible to specify the use of a specific class only (\eg, \code{Signal}, \code{Background}, 
\code{Regression}) by attaching {\tt \_<class name>} to the user option -- 
where {\tt <class name>} has to be replaced by the actual class name 
(\eg, \code{Signal}) -- which defines the transformation (\eg, {\tt VarTransform=G\_Signal}). 
A complex transformation option might hence look like {\tt VarTransform=D,G\_Signal,N}.
The attachment {\tt \_AllClasses} is equivalent to the default, where events from all 
classes are used.

\subsection{Binary search trees\index{Binary search trees}} 
\label{sec:binaryTrees}

When frequent iterations over the training sample need to be performed, it is helpful 
to sort the sample before using it. Event sorting in {\em binary trees} is employed 
by the MVA methods rectangular cut optimisation\index{Cut optimisation},  
PDE-RS\index{PDE-RS} and k-NN\index{kd-tree}. While the former two classifiers rely on 
the simplest possible binary tree implementation, k-NN uses on a better performing 
{\em kd-tree} (\cf\  Ref.~\cite{kd-tree}).\index{kd-tree}

Efficiently searching for and counting events that lie inside a multidimensional
volume spanned by the discriminating input variables is accomplished with the use of a 
binary tree search algorithm~\cite{BinaryTree}.\footnote
{
   The following is extracted from Ref.~\cite{CarliKoblitz} for a two-dimensional 
   range search example.
   Consider a random sequence of signal events $e_i(x_1, x_2)$, $i = 1,2,\dots$, 
   which are to be stored in a binary tree. The first event in the sequence 
   becomes by definition the topmost node of the tree. The second event 
   $e_2(x_1, x_2)$ shall have a larger $x_1$-coordinate than the first event, 
   therefore a new node is created for it and the node is attached to the first
   node as the right child (if the $x_1$-coordinate had been smaller, the 
   node would have become the left child). Event $e_3$ shall have a larger 
   $x_1$-coordinate than event $e_1$, it therefore should be attached to 
   the right branch below $e_1$. Since $e_2$ is already placed at that 
   position, now the $x_2$-coordinates of $e_2$ and $e_3$ are compared, and,
   since $e_3$ has a larger $x_2$, $e_3$ becomes the right child of the node 
   with event $e_2$.
   The tree is sequentially filled by taking every event and, while 
   descending the tree, comparing its $x_1$ and $x_2$ coordinates with 
   the events already in place. Whether $x_1$ or $x_2$ are used for the 
   comparison depends on the level within the tree. On the first level, 
   $x_1$ is used, on the second level $x_2$, on the third again $x_1$ 
   and so on.
}
It is realised in the class \code{BinarySearchTree}, which 
inherits from \code{BinaryTree}, and which is also employed to grow decision 
trees (\cf\  Sec.~\ref{sec:bdt}). The amount of computing time needed to 
sort $N$ events into the tree is~\cite{CarliKoblitz} 
$\propto\sum_{i=1}^N\ln_2(i)=\ln_2(N!)\simeq N\ln_2N$. 
Finding the events within the tree which lie in a given volume is done 
by comparing the bounds of the volume with the coordinates of the events 
in the tree. Searching the tree once requires a CPU time that is
$\propto\ln_2N$, compared to $\propto N^\Nvar$ without prior event sorting.

%% file: CommonTools.tex
\section{Probability Density Functions -- the {\em PDF} Class}
\label{sec:PDF}

Several methods and functionalities in TMVA require the estimation of probability densities
(PDE)\index{PDE} of one or more correlated variables. One may distinguish three conceptually different 
approaches to PDEs: $(i)$ parametric approximation, where the training data are fitted with 
a user-defined parametric function, $(ii)$ nonparametric approximation, where the data 
are fitted piecewise using simple standard functions, such as a polynomial or a Gaussian, and
$(iii)$ nearest-neighbour estimation, where the average of the training data in the 
vicinity of a test event determines the PDF\index{PDF}\index{Probability Density Function}. 

All multidimensional PDEs used in TMVA are based on nearest-neighbour estimation with 
however quite varying implementations. They are described in Secs.~\ref{sec:pders}, 
\ref{sec:pdefoam} and \ref{sec:knn}. 

One-dimensional PDFs in TMVA are estimated by means 
of nonparametric approximation, because parametric functions cannot be generalised to 
a-priori unknown problems. The training data can be in form of binned histograms, or 
unbinned data points (or ``quasi-unbinned'' data, \ie, histograms with very narrow 
bins). In the first case, the bin centres are interpolated with polynomial
spline curves, while in the latter case one attributes a kernel function
to each event such that the PDF is represented by the sum over all kernels. Beside a faithful
representation of the training data, it is important that statistical fluctuations 
are smoothed out as much as possible without destroying significant information. In 
practise, where the true PDFs are unknown, a compromise determines which information
is regarded significant and which is not. Likelihood methods crucially depend on a 
good-quality PDF representation. Since the PDFs are strongly problem dependent, the 
default configuration settings in TMVA will almost never be optimal. The user is 
therefore advised to scrutinise the agreement between training data and PDFs via the 
available plotting macros, and to optimise the settings. 

In TMVA, all PDFs are derived from the \code{PDF} class, which is instantiated via the command
(usually hidden in the MVA methods for normal TMVA usage):
\begin{codeexample}
\begin{tmvacode}
   PDF* pdf = new PDF( "<options>", "Suffix", defaultPDF );
   pdf->BuildPDF( SourceHistogram );
   double p = pdf->GetVal( x );
\end{tmvacode}
\caption[.]{\codeexampleCaptionSize Creating and using a PDF class object.
         The first argument is the configuration options string. Individual options are 
         separated by a ':'. The second optional argument is the suffix appended to the 
         options used in the option string. The suffix is added to the option names 
         given in the Option Table~\ref{opt:pdf} in order to distinguish variables and types. 
         The third (optional) object is a PDF from which default option settings are read. The 
         histogram specified in the second line is a TH1* object from which the PDF is built.
         The third line shows how to retrieve the PDF value at a given test value \code{'x'}.
}
\end{codeexample}
Its configuration options are given in Option Table~\ref{opt:pdf}.
\begin{option}[!t]
\begin{optiontableLong2}
PDFInterpol       & KDE, \hspace{1cm} Spline0, 
                    \hspace{1cm}Spline1, \hspace{1cm}Spline2*, \hspace{1cm}Spline3, \hspace{1cm}Spline5 
                                 & The method of interpolating the reference histograms: either by
                                   using the unbinned kernel density estimator (KDE), or by various degrees
                                   of spline functions (note that currently the KDE characteristics 
                                   cannot be changed individually but apply to all variables that 
                                   select KDE) \\
NSmooth           & 0            & Number of smoothing iterations for the input histograms;
                                   if set, \code{MinNSmooth} and \code{MaxNSmooth} are ignored\\
MinNSmooth        & -1           & Minimum number of smoothing iteration for the input histograms;
                                   for bins with least relative error (see text) \\
MaxNSmooth        & -1           & Maximum number of smoothing iteration for the input histograms;
                                   for bins with most relative error (see text) \\
Nbins             & 0            & Number of bins used to build the reference histogram;
                                   if set to value $>0$, \code{NAvEvtPerBin} is ignored\\
NAvEvtPerBin      & 50           & Average number of events per bin in the reference histogram (see text) \\
KDEtype           & Gauss*       & KDE kernel type (currently only Gauss) \\
KDEiter           & Nonadaptive*, Adaptive
                                 & Non-adaptive or adaptive number of iterations (see text) \\
KDEFineFactor     & 1            & Fine-tuning factor for the adaptive KDE \\
KDEborder         & None*, Renorm, Mirror
                                 & Method for correcting histogram boundary/border effects \\
CheckHist         & False        & Sanity comparison of derived high-binned interpolated PDF histogram 
                                   versus the original PDF function
\end{optiontableLong2}
\caption[.]{\optionCaptionSize 
     Configuration options for class: {\em PDF}. Values given are defaults. If predefined 
     categories exist, the default category is marked by a '$\star$'. In case a suffix is defined 
     for the PDF, it is added in the end of the option name. For example for PDF 
     with suffix \code{MVAPdf} the number of smoothing iterations is given by the 
     option \code{NSmoothMVAPdf}
     (see Option Table~\ref{opt:mva::likelihood} on page~\pageref{opt:mva::likelihood} for a concrete
     use of the PDF options in a MVA method).     
}
\label{opt:pdf}
\end{option}

\subsection{Nonparametric PDF fitting using spline functions
                \index{Splines}
                \index{PDF parameterisation!with splines}}

Polynomial splines of various orders are fitted to one-dimensional (1D) binned histograms 
according to the following procedure.

\begin{enumerate}
\item	The number of bins of the \code{TH1} object representing the distribution of the 
      input variable is driven by the options \code{NAvEvtPerBin} or \code{Nbins} (\cf
      Option Table~\ref{opt:pdf}). Setting \code{Nbins} enforces a fixed number 
		of bins, while \code{NAvEvtPerBin} defines an average number of entries required 
 		per bin. The upper and lower bounds of the histogram coincide with the limits found 
      in the data (or they are $[-1,1]$ if the input variables are normalised).

\item	The histogram is smoothed\index{Histogram smoothing} adaptively between \code{MinNSmooth} 
      and \code{MaxNSmooth} times, using \code{TH1::Smooth(.)} -- an implementation of 
      the 353QH-twice algorithm~\cite{353qh}. The appropriate number of smoothing 
      iterations is derived with the aim to preserve statistically significant structures,
      while smoothing out fluctuations. Bins with the largest (smallest) relative statistical 
      error are maximally (minimally) smoothed. All other bins are smoothed between \code{MaxNSmooth} 
      and \code{MinNSmooth} times according to a linear function of their relative 
      errors. During the smoothing process a histogram with the suffix \code{NSmooth} is 
      created for each variable, where the number of smoothing iterations applied to each 
      bin is stored.

\item	The smoothed \code{TH1} object is internally cloned and the requested polynomial splines
      are used to interpolate adjacent bins. All spline classes are derived from ROOT's 
      \code{TSpline} class. Available are: polynomials of degree 0 (the original smoothed 
      histogram is kept), which is used for discrete variables; degree 1 (linear), 
      2 (quadratic), 3 (cubic) and degree 5. Splines\index{Splines} of degree two or more 
      render the PDF continuous and differentiable in all points excluding the interval 
      borders. In case of a likelihood analysis, this ensures the same property for the 
		likelihood ratio~(\ref{eq:RLik}). Since cubic (and higher) splines equalise
		the first and second derivatives at the spline transitions, the 
		resulting curves, although mathematically smooth, can wiggle in
		quite unexpected ways. Furthermore, there is no local control of 
		the spline: moving one control point (bin) causes the entire curve 
		to change, not just the part near the control point. To ensure 
		a safe interpolation, quadratic splines are used by default.
		
\item To speed up the numerical access to the probability densities,
		the spline functions are stored into a finely binned ($10^4$ bins)
		histogram, where adjacent bins are interpolated by a linear
		function. Only after this step, the PDF is normalised according
		to Eq.~(\ref{eq:pdfNorm}).

\end{enumerate}

\subsection{Nonparametric PDF parameterisation using kernel density estimators
                \index{Kernel density estimators (KDE)}
                \index{PDF parameterisation!with kernel density estimators}}

Another type of nonparametric approximation of a 1D PDF are kernel density
estimators (KDE). As opposed to splines, KDEs are obtained from unbinned data. The idea 
of the approach is to estimate the shape of a PDF by the sum over {\em smeared} 
training events. One then finds for a PDF $p(x)$ of a variable $x$~\cite{scott}
\beq
\label{eq:PDF:KDEf0}
   p(x)=\frac{1}{N\, h} \sum_{i=1}^{N}K\!\left( \frac{x-x_i}{h} \right)
       =\frac{1}{N} \sum_{i=1}^{N}K_h\!\left( x-x_i \right)\,,
\eeq
where $N$ is the number of training events, $K_h(t)=K(t/h)/h$ is the kernel 
function, and $h$ is the {\em bandwidth} of the kernel (also termed the {\em smoothing 
parameter}). Currently, only a Gaussian form of $K$ is implemented in TMVA, where the 
exact form of the kernel function is of minor relevance for the quality of the shape 
estimation. More important is the choice of the bandwidth. 

The KDE smoothing can be applied in either non-adaptive (NA) or adaptive form (A), 
the choice of which is controlled by the option \code{KDEiter}. In the 
non-adaptive case the bandwidth $h_{\rm NA}$ is kept constant for the entire training sample. 
As optimal bandwidth can be taken the one that minimises the {\em asymptotic mean integrated 
square error} (AMISE). For the case of a Gaussian kernel function this leads to~\cite{scott}
\beq
   h_{\rm NA}=\left(\frac{4}{3}\right)^{\!\!\!1/5} \!\!\!\sigma_x N^{-1/5}\,,
\eeq
where $\sigma_x$ is the RMS of the variable $x$.

The so-called {\em sample point adaptive} method uses as input the result of the non-adaptive 
KDE, but also takes into account the local event density. The adaptive bandwidth $h_{\rm A}$
then becomes a function of~$p(x)$~\cite{scott}
\beq
\label{eq:PDF:KDEhx}
   h_{\rm A}(x) = \frac{h_{\rm NA}}{\sqrt{p(x)}}\,.
\eeq
The adaptive approach improves the shape estimation in regions with low event density. However, in regions with high event density it can give rise to ``over-smoothing'' of fine 
structures such as narrow peaks. The degree of smoothing can be tuned by multiplying the 
bandwidth~$h_{\rm A}(x)$ with the user-specified factor \code{KDEFineFactor}. 

For practical reasons, the KDE implementation in TMVA differs somewhat from the procedure 
described above. Instead of unbinned training data a finely-binned histogram is used as 
input, which allows to significantly speed up the algorithm. The calculation of the optimal
bandwidth~$h_{\rm NA}$ is performed in the dedicated class \code{KDEKernel}. If the algorithm 
is run in the adaptive mode, the non-adaptive step is also performed because its output feeds
the computation of $h_{\rm A}(x)$ for the adaptive part. Subsequently, a smoothed high-binned 
histogram estimating the PDF shape is created by looping over the bins of the input histogram 
and summing up the corresponding kernel functions, using $h_{\rm NA}$ ($h_{\rm A}(x)$) in 
case of the non-adaptive (adaptive) mode. This output histogram is returned to the \code{PDF}
class.

Both the non-adaptive and the adaptive methods can suffer from the so-called {\em boundary 
problem} at the histogram boundaries. It occurs for instance if the original distribution is non-zero below a physical 
boundary value and zero above. This property cannot be reproduced by the KDE procedure. 
In general, the stronger the discontinuity the more acute is the boundary problem. TMVA 
provides three options under the term \code{KDEborder} that allow to treat boundary 
problems. 
\begin{itemize}

\item \code{KDEborder=None} \\
      No boundary treatment is performed. The consequence is that 
      close to the boundary the KDE result will be inaccurate: below the boundary it will 
      underestimate the PDF while it will not drop to zero above. In TMVA the PDF resulting from 
      KDE is a (finely-binned) histogram, with bounds equal to the minimum and the maximum 
      values of the input distribution. Hence, the boundary value will be at the edge of the PDF 
      (histogram), and a drop of the PDF due to the proximity of the boundary can be observed
      (while the artificial enhancement beyond the boundary will fall outside of the histogram). 
      In other words, for training events that are close to the boundary some fraction of the 
      probability ``flows'' outside the histogram ({\em probability leakage}). As a consequence, 
      the integral of the kernel function inside the histogram borders is smaller than one.

\item \code{KDEborder=Renorm} \\
      The probability leakage is compensated by renormalising the 
      kernel function such that the integral inside the histogram borders is equal to one.

\item \code{KDEborder=Mirror} \\
      The original distribution is ``mirrored'' around the boundary. The same procedure is 
      applied to the mirrored events and each 
      of them is smeared by a kernel function and its contribution {\em inside the histogram's (PDF) 
      boundaries} is added to the PDF. The mirror copy compensates the probability 
      leakage completely.

\end{itemize}

\section{Optimisation and Fitting}\index{Fitting}
\label{sec:fitting}

Several MVA methods (notably cut optimisation and FDA) require general purpose
parameter fitting to optimise the value of an estimator. For example, an estimator 
could be the sum of the deviations of classifier outputs from 1 for signal events 
and 0 for background events, and the parameters are adjusted so that this sum is as
small as possible. Since the various fitting problems call for dedicated solutions, 
TMVA has a fitter base class, used by the MVA methods, from which all concrete fitters 
inherit. The consequence of this is that the user can choose whatever fitter is
deemed suitable and can configure it through the option string of the MVA method.
At present, four fitters are implemented and described below: Monte Carlo sampling, 
Minuit minimisation, a Genetic Algorithm, Simulated Annealing.
They are selected via the configuration option of the corresponding MVA method for
which the fitter is invoked (see Option Table~\ref{opt:fitter}). Combinations 
of MC and GA with Minuit are available for the FDA method by setting the \code{Converger}
option, as described in Option Table~\ref{opt:mva::fda}.
\begin{option}[t]
\begin{optiontableDescr}
FitMethod         & MC, MINUIT, GA, SA      & Fitter method   \\
Converger         & None*, MINUIT           & Converger which can be combined with MC or GA
                                              (currently only used for FDA) to improve 
                                              finding local minima
\end{optiontableDescr}
\caption[.]{\optionCaptionSize Configuration options for the choice of a fitter.
         The abbreviations stand for Monte Carlo sampling, Minuit, Genetic 
         Algorithm, Simulated Annealing. By setting a Converger (only Minuit is
         currently available) combined use of Monte Carlo sampling and 
         Minuit, and of Genetic Algorithm and Minuit is possible.
         The \code{FitMethod} option can be used in any MVA method that requires fitting. 
         The option \code{Converger} is currently only implemented in FDA.
         The default fitter depends on the MVA method.
         The fitters and their specific options are described below.}
\label{opt:fitter}
\end{option}

\subsection{Monte Carlo sampling\index{Monte Carlo sampling}}
\label{sec:MCsampling}

The simplest and most straightforward, albeit inefficient fitting method is to randomly 
sample the fit parameters and to choose the set of parameters that optimises the estimator. 
The priors
used for the sampling are uniform or Gaussian within the parameter limits. The specific 
configuration options for the MC sampling are given in Option Table~\ref{opt:fitter_mc}. 

For fitting problems
with few local minima out of which one is a global minimum the performance can be enhanced
by setting the parameter \code{Sigma} to a positive value. The newly generated parameters
are then not any more independent of the parameters of the previous samples. The random 
generator will throw random values according to a Gaussian probability density 
with the mean given by the currently known best value for that particular parameter and 
the width in units of the interval size given by the option \code{Sigma}.
Points which are created out of the parameter's interval are mapped back 
into the interval.
% ======= input option table ==========================================
\begin{option}[t]
\input optiontables/Fitter_MC.tex
\caption[.]{\optionCaptionSize 
     Configuration options reference for fitting method: {\em Monte Carlo sampling (MC)}.
}
\label{opt:fitter_mc}
\end{option}
% =====================================================================

\subsection{Minuit minimisation}\index{Minuit!fitter}\index{Minuit!minimisation}
\label{sec:minuit}

Minuit is the standard multivariate minimisation package used in HEP~\cite{Minuit}.
Its purpose is to find the minimum of a multi-parameter estimator function 
and to analyse the shape of the function around the minimum (error analysis). The 
principal application of the TMVA fitters is simple minimisation, while the shape
of the minimum is irrelevant in most cases. The use of Minuit is therefore not 
necessarily the most efficient solution, but because it is a very robust tool we 
have included it here. Minuit searches the solution along the direction of the gradient 
until a minimum or an boundary is reached (MIGRAD algorithm). It 
does not attempt to find the global minimum but is satisfied with local minima.
If during the error analysis with MINOS, the minimum smaller values than the local 
minimum might be obtained. In particular, the use of MINOS may as a side effect of 
an improved error analysis uncover a convergence in a local minimum, in which case 
MIGRAD minimisation is invoked once again. If multiple local and/or global 
solutions exist, it might be preferable to use any of the other fitters which are 
specifically designed for this type of problem. 

The configuration options for Minuit are given in Option Table~\ref{opt:fitter_minuit}.
% ======= input option table ==========================================
\begin{option}[t]
\input optiontables/Fitter_Minuit.tex
\caption[.]{\optionCaptionSize 
     Configuration options reference for fitting method: {\em Minuit}. More information 
     on the Minuit parameters can be found here: \url{http://root.cern.ch/root/html/TMinuit.html}.
}
\label{opt:fitter_minuit}
\end{option}
% =====================================================================

\subsection{Genetic Algorithm}\index{Genetic Algorithm}
\label{sec:geneticAlgorithm}

A Genetic Algorithm is a technique to find approximate solutions to optimisation or 
search problems. The problem is modelled by a group 
({\em population}) of abstract representations ({\em genomes}) of possible solutions 
({\em individuals}). By applying means similar to processes found in biological 
evolution the individuals of the population should evolve towards an optimal 
solution of the problem. Processes which are usually modelled in evolutionary 
algorithms --- of which Genetic Algorithms are a subtype --- are inheritance, mutation 
and ``sexual recombination'' (also termed {\em crossover}). 

Apart from the abstract representation of the solution domain, a {\em fitness}\index{Fitness} 
function must be defined. Its purpose is the evaluation of the goodness of an 
individual. The fitness function is problem dependent. It either returns a value 
representing the individual's goodness or it compares two individuals and indicates 
which of them performs better. 

The Genetic Algorithm proceeds as follows:
\begin{itemize}

\item {\em Initialisation}: A starting population is created. Its size depends 
      on the problem to be solved. Each individual belonging to the population is 
      created by randomly setting the parameters of the abstract 
      representation (variables), thus producing a point in the 
      solution domain of the initial problem.

\item {\em Evaluation}: Each individual is evaluated using the fitness function. 

\item {\em Selection}: Individuals are kept or discarded as a function of their 
      fitness. Several selection procedures are possible. The simplest one is to 
      separate out the worst performing fraction of the population. Another possibility 
      is to decide on the individual's survival by assigning probabilities that 
      depend on the individual's performance compared to the others. 

\item {\em Reproduction}: The surviving individuals are copied, mutated and 
      crossed-over until the initial population size is reached again. 

\item {\em Termination}: The evaluation, selection and reproduction steps are 
      repeated until a maximum number of cycles is reached or an individual 
      satisfies a maximum-fitness criterion. The best individual is selected 
      and taken as solution to the problem. 

\end{itemize}
% ======= input option table ==========================================
\begin{option}[t]
\input optiontables/Fitter_GA.tex
\caption[.]{\optionCaptionSize 
     Configuration options reference for fitting method: {\em Genetic Algorithm (GA)}.
}
\label{opt:fitter_ga}
\end{option}
% =====================================================================
The TMVA Genetic Algorithm provides controls that are set through configuration options 
(\cf\  Table~\ref{opt:fitter_ga}). 
The parameter \code{PopSize} determines the number of individuals 
created at each generation of the Genetic Algorithm. At 
the initialisation, all parameters of all individuals are chosen randomly.  
The individuals are evaluated in terms of their fitness, and each individual 
giving an improvement is immediately stored.

Individuals with a good fitness are selected to engender the next generation. 
The new individuals are created by crossover and mutated afterwards. Mutation 
changes some values of some parameters of some individuals randomly following
a Gaussian distribution function. The width of the Gaussian can be altered
by the parameter \code{SC\_factor}. The current width is multiplied by this factor 
when within the last \code{SC\_steps} generations more than \code{SC\_rate} 
improvements have been obtained. If there were \code{SC\_rate} improvements 
the width remains unchanged. Were there, on the other hand, less than 
\code{SC\_rate} improvements, the width is divided by \code{SC\_factor}. 
This allows to influence the speed of searching through the solution domain.

The cycle of evaluating the fitness of the individuals of a generation and 
producing a new generation is repeated until the improvement of the fitness 
within the last \code{Steps} has been less than \code{ConvCrit}.
The minimisation is then considered to have converged. The whole cycle 
from initialisation over fitness evaluation, selection, reproduction and 
determining the improvement is repeated \code{Cycles} times, before the Genetic 
Algorithm has finished.

\subsubsection*{Guidelines for optimising the GA}

\code{PopSize} is the most important value for enhancing the quality of the results.
This value is by default set to 300, but can be increased to 1000 or more only limited
by the resources available. The calculation time of the GA should increase with O(\code{PopSize}).

\code{Steps} is set by default to 40. This value can be increased moderately to about 60. 
Time consumption increases non linearly but at least with O(\code{Steps}). 

\code{Cycles} is set by default to 3. In this case, the GA is called three times independently
from each other. With \code{SaveBestCycle} and \code{SaveBestGen} it is possible to set
the number of best results which shall be stored each cycle of even each generation. 
These stored results are reused in the last cycle. That way the last cycle is not
independent from the others, but incorporates their best results. The number of 
cycles can be increased moderately to about 10. The time consumption of GA rises
with about O(\code{Cycles}).

\subsection{Simulated Annealing} \index{Simulated Annealing}
\label{sec:simAnnealing}

Simulated Annealing also aims at solving a minimisation problem with several 
discrete or continuous, local or global minima. The algorithm is inspired by 
the process of of annealing which occur in condensed matter physics. When 
first heating and then slowly cooling down a metal (``annealing'') its atoms move 
towards a state of lowest energy, while for sudden cooling the atoms tend to freeze in 
intermediate states higher energy. For infinitesimal annealing activity the system will 
always converge in its global energy minimum (see, \eg, Ref.~\cite{VanLaarhovenEA}).
This physical principle can be converted into an algorithm to achieve slow, but 
correct convergence of an optimisation problem with multiple solutions. Recovery 
out of local minima is achieved by assigning the probability~\cite{MetropolisEA} 
\beq
       p(\Delta E) \propto \exp\left(-\frac{\Delta E}{T}\right)\,,
\eeq
to a perturbation of the parameters leading to a shift $\Delta E$ 
in the energy of the system. The probability of such perturbations to occur 
decreases with the size of a positive energy coefficient of the perturbation,
and increases with the ambient temperature ($T$).

\subsubsection*{Guidelines for optimising SA}

The TMVA implementation of Simulated Annealing includes various different adaptive 
adjustments of the perturbation and temperature gradients. The adjustment procedure 
is chosen by setting \code{KernelTemp} to one of the following values.
\begin{itemize}

\item {\sf\bfseries\small Increasing Adaptive Approach } (\code{IncAdaptive}). 
      The algorithm seeks local minima and explores their neighbourhood, while                
      changing the ambient temperature depending on the number of failures            
      in the previous steps. The performance can be improved by increasing            
      the number of iteration steps (\code{MaxCalls}), or by adjusting the               
      minimal temperature (\code{MinTemp}). Manual adjustments of the             
      speed of the temperature increase (\code{TempScale} and \code{AdaptiveSpeed})  
      for individual data sets might also improve the performance. 

      %Summary: increase \code{MaxCalls}; adjust \code{MinTemp}; adjust \code{TempScale}; adjust \code{AdaptiveSpeed}.

\item {\sf\bfseries\small Decreasing Adaptive Approach} (\code{DecAdaptive}). 
      The algorithm calculates the initial temperature (based on the effectiveness 
      of large steps) and defines a multiplier to ensure that the minimal temperature is reached  
      in the requested number of iteration steps. The performance can be improved 
      by adjusting the minimal temperature (\code{MinTemp}) and by increasing 
      number of steps (\code{MaxCalls}).

\item {\sf\bfseries\small Other Kernels.} 
     Several other procedures to calculate the temperature change are also implemented.
     Each of them starts with the maximum temperature (\code{MaxTemp})
     and decreases while changing the temperature according to : 
     \beqns
     {\rm Temperature}^{(k)} = \left\{
     \begin{array}{ll}
       {\tt Sqrt}: & \frac{{\tt InitialTemp}}{\sqrt{k+2}} \cdot {\tt TempScale} \\[0.25cm]
       {\tt Log}:  & \frac{{\tt InitialTemp}}{\ln(k+2)} \cdot {\tt TempScale} \\[0.25cm]
       {\tt Homo}: & \frac{{\tt InitialTemp}}{k+2} \cdot {\tt TempScale} \\[0.25cm]
       {\tt Sin}:  & \frac{\sin(k/{\tt TempScale}) + 1}{k+1} 
                      \cdot {\tt InitialTemp + \e} \\[0.25cm]
       {\tt Geo}:  & {\tt CurrentTemp} \cdot {\tt TempScale }
      \end{array}
      \right.
      \eeqns
      Their performances can be improved by adjusting the initial
      temperature \code{InitialTemp}\\ (= ${\rm
        Temperature}^{(k=0)}$), the number of iteration steps
      (\code{MaxCalls}), and the multiplier that scales the
      temperature decrease (\code{TempScale}).
     % increase \code{MaxCalls}; adjust   \code{InitialTemp}; adjust   \code{TempScale}; 
     % adjust   \code{KernelTemp}.

\end{itemize}
The configuration options for the Simulated Annealing fitter are given in 
Option Table~\ref{opt:fitter_sa}.

% ======= input option table ==========================================
\begin{option}[t]
\input optiontables/Fitter_SA.tex
\caption[.]{\optionCaptionSize 
     Configuration options reference for fitting method: {\em Simulated Annealing (SA)}.
}
\label{opt:fitter_sa}
\end{option}
% =====================================================================

\subsection{Combined fitters}
           \index{Monte Carlo sampling!combination with Minuit}
           \index{Genetic Algorithm!combination with Minuit}
           \index{Minuit!combination with MC or GA}
\label{sec:converger}

For MVA methods such as FDA, where parameters of a discrimination function
are adjusted to achieve optimal classification performance (\cf\  Sec.~\ref{sec:fda}), 
the user can choose to 
combine Minuit parameter fitting with Monte Carlo sampling or a Genetic Algorithm.
While the strength of Minuit is the speedy detection of a nearby local minimum, 
it might not find a better global minimum. If several local minima exist
Minuit will find different solutions depending on the start values for the fit 
parameters. When combining Minuit with Monte Carlo sampling or a Genetic Algorithm, 
Minuit uses starting values generated by these methods. The subsequent fits then 
converge in local minima. Such a combination is usually more efficient than the
uncontrolled sampling used in Monte Carlo techniques. When combined with a Genetic
Algorithm the user can benefit from the advantages of both methods: the 
Genetic Algorithm to roughly locate the global minimum, and Minuit to find an
accurate solution for it (for an example see the FDA method).

The configuration options for the combined fit methods are the inclusive sum of all 
the individual fitter options. It is recommended to use Minuit in  batch mode 
(option \code{SetBatch}) and without MINOS (option \code{!UseMinos}) to prevent 
TMVA from flooding the output with Minuit messages which cannot be turned off, and 
to speed up the individual fits. It is however important to note that the combination of 
MIGRAD and MINOS together is less susceptible to getting caught in local minima.

\section{Boosting and Bagging}
\label{sec:boost}

Boosting\index{Boosting} is a way of enhancing the classification and regression
performance (and increasing the stability with respect to statistical fluctuations
in the training sample) of typically weak MVA methods by sequentially applying an
MVA algorithm to reweighted ({\em boosted}) versions of the training data and then 
taking a weighted majority vote of the sequence of MVA algorithms thus produced. 
It has been introduced to classification techniques in the early '90s~\cite{Boosting} 
and in many cases this simple strategy results in dramatic performance increases.

Although one my argue that {\em bagging} (\cf\  Sec.~\ref{sec:bagging}) is not
a genuine boosting algorithm, we include it in the same context and typically when 
discussing boosting we also refer to bagging.  The most commonly boosted 
methods are decision trees. However, as described in Sec.~\ref{sec:boosted} on
page~\pageref{sec:boosted}, any MVA method may be boosted with TMVA. Hence, although
the following discussion refers to decision trees, it also applies to other methods. 
(Note however that ``Gradient Boost'' is only available for decision trees and only 
for classification in the present TMVA version).

\subsection{Adaptive Boost (AdaBoost)}\index{Boosting!Adaptive Boost}
\label{sec:adaboost}

The most popular boosting algorithm is the so-called {\em  AdaBoost} 
(adaptive boost)\index{AdaBoost}\index{Adaptive boost}~\cite{AdaBoost}. 
In a {\em classification problem}, events that were misclassified during the
training of a decision tree are given a higher event weight in the training of
the following tree.  Starting with the original event weights 
when training the first decision tree, the subsequent tree is trained 
using a modified event sample where the weights of previously
misclassified events are multiplied by a common {\em boost weight}
$\alpha$. The boost weight is derived from the misclassification rate,
${\rm err}$, of the previous tree\footnote 
{ 
   By construction, the error rate is ${\rm err}\le0.5$ as the same 
   training events used to classify the output nodes of the previous
   tree are used for the calculation of the error rate.  
}, 
\beq
\label{eq:boost}
   \alpha = \frac{1-{\rm err}}{{\rm err}}\,.  
\eeq 
The weights of the entire event sample are then renormalised such that the 
sum of weights remains constant.  
 
We define the result of an individual classifier as $h({\bf x})$, 
with (${\bf x}$ being the tuple of input variables) encoded for signal 
and background as $h({\bf x}) = +1\ \mbox{and }-1$, respectively. 
The boosted event classification $\yBoost(\bf{x})$ is then given by 
\beq
\label{eq:adaboost}
  \yBoost({\bf x}) = \frac{1}{N_{\rm collection }} \cdot {\sum_{i}^{N_{\rm collection}}} 
                     \ln(\alpha_i)\cdot h_i({\bf x})\:,
\eeq
where the sum is over all classifiers in the collection. Small (large)
values for $\yBoost({\bf x})$ indicate a background-like (signal-like)
event. Equation~(\ref{eq:adaboost}) represents the default boosting algorithm.
It can be modified via the configuration option string of the MVA method to 
be boosted (see Option Tables~\ref{opt:mva::bdt_1} and \ref{opt:mva::bdt_2} 
on pages~\pageref{opt:mva::bdt_1} and \pageref{opt:mva::bdt_1} for 
boosted decision trees, and Option Table~\ref{opt:mva::boost} for general
classifier boosting~\ref{sec:boosted}) if one wants to use an unweighted average 
of the boosted decision trees or classifiers instead of the weighted one.

For {\em regression trees}, the AdaBoost algorithm needs to be modified. TMVA uses 
here the so-called AdaBoost.R2 algorithm~\cite{AdaBoostR2}\index{AdaBoost.R2}. The 
idea is similar to AdaBoost albeit with a redefined loss per event to account for the 
the deviation of the estimated target value from the true one. Moreover, as there 
are no longer correctly and wrongly classified events, all events need to be 
reweighted depending on their individual {\em loss}, which -- for event $k$ --
is given  by
\beqn
   Linear: & L(k) = \frac{|y(k) - {\hat y(k)}|}{\max\limits_{{\rm events}\ k^\prime} 
                                  (|y(k^\prime) - {\hat y(k^\prime}|) }\:, \\
   Square: & L(k) = \left[ \frac{|y(k) - {\hat y(k)}|}{\max\limits_{{\rm events}\ k^\prime} 
                                  (|y(k^\prime) - {\hat y(k^\prime}|) } \right]^2\,, \\
   Exponential : & L(k) = 1 - \exp \left[ -\, \frac{|y(k) - {\hat y(k)}|}{\max\limits_{{\rm events}\ k^\prime} 
                                  (|y(k^\prime) - {\hat y(k^\prime}|) }\right]\,.
\eeqn

\newcommand{\Lave}{{\ensuremath \langle L\rangle}\xspace}
The average loss of the classifier $y^{(i)}$ over the whole training sample,
$\Lave^{(i)} = \sum_{{\rm events}\:\:k^\prime} w(k^\prime)L^{(i)}(k^\prime)$, can be 
considered to be the analogon to the error fraction in classification. Given 
$\Lave$, one computes the quantity $ \beta_{(i)} = \Lave^{(i)}/(1-\Lave^{(i)})$, which is used in
the boosting of the events, and for the combination of the regression methods  
belonging to the boosted collection. The boosting weight, $w^{(i+1)}(k)$, for event~$k$ 
and boost step $i+1$ thus reads
\beqn
   w^{(i+1)}(k) = w^{(i)}(k) \cdot \beta_{(i)}^{1-L^{(i)}(k)}\:.
\eeqn
The sum of the event weights is again renormalised to reproduce the original 
overall number of events. The final regressor, $\yBoost$, uses the weighted 
median, $\tilde y_{(i)}$, where $(i)$ is chosen so that it is the minimal $(i)$ 
that satisfies the inequality
\beq
\label{eq:adaboostr2}
  \sum_{t\in {\rm sorted\ collection}\atop t\leq i}\!\!\!\! \ln\frac{1}{\beta_{(t)}}
  \geq \: \frac{1}{2}\!\sum_{t}^{N_{\rm collection}}\!\!\!\!\ln\frac{1}{\beta_{(t)}}  
\eeq

\subsection{Gradient Boost}\index{Boosting!Gradient Boost}
\label{sec:gradientboost}

The idea of function estimation through boosting can be understood by
considering a simple additive expansion approach. The function
$F(\mathbf{x})$ under consideration is assumed to be a weighted sum of
parametrised base functions $f(x;a_m)$, so-called ``weak
learners''. From a technical point of view any TMVA classifier could
act as a weak learner in this approach, but decision trees benefit most
from boosting and are currently the only classifier that implements
GradientBoost (a generalisation may be included in future releases).
Thus each base function in this expansion corresponds to a
decision tree
\beq
   F(\mathbf{x};P)=\sum_{m=0}^{M}\beta_mf(x;a_m); \hspace{0.5cm} 
   P\in \{\beta_m;a_m\}_0^M\:.
\eeq
The boosting procedure is now employed to adjust the parameters $P$
such that the deviation between the model response $F(\mathbf{x})$ and
the true value $y$ obtained from the training sample is minimised. The
deviation is measured by the so-called \textit{loss-function}
$L(F,y)$, a popular choice being squared error loss
$L(F,y)=(F(\mathbf{x})-y)^2$. It can be shown that the loss
function fully determines the boosting procedure.

The most popular boosting method, AdaBoost, is based on exponential loss, 
$L(F,y)=e^{-F(\mathbf{x})y}$, which leads to the well known reweighting 
algorithm described in Sec.~\ref{sec:adaboost}. Exponential loss has the 
shortcoming that it lacks robustness in presence of outliers or mislabelled 
data points. The performance of AdaBoost therefore degrades in noisy settings.

The \textit{GradientBoost} algorithm attempts to cure this weakness by allowing 
for other, potentially more robust, loss functions without giving up on the 
good out-of-the-box performance of AdaBoost. The current TMVA implementation 
of GradientBoost uses the binomial log-likelihood loss
\beq
   L(F,y)=\ln\left(1+e^{-2F(\mathbf{x})y}\right)\:,
\eeq
for classification. As the boosting algorithm corresponding to this
loss function cannot be obtained in a straightforward manner, one has
to resort to a steepest-descent approach to do the minimisation. This
is done by calculating the current gradient of the loss function and
then growing a regression tree whose leaf values are adjusted to match
the mean value of the gradient in each region defined by the tree
structure. Iterating this procedure yields the desired set of
decision trees which minimises the loss function. Note that
GradientBoost can be adapted to any loss
function as long as the calculation of the gradient is feasible.

Giving good results already for small trees (5--10 leaf nodes),
GradientBoost is typically less susceptible to overtraining. Its
robustness can be enhanced by reducing the learning rate of the
algorithm through the \code{Shrinkage} parameter (\cf\  Option 
Table~\ref{opt:mva::bdt_1} on page~\pageref{opt:mva::bdt_1}), which controls 
the weight of the individual trees. A small shrinkage (0.1--0.3) demands
more trees to be grown but can significantly improve the accuracy of
the prediction in difficult settings.

In certain settings GradientBoost may also benefit
from the introduction of a bagging-like resampling procedure using
random subsamples of the training events for growing the trees. This
is called \textit{stochastic gradient boosting} and can be enabled by
selecting the \code{UseBaggedGrad} option. The sample fraction used
in each iteration can be controlled through the parameter
\code{GradBaggingFraction}, where typically the best results are obtained
for values between 0.5 and 0.8.

\subsection{Bagging}
\label{sec:bagging}

The term {\em Bagging}\index{Bagging}\index{Resampling} denotes a resampling 
technique where a classifier is repeatedly trained using resampled training
events such that the combined classifier represents an average of 
the individual classifiers. A priori, bagging does not aim at enhancing a
weak classifier in the way adaptive or gradient boosting does, and is 
thus not a ``boosting'' algorithm in a strict sense. Instead it effectively 
smears over statistical representations of the training data and is hence 
suited to stabilise the response of a classifier. In this context it is 
often accompanied also by a significant performance increase compared to 
the individual classifier.

Resampling includes the possibility of replacement, which means that the 
same event is allowed to be (randomly) picked several times from the parent
sample. This is equivalent to regarding the training sample as being a
representation of the probability density distribution of the parent
sample: indeed, if one draws an event out of the parent sample, it is more
likely to draw an event from a region of phase-space that has a high
probability density, as the original data sample will have more
events in that region. If a selected event is kept in the original
sample (that is when the same event can be selected several times),
the parent sample remains unchanged so that the randomly extracted
samples will have the same parent distribution, albeit statistically
fluctuated.  Training several classifiers with different resampled
training data, and combining them into a collection, results in an
averaged classifier that, just as for boosting, is more stable with
respect to statistical fluctuations in the training sample. 

Technically, resampling is implemented by applying random Poisson 
weights to each event of the parent sample.

%% file: optiontables/Fitter_MC.tex
\begin{optiontableAuto}
               SampleSize  &  \mc{1}{c}{--}  &           100000  &  \mc{1}{l}{--}  &  Number of Monte Carlo events in toy sample \\
                    Sigma  &  \mc{1}{c}{--}  &               -1  &  \mc{1}{l}{--}  &  If $>$ 0: new points are generated according to Gauss around best value and with Sigma in units of interval length \\
                     Seed  &  \mc{1}{c}{--}  &              100  &  \mc{1}{l}{--}  &  Seed for the random generator (0 takes random seeds) 
\end{optiontableAuto}

%% file: optiontables/Fitter_Minuit.tex
\begin{optiontableAuto}
               ErrorLevel  &  \mc{1}{c}{--}  &                1  &  \mc{1}{l}{--}  &  TMinuit: error level: 0.5=logL fit, 1=chi-squared fit \\
               PrintLevel  &  \mc{1}{c}{--}  &               -1  &  \mc{1}{l}{--}  &  TMinuit: output level: -1=least, 0, +1=all garbage \\
              FitStrategy  &  \mc{1}{c}{--}  &                2  &  \mc{1}{l}{--}  &  TMinuit: fit strategy: 2=best \\
            PrintWarnings  &  \mc{1}{c}{--}  &            False  &  \mc{1}{l}{--}  &  TMinuit: suppress warnings \\
               UseImprove  &  \mc{1}{c}{--}  &             True  &  \mc{1}{l}{--}  &  TMinuit: use IMPROVE \\
                 UseMinos  &  \mc{1}{c}{--}  &             True  &  \mc{1}{l}{--}  &  TMinuit: use MINOS \\
                 SetBatch  &  \mc{1}{c}{--}  &            False  &  \mc{1}{l}{--}  &  TMinuit: use batch mode \\
                 MaxCalls  &  \mc{1}{c}{--}  &             1000  &  \mc{1}{l}{--}  &  TMinuit: approximate maximum number of function calls \\
                Tolerance  &  \mc{1}{c}{--}  &              0.1  &  \mc{1}{l}{--}  &  TMinuit: tolerance to the function value at the minimum 
\end{optiontableAuto}

%% file: optiontables/Fitter_GA.tex
\begin{optiontableAuto}
                  PopSize  &  \mc{1}{c}{--}  &              300  &  \mc{1}{l}{--}  &  Population size for GA \\
                    Steps  &  \mc{1}{c}{--}  &               40  &  \mc{1}{l}{--}  &  Number of steps for convergence \\
                   Cycles  &  \mc{1}{c}{--}  &                3  &  \mc{1}{l}{--}  &  Independent cycles of GA fitting \\
                 SC\_steps  &  \mc{1}{c}{--}  &               10  &  \mc{1}{l}{--}  &  Spread control, steps \\
                  SC\_rate  &  \mc{1}{c}{--}  &                5  &  \mc{1}{l}{--}  &  Spread control, rate: factor is changed depending on the rate \\
                SC\_factor  &  \mc{1}{c}{--}  &             0.95  &  \mc{1}{l}{--}  &  Spread control, factor \\
                 ConvCrit  &  \mc{1}{c}{--}  &            0.001  &  \mc{1}{l}{--}  &  Convergence criteria \\
              SaveBestGen  &  \mc{1}{c}{--}  &                1  &  \mc{1}{l}{--}  &  Saves the best n results from each generation. They are included in the last cycle \\
            SaveBestCycle  &  \mc{1}{c}{--}  &               10  &  \mc{1}{l}{--}  &  Saves the best n results from each cycle. They are included in the last cycle. The value should be set to at least 1.0 \\
                     Trim  &  \mc{1}{c}{--}  &            False  &  \mc{1}{l}{--}  &  Trim the population to PopSize after assessing the fitness of each individual \\
                     Seed  &  \mc{1}{c}{--}  &              100  &  \mc{1}{l}{--}  &  Set seed of random generator (0 gives random seeds) 
\end{optiontableAuto}

%% file: optiontables/Fitter_SA.tex
\begin{optiontableAuto}
                 MaxCalls  &  \mc{1}{c}{--}  &           100000  &  \mc{1}{l}{--}  &  Maximum number of minimisation calls \\
              InitialTemp  &  \mc{1}{c}{--}  &            1e+06  &  \mc{1}{l}{--}  &  Initial temperature \\
                  MinTemp  &  \mc{1}{c}{--}  &            1e-06  &  \mc{1}{l}{--}  &  Mimimum temperature \\
                      Eps  &  \mc{1}{c}{--}  &            1e-10  &  \mc{1}{l}{--}  &  Epsilon \\
                TempScale  &  \mc{1}{c}{--}  &                1  &  \mc{1}{l}{--}  &  Temperature scale \\
            AdaptiveSpeed  &  \mc{1}{c}{--}  &                1  &  \mc{1}{l}{--}  &  Adaptive speed \\
         TempAdaptiveStep  &  \mc{1}{c}{--}  &         0.009875  &  \mc{1}{l}{--}  &  Step made in each generation temperature adaptive \\
          UseDefaultScale  &  \mc{1}{c}{--}  &            False  &  \mc{1}{l}{--}  &  Use default temperature scale for temperature minimisation algorithm \\
           UseDefaultTemp  &  \mc{1}{c}{--}  &            False  &  \mc{1}{l}{--}  &  Use default initial temperature \\
               KernelTemp  &  \mc{1}{c}{--}  &      IncAdaptive  &  IncAdaptive, DecAdaptive, Sqrt, Log, Sin, Homo, Geo  &  Temperature minimisation algorithm 
\end{optiontableAuto}

%% file: MethodsIntro.tex
\section{The TMVA Methods}
\label{sec:tmvaClassifiers}

All TMVA classification and regression methods (in most cases, a method serves both 
analysis goals) inherit from \code{MethodBase}, which implements basic 
functionality like the interpretation of common configuration options, the 
interaction with the training and test data sets, I/O operations and common 
performance evaluation calculus. The functionality each MVA method is required 
to implement is defined in the abstract interface \code{IMethod}.\footnote 
{ 
  Two constructors are
  implemented for each method: one that creates the method for
  a first time for training with a configuration (``option'') string
  among the arguments, and another that recreates a method from an
  existing weight file. The use of the first constructor is
  demonstrated in the example macros \code{TMVAClassification.C} and
  \code{TMVARegression.C}, while the second one is employed by the Reader in 
  \code{TMVAClassificationApplication.C} and \code{TMVARegressionApplication.C}.
  Other functions implemented by each methods are: \code{Train}
  (called for training), \code{Write/ReadWeightsToStream} (I/O of
  specific training results), \code{WriteMonitoringHistosToFile}
  (additional specific information for monitoring purposes) and 
  \code{CreateRanking} (variable ranking).  
} 
Each MVA method provides a function that creates a rank object (of 
type \code{Ranking}), which is an ordered list of the input variables 
prioritised according to criteria specific to that method. Also 
provided are brief method-specific help notes (option \code{Help}, 
switched off by default) with information on the adequate usage of 
the method and performance optimisation in case of unsatisfying 
results.

If the option \code{CreateMVAPdfs} is set TMVA creates signal and
background PDFs from the corresponding MVA response
distributions using the training sample (\cf\
Sec.~\ref{sec:usingtmva:training}). The binning and smoothing
properties of the underlying histograms can be customised via controls
implemented in the \code{PDF} class (\cf\ Sec.~\ref{sec:PDF} and Option 
Table~\ref{opt:pdf} on page~\pageref{opt:pdf}). The options specific to
\code{MethodBase} are listed in Option Table~\ref{opt:mva::methodbase}. 
They can be accessed by all MVA methods. 

The following sections describe the methods implemented in TMVA. For each method we 
proceed according to the following scheme: ($i$) a brief introduction, ($ii$) the description 
of the booking options required to configure the method, ($iii$) a description of the 
the method and TMVA implementation specifications for classification and -- where
available -- for regression, ($iv$) the properties of the 
variable ranking, and ($v$) a few comments on performance, favourable (and 
disfavoured) use cases, and comparisons with other methods.
% ======= input option table ==========================================
\begin{option}[!t]
\input optiontables/MVA__MethodBase.tex
\caption[.]{\optionCaptionSize 
   Configuration options that are common for all classifiers (but which can be controlled individually
   for each classifier). Values given are defaults. If predefined categories exist, the default category 
   is marked by a ’*’. The lower options in the table control the PDF fitting of the classifiers (required,
   \eg, for the Rarity calculation).
}
\label{opt:mva::methodbase}
\end{option}
% =====================================================================

%% file: optiontables/MVA__MethodBase.tex
\begin{optiontableAuto}
                        V  &  \mc{1}{c}{--}  &            False  &  \mc{1}{l}{--}  &  Verbose output (short form of VerbosityLevel below - overrides the latter one) \\
           VerbosityLevel  &  \mc{1}{c}{--}  &          Default  &  Default, Debug, Verbose, Info, Warning, Error, Fatal  &  Verbosity level \\
             VarTransform  &  \mc{1}{c}{--}  &             None  &  \mc{1}{l}{--}  &  List of variable transformations performed before training, e.g., D\_Background,P\_Signal,G,N\_AllClasses for: Decorrelation, PCA-transformation, Gaussianisation, Normalisation, each for the given class of events ('AllClasses' denotes all events of all classes, if no class indication is given, 'All' is assumed) \\
                        H  &  \mc{1}{c}{--}  &            False  &  \mc{1}{l}{--}  &  Print method-specific help message \\
            CreateMVAPdfs  &  \mc{1}{c}{--}  &            False  &  \mc{1}{l}{--}  &  Create PDFs for classifier outputs (signal and background) \\
IgnoreNegWeightsInTraining  &  \mc{1}{c}{--}  &            False  &  \mc{1}{l}{--}  &  Events with negative weights are ignored in the training (but are included for testing and performance evaluation) 
\end{optiontableAuto}

%% file: Cuts.tex
\subsection{Rectangular cut optimisation\index{Rectangular cuts}\index{Rectangular cut optimisation}\index{Cut optimisation}}
\label{sec:cuts}

The simplest and most common classifier for selecting signal events from
a mixed sample of signal and background events is the application of an ensemble
of rectangular cuts\index{Cuts} on discriminating variables. Unlike 
all other classifiers in TMVA, the cut classifier only returns a binary 
response (signal {\em or} background)\index{Binary split}.\footnote
{
   Note that cut optimisation is not a {\em multivariate} analyser 
   method but a sequence of univariate ones, because no combination of 
   the variables is achieved. Neither
   does a cut on one variable depend on the value of another variable
   (like it is the case for Decision Trees), nor can a, say, background-like
   value of one variable in a signal event be counterweighed by signal-like
   values of the other variables (like it is the case for the likelihood method).
}
The optimisation of cuts performed by TMVA maximises the background rejection
at given signal efficiency, and scans over the full range of the 
latter quantity. Dedicated analysis optimisation for which, \eg, the signal 
{\em significance} is maximised requires the expected signal and background
yields to be known before applying the cuts. This is not the case for a 
multi-purpose discrimination and hence not used by TMVA.
However, the cut ensemble leading to maximum significance corresponds to 
a particular working point on the efficiency curve, and can hence be easily
derived after the cut optimisation scan has converged.\footnote
{\label{ftn:cutcomp}
  Assuming a large enough number of events so that Gaussian statistics is applicable, the
  significance for a signal is given by $\Signif=\eS \NS/\sqrt{\eS \NS + \eB(\eS) \NS}$,
  where $\e_{\!S(B)}$ and $N_{\!S(B)}$ are the signal and background efficiencies for 
  a cut ensemble and the event yields before applying the cuts, respectively. 
  The background efficiency $\eB$
  is expressed as a function of $\eS$ using the TMVA evaluation curve obtained form 
  the test data sample. The maximum significance is then found at the root of the 
  derivative
  \beq
     \frac{d\Signif}{d\eS} 
       = \NS\frac{2\eB(\eS)\NB + \eS\left(\NS-\frac{d\eB(\eS)}{d\eS}\NB\right)}
                 {2\left(\eS \NS + \eB(\eS)\NB\right)^{3/2}} = 0\,,
  \eeq
  which depends on the problem.
}

TMVA cut optimisation is performed with the use of multivariate parameter fitters 
interfaced by the class \code{FitterBase} (\cf\  Sec.~\ref{sec:fitting}). Any 
fitter implementation can be used, where however because of the peculiar, non-unique 
solution space only Monte Carlo sampling, Genetic Algorithm, and Simulated Annealing 
show satisfying results. Attempts to use Minuit (SIMPLEX or MIGRAD) have not shown 
satisfactory results\index{Minuit!cut optimisation}, with frequently failing fits.

The training events are sorted in {\em binary trees}\index{Binary search trees} 
prior to the optimisation, which significantly reduces the computing time required 
to determine the number of events passing a given cut ensemble (\cf\  
Sec.~\ref{sec:binaryTrees}).

\subsubsection{Booking options}

The rectangular cut optimisation is booked through the Factory via the command:
\begin{codeexample}
\begin{tmvacode}
factory->BookMethod( Types::kCuts, "Cuts", "<options>" );
\end{tmvacode}
\caption[.]{\codeexampleCaptionSize Booking of the cut optimisation classifier: the 
         first argument is a predefined enumerator, the second argument is a 
         user-defined string identifier, and the third argument is the configuration 
         options string. Individual options are separated by a ':'. 
         See Sec.~\ref{sec:usingtmva:booking} for more information on the booking.}
\end{codeexample}

The configuration options for the various cut optimisation techniques 
are given in Option Table~\ref{opt:mva::cuts}.

% ======= input option table ==========================================
\begin{option}[t]
\input optiontables/MVA__Cuts.tex
\caption[.]{\optionCaptionSize 
     Configuration options reference for MVA method: {\em Cuts}.
     Values given are defaults. If predefined categories exist, the default category is marked by a '$\star$'. 
     The options in Option Table~\ref{opt:mva::methodbase} on page~\pageref{opt:mva::methodbase} can also be configured.     
}
\label{opt:mva::cuts}
\end{option}
% =====================================================================

\subsubsection{Description and implementation}

The cut optimisation analysis proceeds by first building binary search trees for 
signal and background. For each variable, statistical properties like mean, 
root-mean-squared (RMS), variable ranges are computed to guide the search 
for optimal cuts. Cut optimisation requires an estimator that quantifies the goodness
of a given cut ensemble. Maximising this estimator minimises (maximises)
the background efficiency, $\eB$ (background rejection, $\rB=1-\eB$) for 
each signal efficiency $\eS$. 

All optimisation methods (fitters) act on the assumption
that one minimum and one maximum requirement on each variable is sufficient 
to optimally discriminate signal from background (\ie, the signal is clustered).
If this is not the case, the variables must be transformed prior to the cut 
optimisation to make them compliant with this assumption.

For a given cut ensemble the signal and background efficiencies 
are derived by counting the training events that pass the cuts and dividing
the numbers found by the original sample sizes.
The resulting efficiencies are therefore rational numbers that may exhibit
visible discontinuities when the number of training events is small and an
efficiency is either very small or very large. Another way to compute efficiencies
is to parameterise the probability density functions of all input variables
and thus to achieve continuous efficiencies for any cut value. Note however
that this method expects the input variables to be uncorrelated! Non-vanishing
correlations would lead to incorrect efficiency estimates and hence to underperforming
cuts. The two methods are chosen with the option \code{EffMethod} set to \code{EffSel}
and \code{EffPDF}, respectively.

\subsubsection*{Monte Carlo sampling\index{Monte Carlo sampling}}

Each generated cut sample (\cf\  Sec.~\ref{sec:MCsampling}) corresponds to a 
point in the $(\eS,\rB)$ plane. The $\eS$ dimension is (finely) binned
and a cut sample is retained if its $\rB$ value is larger than the value already
contained in that bin. This way a reasonably smooth efficiency curve can be obtained
if the number of input variables is not too large (the required number of MC samples
grows with powers of $2\Nvar$). 

Prior information on the variable distributions can be used to reduce the 
number of cuts that need to be sampled. For example, if a discriminating variable 
follows Gaussian distributions for signal and background, with equal width but 
a larger mean value for the background distribution, there is no useful minimum
requirement (other than $-\infty$) so that a single maximum requirement is 
sufficient for this variable. To instruct TMVA to remove obsolete requirements,
the option \code{VarProp[i]} must be used, where \code{[i]} indicates the counter 
of the variable (following the order in which they have been registered with the 
Factory, beginning with 0) must be set to either \code{FMax} or \code{FMin}.
TMVA is capable of automatically detecting which of the requirements should be 
removed. Use the option \code{VarProp[i]=FSmart} (where again \code{[i]} must
be replaced by the appropriate variable counter, beginning with 0). Note that in 
many realistic use cases the mean values between signal and background of a 
variable are indeed distinct, but the background can have large tails. In such a 
case, the removal of a requirement is inappropriate, and would lead to 
underperforming cuts.

\subsubsection*{Genetic Algorithm\index{Genetic Algorithm}}

Genetic Algorithm (\cf\  Sec.~\ref{sec:geneticAlgorithm}) is a technique to 
find approximate solutions to optimisation or search problems. Apart from the 
abstract representation of the solution domain, 
a {\em fitness}\index{Fitness! for cut optimisation} 
function must be defined. In cut optimisation, the fitness of a rectangular cut 
is given by good background rejection combined with high signal efficiency. 

At the initialization step, all parameters of all individuals (cut ensembles) 
are chosen randomly. The individuals are evaluated in terms of their background 
rejection and signal efficiency. Each cut ensemble giving an improvement 
in the background rejection for a specific signal efficiency bin is immediately 
stored. Each individual's fitness is assessed, where the fitness is 
largely determined by the difference of the best found background rejection 
for a particular bin of signal efficiency and the value produced by the 
current individual. The same individual that has at one generation a very 
good fitness will have only average fitness at the following generation. This 
forces the algorithm to focus on the region where the potential of improvement 
is the highest. 
Individuals with a good fitness are selected to produce the next generation. 
The new individuals are created by crossover and mutated afterwards. Mutation 
changes some values of some parameters of some individuals randomly following
a Gaussian distribution function, etc. This process can be controlled with the 
parameters listed in Option Table~\ref{opt:fitter_ga}, page~\pageref{opt:fitter_ga}.

\subsubsection*{Simulated Annealing\index{Simulated Annealing}}

Cut optimisation using Simulated Annealing works similarly as for the Genetic 
Algorithm and achieves comparable performance. In particular, the same fitness 
function is used to estimator the goodness of a given cut ensemble. Details 
on the algorithm and the configuration options can be found in 
Sec.~\ref{sec:simAnnealing} on page~\pageref{sec:simAnnealing}.

\subsubsection{Variable ranking}

The present implementation of Cuts does not provide a ranking of 
the input variables.

\subsubsection{Performance}

The Genetic Algorithm currently provides the best cut optimisation convergence. 
However, it is found that with rising number of discriminating input variables the 
goodness of the solution found (and hence the smoothness of the background-rejections 
versus signal efficiency plot) deteriorates quickly. Rectangular cut optimisation
should therefore be reduced to the variables that have the largest discriminating
power. 

If variables with excellent signal from background separation exist, applying cuts
can be quite competitive with more involved classifiers. Cuts are known to underperform
in presence of strong nonlinear correlations and/or if several weakly discriminating 
variables are used. In the latter case, a true multivariate combination of the information
will be rewarding.

%% file: optiontables/MVA__Cuts.tex
\begin{optiontableAuto}
                FitMethod  &  \mc{1}{c}{--}  &               GA  &  GA, SA, MC, MCEvents, MINUIT, EventScan  &  Minimisation Method (GA, SA, and MC are the primary methods to be used; the others have been introduced for testing purposes and are depreciated) \\
                EffMethod  &  \mc{1}{c}{--}  &           EffSel  &  EffSel, EffPDF  &  Selection Method \\
              CutRangeMin  &  Yes  &               -1  &  \mc{1}{l}{--}  &  Minimum of allowed cut range (set per variable) \\
              CutRangeMax  &  Yes  &               -1  &  \mc{1}{l}{--}  &  Maximum of allowed cut range (set per variable) \\
                  VarProp  &  Yes  &      NotEnforced  &  NotEnforced, FMax, FMin, FSmart, FVerySmart  &  Categorisation of cuts 
\end{optiontableAuto}

%% file: Likelihood.tex
\subsection{Projective likelihood estimator (PDE approach)}
\label{sec:likelihood}

The method of maximum likelihood\index{Likelihood} consists of building a 
model out of probability density functions (PDF)\index{PDF}\index{Probability Density Function} that reproduces the input 
variables for signal and background. For a given event, the likelihood for 
being of signal type is obtained by multiplying the signal probability densities 
of all input variables, which are assumed to be independent, and normalising this 
by the sum of the signal and background likelihoods. Because correlations among 
the variables are ignored, this PDE approach is also called ``naive Bayes estimator'', 
unlike the full multidimensional PDE approaches such as PDE-range search, 
PDE-foam and k-nearest-neighbour discussed in the subsequent sections, which 
approximate the true Bayes limit. 

\subsubsection{Booking options}

The likelihood classifier is booked via the command:
\begin{codeexample}
\begin{tmvacode}
factory->BookMethod( Types::kLikelihood, "Likelihood", "<options>" );
\end{tmvacode}
\caption[.]{\codeexampleCaptionSize Booking of the (projective) likelihood classifier: 
         the first argument is the predefined enumerator, the 
         second argument is a user-defined string identifier, and the third argument 
         is the configuration options string. Individual options are separated by a ':'. 
         See Sec.~\ref{sec:usingtmva:booking} for more information on the booking.}
\end{codeexample}

The likelihood configuration options are given in Option 
Table~\ref{opt:mva::likelihood}.

% ======= input option table ==========================================
\begin{option}[!t]
\input optiontables/MVA__Likelihood.tex
\caption[.]{\optionCaptionSize 
     Configuration options reference for MVA method: {\em Likelihood}.
     Values given are defaults. If predefined categories exist, the default category 
     is marked by a '$\star$'. The options in Option Table~\ref{opt:mva::methodbase} on 
     page~\pageref{opt:mva::methodbase} can also be configured.
     
}
\label{opt:mva::likelihood}
\end{option}
% =====================================================================

\subsubsection{Description and implementation}
\label{sec:likelihood:description}

The likelihood\index{Likelihood} ratio $\RLik(i)$ for event $i$ is defined by
\beq
\label{eq:RLik}
	\RLik(i) = \frac{\Lik_S(i)}{\Lik_S(i) + \Lik_B(i)}\,,
\eeq
where 
\beq
\label{eq:Likelihood}
	\Lik_{S(B)}(i) = \prod_{k=1}^\Nvar p_{S(B),k}(x_k(i))\,,
\eeq
and where $p_{S(B),k}$ is the signal (background) PDF for the $k$th
input variable $x_k$. The PDFs are normalised 
\beq
\label{eq:pdfNorm}
	\intl_{-\infty}^{+\infty}p_{S(B),k}(x_k) dx_k = 1\,, \hspace{0.5cm}\forall k.
\eeq
It can be shown that in absence of model inaccuracies (such as correlations
between input variables not removed by the de-correlation procedure, or an inaccurate
probability density model), the ratio~(\ref{eq:RLik}) provides optimal signal from 
background separation for the given set of input variables. 

Since the parametric form of the PDFs is generally unknown, the PDF shapes are empirically
approximated from the training data by nonparametric functions, which can be 
chosen individually for each variable and are either  polynomial 
splines of various degrees fitted to histograms or unbinned kernel density estimators (KDE),
as discussed in Sec.~(\ref{sec:PDF}). 

A certain number of primary validations are performed during the PDF
creation, the results of which are printed to standard output.
Among these are the computation
of a $\chi^2$ estimator between all nonzero bins of the original 
histogram and its PDF, and a comparison of the number of outliers 
(in sigmas) found in the original histogram with respect to the 
(smoothed) PDF shape, with the statistically expected one. The 
fidelity of the PDF estimate can be also inspected visually by executing 
the macro \code{likelihoodrefs.C} (\cf\  Table~\ref{pgr:scripttable2}).

\subsubsection*{Transforming the likelihood output\index{Likelihood output transformation}}

\begin{figure}[t]
  \begin{center}
	  \includegraphics[width=0.53\textwidth]{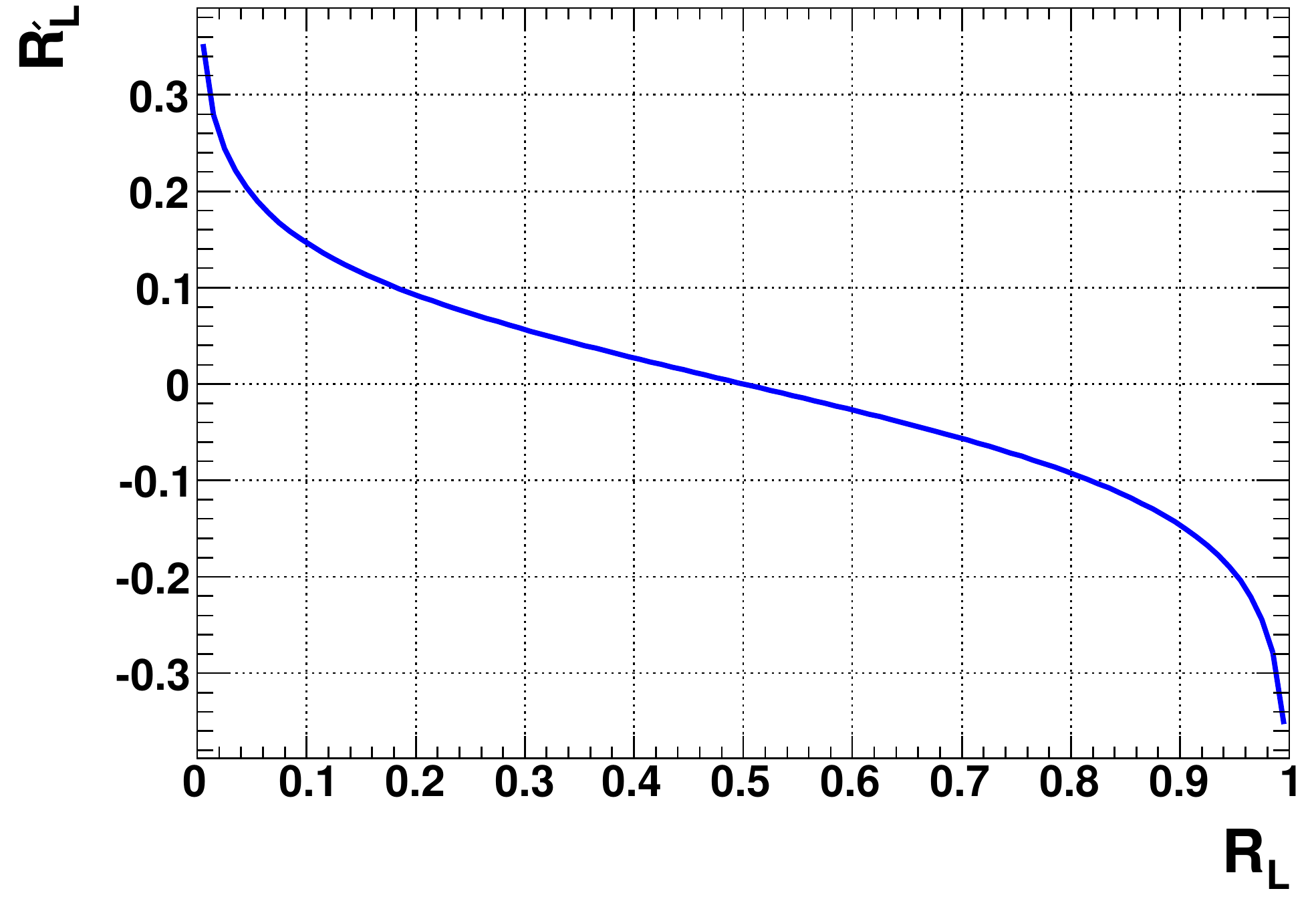}
  \end{center}
  \vspace{-0.5cm}
  \caption[.]{Transformation~(\ref{eq:RLikTransformed}) of the likelihood output. }
\label{fig:liktransform}
\end{figure}

If a data-mining problem offers a large number of input variables, or 
variables with excellent separation power, the likelihood response $\RLik$ 
is often strongly peaked at 0 (background) and 1 (signal). Such a response
is inconvenient for the use in subsequent analysis steps. TMVA therefore 
allows to transform the likelihood output by an inverse sigmoid function
that zooms into the peaks
\beq
\label{eq:RLikTransformed}
		\RLik(i) \longrightarrow \RLik^\prime(i) = 
		-\tau^{-1}\ln\!\left(\RLik^{-1}-1\right)\,,
\eeq
where $\tau=15$ is used. Note that $\RLik^\prime(i)$ is no 
longer contained within $[0,1]$ (see Fig.~\ref{fig:liktransform}).
The transformation~(\ref{eq:RLikTransformed}) is enabled (disabled) with 
the booking option \code{TransformOutput=True(False)}. 

\subsubsection{Variable ranking}

The present likelihood implementation does not provide a ranking of 
the input variables.

\subsubsection{Performance}

Both the training and the application of the likelihood 
classifier are very fast operations that are suitable for large data sets. 

The performance of the classifier relies on the accuracy of the likelihood model.
Because high fidelity PDF estimates are mandatory, sufficient training statistics 
is required to populate the tails of the distributions.
The neglect of correlations between input variables in the model~(\ref{eq:Likelihood}), 
often leads to a diminution of the discrimination performance. While linear 
Gaussian correlations can be rotated away (see Sec.~\ref{sec:variableTransform}),
such an ideal situation is rarely given. Positive correlations lead to 
peaks at both $\RLik\to 0,1$. Correlations can be reduced by categorising 
the data samples and building an independent likelihood classifier for each 
event category. Such categories could be geometrical regions in the detector,
kinematic properties, etc. In spite of this, realistic applications with a 
large number of input variables are often plagued by irreducible correlations, 
so that projective likelihood approaches like the one discussed here are 
under-performing. This finding led to the development of the many 
alternative classifiers that exist in statistical theory today.

%% file: optiontables/MVA__Likelihood.tex
\begin{optiontableAuto}
          TransformOutput  &  \mc{1}{c}{--}  &            False  &  \mc{1}{l}{--}  &  Transform likelihood output by inverse sigmoid function 
\end{optiontableAuto}

%% file: PDERS.tex
\subsection{Multidimensional likelihood estimator (PDE range-search approach)}
\label{sec:pders}

This is a generalization of the projective likelihood classifier described in
Sec.~\ref{sec:likelihood} to $\Nvar$ dimensions, where $\Nvar$ is
the number of input variables used. If the multidimensional 
PDF\index{PDE-RS, multidimensional likelihood} 
for signal and background (or regression data) were known, this classifier 
would exploit the full information contained in the input variables, and  
would hence be optimal. In practice however, huge training samples are necessary 
to sufficiently populate the multidimensional phase space.\footnote
{
	Due to correlations between the input variables, only a sub-space of 
	the full phase space may be populated.
} 
Kernel estimation methods may be used to approximate the shape of the PDF
for finite training statistics.

A simple probability density estimator\index{PDE} denoted {\em PDE range search}, or {\em PDE-RS}, 
has been suggested in Ref.~\cite{CarliKoblitz}. The PDE for a given test event (discriminant) 
is obtained by counting the (normalised) number of 
training events that occur in the "vicinity" of the test event. The 
classification of the test event may then be conducted on the basis of the 
majority of the nearest training events. The $\Nvar$-dimensional
volume that encloses the "vicinity" is user-defined and can be adaptive. A search method 
based on sorted binary trees is used to reduce the computing time for the 
range search. To enhance the sensitivity within the volume, kernel 
functions are used to weight the reference events according to their 
distance from the test event. PDE-RS is a variant of the k-nearest neighbour
classifier described in Sec.~\ref{sec:knn}.

\subsubsection{Booking options}

The PDE-RS classifier is booked via the command:
\begin{codeexample}
\begin{tmvacode}
factory->BookMethod( Types::kPDERS, "PDERS", "<options>" );
\end{tmvacode}
\caption[.]{\codeexampleCaptionSize Booking of PDE-RS: the first argument is a predefined 
		   enumerator, the second argument is a user-defined 
		   string identifier, and the third argument is the configuration options string.
         Individual options are separated by a ':'. 
         See Sec.~\ref{sec:usingtmva:booking} for more information on the booking.}
\end{codeexample}

The configuration options for the PDE-RS classifier are given in 
Option Table~\ref{opt:mva::pders}.

% ======= input option table ==========================================
\begin{option}[t]
\input optiontables/MVA__PDERS.tex
\caption[.]{\optionCaptionSize 
     Configuration options reference for MVA method: {\em PDE-RS}.
     Values given are defaults. If predefined categories exist, the default category 
     is marked by a '$\star$'. The options in Option Table~\ref{opt:mva::methodbase} on 
     page~\pageref{opt:mva::methodbase} can also be configured.     
}
\label{opt:mva::pders}
\end{option}
% =====================================================================
\subsubsection{Description and implementation}

\subsubsection*{Classification}

To classify an event as being either of signal or of background type, a {\em local}
estimate of the probability density of it belonging to either class is 
computed. The method of PDE-RS provides such an estimate by defining a 
volume ($V$) around the test event ($i$), and by counting the number of signal 
($n_S(i,V)$) and background events ($n_B(i,V)$) obtained from the training 
sample in that volume. The ratio 
\beq
\label{eq:PDERSratio}
	\RPDERS(i,V) = \frac{1}{1 + r(i,V)}\,,
\eeq
is taken as the estimate, where $r(i,V)=(n_B(i,V)/N_B)\cdot(N_S/n_S(i,V))$, 
and $N_{S(B)}$ is the total number of signal (background) events in the 
training sample. The estimator $\RPDERS(i,V)$ peaks at 1 (0) for signal 
(background) events. The counting method averages over the PDF within $V$, 
and hence ignores the available shape information inside (and outside) that 
volume.

\subsubsection*{Binary tree search}

Efficiently searching for and counting the events that lie inside the
volume is accomplished with the use of a $\Nvar$-variable binary tree 
search algorithm~\cite{BinaryTree} (\cf\  
Sec.~\ref{sec:binaryTrees}).

\subsubsection*{Choosing a volume}

The TMVA implementation of PDE-RS optionally provides four different 
volume definitions selected via the configuration option \code{VolumeRangeMode}.
\begin{itemize}

\item	\code{Unscaled} \\
		The simplest volume definition consisting of a rigid box of size \code{DeltaFrac},
      in units of the variables.
		This method was the one originally used by the developers of 	
		PDE-RS~\cite{CarliKoblitz}.

\item \code{MinMax} \\ 
		The volume is defined in each dimension (\ie, input variable) 
      with respect to the full range of values found for that dimension
		in the training sample. The fraction of this volume used 
		for the range search is defined by the option \code{DeltaFrac}.

\item \code{RMS} \\
		The volume is defined in each dimension with respect 
		to the RMS of that dimension (input variable), estimated from the 
		training sample. The fraction of this volume used 
		for the range search is defined by the option \code{DeltaFrac}.

\item	\code{Adaptive} \\ 
		A volume is defined in each dimension
		with respect to the RMS of that dimension, estimated from the 
		training sample. The overall scale of the volume is  
		adjusted individually for each test event such 
		that the total number of events confined in the volume lies within 
		a user-defined range (options \code{NEventsMin/Max}). The adjustment
		is performed by the class \code{RootFinder}, which is a C++ implementation
		of Brent's algorithm (translated from the CERNLIB function RZERO).
		The maximum initial volume (fraction of the RMS) and the maximum
		number of iterations for the root finding is set by the options 
		\code{InitialScale} and \code{MaxVIterations}, respectively. The
		requirement to collect a certain number of events in the volume 
		automatically leads to small volume sizes in strongly populated 
		phase space regions, and enlarged volumes in areas where the
		population is scarce.			

\end{itemize}
Although the adaptive volume adjustment is more flexible and should perform
better, it significantly increases the computing time of the PDE-RS discriminant.
If found too slow, one can reduce the number of necessary iterations by 
choosing a larger \code{NEventsMin/Max} interval.

\subsubsection*{Event weighting with kernel functions\index{Kernel estimation}}

One of the shortcomings of the original PDE-RS implementation is its 
sensitivity to the exact location of the sampling volume boundaries: an
infinitesimal change in the boundary placement can include or exclude a 
training event, thus changing $r(i,V)$ by a finite amount.\footnote
{
	Such an introduction of artefacts by having sharp boundaries in the 
	sampled space is an example of Gibbs's phenomenon, and is commonly 
   referred to as {\em ringing} or {\em aliasing}. 
} 
In addition, the shape information within the volume is ignored.

\begin{figure}[t]
\begin{center}
   \includegraphics[width=0.49\textwidth]{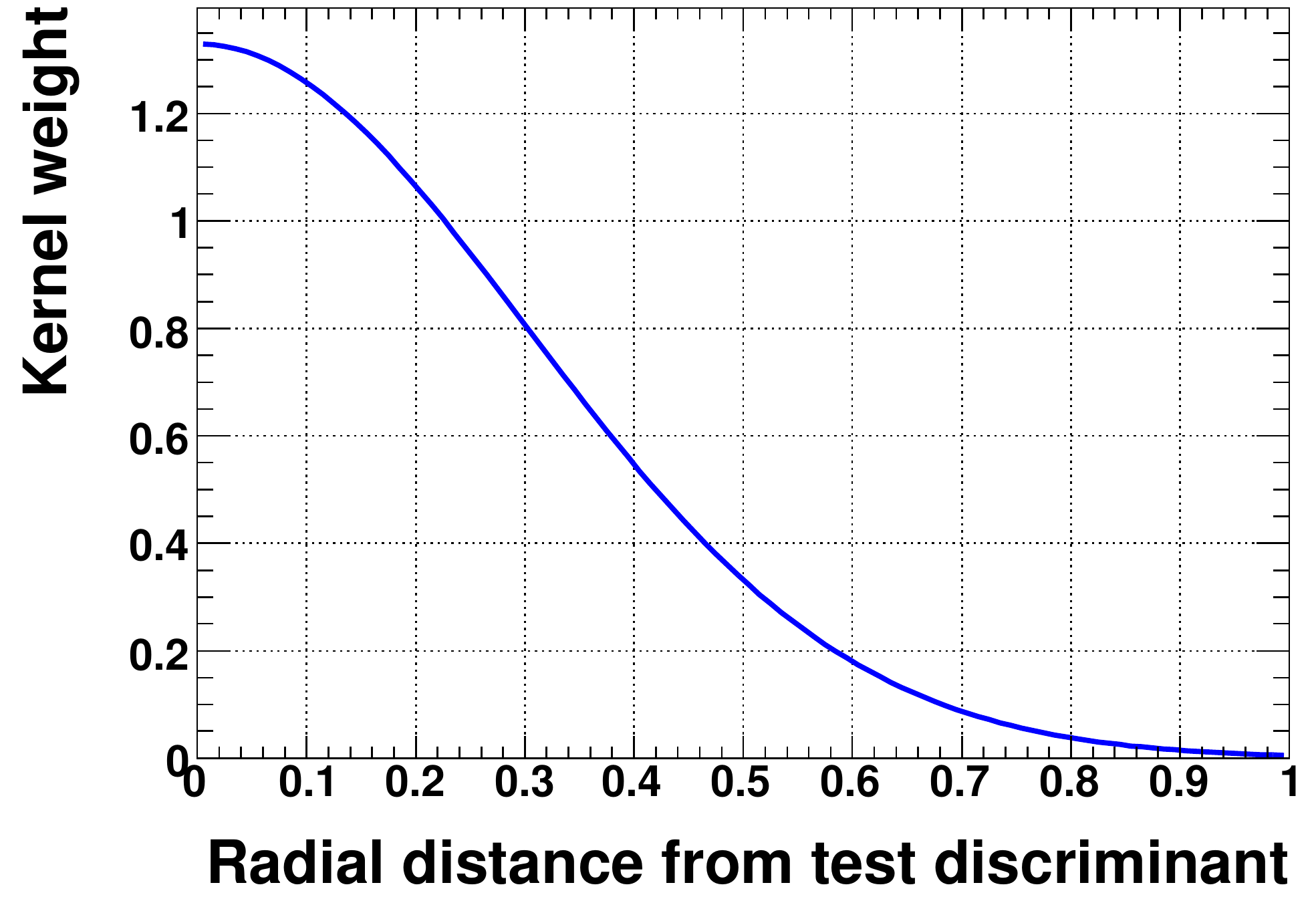}
   \includegraphics[width=0.49\textwidth]{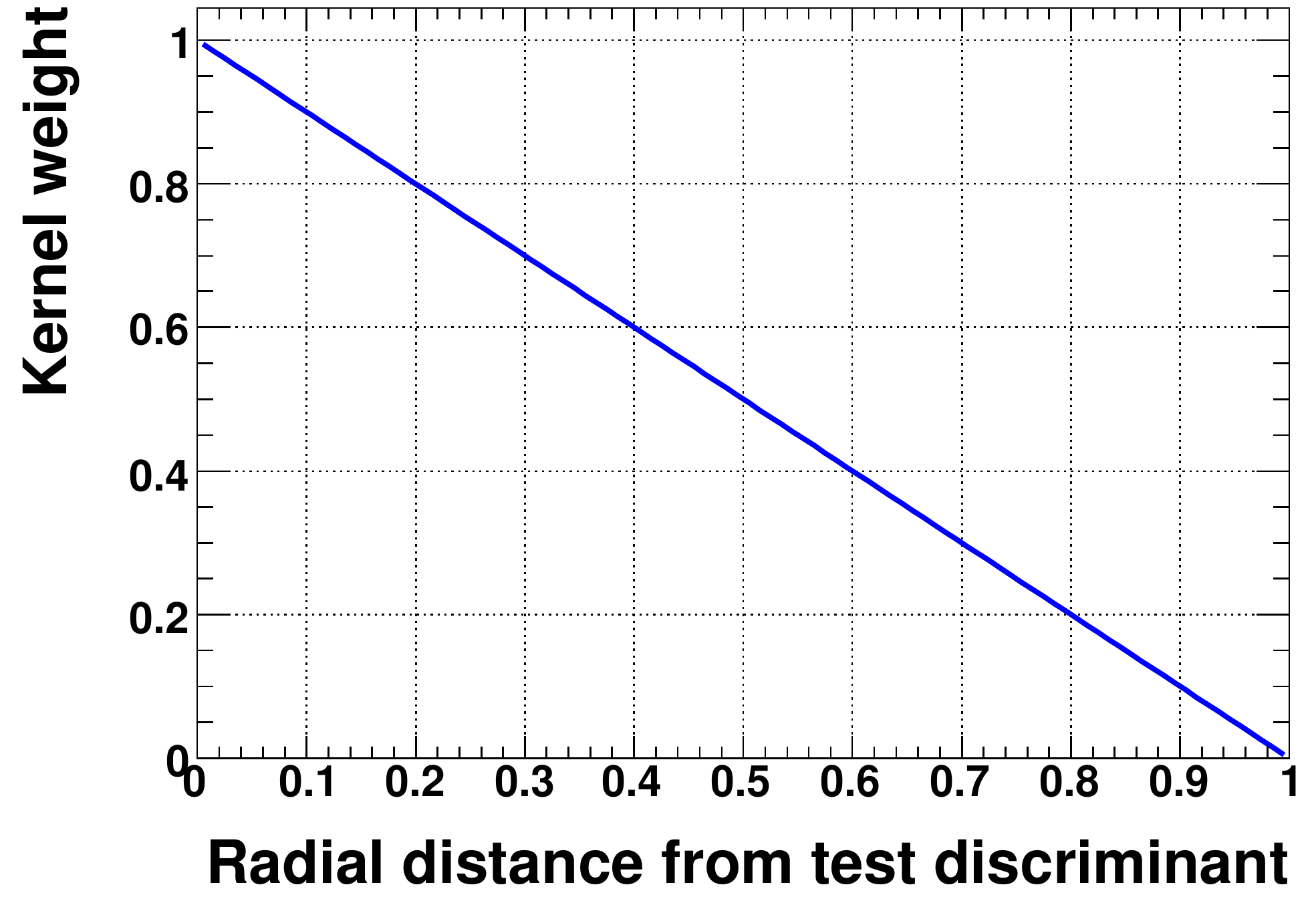}
\end{center}
\vspace{-0.5cm}
\caption[.]{Kernel functions (left: Gaussian, right: Teepee) 
 		   used to weight the events that are found 
	      inside the reference volume of a test event.}
\label{fig:pdersKernels}
\end{figure}

Kernel functions mitigate these problems by weighting each event within
the volume as a function of its distance to the test event. The 
farer it is away, the smaller is its weight. The following kernel functions are
implemented in TMVA, and can be selected with the option \code{KernelEstimator}.
\begin{itemize}

\item	\code{Box} \\
		Corresponds to the original rectangular volume element 
		without application of event weights.

\item \code{Sphere} \\
		A hyper-elliptic volume element is used
		without application of event weights. The hyper-ellipsoid corresponds 
		to a sphere of constant fraction in the \code{MinMax} or \code{RMS}
		metrics. The size of the sphere can be chosen adaptive, just as 
		for the rectangular volume.

\item	\code{Teepee} \\
		The simplest linear interpolation that eliminates the discontinuity 
      problem of the box. The training events are 
      given a weight that decreases linearly with their distance
      from the centre of the volume (the position of the test event).
		In other words: these events are convolved with the triangle 
      or tent function, becoming a sort of teepee in multi-dimensions.

\item \code{Trim} \\ 
      Modified Teepee given by the function $(1-\hat x^3)^3$, where $\hat x$
      is the the normalised distance between test event and reference. 

\item \code{Gauss} \\ 
		The simplest well behaved convolution kernel.
		The width of the Gaussian (fraction of the volume size) can be 
		set by the option \code{GaussSigma}.

\end{itemize}
Other kernels implemented for test purposes are ``Sinc'' and 
''Lanczos'' functions $\propto \sin x/x$ of different (symmetric) orders.
They exhibit strong peaks at zero and oscillating tails. The Gaussian and 
Teepee kernel functions are shown in Fig.~\ref{fig:pdersKernels}.

\subsubsection*{Regression}

Regression with PDE-RS proceeds  similar to classification. The difference 
lies in the replacement of Eq.~(\ref{eq:PDERSratio}) by the average target value 
of all events belonging to the volume $V$ defined by event $i$ (the {\em test} event)
\beq
\label{eq:PDERSregratio}
	\RPDERS(i,V) = \langle t(i,V)\rangle = 
                   \frac{\sum_{j \in V}{w_{j} t_{j} f(\operatorname{dis}(i,j))}} 
                       {\sum_{j \in V}{w_{j} f(\operatorname{dis}(i,j))}}\,,
\eeq
where the sum is over all training events in $V$, $w_j$ and $t_j$ are the weight and 
target value of event $j$ in $V$, $\operatorname{dis}(i,j)$ is a measure of the distance
between events $i$ and $j$, and $f(\dots)$ is a kernel function.

\subsubsection{Variable ranking}

The present implementation of PDE-RS does not provide a ranking of 
the input variables.

\subsubsection{Performance}

As opposed to many of the more sophisticated data-mining approaches, which 
tend to present the user with a ''black box'', PDE-RS is simple enough that 
the algorithm can be easily traced and tuned by hand. PDE-RS can yield competitive 
performance if the number of input variables is not too large and the statistics
of the training sample is ample. In particular, it naturally deals with complex
nonlinear variable correlations, the reproduction of which may, for example, require
involved neural network architectures.

PDE-RS is a slowly responding classifier. Only the training, \ie, the fabrication 
of the binary tree is fast, which is usually not the critical part. 
The necessity to store the entire binary tree in memory to avoid accessing
virtual memory limits the number of training events that can effectively be 
used to model the multidimensional PDF. This is not the case for the other 
classifiers implemented in TMVA (with some exception for Boosted Decision Trees).

%% file: optiontables/MVA__PDERS.tex
\begin{optiontableAuto}
          VolumeRangeMode  &  \mc{1}{c}{--}  &         Adaptive  &  Unscaled, MinMax, RMS, Adaptive, kNN  &  Method to determine volume size \\
          KernelEstimator  &  \mc{1}{c}{--}  &              Box  &  Box, Sphere, Teepee, Gauss, Sinc3, Sinc5, Sinc7, Sinc9, Sinc11, Lanczos2, Lanczos3, Lanczos5, Lanczos8, Trim  &  Kernel estimation function \\
                DeltaFrac  &  \mc{1}{c}{--}  &                3  &  \mc{1}{l}{--}  &  nEventsMin/Max for minmax and rms volume range \\
               NEventsMin  &  \mc{1}{c}{--}  &              100  &  \mc{1}{l}{--}  &  nEventsMin for adaptive volume range \\
               NEventsMax  &  \mc{1}{c}{--}  &              200  &  \mc{1}{l}{--}  &  nEventsMax for adaptive volume range \\
           MaxVIterations  &  \mc{1}{c}{--}  &              150  &  \mc{1}{l}{--}  &  MaxVIterations for adaptive volume range \\
             InitialScale  &  \mc{1}{c}{--}  &             0.99  &  \mc{1}{l}{--}  &  InitialScale for adaptive volume range \\
               GaussSigma  &  \mc{1}{c}{--}  &              0.1  &  \mc{1}{l}{--}  &  Width (wrt volume size) of Gaussian kernel estimator \\
                 NormTree  &  \mc{1}{c}{--}  &            False  &  \mc{1}{l}{--}  &  Normalize binary search tree 
\end{optiontableAuto}

%% file: PDEFoam.tex
\subsection{Likelihood estimator using self-adapting phase-space binning (PDE-Foam)}
\label{sec:pdefoam}

The PDE-Foam\index{PDE-Foam, multi-dimensional likelihood} 
method~\cite{pdefoam} is an extension of PDE-RS, which
divides the multi-dimensional phase space in a finite number of
hyper-rectangles (cells) of constant event density.  This ``foam'' of
cells is filled with averaged probability density information sampled
from the training data.  For a given number of cells, the
binning algorithm adjusts the size and position of the cells inside
the multi-dimensional phase space based on a binary split algorithm that
minimises the variance of the event density in the cell.  The binned
event density information of the final foam is stored in cells,
organised in a binary tree, to allow a fast and memory-efficient
storage and retrieval of the event density information necessary for
classification or regression.  The implementation of PDE-Foam is based on the
Monte-Carlo integration package TFoam~\cite{tfoam} included in ROOT.

In classification mode PDE-Foam forms bins of similar density of signal 
and background events or the ratio of signal to background. In regression 
mode the algorithm determines cells with small varying regression targets. 
In the following, we use the term density ($\rho$) for the event density in
case of classification or for the target variable density in case of
regression.

\subsubsection{Booking options}

The PDE-Foam classifier is booked via the command:
\begin{codeexample}
\begin{tmvacode}
factory->BookMethod( Types::kPDEFoam, "PDEFoam", "<options>" );
\end{tmvacode}
\caption[.]{\codeexampleCaptionSize Booking of PDE-Foam: the first 
         argument is a predefined enumerator, the second argument is a
	 user-defined string identifier, and the third argument is the
	 configuration options string.  Individual options are
	 separated by a ':'.  See Sec.~\ref{sec:usingtmva:booking} for
	 more information on the booking.}
\end{codeexample}
The configuration options for the PDE-Foam method are summarised in 
Option Table~\ref{opt:mva::pdefoam}.

% ======= input option table ==========================================
\begin{option}[t]
\input optiontables/MVA__PDEFoam.tex
\caption[.]{\optionCaptionSize Configuration options reference for MVA
  method: {\em PDE-Foam}.  The options in Option
  Table~\ref{opt:mva::methodbase} on
  page~\pageref{opt:mva::methodbase} can also be configured.  }
\label{opt:mva::pdefoam}
\end{option}
% =====================================================================

Table~\ref{tab:opt-combinations} gives an overview of supported
combinations of configuration options.

\begin{table}[th]
\centering
\renewcommand{\checkmark}{\YES}
\setlength{\tabcolsep}{0.0pc}
{\small
\begin{tabular*}{\textwidth}{@{\extracolsep{\fill}}lcccc} 
\hline
&&&&\\[\BD]
                 &    \multicolumn{2}{c}{Classification}     & \multicolumn{2}{c}{Regression} \\
  Option         & Separated foams & Single foam             & Mono target  & Multi target \\
&&&&\\[\BD]                                       
\hline                                                 
&&&&\\[\BD]                                       
  \code{SigBgSeparate}  & True            & False                   & --           & -- \\
  \code{MultiTargetRegression} & --       & --                      & False        & True \\
&&&&\\[\BD]                                       
\hline                                                 
&&&&\\[\BD]                                       
  \code{CutNmin}        & \checkmark      & \checkmark              & \checkmark   & \checkmark \\
%%  \code{CutRMSmin}      & \checkmark      & \checkmark              & \checkmark   & \checkmark \\
  \code{Kernel}         & \checkmark      & \checkmark              & \checkmark   & \checkmark \\
  \code{TargetSelection} & \NO             & \NO                      & \NO           & \checkmark \\
  \code{TailCut}        & \checkmark      & \checkmark              & \checkmark   & \checkmark \\
%%  \code{DiscrErrCut}    & \checkmark      & \checkmark              & \checkmark   & \checkmark \\\hline
&&&&\\[\BD]                                       
\hline                                                 
\end{tabular*}
}
\caption[.]{\captionfont Availability of options for the two classification and 
            two regression modes of PDE-Foam. Supported options are marked by a 
            '\checkmark', while disregarded ones are marked by a '\NO'.}
\label{tab:opt-combinations}
\end{table}

\subsubsection{Description and implementation of the foam algorithm}

Foams for an arbitrary event sample are formed as follows.

\begin{figure}[t]
\centering
   \subfigure[foam projection without kernel]{%
      \label{fig:pdefoam-foam-nokernel}
      \includegraphics[width=0.48\textwidth,clip]{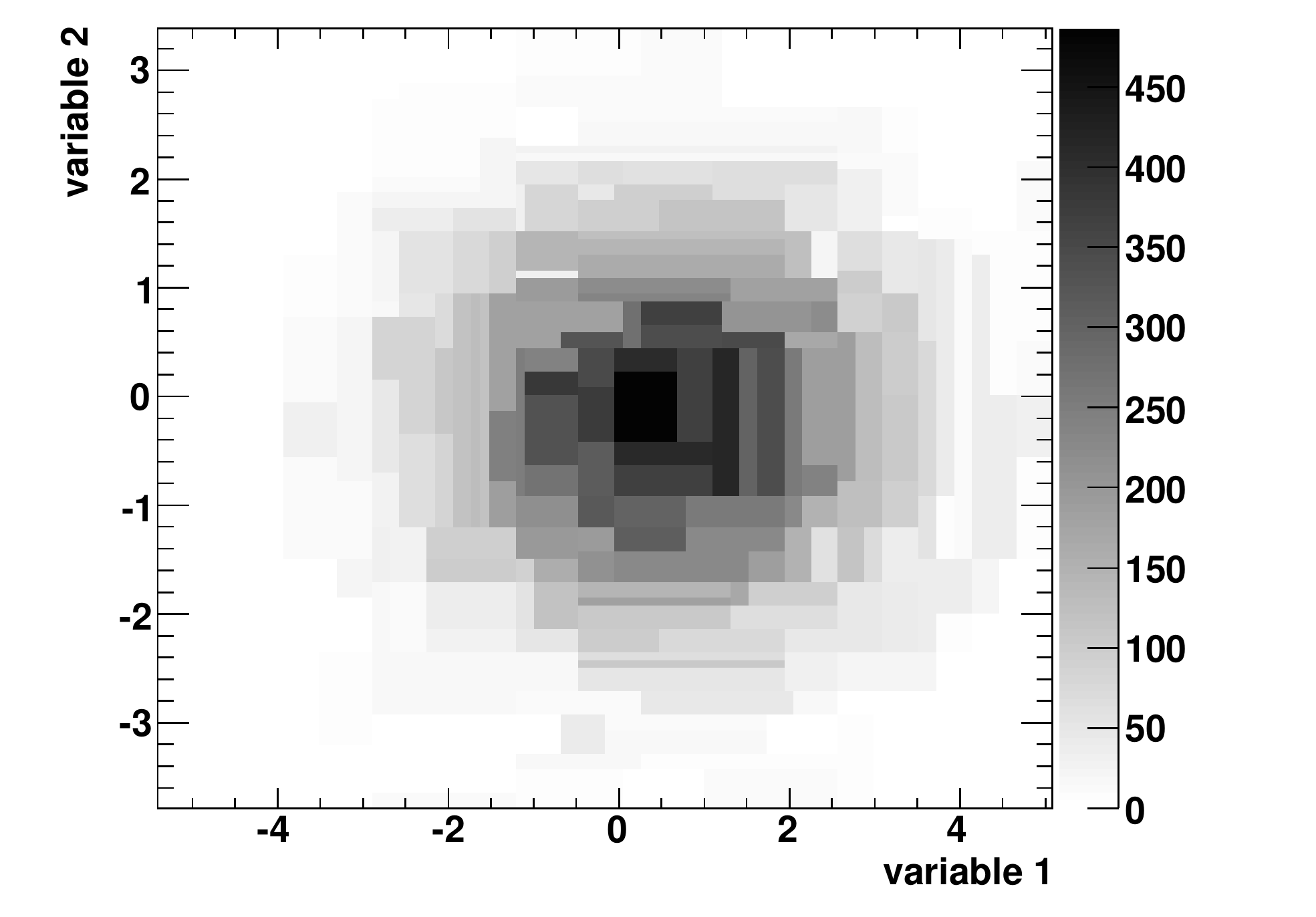}}
   \subfigure[foam projection with Gaussian kernel]{%
      \label{fig:pdefoam-foam-kernel}
      \includegraphics[width=0.48\textwidth,clip]{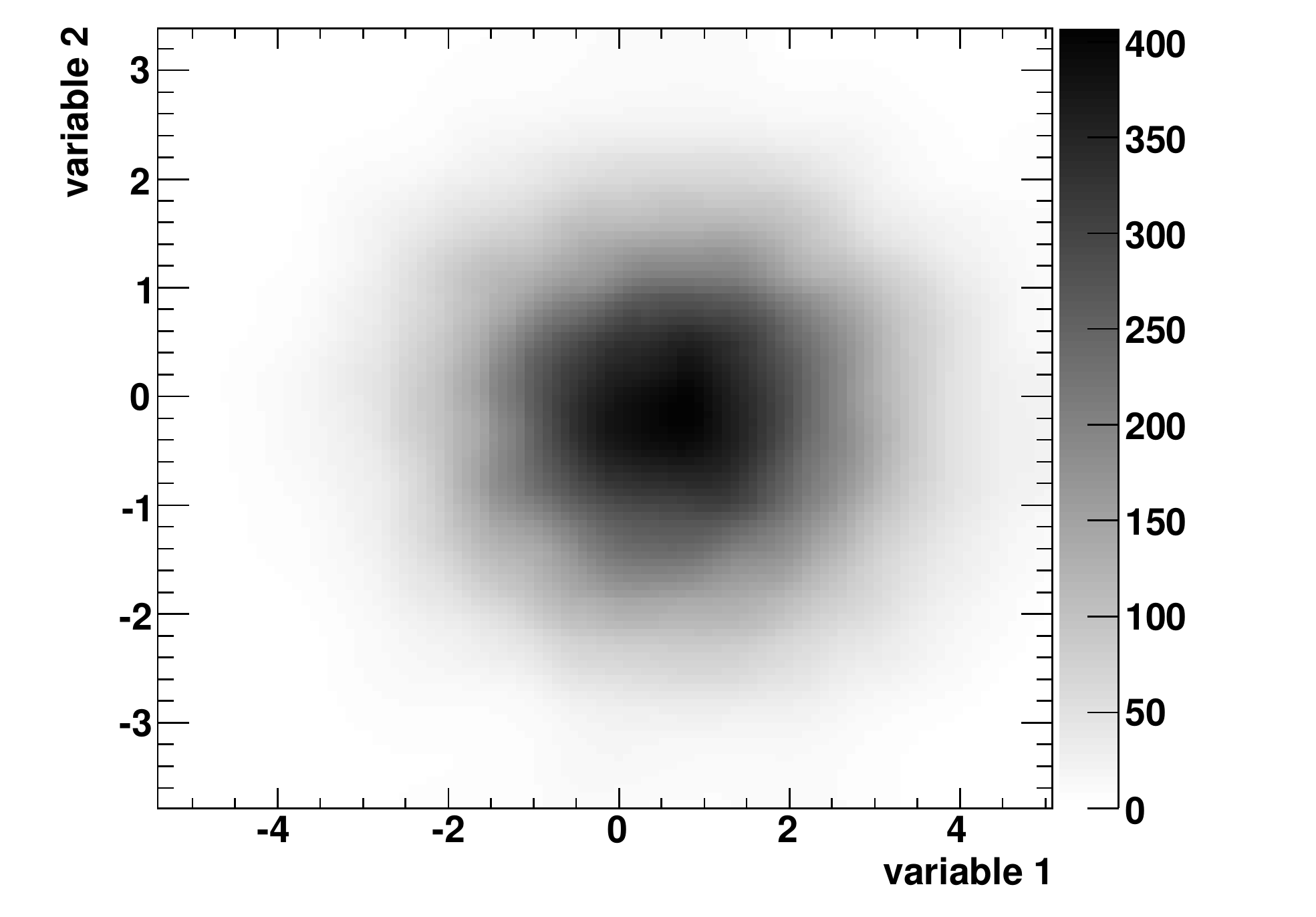}}
\caption[.]{Projections of a two-dimensional foam with 500 cells 
    for a Gaussian distribution on a two-dimensional histogram. The foam
    was created with 5000 events from the input tree. (a) shows the
    reconstructed distribution without kernel weighting and (b)
    shows the distribution weighted with a Gaussian kernel.  The grey shades
    indicate the event density of the drawn cell.  For more details
    about the projection function see the description on page
    \pageref{sec:pdefoam-visualise}.}
\label{fig:pdefoam-foam}
\end{figure}

\begin{enumerate}
  \item \emph{Setup of binary search trees}. A binary search tree is
    created and filled with the $d$-dimensional event tuples form the 
    training sample as for the PDE-RS method (\cf Sec.~\ref{sec:pders}).

  \item \emph{Initialisation phase}. A foam is created, which at first
    consists of one $d$-dimensional hyper-rectangle (base cell).  The
    coordinate system of the foam is normalised such that the base
    cell extends from 0 to 1 in each dimension. The coordinates of the
    events in the corresponding training tree are linearly transformed
    into the coordinate system of the foam.

  \item \emph{Growing phase}. A binary splitting algorithm iteratively
    splits cells of the foam along axis-parallel hyperplanes until the maximum
    number of active cells (set by \code{nActiveCells}) is reached.
    The splitting algorithm minimises the relative variance of the
    density $\sigma_{\rho}/\langle\rho\rangle$ across each cell (\cf
    Ref.~\cite{tfoam}).  For each cell \code{nSampl} random points
    uniformly distributed over the cell volume are generated. For each
    of these points a small box centred around this point is defined.
    The box has a size of \code{VolFrac} times the size of the base
    cell in each dimension.  The density is estimated as the number of
    events contained in this box divided by the volume of the box.\footnote
    {
       In case of regression this is the average target value computed 
       according to Eq.~(\ref{eq:PDERSregratio}), page~\pageref{eq:PDERSregratio}.
    }  
    The densities obtained for all sampled points in the cell are
    projected onto the $d$ axes of the cell and the projected values are
    filled in histograms with \code{nBin} bins.  The next cell to be
    split and the corresponding division edge (bin) for the split
    are selected as the ones that have the largest relative variance.
    The two new daughter cells are marked as `active' cells and the
    old mother cell is marked as `inactive'.  A detailed description
    of the splitting algorithm can be found in Ref.~\cite{tfoam}.  The
    geometry of the final foam reflects the distribution of the
    training samples: phase-space regions where the density is
    constant are combined in large cells, while in regions with large
    density gradients many small cells are created. 
    Figure~\ref{fig:pdefoam-foam-nokernel} displays a foam obtained 
    from a two-dimensional Gaussian-distributed training sample.

  \item \emph{Filling phase}. Each active cell is filled with values
    that classify the event distribution within this cell and which permit
    the calculation of the classification or regression discriminator. 

  \item \emph{Evaluation phase}. The estimator for a given event is
    evaluated based on the information stored in the foam cells. The
    corresponding foam cell, in which the event variables
    ($d$-dimensional vector) of a given event is contained, is determined
    with a binary search algorithm.\footnote
    {
       For an event that falls outside the foam boundaries, the cell with the smallest 
       Cartesian distance to the event is chosen.
    }
\end{enumerate}

The initial trees which contain the training events and which are needed to
evaluate the densities for the foam build-up, are discarded after the
training phase.  The memory consumption for the foam is $160$ bytes
per foam cell plus an overhead of $1.4$ kbytes for the PDE-Foam object
on a $64$ bit architecture.  Note that in the foam all cells created
during the growing phase are stored within a binary tree
structure. Cells which have been split are marked as inactive and
remain empty. To reduce memory consumption, the cell geometry is
not stored with the cell, but rather obtained recursively from the
information about the division edge of the corresponding mother
cell. This way only two short integer numbers per cell represent the
information of the entire foam geometry: the division coordinate
and the bin number of the division edge.

%%%%%%%%%%%%%%%%%%%%%%%%%%%%%%%%%%%%%%%%%%%%%%%%%%%%%%%%%%%%%%%
\subsubsection*{PDE-Foam options}

\begin{itemize}
\item \code{TailCut} -- {\em boundaries for foam geometry}\\
A parameter
\code{TailCut} is used to define the geometry of the base
cell(s) such that outliers of the distributions in the training ensemble
are not included in the volume of the base cell(s). In a first
step, the upper and lower limits for each input variable are determined
from the training sample.  Upper and a lower bounds are then
determined for each input variable, such that on both sides a fraction 
\code{TailCut} of all events is excluded.
The default value of \code{TailCut = 0.001} ensures a sufficient
suppression of outliers for most input distributions. For cases where
the input distributions have a fixed range or where they are
discontinuous and/or have peaks towards the edges, it can be 
appropriate to set \code{TailCut} to 0.
Note that for event classification it is guaranteed that 
the foam has an infinite coverage: events outside the foam volume 
are assigned to the cell with the smallest Cartesian distance to the event.

\item \code{nActiveCells} -- {\em maximum number of active foam cells }\\
In most cases larger \code{nActiveCells} values result in more accurate 
foams, but also lead to longer computation time during the foam formation, 
and require more storage memory. The default value of 500
was found to be a good compromise for most practical applications if
the size of the training samples is of the order of $10^4$ events.
Note that the total number of cells, \code{nCells}, is given
as $\texttt{nCells}=\texttt{nActiveCells}\cdot 2-1$,
since the binary-tree structure of the foam requires all inactive
cells to remain part of the foam (see \emph{Growing phase}).

\item \code{VolFrac} -- {\em size of the probe volume for the density sampling of the training data}\\
The volume is defined as a $d$-dimensional box with edge length
\code{VolFrac} times the extension of the base cell in each dimension. 
The default of $1/30$ results in a box with volume $1/30^{d}$ times
the volume of the base cell. Smaller values of \code{VolFrac}
increase the sampling speed, but also make the algorithm more
vulnerable to statistical fluctuations in the training sample (overtraining).
In cases with many observables ($>5$) and small training samples ($<10^4$),
\code{VolFrac} should be increased for better classification results. 

\item \code{nSampl} -- {\em number of samplings per cell and per cell-division step }\\
The computation time for the foam formation scales
linearly with the number of samplings. The default of 2000 was found
to give sufficiently accurate estimates of the density distributions
with an acceptable computation time.\footnote
{
   Contrary to the original application where an analytical function is 
   used, increasing the number of samplings does not automatically lead to 
   better accuracy. Since the density is defined by a limited event sample, 
   it may occur that always the same event is found for all sample points.
}.

\item \code{nBin} -- {\em number of bins for edge histograms }\\
The number of bins for the edge histograms used to determine the
variance of the sampled density distributions within one cell are set
with the parameter \code{nBin}. The performance in typical
applications was found to be rather insensitive to the number of bins.
The default value of 5 gives the foam algorithm sufficient flexibility
to find the division points, while maintaining sufficient statistical
accuracy also in case of small event samples.

\item \code{Nmin} -- {\em Minimal number of events for cell split}\\
If the option \code{CutNmin = T} is set, the foam will only consider
cells with a number of events greater than \code{Nmin} to be split.
If no more cells are found during the growing phase for which this
condition is fulfilled, the foam will stop splitting cells, even if
the target number of cells is not yet reached. 
This option prevents
the foam from adapting to statistical fluctuations in the training
samples (overtraining). 
Note that event 
weights are not taken into account for evaluating the number of events 
in the cell. 

In particular for training samples with small event numbers of less
than $10^4$ events this cut improves the performance. The default
value of (\code{Nmin = 100}) was found to give good results in most
cases.  It should be reduced in applications
with very small training samples (less than 200 training events) and
with many cells. 

\item \texttt{Kernel} -- {\em cell weighting with kernel functions:\index{Kernel estimation}}\\
\label{sec:PDEFoam-kernel}%
A Gaussian Kernel smoothing is applied during the evaluation
phase, if the option \code{Kernel} is set to ``\code{Gauss}''.  In this
case all cells contribute to the calculation of the discriminant for a
given event, weighted with their Gaussian distance to the
event.  The average cell value $v$ (event density in case of
separated foams, and the ratio
$n_S/(n_S+n_B)$ in case of a single foam)
for a given event ${\mathbf x}=(x_1, \ldots, x_\Nvar)$ is calculated by
\beq
  v = \frac{\sum_{\text{all cells } i}G(D(i, \mathbf{x}), 0, \texttt{VolFrac})\cdot v_i}%
           {\sum_{\text{all cells } i}G(D(i, \mathbf{x}), 0, \texttt{VolFrac})}\,,
\eeq
where $v_i$ is the output value in cell $i$, 
$G(x, x_0, \sigma) = \exp(-(x-x_0)^2/2\sigma^2)$, and $D(i,\mathbf{x})$ 
is the minimal Euclidean distance between $\mathbf{x}$ and any point
$\mathbf{y}$ in cell $i$
\beq
  D(i, \mathbf{x}) = \min_{\mathbf{y}\in\text{cell }i}|\mathbf{x}-\mathbf{y}|\,.
\eeq
The Gaussian kernel avoids discontinuities of the discriminant
values at the cell boundaries.  In most cases it results in an
improved separation power between signal and background.  However, the
time needed for classification increases due to the larger number of
computations performed.  A comparison between foams with and without
Gaussian kernel can be seen in Fig.~\ref{fig:pdefoam-foam}.

A linear interpolation with adjacent cells in each dimension is applied during 
the classification phase, if the option \code{Kernel} is set to ``\code{LinNeighbors}''.
This results in faster classification than the Gauss weighting of all cells
in the foam.
\end{itemize}

%%%%%%%%%%%%%%%%%%%%%%%%%%%%%%%%%%%%%%%%%%%%%%%%%%%%%%%%%%%%%%%

The PDE-Foam algorithm exhibits stable performance with respect to variations
in most of the parameter settings. However, optimisation of the parameters is required
for small training samples ($<10^4$ events) in combination with many
observables ($>5$). In such cases, \code{VolFrac} should be increased until 
an acceptable performance is reached. Moreover, in cases where the classification 
time is not critical, one of the \code{Kernel} methods should be applied
to further improve the classification performance. For large training samples ($>10^5$) 
and if the training time is not critical,
\code{nActiveCells} should be increased to improve the classification performance.

%%%%%%%%%%%%%%%%%%%%%%%%%%%%%%%%%%%%%%%%%%%%%%%%%%%%%%%%%%%

\subsubsection{Classification}

To classify an event in a $d$-dimensional phase space as being either
of signal or of background type, a local estimator of the probability
that this event belongs to either class can be obtained from the
foam's hyper-rectangular cells. The foams are created and filled based
on samples of signal and background training events.  For
classification two possibilities are implemented. One foam can be used
to separate the S/B probability density or two separate foams are
created, one for the signal events and one for background events.

\subsubsection*{1)~Separate signal and background foams} 

If the option \code{SigBgSeparate = True} is set, the
method PDE-Foam treats the signal- and background distributions
separately and performs the following steps to form the two foams and
to calculate the classifier discriminator for a given event:

\begin{enumerate}
  \item \emph{Setup of training trees}. Two binary search trees are
    created and filled with the $d$-dimensional observable vectors of
    all signal and background training events, respectively.

  \item \emph{Initialisation phase}. Two independent foams for signal
    and background are created.

  \item \emph{Growing phase}. The growing is performed independently
    for the two foams.  The density of events is estimated as the
    number of events found in the corresponding tree that are
    contained in the sampling box divided by the volume of the box
    (see \code{VolFrac} option).  The geometries of the final foams
    reflect the distribution of the training samples: phase-space
    regions where the density of events is constant are combined in
    large cells, while in regions with large gradients in density many
    small cells are created.

  \item \emph{Filling phase}. Both for the signal and background foam
    each active cell is filled with the number of training
    events, $n_S$ (signal) or $n_B$ (background), contained in the 
    corresponding cell volume, taking into account
    the event weights $w_i$: $n_S=\sum_{\text{sig. cell}}w_i$, 
    $n_B=\sum_{\text{bg. cell}}w_i$. 

  \item \emph{Evaluation phase}. The estimator for a given event is
    evaluated based on the number of events stored in the foam
    cells. The two corresponding foam cells that contain the event are
    found.  The number of events ($n_S$ and $n_B$) is
    read from the cells.  The estimator $\RPDEFoam(i)$ is then given by
    \beq
    \label{eq:PDEFoamSeparatedRatio}
    \RPDEFoam(i) = \frac{n_{S}/V_{S}}
    {\frac{n_B}{V_B} + 
      \frac{n_S}{V_S}} \, , 
    \eeq
    where $V_S$ and $V_B$ are the respective
    cell volumes.
    The statistical error of the estimator is calculated as:
    \beq
    \label{eq:PDEFoamSeparatedError}
    \sigma_{\RPDEFoam}(i) = \sqrt{ \left(\frac{n_S\sqrt{n_B}}{(n_S+ n_B)^2}\right)^2 +
                                   \left(\frac{n_B\sqrt{n_S}}{(n_S+ n_B)^2}\right)^2 }.
    \eeq
    Note that the so defined discriminant approximates the probability for an event
    from within the cell to be of class signal, if the weights are normalised such
    that the total number of weighted signal events is equal to the total number of weighted
    background events. This can be enforced with the normalisation mode ``EqualNumEvents''
    (\cf Sec.~\ref{sec:PreparingTrainingTestData}).
\end{enumerate}

Steps 1-4 correspond to the training phase of the method. Step 5 is
performed during the testing phase. In contrast to the PDE-RS method
the memory consumption and computation time for the testing phase does
not depend on the number of training events.

Two separate foams for signal and background allow the foam algorithm to 
adapt the foam geometries to the individual shapes of the signal and 
background event distributions. It is therefore well suited for cases 
where the shapes of the two distributions are very different.

\subsubsection*{2)~Single signal and background foam} 

If the option \code{SigBgSeparate = False} is set (default), the PDE-Foam
method creates only one foam, which holds directly the estimator
instead of the number of signal and background events.  The
differences with respect to separate signal and backgrounds foams are:

\begin{enumerate}
  \item \emph{Setup of training trees}. Fill only one binary search
    tree with both signal events and background events.  This is
    possible, since the binary search tree has the information whether
    a event is of signal or background type.
    
  \item \emph{Initialisation phase}. Only one foam is created. The
    cells of this foam will contain the estimator $\RPDEFoam(i)$ (see
    eq. \eqref{eq:PDEFoamSeparatedRatio}).

  \item \emph{Growing phase}. The splitting algorithm in this case
    minimises the variance of the estimator density
    $\sigma_{\rho}/\langle \rho\rangle$ across each cell.  The
    estimator density $\rho$ is sampled as the number of weighted signal events
    $n_S$ over the total number of weighted events
    $n_S+n_B$ in a small box around the sampling
    points:
    \beq
    \rho = \frac{n_S}{n_S+n_B} \, 
    \frac{1}{\texttt{VolFrac}}
    \eeq
    In this case the geometries of the final foams reflect 
    the distribution of the estimator density in the training sample:
    phase-space regions where the signal to background ratio
    is constant are combined in large cells, while in regions where
    the signal-to-background ratio changes rapidly many small
    cells are created. 

  \item \emph{Filling phase}. Each active cell is filled with the estimator
    given as the ratio of weighted signal events to the total number of 
    weighted events
    in the cell:
    \beq
    \RPDEFoam(i) = \frac{n_S}{n_S+n_B}.
    \eeq
    The statistical error of the estimator \eqref{eq:PDEFoamSeparatedError}
    also is stored in the cell.

  \item \emph{Evaluation phase}. The estimator for a given event is
    directly obtained as the discriminator that is stored in the cell
    which contains the event. 

\end{enumerate}

For the same total number of foam cells, the performance of the
two implementations was found to be similar. 

%%%%%%%%%%%%%%%%%%%%%%%%%%%%%%%%%%%%%%%%%%%%%%%%%%%%%%%%%%%%%%%
\subsubsection{Regression}

Two different methods are implemented for regression. In the first method,
applicable for single targets only ({\em mono-target regression}), the target
value is stored in each cell of the foam. In the second method, applicable 
to any number of targets ({\em multi-target regression}), the target 
values are stored in higher foam dimensions. 

In {\bf mono-target regression} the density used to form the foam is given by
the mean target density in a given box. The procedure is as follows. 

\begin{enumerate}
  \item \emph{Setup of training trees}. A binary search tree is filled
        with all training events.

  \item \emph{Growing phase}. One $\Nvar$-dimensional foam is
        formed: the density $\rho$ is given by the mean target value
        $\langle t\rangle$ within the sampling box, divided by the box
        volume (given by the \code{VolFrac} option):
	\beq
        \rho = \frac{\langle t\rangle}{\texttt{VolFrac}} \equiv
        \frac{\sum_{i=1}^{N_\text{box}}t_i}{N_\text{box}\cdot
        \texttt{VolFrac}} \, , 
	\eeq 
	where the sum is over all $N_\text{box}$ events within the
        sampling box, and $t_i$ is the target value of event $i$.

  \item \emph{Filling phase}. The cells are filled with their
        average target values, $\langle t\rangle =
        \sum_{i=1}^{N_\text{box}}t^{(i)}/N_\text{box}$.

  \item \emph{Evaluation phase}. Estimate the target value for a
        given test event: find the corresponding foam cell in which
        the test event is situated and read the average target value
        $\langle t\rangle$ from the cell.
\end{enumerate}

For {\bf multi-target regression} the target information is stored in additional 
foam dimensions. For a training sample with $\Nvar$ ($\Ntar$) input variables 
(regression targets), one would form a ($\Nvar+\Ntar$)-dimensional foam.
To compute a target estimate for event $i$, one needs the coordinates of 
the cell centre $C(i, k)$ in each foam dimension $k$. The procedure is 
as follows.

\begin{enumerate}
  \item \emph{Setup of training trees}. A binary search tree is filled
        with all training events.

  \item \emph{Growing phase}. A ($\Nvar+n_\text{tar}$)-dimensional
        foam is formed: the event density $\rho$ is estimated by the
        number of events $N_\text{box}$ within a predefined box of the
        dimension ($\Nvar+n_\text{tar}$), divided by the box volume,
        whereas the box volume is given by the \code{VolFrac} option
	\beq
        \rho = \frac{N_\text{box}}{\texttt{VolFrac}} \, .
	\eeq

  \item \emph{Filling phase}. Each foam cell is filled with the
        number of corresponding training events.

  \item \emph{Evaluation phase}. Estimate the target value for a
        given test event: find the $N_\text{cells}$ foam cells in
        which the coordinates $(x_1, \ldots, x_\Nvar)$ of the event
        vector are situated.  Depending on the \code{TargetSelection}
        option, the target value $t_k$ ($k=1, \ldots, \Ntar$) is
	\begin{enumerate}
	  \item the coordinate of the cell centre in direction of the
	        target dimension $\Nvar+k$ of the cell $j$ ($j=1,
	        \ldots, N_\text{cells}$), which has the maximum event
	        density
		\beq
		  t_k  = C(j, \Nvar+k) \, ,
		\eeq
		if \code{TargetSelection = Mpv}.
	  \item the mean cell centre in direction of the target
	        dimension $\Nvar+k$ weighted by the event densities
	        $d_\text{ev}(j)$ ($j=1, \ldots, N_\text{cells}$) of
	        the cells
		\beq
		  t_k = \frac{\sum_{j=1}^{N_\text{cells}}\, C(j, \Nvar+k) 
		        \cdot d_\text{ev}(j)}{\sum_{j=1}^{N_\text{cells}} d_\text{ev}(j)}
		\eeq
		if \code{TargetSelection = Mean}.  
	\end{enumerate}
\end{enumerate}

\subsubsection*{Kernel functions for regression}

The kernel weighting methods have been implemented also for regression,
taking into account the modified structure of the foam in case 
of multi-target regression.

%%%%%%%%%%%%%%%%%%%%%%%%%%%%%%%%%%%%%%%%%%%%%%%%%%%%%%%%%%%%%%%
\subsubsection{Visualisation of the foam via projections to 2 dimensions}
\label{sec:pdefoam-visualise}

A projection function is used to visualise the foam in two dimensions. 
It is called via:

\begin{codeexample}
\begin{tmvacode}
TH2D *proj = foam->Project2( dim1, dim2, "<options>", "<kernel>" );
\end{tmvacode}
\caption[.]{\codeexampleCaptionSize Call of the projection function.
   The first two arguments are the dimensions one wishes to project
   on, the third is a string identifier to specify quantity to plot
   (\code{nev}, \code{discr}, \code{rms}, \code{rms_ov_mean},
   \code{MonoTargetRegression}, \code{MultiTargetRegression}), and
   the fourth argument chooses the kernel (\code{kNone},
   \code{kGaus}).}
\end{codeexample}

For each active cell $i$ the two-dimensional rectangular sub-space
(dimensions \code{dim1} and \code{dim2}) is calculated and all
bins contained in this sub-space are filled with the value $v(i)$ of
cell $i$.  This implies that in the case of more than two dimensions
the values $v(i)$ in the dimensions that are not visible are summed.
The filled value $v(i)$ depends on the given \code{<option>}, which
allows one to display all variables stored in the foam cells.

For the following description we define by $L(i, k)$ 
the length of the foam cell $i$ in the dimension $k$ of a
$d$-dimensional foam.  Symbolic, may $L(\text{Foam},
k)$ be the scaled length of the entire foam in dimension $k$.

\begin{itemize}
\item \code{MultiTargetRegression}, \code{nev} -- {\em projecting the number of events} \\
These options apply to classification with separate signal and
background foams and multi-target regression.
The value $v(i)$ filled in the histogram is equal to
the number of events $N_\text{ev}(i)$ stored in the foam cell $i$
divided by the scaled two-dimensional cell area in dimension
\code{dim1} and \code{dim2}
\beq
  v(i) = \frac{N_\text{ev}(i)}{L(i, \texttt{dim1})\cdot L(i, \texttt{dim2})\cdot
                               L({\rm Foam}, \texttt{dim1})\cdot L({\rm Foam}, \texttt{dim2})}
\eeq

\item \code{discr} -- {\em projecting the discriminator}\\
If the foam cells are filled with the discriminator, which is the case
for classification with a single foam (\code{SigBgSeparate = False}), one
can use this option.  Here the value $v(i)$, filled in the histogram
is equal to the discriminator $\operatorname{Discr}(i)$ saved in cell
$i$ multiplied by the cell volume excluding the dimensions \code{dim1}
and \code{dim2}
\beq
  v(i) = \operatorname{Discr}(i)\!\!\!
         \prod_{\substack{k=1\\k\neq\texttt{dim1}\\k\neq\tt{dim2}}}^{d}\!\!\!\!L(i, k)
\eeq
This means that the average discriminator weighted with the cell size
of the non-visualised dimensions is filled.

\item \code{rms}, \code{rms_ov_mean} -- {\em projection of cell variances}\\
The variance (RMS) and the mean of the event distribution are saved in
every foam cell.  If the option \code{rms} is used, the plotted cell
value $v(i)$ is equal to the cell RMS.  If the option
\code{rms_ov_mean} is given, $v(i)=\text{RMS}/\text{Mean}$ is
filled into the histogram.

\item \code{MonoTargetRegression} -- {\em projection of targets}\\
If the foam cells are filled with targets by using the mono-target
regression option in order to do regression
(\code{MultiTargetRegression = False}), one can use this option.
Here the value $v(i)$, filled in the histogram is equal to the target
saved in cell $i$.

\item \code{kNone}, \code{kGaus} -- {\em Using kernels for projecting}\\
Instead of filling the rectangle shaped cell areas into the histogram
one can use the build-in kernel estimator to interpolate between the
cell mean values in order to visualise the effect of the kernel to the
foam.  In this case the function performs a loop over all cells and
fills the sum of the weighted cell values $v(i)$ into the histogram.
See page~\pageref{sec:PDEFoam-kernel} for more details on kernels
included in PDE-Foam.
\end{itemize}

%%%%%%%%%%%%%%%%%%%%%%%%%%%%%%%%%%%%%%%%%%%%%%%%%%%%%%%%%%%%%%%
\subsubsection{Performance}

Like PDE-RS (see Sec.~\ref{sec:pders}), this method is a powerful
classification tool for problems with highly nonlinearly correlated
observables.  Furthermore PDE-Foam is a fast responding classifier,
because of its limited number of cells, independent of the size of the
training samples.

An exception is the multi-target regression with Gauss kernel because
the time scales with the number of cells squared.  Also the training
can be slow, depending on the number of training events and number of
cells one wishes to create.

%% file: optiontables/MVA__PDEFoam.tex
\begin{optiontableAuto}
            SigBgSeparate  &  \mc{1}{c}{--}  &            False  &  \mc{1}{l}{--}  &  Separate foams for signal and background \\
                  TailCut  &  \mc{1}{c}{--}  &            0.001  &  \mc{1}{l}{--}  &  Fraction of outlier events that are excluded from the foam in each dimension \\
                  VolFrac  &  \mc{1}{c}{--}  &        0.0333333  &  \mc{1}{l}{--}  &  Size of sampling box, used for density calculation during foam build-up (maximum value: 1.0 is equivalent to volume of entire foam) \\
             nActiveCells  &  \mc{1}{c}{--}  &              500  &  \mc{1}{l}{--}  &  Maximum number of active cells to be created by the foam \\
                   nSampl  &  \mc{1}{c}{--}  &             2000  &  \mc{1}{l}{--}  &  Number of generated MC events per cell \\
                     nBin  &  \mc{1}{c}{--}  &                5  &  \mc{1}{l}{--}  &  Number of bins in edge histograms \\
                 Compress  &  \mc{1}{c}{--}  &             True  &  \mc{1}{l}{--}  &  Compress XML file \\
    MultiTargetRegression  &  \mc{1}{c}{--}  &            False  &  \mc{1}{l}{--}  &  Do regression with multiple targets \\
                  CutNmin  &  \mc{1}{c}{--}  &             True  &  \mc{1}{l}{--}  &  Requirement for minimal number of events in cell \\
                     Nmin  &  \mc{1}{c}{--}  &              100  &  \mc{1}{l}{--}  &  Number of events in cell required to split cell \\
                   Kernel  &  \mc{1}{c}{--}  &             None  &  None, Gauss, LinNeighbors  &  Kernel type used \\
          TargetSelection  &  \mc{1}{c}{--}  &             Mean  &  Mean, Mpv  &  Target selection method 
\end{optiontableAuto}

%% file: KNN.tex
\subsection{k-Nearest Neighbour (k-NN) Classifier}\index{k-Nearest Neighbour classifier}\index{k-NN}
\label{sec:knn}

Similar to PDE-RS (\cf\  Sec.~\ref{sec:pders}), the k-nearest neighbour method compares 
an observed (test) event to reference events from a training data set~\cite{FriedmanBook}. 
However, unlike PDE-RS, which in its original form uses a fixed-sized multidimensional volume 
surrounding the test event, and in its augmented form resizes the volume as a function of 
the local data density, the k-NN algorithm is intrinsically adaptive. It searches for a 
fixed number of adjacent events, which then define a volume for the metric used. The k-NN 
classifier has best performance when the boundary that separates signal and background 
events has irregular features that cannot be easily approximated by parametric learning 
methods. 

\subsubsection{Booking options}

The k-NN classifier is booked via the command:
\begin{codeexample}
\begin{tmvacode}
factory->BookMethod( Types::kKNN, "kNN", "<options>" );
\end{tmvacode}
\caption[.]{\codeexampleCaptionSize Booking of the k-NN classifier: the first argument is 
		   a predefined enumerator, the second argument is a user-defined 
		   string identifier, and the third argument is the configuration options string.
         Individual options are separated by a ':'. 
         See Sec.~\ref{sec:usingtmva:booking} for more information on the booking.}
\end{codeexample}

The configuration options for the k-NN classifier are listed in Option Table~\ref{opt:mva::knn}
(see also Sec.~\ref{sec:fitting}).

% ======= input option table ==========================================
\begin{option}[t]
\input optiontables/MVA__KNN.tex
\caption[.]{\optionCaptionSize 
     Configuration options reference for MVA method: {\em k-NN}.
     Values given are defaults. If predefined categories exist, the default category 
     is marked by a '$\star$'. The options in Option Table~\ref{opt:mva::methodbase} on 
     page~\pageref{opt:mva::methodbase} can also be configured.     
}
\label{opt:mva::knn}
\end{option}
% =====================================================================

\subsubsection{Description and implementation}

The k-NN algorithm searches for $k$ events that are closest to the test event. Closeness 
is thereby measured using a metric function. The simplest metric choice is the Euclidean 
distance
\beq
   R = \left(\sum_{i = 1}^{\Nvar} |x_{i} - y_{i}|^{2}\right)^{\!\!\frac{1}{2}}\;.
\eeq
where $\Nvar$ is the number of input variables used for the classification, $x_{i}$ are 
coordinates of an event from a training sample and $y_{i}$ are variables of an observed 
test event. The $k$ events with the smallest values of $R$ are the {\em k-nearest neighbours}. 
The value of $k$ determines the size of the neighbourhood for which a probability density 
function is evaluated. Large values of $k$ do not capture the local behavior of the 
probability density function. On the other hand, small values of $k$ cause statistical
fluctuations in the probability density estimate. A case study with real data suggests 
that values of $k$ between 10 and 100 are appropriate and result in similar classification 
performance when the training sample contains hundreds of thousands of events (and $\Nvar$
is of the order of a few variables).

The classification algorithm finds k-nearest training events around a query point
\beq
   k = k_{S} + k_{B}\;,
\eeq
where $k_{S(B)}$ is number of the signal (background) events in the training sample. 
The relative probability that the test event is of signal type is given by
\beq
   P_{S} = \frac{k_{S}}{k_{S} + k_{B}} = \frac{k_{S}}{k}\;.
\eeq
The choice of the metric governs the performance of the nearest neighbour algorithm. 
When input variables have different units a variable that has a wider distribution 
contributes with a greater weight to the Euclidean metric. This feature is compensated 
by rescaling the variables using a scaling fraction determined by the option \code{ScaleFrac}.
Rescaling can be turned off by setting \code{ScaleFrac} to 0. The scaling factor applied
to variable $i$ is determined by the width $w_{i}$ of the $x_{i}$ distribution for the 
combined sample of signal and background events: $w_{i}$ is the interval that contains 
the fraction \code{ScaleFrac} of $x_{i}$ training values. The input variables are 
then rescaled by $1/w_{i}$, leading to the rescaled metric
\beq
   R_{\rm rescaled} = 
   \left(\sum_{i = 1}^{d} \frac{1}{w_{i}^{2}}|x_{i} - y_{i}|^{2}\right)^{\!\!\frac{1}{2}}\;. 
\eeq
\begin{figure}[t]
  \begin{center}
    \includegraphics[width=0.325\textwidth]{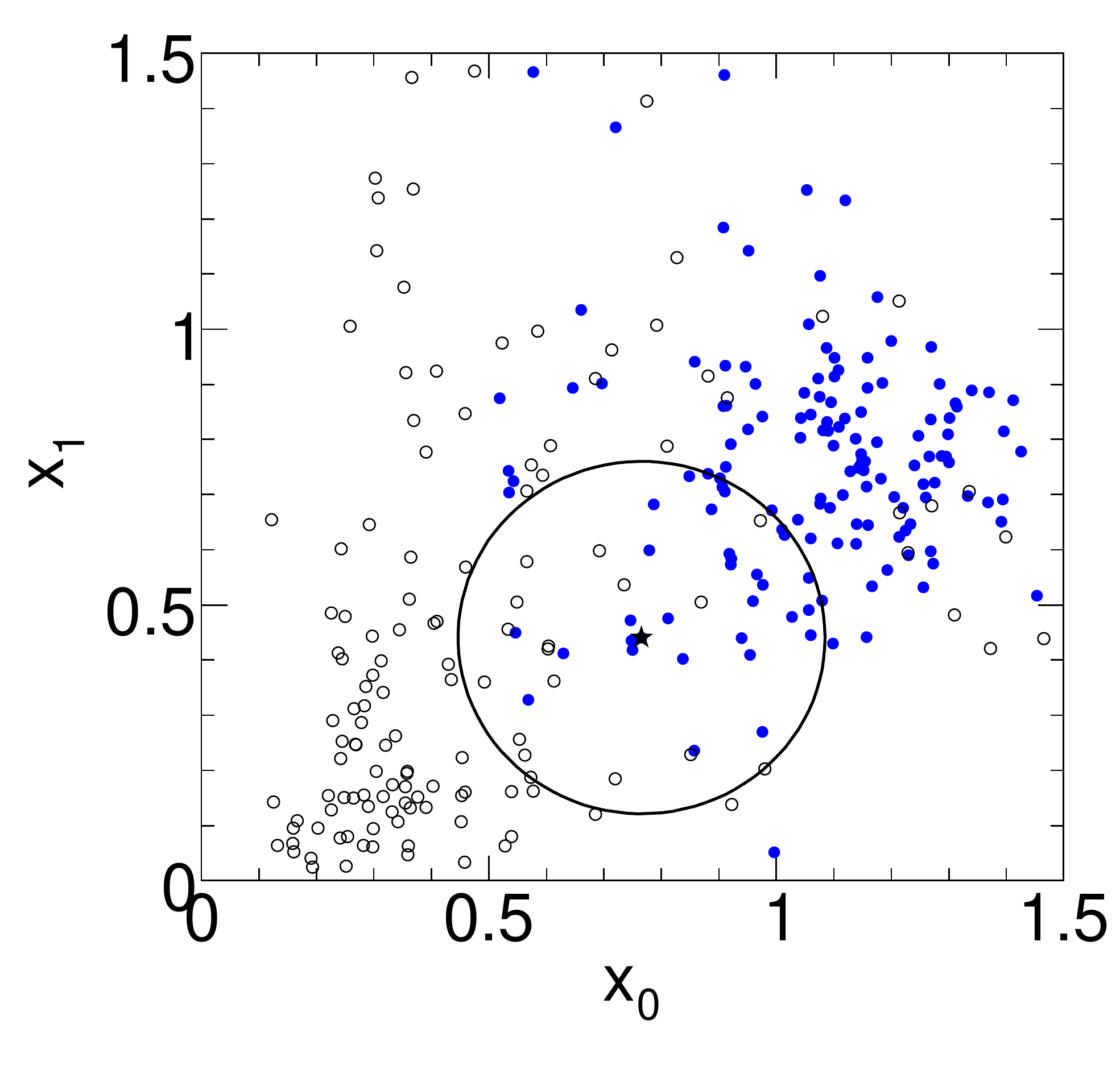}
    \includegraphics[width=0.325\textwidth]{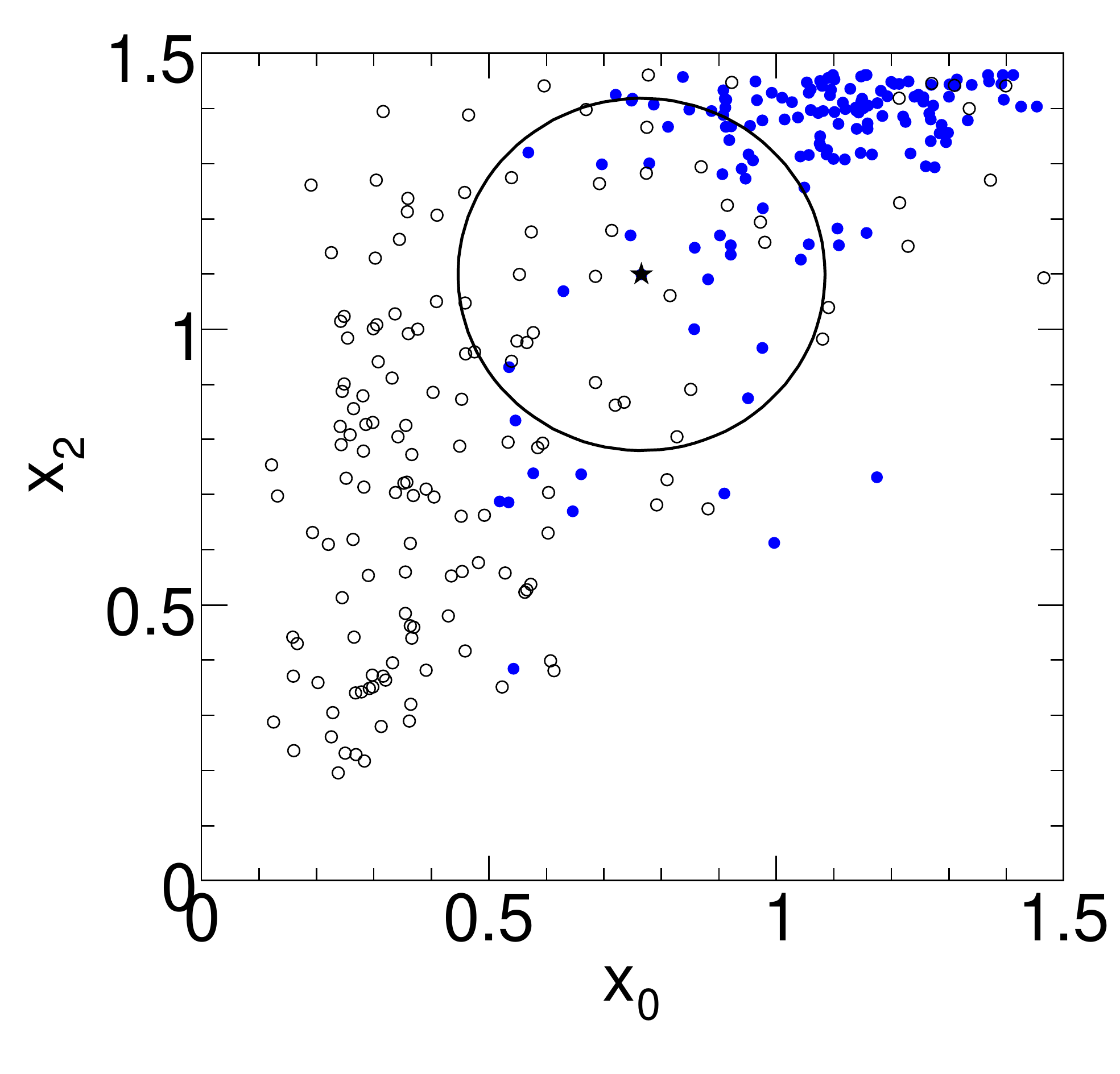}
    \includegraphics[width=0.325\textwidth]{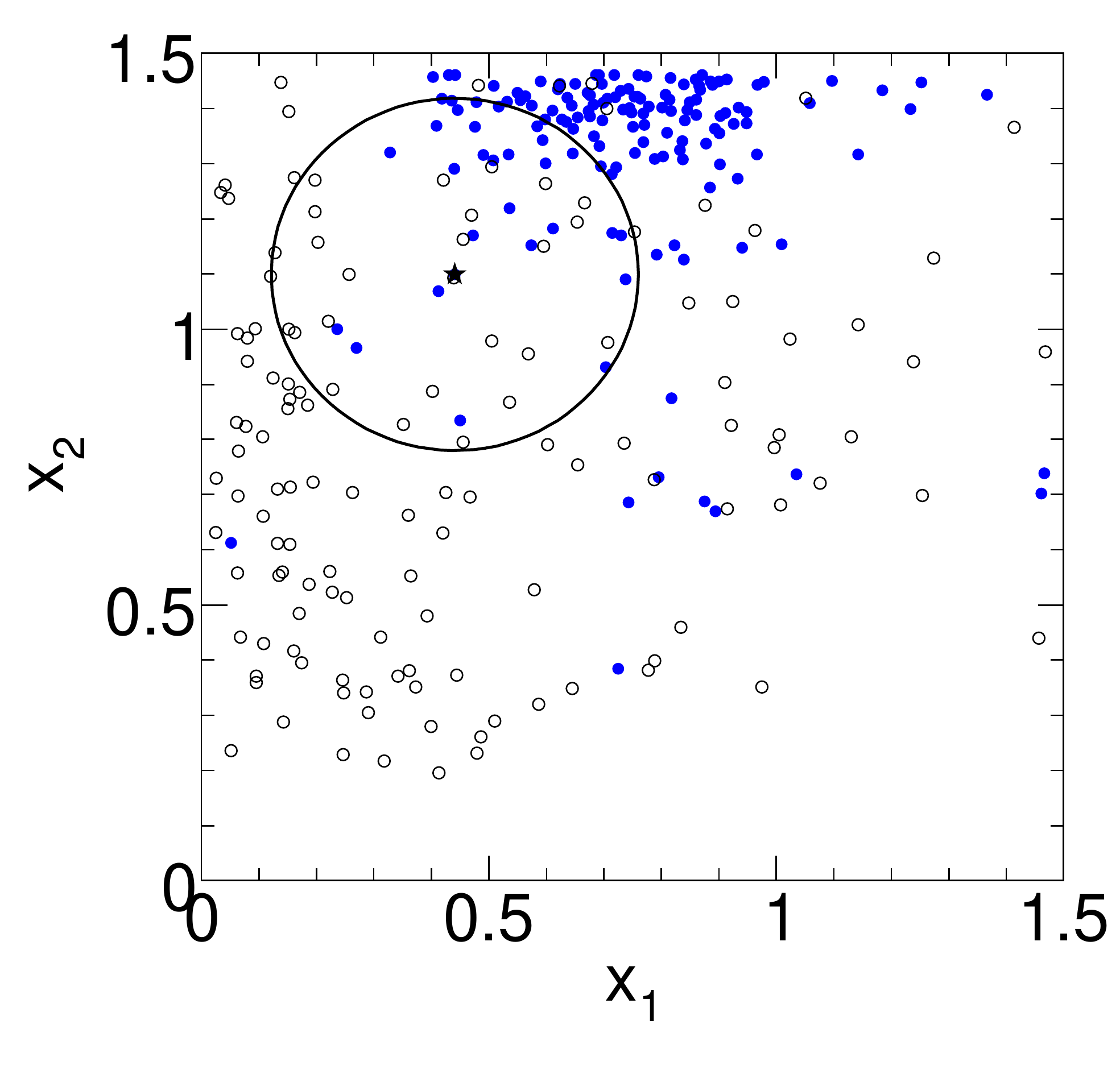}
  \end{center}
  \vspace{-0.7cm}
  \caption[.] { Example for the k-nearest neighbour algorithm in a three-dimensional space
                (\ie, for three discriminating input variables). The three plots are 
                projections upon the two-dimensional coordinate planes. The full (open) circles 
                are the signal (background) events. The k-NN algorithm searches for 20 nearest 
                points in the {\em nearest neighborhood} (circle) of the query event, shown as 
                a star. The nearest neighborhood counts 13 signal and 7 background points 
                so that query event may be classified as a signal candidate. }
\label{fig:knn-3d} 
\end{figure}
Figure~\ref{fig:knn-3d} shows an example of event classification with the k-nearest 
neighbour algorithm.\footnote
{
  The number of training events shown has been greatly reduced to illustrate the 
  principle of the algorithm. In a real application a typical k-NN training sample 
  should be ample.
}

The output of the k-nearest neighbour algorithm can be interpreted as a probability 
that an event is of signal type, if the numbers (better: sum of event weights) of 
signal and background events in the training sample are equal. This can be enforced
via the \code{Trim} option. If set training events of the overabundant type are 
randomly removed until parity is achieved.
% Using equal number of signal and background events for the classification allows 
% to measure the relative fractions of signal and background events in the observed 
% data. This way the dependency on the underlying model used to generated training 
% events is removed.

Like (more or less) all TMVA classifiers, the k-nearest neighbour estimate suffers 
from statistical fluctuations in the training data. The typically high variance of the 
k-NN response is mitigated by adding a weight function that depends smoothly on the 
distance from a test event. The current k-NN implementation uses a polynomial kernel
\beq
  W(x) = \left\{\begin{array}{ll}
                  (1 - |x|^{3})^{3} & \text{ if } |x| < 1\;, \\[0.1cm]
                  0 & \text{ otherwise}\;.
                \end{array}     
         \right. 
\eeq
If $R_{k}$ is the distance between the test event and the $k$th neighbour, the events 
are weighted according to the formula:
\beq
  W_{S(B)} = \sum_{i = 1}^{k_{S(B)}} W\left(\frac{R_{i}}{R_{k}}\right)\;,
\eeq
where $k_{S(B)}$ is number of the signal (background) events in the neighbourhood. 
The weighted signal probability for the test event is then given by
\beq
  P_{S} = \frac{W_{S}}{W_{S} + W_{B}}\;.
\eeq
The kernel use is switched on/off by the option \code{UseKernel}.

\subsubsection*{Regression}

The k-NN algorithm in TMVA also implements a simple multi-dimensional (multi-target)
regression model. For a test event, the algorithm finds the k-nearest neighbours 
using the input variables, where each training event contains a regression value. 
The predicted regression value for the test event is the weighted average of the 
regression values of the k-nearest neighbours, \cf Eq.~(\ref{eq:PDERSregratio})
on page~\pageref{eq:PDERSregratio}.

\subsubsection{Ranking}

The present implementation of k-NN does not provide a ranking of the input variables.

\subsubsection{Performance}

The simplest implementation of the k-NN algorithm would store all training events in an array. 
The classification would then be performed by looping over all stored events and finding 
the k-nearest neighbours. As discussed in Sec.~\ref{sec:binaryTrees}, such an implementation
is impractical for large training samples. The k-NN algorithm therefore uses a {\em kd-tree} 
structure~\cite{kd-tree} that significantly improves the performance.

The TMVA implementation of the k-NN method is reasonably fast to allow classification of large 
data sets. In particular, it is faster than the adaptive PDE-RS method (\cf\  Sec.~\ref{sec:pders}).
Note that the k-NN method is not appropriate for problems where the number of input 
variables exceeds $\Nvar\gtrsim 10$. The neighbourood size $R$ depends on $\Nvar$
and the size of the training sample $\Ntrain$ as
\beq
  R_{N} \propto \frac{1}{\sqrt[\Nvar]N}\;.
\eeq
A large training set allows the algorithm to probe small-scale features that distinguish 
signal and background events.

%%% Local Variables: 
%%% mode: latex
%%% TeX-master: "TMVAUsersGuide"
%%% End: 

%% file: optiontables/MVA__KNN.tex
\begin{optiontableAuto}
                     nkNN  &  \mc{1}{c}{--}  &               20  &  \mc{1}{l}{--}  &  Number of k-nearest neighbors \\
             BalanceDepth  &  \mc{1}{c}{--}  &                6  &  \mc{1}{l}{--}  &  Binary tree balance depth \\
                ScaleFrac  &  \mc{1}{c}{--}  &              0.8  &  \mc{1}{l}{--}  &  Fraction of events used to compute variable width \\
                SigmaFact  &  \mc{1}{c}{--}  &                1  &  \mc{1}{l}{--}  &  Scale factor for sigma in Gaussian kernel \\
                   Kernel  &  \mc{1}{c}{--}  &             Gaus  &  \mc{1}{l}{--}  &  Use polynomial (=Poln) or Gaussian (=Gaus) kernel \\
                     Trim  &  \mc{1}{c}{--}  &            False  &  \mc{1}{l}{--}  &  Use equal number of signal and background events \\
                UseKernel  &  \mc{1}{c}{--}  &            False  &  \mc{1}{l}{--}  &  Use polynomial kernel weight \\
                UseWeight  &  \mc{1}{c}{--}  &             True  &  \mc{1}{l}{--}  &  Use weight to count kNN events \\
                   UseLDA  &  \mc{1}{c}{--}  &            False  &  \mc{1}{l}{--}  &  Use local linear discriminant - experimental feature 
\end{optiontableAuto}

%% file: HMatrix.tex
\subsection{H-Matrix discriminant}
\label{sec:hmatrix}

The origins of the H-Matrix\index{H-Matrix} approach dates back to works 
of Fisher and Mahalanobis in the context of Gaussian 
classifiers~\cite{Fisher,Mahalanobis}. 
It discriminates one class (signal) of a feature vector from another 
(background). The correlated elements of the vector are assumed to be 
Gaussian distributed, and the inverse of the covariance matrix is 
the {\em H-Matrix}. A multivariate $\chi^2$ estimator\index{Chi-squared estimator} 
is built that exploits differences in the mean values of the vector elements 
between the two classes for the purpose of discrimination.

The H-Matrix classifier as it is implemented in TMVA is equal or less performing
than the Fisher discriminant (see Sec.~\ref{sec:fisher}), and has been only  
included for completeness. 

\subsubsection{Booking options}

The H-Matrix discriminant is booked via the command:
\begin{codeexample}
\begin{tmvacode}
factory->BookMethod( Types::kHMatrix, "HMatrix", "<options>" );
\end{tmvacode}
\caption[.]{\codeexampleCaptionSize Booking of the H-Matrix classifier: the first argument is 
		   a predefined enumerator, the second argument is a user-defined 
		   string identifier, and the third argument is the configuration options string.
         Individual options are separated by a ':'. 
         See Sec.~\ref{sec:usingtmva:booking} for more information on the booking.}
\end{codeexample}

No specific options are defined for this method beyond those shared with all the other 
methods (\cf Option Table~\ref{opt:mva::methodbase} on page~\pageref{opt:mva::methodbase}).

\subsubsection{Description and implementation}

For an event $i$, each one $\chi^2$ estimator ($\chi^2_{S(B)}$) is computed for 
signal ($S$) and background ($B$), using estimates for the sample means 
($\overline x_{S(B),k}$) and covariance matrices ($C_{S(B)}$) obtained 
from the training data
\beq
  \chi^2_U(i)=\sum_{k,\ell=1}^{\Nvar}
            \left(x_k(i) - \overline x_{U,k}\right)
            C^{-1}_{U,k\ell}
            \left(x_\ell(i) - \overline x_{U,\ell}\right)\,,
\eeq
where $U=S,B$. From this, the discriminant
\beq
   \HMATRIX(i) = \frac{\chi^2_B(i)-\chi^2_S(i)}{\chi^2_B(i)+\chi^2_S(i)}\,,
\eeq
is computed to discriminate between the signal and background classes.

\subsubsection{Variable ranking}

The present implementation of the H-Matrix discriminant does not provide a ranking 
of the input variables.

\subsubsection{Performance}

The TMVA implementation of the H-Matrix classifier has been shown to underperform 
in comparison with the corresponding Fisher discriminant (\cf\  Sec.~\ref{sec:fisher}),
when using similar assumptions and complexity. It might therefore be considered to be depreciated.

%% file: Fisher.tex
\subsection{Fisher discriminants (linear discriminant 
            analysis\index{Linear Discriminant Analysis})}
\label{sec:fisher}

In the method of Fisher discriminants\index{Fisher discriminant}~\cite{Fisher} 
event selection is performed in a transformed variable space with zero
linear correlations, by distinguishing the mean values of the signal
and background distributions.
The linear discriminant analysis\index{Linear Discriminant Analysis}
determines an axis in the (correlated) hyperspace of the input
variables such that, when projecting the output classes (signal and
background) upon this axis, they are pushed as far as possible away
from each other, while events of a same class are confined in a close
vicinity. The linearity property of this classifier is reflected in the
metric with which "far apart" and "close vicinity" are determined: the
covariance matrix of the discriminating variable space.

\subsubsection{Booking options}

The Fisher discriminant is booked via the command:
\begin{codeexample}
\begin{tmvacode}
factory->BookMethod( Types::kFisher, "Fisher", "<options>" );
\end{tmvacode}
\caption[.]{\codeexampleCaptionSize Booking of the Fisher discriminant: the first 
		   argument is a predefined enumerator, the second argument is a user-defined 
		   string identifier, and the third argument is the configuration options string.
         Individual options are separated by a ':'. 
         See Sec.~\ref{sec:usingtmva:booking} for more information on the booking.}
\end{codeexample}
The configuration options for the Fisher discriminant are given in Option Table~\ref{opt:mva::fisher}.

% ======= input option table ==========================================
\begin{option}[t]
\input optiontables/MVA__Fisher.tex
\caption[.]{\optionCaptionSize 
     Configuration options reference for MVA method: {\em Fisher}.
     Values given are defaults. If predefined categories exist, the default category is marked by a '$star$'. The options in Option Table~ref{opt:mva::methodbase} on page~pageref{opt:mva::methodbase} can also be configured.
     
}
\label{opt:mva::fisher}
\end{option}
% =====================================================================

\subsubsection{Description and implementation}

The classification of the events in signal and background classes relies 
on the following characteristics: the overall sample means $\overline x_k$
for each input variable $k=1,\dots,\Nvar$, the class-specific sample means
$\overline x_{S(B),k}$, and total covariance matrix $C$ of the 
sample. The covariance matrix can be decomposed into the sum of a {\em within-}
($W$) and a {\em between-class matrix} ($B$). They respectively describe 
the dispersion of events relative to the means of their own 
class (within-class matrix), and relative to the overall sample means 
(between-class matrix)\footnote
{
	The within-class matrix is given by
	\beqns
		W_{k\ell} = \sum_{U=S,B}\langle x_{U,k}-\overline x_{U,k}\rangle
                                \langle x_{U,\ell}-\overline x_{U,\ell}\rangle
					 = C_{S,k\ell} + C_{B,k\ell}\,,
	\eeqns
	where $C_{S(B)}$ is the covariance matrix of the signal (background)
	sample. The between-class matrix is obtained by
	\beqns
		B_{k\ell} = \frac{1}{2}\sum_{U=S,B}
                  \left(\overline x_{U,k} - \overline x_{k}\right)
                  \left(\overline x_{U,\ell} - \overline x_{\ell}\right)\,,
	\eeqns
	where $\overline x_{S(B),k}$ is the average of variable $x_{k}$ for the 
   signal (background) sample, and $\overline x_{k}$ denotes the average for 
   the entire sample.
}. 

The {\em Fisher coefficients}, $F_k$, are then given by 
\beq
\label{eq:Fisher}
	F_k = \frac{\sqrt{\NS \NB}}{\NS+\NB}
		   \sum_{\ell=1}^{\Nvar}W_{k\ell}^{-1}
         \left(\overline x_{S,\ell} - \overline x_{B,\ell}\right)\,,
\eeq
where $\NSB$ are the number of signal (background) events in the 
training sample. The Fisher discriminant $\Fisher(i)$ for event $i$ 
is given by
\beq
	\Fisher(i) = F_0 + \sum_{k=1}^{\Nvar}F_k x_k(i)\,.
\eeq
The offset $F_0$ centers the sample mean $\FisherMean$ of
all $\NS+\NB$ events at zero. 

Instead of using the within-class matrix, the Mahalanobis\index{Mahalanobis distance} 
variant determines the Fisher coefficients as follows~\cite{Mahalanobis}
\beq
\label{eq:Mahalanobis}
	F_k = \frac{\sqrt{\NS \NB}}{\NS+\NB}
		   \sum_{\ell=1}^{\Nvar}C_{k\ell}^{-1}
         \left(\overline x_{S,\ell} - \overline x_{B,\ell}\right)\,,
\eeq
where $C_{k\ell}=W_{k\ell}+B_{k\ell}$.

\subsubsection{Variable ranking}

The Fisher discriminant analysis aims at simultaneously maximising the 
between-class separation while minimising the within-class dispersion. 
A useful measure of the discrimination power of a variable is therefore
given by the diagonal quantity $B_{kk}/C_{kk}$, which is used for the 
ranking of the input variables.

\subsubsection{Performance}

In spite of the simplicity of the classifier, Fisher discriminants can 
be competitive with likelihood and nonlinear discriminants
in certain cases. In particular, Fisher discriminants are 
optimal for Gaussian distributed variables with linear correlations
(\cf\  the standard toy example that comes with TMVA). 

On the other hand, no discrimination at all is achieved when a variable 
has the same sample mean for signal and background, even if the shapes
of the distributions are very different. Thus, Fisher discriminants 
often benefit from suitable transformations 
of the input variables. For example, if a variable $x\in[-1,1]$ has a 
a signal distributions of the form $x^2$, and a uniform background
distributions, their mean value is zero in both cases, leading to no 
separation. The simple transformation $x\to |x|$ renders this variable 
powerful for the use in a Fisher discriminant.

%% file: optiontables/MVA__Fisher.tex
\begin{optiontableAuto}
                   Method  &  \mc{1}{c}{--}  &           Fisher  &  Fisher, Mahalanobis  &  Discrimination method 
\end{optiontableAuto}

%% file: LD.tex
\subsection{Linear discriminant analysis (LD)}\index{Linear Discriminant}\index{LD}
\label{sec:ld}

The linear discriminant analysis provides data classification using a linear model, 
where \textit{linear} refers to the discriminant function $y(\mathbf{x})$ being 
linear in the parameters $\mathbf{\beta}$
\beq
	y(\mathbf{x})=\mathbf{x}^\top\beta + \beta_0\;,
\eeq
where $\beta_0$ (denoted the {\em bias}) is adjusted so that $y(\mathbf{x})\geq0$ 
for signal and $y(\mathbf{x})<0$ for background. It can be shown that this is equivalent 
to the Fisher discriminant, which seeks to maximise the ratio of between-class 
variance to within-class variance by projecting the data onto a linear subspace.

\subsubsection{Booking options}

The LD is booked via the command:
\begin{codeexample}
\begin{tmvacode}
factory->BookMethod( Types::kLD, "LD" );
\end{tmvacode}
\caption[.]{\codeexampleCaptionSize Booking of the linear discriminant: the first argument is 
		   a predefined enumerator, the second argument is a user-defined 
	   	string identifier. No method-specific options are available.
        See Sec.~\ref{sec:usingtmva:booking} for more information on the booking.}
\end{codeexample}

No specific options are defined for this method beyond those shared with all the other 
methods (\cf Option Table~\ref{opt:mva::methodbase} on page~\pageref{opt:mva::methodbase}).

\subsubsection{Description and implementation}

Assuming that there are $m+1$ parameters $\beta_0, \cdots ,\beta_m$ to be estimated using 
a training set comprised of $n$ events, the defining equation for $\mathbf{\beta}$ is
\beq
	Y=X\mathbf{\beta}\;,
\eeq
where we have absorbed $\beta_0$ into the vector $\beta$ and introduced the matrices
\beq
	Y=\left( \begin{array}{c}
	y_1\\
	y_2\\
	\vdots\\
	y_n \end{array} \right) \mbox{  and  } X=\left( \begin{array}{cccc}
							1 & x_{11} & \cdots & x_{1m} \\
							1 & x_{21} & \cdots & x_{2m} \\
							\vdots & \vdots & \ddots & \vdots \\
							1 & x_{n1} & \cdots & x_{nm} \end{array} \right)\;,
\eeq
where the constant column in $X$ represents the bias $\beta_0$ and $Y$ is composed of 
the target values with $y_i=1$ if the $i$th event belongs to the signal class and $y_i=0$ 
if the $i$th event belongs to the background class. Applying the method of least squares, 
we now obtain the {\em normal equations} for the classification problem, given by
\beq
	X^TX\beta=X^TY \Longleftrightarrow \beta=(X^TX)^{-1}X^TY\;.
\eeq
The transformation $(X^TX)^{-1}X^T$ is known as the \textit{Moore-Penrose pseudo inverse} 
of $X$ and can be regarded as a generalisation of the matrix inverse to non-square 
matrices. It requires that the matrix $X$ has full rank.

If weighted events are used, this is simply taken into account by introducing a diagonal
weight matrix $W$ and modifying the normal equations as follows:
\beq
	\beta=(X^TWX)^{-1}X^TWY\;.
\eeq
Considering two events $\mathbf{x}_1$ and $\mathbf{x}_2$ on the decision boundary, we 
have $y(\mathbf{x}_1)=y(\mathbf{x}_2)=0$ and hence $(\mathbf{x}_1-\mathbf{x}_2)^T\beta=0$. 
Thus we see that the LD can be geometrically interpreted as determining the decision 
boundary by finding an orthogonal vector $\beta$.

\subsubsection{Variable ranking}

The present implementation of LD provides a ranking of the input variables based on the 
coefficients of the variables in the linear combination that forms the decision boundary. 
The order of importance of the discriminating variables is assumed to agree with the 
order of the absolute values of the coefficients.

\subsubsection{Regression with LD}

It is straightforward to apply the LD algorithm to linear regression by replacing the 
binary targets $y_i \in {0,1}$ in the training data with the measured values of the 
function which is to be estimated. The resulting function $y(\mathbf{x})$ is then 
the best estimate for the data obtained by least-squares regression.

\subsubsection{Performance}

The LD is optimal for Gaussian distributed variables with linear correlations (\cf the 
standard toy example that comes with TMVA) and can be competitive with likelihood and 
nonlinear discriminants in certain cases. No discrimination is achieved when a variable 
has the same sample mean for signal and background, but the LD can often benefit from 
suitable transformations of the input variables. For example, if a variable 
$x \in [-1,1]$ has a signal distribution of the form $x^2$ and a uniform background 
distribution, their mean value is zero in both cases, leading to no separation. The 
simple transformation $x \rightarrow |x|$ renders this variable powerful for the use 
with LD.

%%% Local Variables: 
%%% mode: latex
%%% TeX-master: "TMVAUsersGuide"
%%% End: 

%% file: FDA.tex
\subsection{Function discriminant analysis (FDA)}\index{Function Discriminant Analysis}\index{FDA}
\label{sec:fda}

The common goal of all TMVA discriminators is to determine an optimal separating
function in the multivariate space of all input variables. The Fisher 
discriminant solves this analytically for the linear case, while artificial neural
networks, support vector machines or boosted decision trees provide nonlinear 
approximations with -- in principle -- arbitrary precision if enough training 
statistics is available and the chosen architecture is flexible enough. 

The function discriminant analysis (FDA) provides an intermediate solution to the 
problem with the aim to solve relatively simple or partially nonlinear problems. 
The user provides the desired function with adjustable parameters via the configuration 
option string, and FDA fits the parameters to it, requiring the function value to be as 
close as possible to the real value (to 1 for signal and 0 for background in classification).
Its advantage over the more involved and automatic nonlinear discriminators is the simplicity 
and transparency of the discrimination expression. A shortcoming is that FDA will underperform 
for involved problems with complicated, phase space dependent nonlinear correlations.

\subsubsection{Booking options}

FDA is booked via the command:
\begin{codeexample}
\begin{tmvacode}
factory->BookMethod( Types::kFDA, "FDA", "<options>" );
\end{tmvacode}
\caption[.]{\codeexampleCaptionSize Booking of the FDA classifier: the first argument is 
		   a predefined enumerator, the second argument is a user-defined 
		   string identifier, and the third argument is the configuration options string.
         Individual options are separated by a ':'. 
         See Sec.~\ref{sec:usingtmva:booking} for more information on the booking.}
\end{codeexample}

The configuration options for the FDA classifier are listed in Option Table~\ref{opt:mva::fda}
(see also Sec.~\ref{sec:fitting}).

% ======= input option table ==========================================
\begin{option}[t]
\input optiontables/MVA__FDA.tex
\caption[.]{\optionCaptionSize 
     Configuration options reference for MVA method: {\em FDA}.
     Values given are defaults. If predefined categories exist, the default category 
     is marked by a '$\star$'. The options in Option Table~\ref{opt:mva::methodbase} on 
     page~\pageref{opt:mva::methodbase} can also be configured.     
     The input variables In the discriminator expression are denoted 
     \code{x0}, \code{x1}, \dots (until $\Nvar-1$), where the 
     number follows the order in which the variables have been 
     registered with the Factory; coefficients to be determined 
     by the fit must be denoted \code{(0)}, \code{(1)}, \dots (the 
     number of coefficients is free) in the formula; allowed is  
     any functional expression that can be interpreted by a ROOT 
     \href{http://root.cern.ch/root/html/TFormula.html}{TFormula}. 
     See Code Example~\ref{ce:fdaexample} for an example expression.
     The limits for the fit parameters (coefficients) defined in the 
     formula expression are given with the syntax:  "\code{(-1,3);(2,10);...}", 
     where the first interval corresponds to parameter \code{(0)}.
     The converger allows to use (presently only) Minuit 
     fitting in addition to Monte Carlo sampling or a Genetic Algorithm. More
     details on this combination are given in Sec.~\ref{sec:converger}. The 
     various fitters are configured using the options given in Tables~\ref{opt:fitter_mc}, 
     \ref{opt:fitter_minuit}, \ref{opt:fitter_ga} and \ref{opt:fitter_sa}, for MC, Minuit, GA and SA, 
     respectively.
}
\label{opt:mva::fda}
\end{option}
% =====================================================================

A typical option string could look as follows:
\begin{codeexample}
\begin{tmvacode}
"Formula=(0)+(1)*x0+(2)*x1+(3)*x2+(4)*x3:\ 
 ParRanges=(-1,1);(-10,10);(-10,10);(-10,10);(-10,10):\ 
 FitMethod=MINUIT:\
 ErrorLevel=1:PrintLevel=-1:FitStrategy=2:UseImprove:UseMinos:SetBatch"
\end{tmvacode}
\caption[.]{\codeexampleCaptionSize FDA booking option example simulating a linear Fisher
            discriminant (\cf\  Sec.~\ref{sec:fisher}). The 
            top line gives the discriminator expression, where the $xi$ denote the 
            input variables and the $(j)$ denote the coefficients to be determined by the 
            fit. Allowed are all standard functions and expressions, including the functions 
            belonging to the ROOT \href{http://root.cern.ch/root/html/TMath.html}{TMath} library.
            The second line determines the limits for the fit parameters, where the 
            numbers of intervals given must correspond to the number of fit parameters defined.
            The third line defines the fitter to be used (here Minuit), and the last line 
            is the fitter configuration. }
\label{ce:fdaexample}
\end{codeexample}

\subsubsection{Description and implementation}

The parsing of the discriminator function employs ROOT's  
\href{http://root.cern.ch/root/html/TFormula.html}{TFormula} class, which requires that the 
expression complies with its rules (which are the same as those that apply for the 
\code{TTree::Draw} command). For simple formula with a single global fit solution, Minuit will 
be the most efficient fitter. However, if the problem is complicated, highly nonlinear, and/or 
has a non-unique solution space, more involved fitting algorithms may be required. In that 
case the Genetic Algorithm combined or not with a Minuit converger should lead to the 
best results. After fit convergence, FDA prints the fit results (parameters and estimator 
value) as well as the discriminator expression used on standard output. The smaller the 
estimator value, the better the solution found. The normalised estimator is given by 
\beq
\begin{array}{lrcl}
\mbox{\em For classification:} &
      {\cal E} &=& \frac{1}{\WS}\sum_{i=1}^{\NS} \left(F({\bf x}_i)-1\right)^2w_i +
                 \frac{1}{\WB}\sum_{i=1}^{\NB} F^2({\bf x}_i) w_i\;, \\[0.3cm]
\mbox{\em For regression:} &
      {\cal E} &=& \frac{1}{W}\sum_{i=1}^{N} \left(F({\bf x}_i)-{\bf t}_i\right)^2w_i\;,

\end{array}
\eeq
where for classification the first (second) sum is over the signal (background) 
training events, and for regression it is over all training events, 
$F({\bf x}_i)$ is the discriminator function, ${\bf x}_i$ is the tuple of the 
$\Nvar$ input variables for event $i$, $w_i$ is the event weight, ${\bf t}_i$ the 
tuple of training regression targets,  $\WSB$ is the sum of all signal (background) 
weights in the training sample, and $W$ the sum over all training weights.

\subsubsection{Variable ranking}

The present implementation of FDA does not provide a ranking 
of the input variables.

\subsubsection{Performance}

The FDA performance depends on the complexity and fidelity of the user-defined
discriminator function. As a general rule, it should be able to reproduce the 
discrimination power of any linear discriminant analysis. To reach into the nonlinear 
domain, it is useful to inspect the correlation profiles of the input variables, and 
add quadratic and higher polynomial terms between variables as necessary. Comparison
with more involved nonlinear classifiers can be used as a guide.

%%% Local Variables: 
%%% mode: latex
%%% TeX-master: "TMVAUsersGuide"
%%% End: 

%% file: optiontables/MVA__FDA.tex
\begin{optiontableAuto}
                  Formula  &  \mc{1}{c}{--}  &              (0)  &  \mc{1}{l}{--}  &  The discrimination formula \\
                ParRanges  &  \mc{1}{c}{--}  &               ()  &  \mc{1}{l}{--}  &  Parameter ranges \\
                FitMethod  &  \mc{1}{c}{--}  &           MINUIT  &  MC, GA, SA, MINUIT  &  Optimisation Method \\
                Converger  &  \mc{1}{c}{--}  &             None  &  None, MINUIT  &  FitMethod uses Converger to improve result 
\end{optiontableAuto}

%% file: MLPs.tex
\subsection{Artificial Neural Networks
 (nonlinear discriminant analysis\index{Nonlinear discriminant analysis})}
\label{sec:ann}

An Artificial Neural Network \index{Artificial neural networks} (ANN)
is most generally speaking any simulated collection of interconnected
neurons, with each neuron producing a certain response at a given set
of input signals. By applying an external signal to some (input)
neurons the network is put into a defined state that can be measured
from the response of one or several (output) neurons. One can therefore
view the neural network as a mapping from a space of input variables
$x_1,\dots, x_{\rm n_{var}}$ onto a one-dimensional (e.g. in case of a 
signal-versus-background discrimination problem) or multi-dimensional 
space of output variables $y_1,\dots, y_{\rm m_{var}}$. The mapping is 
nonlinear if at least one neuron has a nonlinear response to its input.

In TMVA three neural network implementations are available to the user. The
first was adapted from a FORTRAN code developed at the Universit\'e Blaise
Pascal in Clermont-Ferrand,\footnote
{
   The original Clermont-Ferrand neural network has been used for Higgs 
   search analyses in ALEPH, and background fighting in rare $B$-decay
   searches by the BABAR Collaboration. For the use in TMVA the FORTRAN 
   code has been converted to C++.
} 
the second is the ANN implementation that comes with ROOT. The third is 
a newly developed neural network (denoted {\em MLP}) that is faster and more 
flexible than the other two and is the recommended neural network to use with 
TMVA. All three neural networks are feed-forward multilayer perceptrons.
\index{Multilayer perceptron (MLP)}\index{Feed-forward MLP}

\subsubsection{Booking options}

\subsubsection*{The Clermont-Ferrand neural network\index{Clermont-Ferrand neural network}}

The Clermont-Ferrand neural network is booked via the command:
\begin{codeexample}
\begin{tmvacode}
factory->BookMethod( Types::kCFMlpANN, "CF_ANN", "<options>" );
\end{tmvacode}
\caption[.]{\codeexampleCaptionSize Booking of the Clermont-Ferrand neural network: the first 
            argument is a predefined enumerator, the second argument is a user-defined string 
            identifier, and the third argument is the options string.
            Individual options are separated by a ':'. 
            See Sec.~\ref{sec:usingtmva:booking} for more information on the booking.}
\end{codeexample}

The configuration options for the Clermont-Ferrand neural net are given
in Option Table~\ref{opt:mva::cfmlpann}.

% ======= input option table ==========================================
\begin{option}[t]
\input optiontables/MVA__CFMlpANN.tex
\caption[.]{\optionCaptionSize 
     Configuration options reference for MVA method: {\em CFMlpANN}.
     Values given are defaults. If predefined categories exist, the default category 
     is marked by a '$\star$'. The options in Option Table~\ref{opt:mva::methodbase} on 
     page~\pageref{opt:mva::methodbase} can also be configured.     
     See Sec.~\ref{sec:MLP:hiddenLayers} for a description of the 
     network architecture configuration.
}
\label{opt:mva::cfmlpann}
\end{option}

\subsubsection*{The ROOT neural network 
                (class TMultiLayerPerceptron)\index{ROOT neural network}\index{TMultiLayerPerceptron}}

This neural network interfaces the ROOT class \code{TMultiLayerPerceptron} and is
booked through the Factory via the command line:
\begin{codeexample}
\begin{tmvacode}
factory->BookMethod( Types::kTMlpANN, "TMlp_ANN", "<options>" );
\end{tmvacode}
\caption[.]{\codeexampleCaptionSize Booking of the ROOT neural network: the 
            first argument is a predefined enumerator, the second argument is a 
            user-defined string identifier, and the third argument is the configuration 
            options string. See Sec.~\ref{sec:usingtmva:booking} for more information on 
            the booking.}
\end{codeexample}

The configuration options for the ROOT neural net are given in Option Table~\ref{opt:mva::tmlpann}.
% ======= input option table ==========================================
\begin{option}[t]
\input optiontables/MVA__TMlpANN.tex
\caption[.]{\optionCaptionSize 
     Configuration options reference for MVA method: {\em TMlpANN}.
     Values given are defaults. If predefined categories exist, the default category 
     is marked by a '$\star$'. The options in Option Table~\ref{opt:mva::methodbase} on 
     page~\pageref{opt:mva::methodbase} can also be configured.     
     See Sec.~\ref{sec:MLP:hiddenLayers} for a description of the 
     network architecture configuration.
}
\label{opt:mva::tmlpann}
\end{option}
% =====================================================================

\subsubsection*{The MLP neural network \index{MLP neural network}}

The MLP neural network is booked through the Factory via the command line:
\begin{codeexample}
\begin{tmvacode}
factory->BookMethod( Types::kMLP, "MLP_ANN", "<options>" );
\end{tmvacode}
\caption[.]{\codeexampleCaptionSize Booking of the MLP neural network: the first argument 
            is a predefined enumerator, the second argument is a user-defined string 
            identifier, and the third argument is the options string.
            See Sec.~\ref{sec:usingtmva:booking} for more information on the booking.}
\end{codeexample}

The configuration options for the MLP neural net are given in Option Table~\ref{opt:mva::mlp}.
% ======= input option table ==========================================
\begin{option}[p]
\input optiontables/MVA__MLP.tex
\caption[.]{\optionCaptionSize 
     Configuration options reference for MVA method: {\em MLP}.
     Values given are defaults. If predefined categories exist, the default category 
     is marked by a '$\star$'. The options in Option Table~\ref{opt:mva::methodbase} on 
     page~\pageref{opt:mva::methodbase} can also be configured.     
     See Sec.~\ref{sec:MLP:hiddenLayers} for a description of the 
     network architecture configuration.
}
\label{opt:mva::mlp}
\end{option}
% =====================================================================
The TMVA implementation of MLP supports random and importance event sampling. With 
event sampling enabled, only a fraction (set by the option \code{Sampling}) 
of the training events is used for the training of the MLP. Values in the interval 
$[0,1]$ are possible. If the  option \code{SamplingImportance} is set to 1, the 
events are selected randomly, while for a value below 1 the probability for the same 
events to be sampled again depends on the training performance achieved for 
classification or regression. If for a given set of events the training leads to a 
decrease of the error of the test sample, the probability for the events of being 
selected again is multiplied with the factor given in \code{SamplingImportance} 
and thus decreases. In the case of an increased error of the test sample, the 
probability for the events to be selected again is divided by the factor 
\code{SamplingImportance} and thus increases. The probability for an event to 
be selected is constrained to the interval $[0,1]$. For each set of events, 
the importance sampling described above is performed together with the overtraining 
test.

Event sampling is performed until the fraction specified by the option 
\code{SamplingEpoch} of the total number of epochs (\code{NCycles}) has been 
reached. Afterwards, all available training events are used for the training. 
Event sampling can be turned on and off for training and testing events 
individually with the options \code{SamplingTraining} and \code{SamplingTesting}.

The aim of random and importance sampling is foremost to speed-up the training 
for large training samples. As a side effect, random or importance sampling may
also increase the robustness of the training algorithm with respect to convergence
in a local minimum.

Since it is typically not known beforehand how many epochs are necessary to 
achieve a sufficiently good training of the neural network, a convergence test 
can be activated by setting \code{ConvergenceTests} to a value above 0. This 
value denotes the number of subsequent convergence tests which have to fail 
(\ie no improvement of the estimator larger than \code{ConvergenceImprove}) to 
consider the training to be complete. Convergence tests are performed at the 
same time as overtraining tests. The test frequency is given by the parameter 
\code{TestRate}.

It is recommended to set the option \code{VarTransform=Norm}, such that 
the input (and in case of regression the output as well) is normalised to the 
interval $[-1,1]$.

\subsubsection{Description and implementation}
\label{sec:mlp:impl}

The behaviour of an artificial neural network is determined by the layout of
the neurons, the weights of the inter-neuron connections, and by the
response of the neurons to the input, described by the {\em neuron
  response function} $\rho$.

\subsubsection*{Multilayer Perceptron}

\begin{figure}[t]
  \centering
  \includegraphics[width=0.62\textwidth]{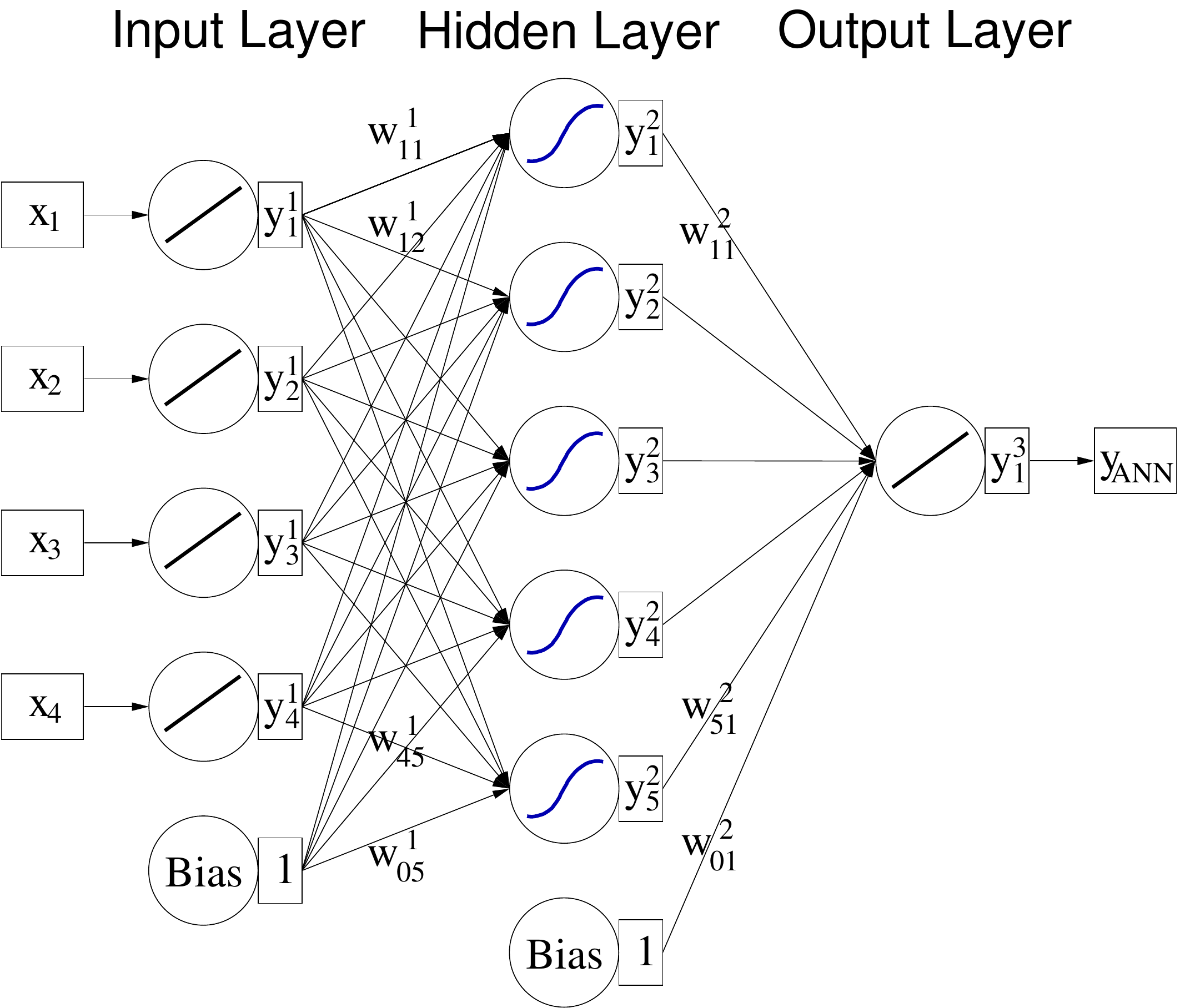}
  \caption[.]{Multilayer perceptron with one hidden layer.}
  \label{fig:mlp:nw}
\end{figure}

While in principle a neural network with $n$ neurons can have $n^2$
directional connections, the complexity can be reduced by organising the
neurons in layers and only allowing direct connections from a given layer 
to the following layer (see Fig.~\ref{fig:mlp:nw}). This kind of
neural network is termed {\em multi-layer perceptron}; all neural net
implementations in TMVA are of this type. The first layer of a multilayer perceptron 
is the input layer, the last one the output layer, and all others are 
{\em hidden} layers.  For a classification problem with \Nvar input
variables the input layer consists of \Nvar
neurons that hold the input values, $x_1,\dots,x_\Nvar$, and one
neuron in the output layer that holds the output variable, the neural
net estimator $\yANN$.

For a regression problem the network structure is similar, except that 
for multi-target regression each of the targets is represented by one
output neuron.
A weight is associated to each directional connection between the output 
of one neuron and the input of another neuron.  When calculating the input 
value to the response function of a neuron, the output values of all 
neurons connected to the given neuron are multiplied with theses weights.
\begin{figure}[t]
  \centering
  \includegraphics[width=.30\textwidth]{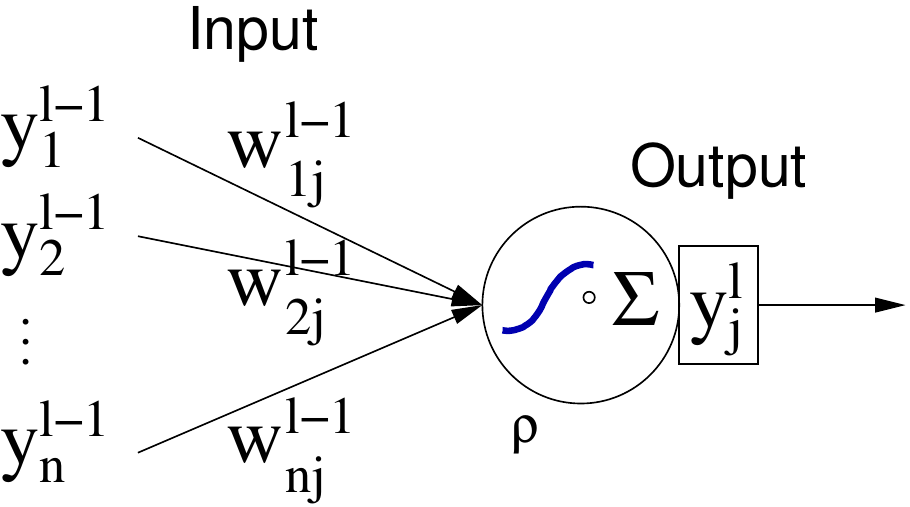}
  \caption[.]{Single neuron $j$ in layer $\ell$ with $n$ input connections. The
    incoming connections carry a weight of $w_{ij}^{(l-1)}$.}
  \label{fig:mlp:node}
\end{figure}

\subsubsection*{Neuron response function}

The neuron response function $\rho$ maps the neuron input
$i_1,\dots,i_n$ onto the neuron output (Fig.~\ref{fig:mlp:node}).
Often it can be separated into a ${\cal R}^n\mapsto{\cal R}$ 
{\em synapse function} $\kappa$, and a ${\cal R}\mapsto{\cal R}$
{\em neuron activation function} $\alpha$, so that $\rho=\alpha\circ\kappa$.
The functions $\kappa$ and $\alpha$ can have the following forms:
\beq
  \label{eq:mlp:synfnc}
  \kappa:~(y_1^{(\ell)},..,y_n^{(\ell)}|w_{0j}^{(\ell)},..,w_{nj}^{(\ell)})\rightarrow 
  \begin{cases}
    w_{0j}^{(\ell)}+\sum\limits_{i=1}^n y_{i}^{(\ell)} w_{ij}^{(\ell)}   
         &\text{\em Sum},\\[0.3cm]
    w_{0j}^{(\ell)}+\sum\limits_{i=1}^n \left(y_{i}^{(\ell)} w_{ij}^{(\ell)}\right)^2 
         &\text{\em Sum of squares},\\[0.3cm]
    w_{0j}^{(\ell)}+\sum\limits_{i=1}^n |y_{i}^{(\ell)} w_{ij}^{(\ell)}| 
         &\text{\em Sum of absolutes},
  \end{cases}  
\eeq
\beq
  \label{eq:mlp:actfnc}
  \alpha:~x\rightarrow
  \begin{cases}
    \ x                                   & \text{\em Linear},\\[0.2cm]
    \ \Dfrac{1}{1+e^{-kx}}                & \text{\em Sigmoid},\\[0.3cm]
    \ \Dfrac{e^{x}-e^{-x}}{e^{x}+e^{-x}}  & \text{\em Tanh},\\[0.3cm]
    \ e^{-x^2/2}                          & \text{\em Radial}.
  \end{cases}
\eeq

\subsubsection{Network architecture}
\label{sec:MLP:hiddenLayers}

The number of hidden layers in a network and the number of neurons in these
layers are configurable via the option \code{HiddenLayers}. For example the 
configuration \code{"HiddenLayers=N-1,N+10,3"} creates a network with three 
hidden layers, the first hidden layer with $\Nvar-1$ neurons, the second with 
$\Nvar+10$ neurons, and the third with 3 neurons.

When building a network two rules should be kept in mind. The first is the 
theorem by Weierstrass, which if applied to neural nets, ascertains 
that for a multilayer perceptron a single 
hidden layer is sufficient to approximate a given continuous correlation function
to any precision, provided that a sufficiently large number of neurons is used 
in the hidden layer. If the available computing power and the size of the training 
data sample suffice, one can increase the number of neurons in the hidden layer 
until the optimal performance is reached.

It is likely that the same performance can be achieved with a
network of more than one hidden layer and a potentially much smaller
total number of hidden neurons. This would lead to a shorter training
time and a more robust network.

\subsubsection{Training of the neural network}

\subsubsection*{Back-propagation (BP)}

The most common algorithm for adjusting the weights that optimise the 
classification performance of a neural network is the so-called 
{\em back propagation.}\index{Back propagation} 
It belongs to the family of supervised learning methods, where the desired output for 
every input event is known. Back propagation is used by all neural networks in TMVA.  
The output of a network (here for simplicity assumed to have a single hidden layer with 
a Tanh activation function, and a linear activation function in the output layer) is 
given by
\beq
  \label{eq:mlp:mvacalc}
  \yANN
  =
  \sum_{j=1}^{n_\text{h}} y_j^{(2)}w_{j1}^{(2)}
  =
  \sum_{j=1}^{n_\text{h}}\tanh\!\left(\sum_{i=1}^\Nvar x_iw_{ij}^{(1)}\right)\cdot w_{j1}^{(2)}\,,
\eeq
where \Nvar and $n_\text{h}$ are the number of neurons in the input
layer and in the hidden layer, respectively, $w^{(1)}_{ij}$ is the
weight between input-layer neuron $i$ and hidden-layer neuron $j$,
and $w^{(2)}_{j1}$ is the weight between the hidden-layer neuron $j$ and the
output neuron. A simple sum was used in Eq.~(\ref{eq:mlp:mvacalc}) 
for the synapse function $\kappa$.

During the learning process the network is supplied with $\Ntrain$ training
events ${\bf x}_a = (x_1,\dots,x_\Nvar)_a$, $a=1,\dots,\Ntrain$. For each 
training event $a$ the neural network output $\yANNa$ is computed
and compared to the desired output $\hat y_a\in\{1,0\}$ (in classification 1 for signal
events and 0 for background events). An {\em error function} $E$, measuring 
the agreement of the network response with the desired one, is defined by
\beq
  E({\bf x}_1,\dots,{\bf x}_{\Ntrain} | {\bf w})
  = 
  \sum_{a=1}^{\Ntrain} E_a({\bf x}_a | {\bf w})
  =
  \sum_{a=1}^{\Ntrain} \frac12 \left(\yANNa - \hat y_{a}\right)^{2}\,,
\eeq
where ${\bf w}$ denotes the ensemble of adjustable weights in the network.
The set of weights that minimises the error function can be found using 
the method of {\em steepest} or {\em gradient descent}, provided that the neuron
response function is differentiable with respect to the input weights. Starting 
from a random set of weights ${\bf w}^{(\rho)}$ the weights are updated by moving a 
small distance in ${\bf w}$-space into the direction $-{\boldsymbol\nabla}_{\bf w} E$
where $E$ decreases most rapidly
\beq
  \label{eq:mlp:weightIter}
  {\bf w}^{(\rho+1)} = {\bf w}^{(\rho)} - \eta {\boldsymbol\nabla}_{\bf w} E\,,
\eeq
where the positive number $\eta$ is the {\em learning rate}.

The weights connected with the output layer are updated by
\beq
\label{eq:mlp:ouputupdate}
  \Delta w_{j1}^{(2)}
  =
  -\eta\sum_{a=1}^{\Ntrain} \frac{\partial E_a}{\partial w_{j1}^{(2)}}
  =
  -\eta\sum_{a=1}^{\Ntrain}\left(\yANNa - \hat y_{a}\right) y^{(2)}_{j,a}\,,
\eeq
and the weights connected with the hidden layers are updated by
\beq
\label{eq:mlp:hiddenupdate}
  \Delta w_{ij}^{(1)}
  = 
  -\eta\sum_{a=1}^{\Ntrain} \frac{\partial E_a}{\partial w_{ij}^{(1)}}
  =
  -\eta\sum_{a=1}^{\Ntrain} \left(\yANNa - \hat y_{a}\right) 
                                   y^{(2)}_{j,a}(1-y^{(2)}_{j,a}) w_{j1}^{(2)} x_{i,a}\,,
\eeq
where we have used $\tanh^\prime x = \tanh x(1-\tanh x)$. This method of training the 
network is denoted {\em bulk learning}, since the sum of errors of all training 
events is used to update the weights. An alternative choice is the so-called
{\em online learning}, where the update of the weights occurs at each event. 
The weight updates are obtained from Eqs.~(\ref{eq:mlp:ouputupdate}) and 
(\ref{eq:mlp:hiddenupdate}) by removing the event summations.
In this case it is important to use a well randomised training sample. 
Online learning is the learning method implemented in TMVA.

\subsubsection*{BFGS}

The Broyden-Fletcher-Goldfarb-Shannon (BFGS)\index{BFGS} method~\cite{BFGS} differs from 
back propagation by the use of second derivatives of the error function to 
adapt the synapse weight by an algorithm which is composed of four main steps.
\begin{enumerate}

\item Two vectors, $D$ and $Y$ are calculated. The vector of weight changes 
      $D$ represents the evolution between one iteration of the algorithm $(k-1)$ 
      to the next $(k)$. Each synapse weight corresponds to one element of the 
      vector. The vector $Y$ is the vector of gradient errors. 
      \beqn
        D_{i}^{(k)} &=& w_i^{(k)} - w_i^{(k-1)}\:,  \\
        Y_{i}^{(k)} &=& g_i^{(k)} - g_i^{(k-1)}\:,
      \eeqn
      where $i$ is the synapse index, $g_i$ is the $i$-th synapse gradient,\footnote
      {
         The synapse gradient is estimated in the same way as in the BP method 
         (with initial gradient and weights set to zero).
      } 
      $w_i$ is the weight of the $i$-th synapse, and $k$ denotes the iteration counter.

\item Approximate the inverse of the Hessian matrix, \IHessian, at iteration $k$ by
      \beq
        \IHessiank = \frac{D\cdot D^{T}\cdot (1+Y^{T}\cdot \IHessiankmone\cdot Y)}{Y^{T}\cdot D} 
                       - D\cdot Y^{T}\cdot H + H\cdot Y\cdot D^{T} + \IHessiankmone\:,
      \eeq
      where superscripts $(k)$ are implicit for $D$ and $Y$. 

\item Estimate the vector of weight changes by
      \beq 
        D^{(k)} = -\IHessiank\cdot Y^{(k)}\:.
      \eeq

\item Compute a new vector of weights by applying a {\em line search} algorithm.
In the line search the error function is locally approximated 
by a parabola. The algorithm evaluates the second derivatives and 
determines the point where the minimum of the parabola 
is expected. The total error is evaluated for this point. The algorithm then 
evaluates points along the line defined by the direction of the gradient 
in weights space to find the absolute minimum. The weights at the 
minimum are used for the next iteration. The learning rate can be set
With the option \code{Tau}. The learning
parameter, which defines by how much the weights are changed in one epoch
along the line where the minimum is suspected, is multiplied with the
 learning rate as long as the training error of the 
neural net with the changed weights is below the one with unchanged weights.
If the training error of the changed neural net were already larger
for the initial learning parameter, it is divided by the learning rate
until the training error becomes smaller. 
The iterative and approximate calculation of $\IHessiank$ turns less 
accurate with an increasing number of iterations. The 
matrix is therefore reset to the unit matrix every \code{ResetStep} steps. 
\end{enumerate}

The advantage of the BFGS method compared to BG is the smaller number of 
iterations. However, because the computing time for one iteration is 
proportional to the squared number of synapses, large networks are 
particularly penalised.

\subsubsection{Variable ranking}
\label{sec:ann:ranking}

The MLP neural network implements a variable ranking that uses the sum of the 
weights-squared of the connections between the variable's neuron in the input 
layer and the first hidden layer. The importance $I_i$ of the input variable 
$i$ is given by
\beq
  \label{eq:mlp:ranking}
  I_i = \overline x_i^2 \sum_{j=1}^{n_\text{h}} \left(w^{(1)}_{ij}\right)^2, \qquad i=1,\dots,\Nvar\,,
\eeq
where $\overline x_{i}$ is the sample mean of input variable $i$.

\subsubsection{Performance}
\label{sec:ann:perf}

In the tests we have carried out so far, the MLP and ROOT networks performed equally well, 
however with a clear speed advantage for the MLP. The Clermont-Ferrand neural net
exhibited worse classification performance in these tests, which is partly due to the slow 
convergence of its training (at least 10k training cycles are required to achieve 
approximately competitive results). 

%%% Local Variables: 
%%% mode: latex
%%% TeX-master: "TMVAUsersGuide"
%%% End: 

%% file: optiontables/MVA__CFMlpANN.tex
\begin{optiontableAuto}
                  NCycles  &  \mc{1}{c}{--}  &             3000  &  \mc{1}{l}{--}  &  Number of training cycles \\
             HiddenLayers  &  \mc{1}{c}{--}  &            N,N-1  &  \mc{1}{l}{--}  &  Specification of hidden layer architecture 
\end{optiontableAuto}

%% file: optiontables/MVA__TMlpANN.tex
\begin{optiontableAuto}
                  NCycles  &  \mc{1}{c}{--}  &              200  &  \mc{1}{l}{--}  &  Number of training cycles \\
             HiddenLayers  &  \mc{1}{c}{--}  &            N,N-1  &  \mc{1}{l}{--}  &  Specification of hidden layer architecture (N stands for number of variables; any integers may also be used) \\
       ValidationFraction  &  \mc{1}{c}{--}  &              0.5  &  \mc{1}{l}{--}  &  Fraction of events in training tree used for cross validation \\
           LearningMethod  &  \mc{1}{c}{--}  &       Stochastic  &  Stochastic, Batch, SteepestDescent, RibierePolak, FletcherReeves, BFGS  &  Learning method 
\end{optiontableAuto}

%% file: optiontables/MVA__MLP.tex
\begin{optiontableAuto}
                  NCycles  &  \mc{1}{c}{--}  &              500  &  \mc{1}{l}{--}  &  Number of training cycles \\
             HiddenLayers  &  \mc{1}{c}{--}  &            N,N-1  &  \mc{1}{l}{--}  &  Specification of hidden layer architecture \\
               NeuronType  &  \mc{1}{c}{--}  &          sigmoid  &  linear, sigmoid, tanh, radial  &  Neuron activation function type \\
          NeuronInputType  &  \mc{1}{c}{--}  &              sum  &  sum, sqsum, abssum  &  Neuron input function type \\
           TrainingMethod  &  \mc{1}{c}{--}  &               BP  &  BP, GA, BFGS  &  Train with Back-Propagation (BP), BFGS Algorithm (BFGS), or Genetic Algorithm (GA - slower and worse) \\
             LearningRate  &  \mc{1}{c}{--}  &             0.02  &  \mc{1}{l}{--}  &  ANN learning rate parameter \\
                DecayRate  &  \mc{1}{c}{--}  &             0.01  &  \mc{1}{l}{--}  &  Decay rate for learning parameter \\
                 TestRate  &  \mc{1}{c}{--}  &               10  &  \mc{1}{l}{--}  &  Test for overtraining performed at each \#th epochs \\
                 Sampling  &  \mc{1}{c}{--}  &                1  &  \mc{1}{l}{--}  &  Only 'Sampling' (randomly selected) events are trained each epoch \\
            SamplingEpoch  &  \mc{1}{c}{--}  &                1  &  \mc{1}{l}{--}  &  Sampling is used for the first 'SamplingEpoch' epochs, afterwards, all events are taken for training \\
       SamplingImportance  &  \mc{1}{c}{--}  &                1  &  \mc{1}{l}{--}  &   The sampling weights of events in epochs which successful (worse estimator than before) are multiplied with SamplingImportance, else they are divided. \\
         SamplingTraining  &  \mc{1}{c}{--}  &             True  &  \mc{1}{l}{--}  &  The training sample is sampled \\
          SamplingTesting  &  \mc{1}{c}{--}  &            False  &  \mc{1}{l}{--}  &  The testing sample is sampled \\
                ResetStep  &  \mc{1}{c}{--}  &               50  &  \mc{1}{l}{--}  &  How often BFGS should reset history \\
                      Tau  &  \mc{1}{c}{--}  &                3  &  \mc{1}{l}{--}  &  LineSearch size step \\
                   BPMode  &  \mc{1}{c}{--}  &       sequential  &  sequential, batch  &  Back-propagation learning mode: sequential or batch \\
                BatchSize  &  \mc{1}{c}{--}  &               -1  &  \mc{1}{l}{--}  &  Batch size: number of events/batch, only set if in Batch Mode, -1 for BatchSize=number\_of\_events \\
       ConvergenceImprove  &  \mc{1}{c}{--}  &                0  &  \mc{1}{l}{--}  &  Minimum improvement which counts as improvement ($<$0 means automatic convergence check is turned off) \\
         ConvergenceTests  &  \mc{1}{c}{--}  &               -1  &  \mc{1}{l}{--}  &  Number of steps (without improvement) required for convergence ($<$0 means automatic convergence check is turned off) 
\end{optiontableAuto}

%% file: SVM.tex
\subsection{Support Vector Machine (SVM)}
\label{sec:SVM}

In the early 1960s a linear support vector method\index{Support vector machine, SVM}
has been developed for the construction of
separating hyperplanes for pattern recognition problems~\cite{Vapnik1963,Vapnik1964}. 
It took 30 years before the method was generalised to nonlinear separating 
functions~\cite{Vapnik1992,Vapnik1995a} and for estimating real-valued functions 
(regression)~\cite{Vapnik1995b}. At that moment it became a general purpose algorithm, 
performing classification and regression tasks which can compete with neural networks 
and probability density estimators. Typical applications of SVMs include text 
categorisation, character recognition, bio-informatics and face detection. 

The main idea of the SVM approach to classification problems is to build a hyperplane 
that separates signal and background {\em vectors} (events) using only a minimal subset 
of all training vectors ({\em support vectors}). The position of the hyperplane is 
obtained by maximizing the margin (distance) between it and the support vectors.  
The extension to nonlinear SVMs is performed by mapping the input vectors onto 
a higher dimensional feature space in which signal and background events can be 
separated by a linear procedure using an optimally separating hyperplane. The use 
of kernel functions eliminates thereby the explicit transformation to the feature 
space and simplifies the computation.

The implementation of the newly introduced regression is similar to the approach in classification.
It also maps input data into higher dimensional space using previously chosen support 
vectors. Instead of separating events of two types, it determines the hyperplane with 
events of the same value (which is equal to the {\em mean} from all training events). 
The final value is estimated based on the distance to the hyperplane which is computed by the selected kernel function.% Comment: currently we can only select Gaus, soon we will have most of the old kernel options back
\subsubsection{Booking options}

The SVM classifier is booked via the command:
\begin{codeexample}
\begin{tmvacode}
factory->BookMethod( TMVA::Types::kSVM, "SVM", "<options>" ); 
\end{tmvacode}
\caption[.]{\codeexampleCaptionSize Booking of the SVM classifier: the first  
            argument is a unique type enumerator, the second is a user-defined  
            name which must be unique among all booked classifiers, and the third argument 
            is the configuration option string. Individual options are separated by a ':'. 
            For options that are not set in the string default values are used. 
            See Sec.~\ref{sec:usingtmva:booking} for more information on the booking.}
\end{codeexample}

The configuration options for the SVM classifier are given in Option Table~\ref{opt:mva::svm}.

% ======= input option table ==========================================
\begin{option}[t]
\input optiontables/MVA__SVM.tex
\caption[.]{\optionCaptionSize 
     Configuration options reference for MVA method: {\em SVM}.
     Values given are defaults. If predefined categories exist, the default category 
     is marked by a '$\star$'. The options in Option Table~\ref{opt:mva::methodbase} on 
     page~\pageref{opt:mva::methodbase} can also be configured.     
     Definition of the kernel function: \code{Linear} is 
     $K(\vec x, \vec y) = \vec x \cdot \vec y $ (no extra parameters),
     \code{Polynomial} is $K(\vec x, \vec y) = (\vec x \cdot \vec y +\theta)^d$,
     \code{Gauss} is 
     $K(\vec x, \vec y) = \exp\big(-\left|\vec x - \vec y \right|^2 /2\sigma^2\big)$,
     and \code{Sigmoid} corresponds to
     $K(\vec x, \vec y) = \tanh\left(\kappa (\vec x \cdot \vec y) +\theta \right)$
}
\label{opt:mva::svm}
\end{option}
% =====================================================================

\subsubsection{Description and implementation}

A detailed  description of the SVM formalism can be found, for example,
in Ref.~\cite{Burges}. Here only a brief introduction of the TMVA 
implementation is given. 

\subsubsection*{Linear SVM}
\index{Support vector machine, SVM!linear}

\begin{figure}[t]
\centering
	  \includegraphics[width=0.47\textwidth]{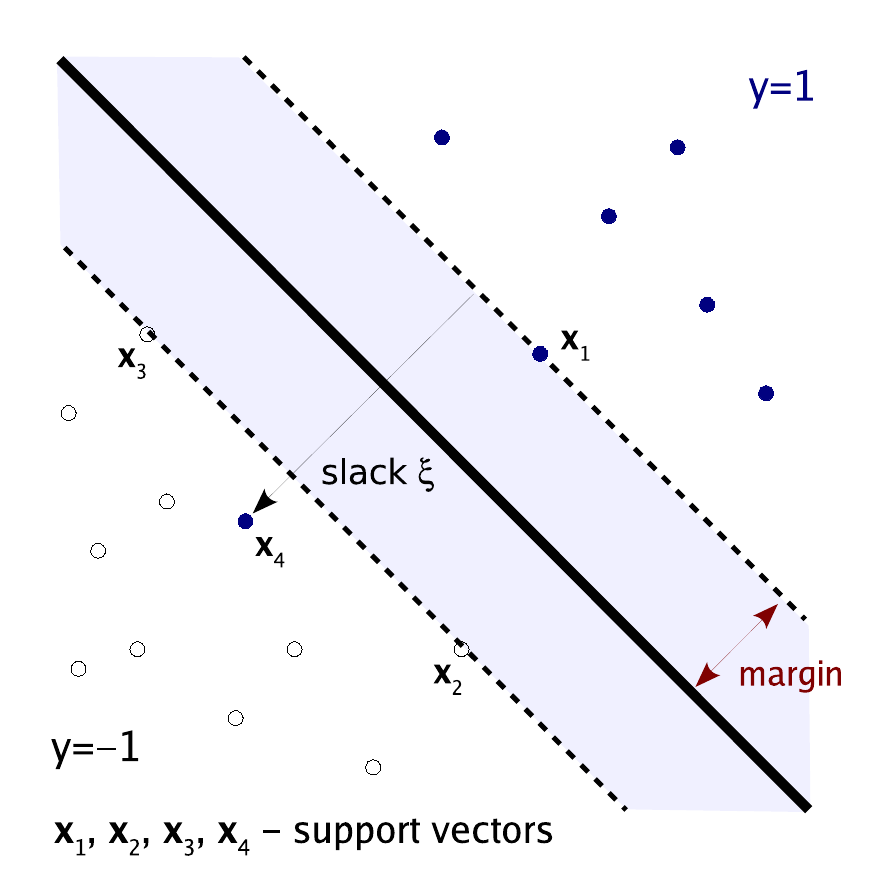}
\caption{ Hyperplane classifier in two dimensions. The vectors (events) ${\bf x}_{1-4}$          
          define the hyperplane and margin, \ie, they are the support vectors.
}
\label{fig:classifier}
\end{figure}
Consider a simple two-class classifier  with oriented hyperplanes.  If the training 
data is linearly separable, a vector-scalar pair $(\vec w, b)$ can be found that 
satisfies the constraints
\beq
   \label{eq:SVM1}
   y_i (\vec {x_i}\cdot \vec w+ b)-1 \ge 0\,, \space \mbox{\hspace{0.5cm}} \forall_i\,,
\eeq
where $\vec x_i$ are the input vectors, $y_i$ the desired outputs ($y_i=\pm 1$),  
and where the pair $(\vec w,b)$ defines a hyperplane. The decision function of 
the classifier is  
$f(\vec x_i)= {\rm sign} (\vec x_i\cdot \vec w+ b)$, which is $+1$ for all points on   
one side of the hyperplane and $-1$ for the points on the other side. 

Intuitively, the classifier with the largest margin will give better separation.
The margin for this linear classifier is just $2/|\vec w|$.  Hence to maximise the 
margin, one needs to minimise the {\em cost function} $W = |\vec w|^2/w$ with the 
constraints from Eq.~(\ref{eq:SVM1}). 

At this point it is beneficial to consider the significance of different input  
vectors $\vec x_i$.  The training events lying on the margins, which are called  
the support vectors (SV), are the events that contribute to defining the decision boundary  
(see Fig.~\ref{fig:classifier}). Hence if the other events are removed from the training 
sample and the classifier is retrained on the remaining events, the training will result 
in the same decision boundary. To solve the constrained quadratic optimisation problem, 
we first reformulate it in terms of a Lagrangian
\beq
   {\cal L}( \vec w, b, \vec \alpha) = \frac{1}{2}\left| \vec w \right|^2-
   \sum_i \alpha_i \left( y_i \left( ( \vec x_i \cdot \vec w ) + b\right)-1\right)
\eeq
where $\alpha_i \ge 0$ and the condition from Eq.~(\ref{eq:SVM1}) must be fulfilled. 
The Lagrangian ${\cal L}$ is minimised with respect to $\vec w$ and $b$  and  
maximised  with respect to $\vec \alpha$. The solution has an expansion in terms  
of a subset of input vectors for which $\alpha_i \ne 0$ (the support vectors):
\beq
   \vec w = \sum_i \alpha_i y_i \vec x_i\,,
\eeq
because $\partial {\cal L} / \partial b =0$ and $\partial {\cal L}/\partial\vec w=0$
hold at the extremum. The optimisation problem translates to finding the vector 
$\vec \alpha$ which maximises
\beq
\label{eq:SVMdot}
   {\cal L}( \vec \alpha )=  \sum_i \alpha_i - 
   \frac{1}{2}  \sum_{ij} \alpha_i \alpha_j y_i y_j \vec x_i \cdot \vec x_j \,.
\eeq
Both the optimisation problem and the final decision function depend only on scalar  
products between input vectors, which is a crucial property for the generalisation 
to the nonlinear case.

\subsubsection*{Nonseparable data}
\label{SVM:nonseparable}
\index{Nonseparable data}

The above algorithm can be extended to non-separable data. The classification  
constraints in Eq.~(\ref{eq:SVM1}) are modified by adding a ``slack'' variable $\xi_i$ to  
it ($\xi_i=0$ if the vector is properly classified, otherwise $\xi_i$ is the distance  
to the decision hyperplane)
\beq
\label{eq:SVM2}
   y_i (\vec {x_i}\cdot \vec w+ b)-1 +\xi_i \ge 0,\space 
   \mbox{\hspace{1cm}} \xi_i\ge 0\,, \mbox{\hspace{0.5cm}} \forall_i\,.
\eeq
This admits a certain amount of misclassification. The training algorithm thus
minimises the modified cost function 
\beq
\label{eq:SVM3}
   W = \frac{1}{2}\left| \vec w \right|^2 + C \sum_i \xi_i\,,
\eeq
describing a trade-off between margin and misclassification. The cost parameter \code{C} 
sets the scale by how much misclassification increases the cost function (see Tab.~\ref{opt:mva::svm}).

\subsubsection*{Nonlinear SVM}
\index{Support vector machine, SVM!nonlinear}

The SVM formulation given above can be further extended to build a nonlinear  
SVM which can classify nonlinearly separable data.
Consider a function $\Phi:\;{\rm R}^\Nvar\to\cal H$, which maps the training data 
from ${\rm R}^\Nvar$, where $\Nvar$ is the number of discriminating input variables, 
to some higher dimensional space $\cal H$. In the $\cal H$ space the signal and background 
events can be linearly separated so that the linear SVM formulation can be applied. We have 
seen in Eq.~(\ref{eq:SVMdot}) that event variables only appear in the form of 
scalar products  $\vec x_i\cdot \vec x_j$, which become $\Phi(\vec x_i)\cdot \Phi(\vec x_j)$ 
in the higher dimensional feature space $\cal H$. The latter scalar product can be approximated 
by a kernel function
\beq
   K(\vec x_i, \vec x_j)\approx\Phi(\vec x_i)\cdot \Phi(\vec x_j)\,,
\eeq
which avoids the explicit computation of the mapping function $\Phi(\vec x)$. 
This is desirable because the exact form of $\Phi(\vec x)$ is hard to derive from
the training data. Most frequently used kernel functions are
\beq
\label{SMV:kernels}
\def\smallEq{\hspace{-0.15cm}=}
\begin{array}{rcll}
  K(\vec x, \vec y) &\smallEq
                      &\hspace{-0.15cm} (\vec x \cdot \vec y +\theta)^d 
                      &\hspace{0.2cm}\text{\em Polynomial}, \\ 
  K(\vec x, \vec y) &\smallEq
                      &\hspace{-0.15cm} \exp\left(-\left|\vec x - \vec y \right|^2 /2\sigma^2\right) 
                      &\hspace{0.2cm} \text{\em Gaussian}, \\
  K(\vec x, \vec y) &\smallEq
                      &\hspace{-0.15cm} \tanh\left(\kappa (\vec x \cdot \vec y) +\theta\right) 
                      &\hspace{0.2cm} \text{\em Sigmoidal}.
\end{array}
\eeq
It was shown in Ref.~\cite{Vapnik1995b} that a suitable function kernel must fulfill  
Mercer's condition
\beq
   \int K(\vec x, \vec y)g(\vec x)g(\vec y)d\vec x d\vec y \ge 0\,,
\eeq
for any function $g$ such that $\int g^2(\vec x) d\vec x$ is finite.
While Gaussian and polynomial kernels are known to comply with Mercer's condition,  
this is not strictly the case for sigmoidal kernels. To extend the linear 
methodology to nonlinear problems one substitutes  $\vec x_i \cdot \vec x_j$ 
by $K(\vec x_i, \vec x_j)$ in Eq.~(\ref{eq:SVMdot}).
Due to Mercer's conditions on the kernel, the corresponding optimisation problem  
is a well defined convex quadratic programming problem with a global minimum. 
This is an advantage of SVMs compared to neural networks where local minima occur.

For regression problems, the same algorithm is used as for classification with the exception that instead of dividing 
events based on their type (signal/background), it separates them based on the value
(larger/smaller than average). In the end, it does not return the sigmoid of the distance 
between the event and the hyperplane, but the distance itself -- increased by the average target value. 

\subsubsection*{Implementation}

The TMVA implementation of the Support Vector Machine follows closely the description  
given in the literature. It employs a sequential minimal optimisation (SMO)~\cite{Platt}  
to solve the quadratic problem. Acceleration of the minimisation is achieved by dividing
a set of vectors into smaller subsets~\cite{Keerthi}. The number of training subsets is 
controlled by option \code{NSubSets}. The SMO method drives the subset 
selection to the extreme by selecting subsets of two vectors (for details see 
Ref.~\cite{Burges}). The pairs of vectors are chosen, using heuristic rules, to achieve
the largest possible improvement (minimisation) per step. Because the working set is of 
size two, it is straightforward to write down the analytical solution. The minimisation 
procedure is repeated recursively until the minimum is found. The SMO algorithm has proven 
to be significantly faster than other methods and has become the most common  
minimisation method used in SVM implementations. The precision of the minimisation 
is controlled by the tolerance parameter \code{Tol} (see Tab.~\ref{opt:mva::svm}). The 
SVM training time can be reduced by increasing the tolerance. Most classification problems 
should be solved with less then 1000 training iterations. Interrupting the SVM algorithm 
using the option \code{MaxIter} may thus be helpful when optimising the SVM training 
parameters.  \code{MaxIter} can be released for the final classifier training.

\subsubsection{Variable ranking}

The present implementation of the SVM classifier does not provide a ranking 
of the input variables.

\subsubsection{Performance}
\label{SVM:performance}
\index{Support vector machine, SVM!performance of}

The TMVA SVM algorithm comes with linear, polynomial, Gaussian and sigmoidal kernel 
functions. With sufficient training statistics, the Gaussian kernel allows to approximate 
any separating function in the input space. It is crucial for the performance of the
SVM to appropriately tune the kernel parameters and the cost parameter \code{C}. 
In case of a Gaussian, the kernel is tuned via option \code{Gamma} which is related to 
the width $\sigma$ by $\Gamma=1/(2 \sigma^2)$.
The optimal tuning of these parameters is specific to the problem and must be done by the user. 

The SVM training time scales with $n^2$, where $n$ is the number of vectors (events) in   
the training data set. The user is therefore advised to restrict the sample size
during the first rough scan of the kernel parameters. Also increasing the minimisation
tolerance helps to speed up the training.

SVM is a nonlinear general purpose classification and regression algorithm with 
a performance similar to neural networks (Sec.~\ref{sec:ann}) or to a multidimensional 
likelihood estimator (Sec.~\ref{sec:pders}).

%% file: optiontables/MVA__SVM.tex
\begin{optiontableAuto}
                        C  &  \mc{1}{c}{--}  &                1  &  \mc{1}{l}{--}  &  Cost parameter \\
                      Tol  &  \mc{1}{c}{--}  &             0.01  &  \mc{1}{l}{--}  &  Tolerance parameter \\
                  MaxIter  &  \mc{1}{c}{--}  &             1000  &  \mc{1}{l}{--}  &  Maximum number of training loops \\
                 NSubSets  &  \mc{1}{c}{--}  &                1  &  \mc{1}{l}{--}  &  Number of training subsets \\
                    Gamma  &  \mc{1}{c}{--}  &                1  &  \mc{1}{l}{--}  &  RBF kernel parameter: Gamma 
\end{optiontableAuto}

%% file: BDTs.tex
\subsection{Boosted Decision and Regression Trees}
\label{sec:bdt}

\begin{figure}[t]
  \begin{center}
          \includegraphics[width=0.60\textwidth]{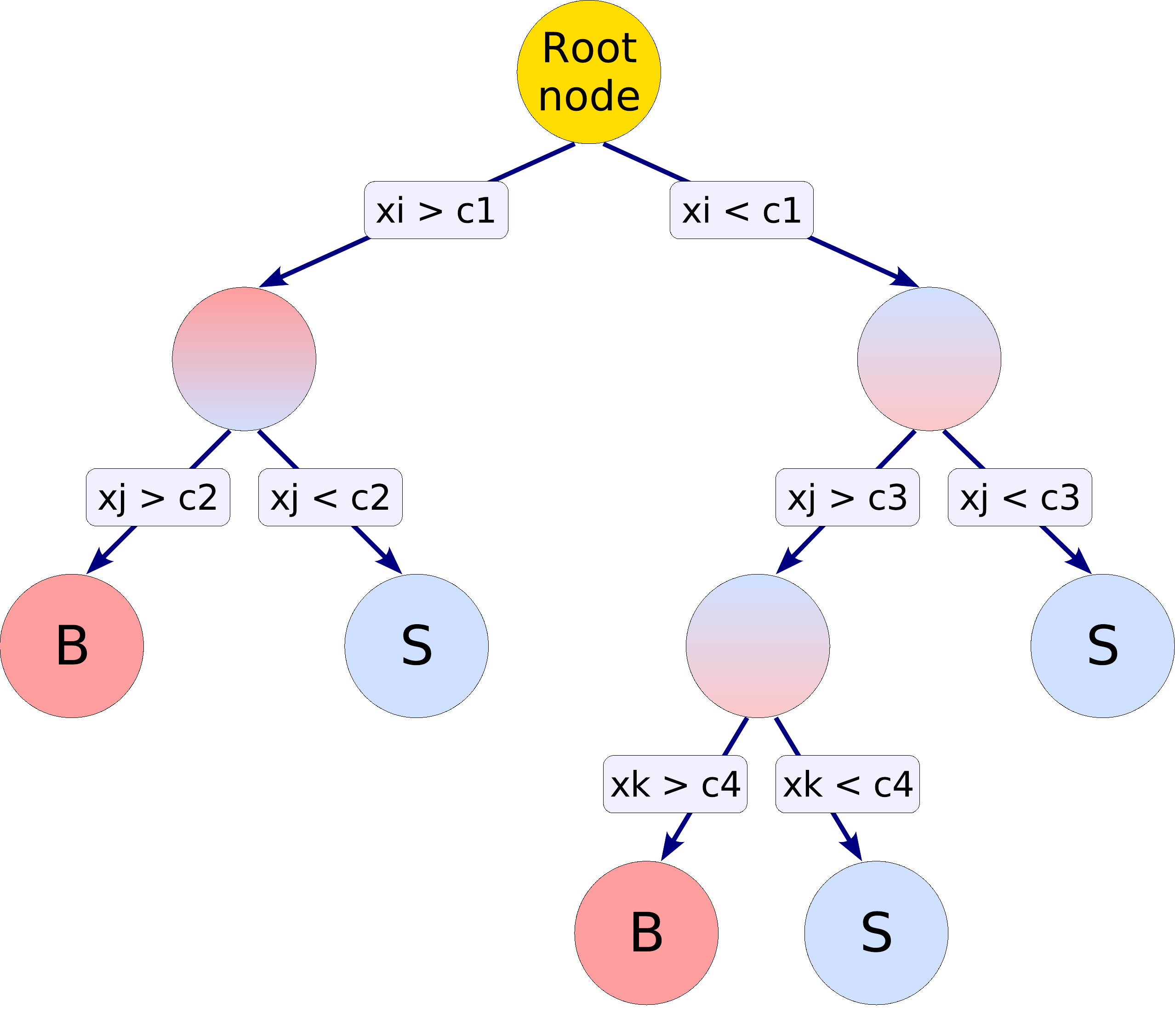}
  \end{center}
  \vspace{-0.3cm}
  \caption[.]{Schematic view of a decision tree.  Starting from the
    root node, a sequence of binary splits using the discriminating
    variables $x_i$ is applied to the data. Each split uses the
    variable that at this node gives the best separation between
    signal and background when being cut on.  The same variable may
    thus be used at several nodes, while others might not be used at
    all.  The leaf nodes at the bottom end of the tree are labeled
    ``S'' for signal and ``B'' for background depending on the
    majority of events that end up in the respective nodes. For
    regression trees, the node splitting is performed on the variable
    that gives the maximum decrease in the average squared error when
    attributing a constant value of the target variable as
    output of the node, given by the average of the training events in
    the corresponding (leaf) node (see Sec.~\ref{sec:treebuilding}). }
\label{fig:decisiontree}
\end{figure}

A {\em decision (regression) tree} (BDT)\footnote{We use the acronym
  BDT for {\em decision} as well as {\em regression} trees.} is a
binary tree structured classifier (regressor) similar to the one sketched in
Fig.~\ref{fig:decisiontree}. Repeated left/right (yes/no) decisions
are taken on one single variable at a time until a stop criterion
is fulfilled.  The phase space is split this way into many regions that are
eventually classified as signal or background, depending on the
majority of training events that end up in the final {\em leaf}
node. In case of {\em regression trees}, each output node represents 
a specific value of the target variable.\footnote{The target variable 
is the variable the regression ``function'' is trying to estimate.} The
boosting (see Sec.~~\ref{sec:boost}) of a decision (regression) tree extends 
this concept from one tree to several trees which form a 
{\em forest}\index{Forest}. The trees are derived from the same 
training ensemble by reweighting events, and are finally combined into a 
single classifier (regressor)
which is given by a (weighted) average of the individual decision
(regression) trees. Boosting stabilizes the response of the decision
trees with respect to fluctuations in the training sample and is able
to considerably enhance the performance w.r.t. a single tree.  In the
following, we will use the term {\em decision tree } for both, {\em
  decision-} and {\em regression trees} and we refer to regression
trees only if both types are treated differently.

\subsubsection{Booking options}

The BDT classifier is booked via the command:
\begin{codeexample}
\begin{tmvacode}
factory->BookMethod( Types::kBDT, "BDT", "<options>" );
\end{tmvacode}
\caption[.]{\codeexampleCaptionSize Booking of the BDT classifier: 
         the first argument is a predefined enumerator, the
         second argument is a user-defined string identifier, and the third
         argument is the configuration options string. 
         Individual options are separated by a ':'. See 
         Sec.~\ref{sec:usingtmva:booking} for more information on the booking.}
\end{codeexample}
% ======= input option table ==========================================
\begin{option}[h!]
\input optiontables/MVA__BDT_1.tex
\caption[.]{\optionCaptionSize Configuration options reference for MVA
  method: {\em BDT}.  Values given are defaults. If predefined
  categories exist, the default category is marked by a '$\star$'. The
  options in Option Table~\ref{opt:mva::methodbase} on
  page~\pageref{opt:mva::methodbase} can also be configured. The table
  is continued in Option Table~\ref{opt:mva::bdt_2}.  }
\label{opt:mva::bdt_1}
\end{option}
\begin{option}[t]
\input optiontables/MVA__BDT_2.tex
\caption[.]{\optionCaptionSize 
     Continuation of Option Table~\ref{opt:mva::bdt_1}.     
}
\label{opt:mva::bdt_2}
\end{option}

Several configuration options are available to customize the BDT
classifier. They are summarized in Option Tables~\ref{opt:mva::bdt_1}
and \ref{opt:mva::bdt_2} and described in more detail in
Sec.~\ref{sec:bdt_descr}.

\subsubsection{Description and implementation}
\label{sec:bdt_descr}

Decision trees are well known classifiers that allow a straightforward
interpretation as they can be visualized by a simple two-dimensional
tree structure. They are in this respect similar to rectangular cuts.
However, whereas a cut-based analysis is able to select only {\em one}
hypercube as region of phase space, the decision tree is able to split
the phase space into a large number of hypercubes, each of which is
identified as either ``signal-like'' or ``background-like'', or
attributed a constant event (target) value in case of a regression
tree. For classification trees, the path down the tree to each leaf
node represents an individual cut sequence that selects signal or
background depending on the type of the leaf node.
 
A shortcoming of decision trees is their instability with respect to 
statistical fluctuations in the training sample from which the tree
structure is derived. For example, if two input variables exhibit
similar separation power, a fluctuation in the training sample may cause the 
tree growing algorithm to decide to split on one variable, while the other 
variable could have been selected without that fluctuation. In
such a case the whole tree structure is altered below this node,
possibly resulting also in a substantially different classifier response.

This problem is overcome by constructing a forest\index{Forest} of decision trees
and classifying an event on a majority vote of the classifications
done by each tree in the forest. All trees in the forest are derived
from the same training sample, with the events being subsequently
subjected to so-called boosting (see~\ref{sec:boost}), a procedure
which modifies their weights in the sample. Boosting increases the
statistical stability of the classifier and typically also improves
the separation performance compared to a single decision tree.
However, the advantage of the straightforward interpretation of the
decision tree is lost. While one can of course still look at a limited
number of trees trying to interpret the training result, one will
hardly be able to do so for hundreds of trees in a
forest. Nevertheless, the general structure of the selection can
already be understood by looking at a limited number of individual
trees.  In many cases, the boosting performs best if applied to
trees (classifiers) that, taken individually, have not much
classification power, i.e. small trees.

\subsubsection{Boosting, Bagging and Randomising}

The different ``boosting'' algorithms (in the following we will 
call also bagging or randomised trees ``boosted'') available
for decision trees in TMVA are currently:
\begin{itemize}
\item AdaBoost (see Sec.~\ref{sec:adaboost}) and AdaBoostR2(see Sec.~\ref{eq:adaboostr2}) for regression
\item Gradient Boost (see Sec.~\ref{sec:gradientboost}) (not for regression)
\item Bagging (see Sec.~\ref{sec:bagging})
\item Randomised Trees\index{Randomising}, like the Random Forests of L.~Breiman~\cite{Breiman2001}. 
  Each tree is grown in such a way that at each split only a random
  subset of all variables is considered. Moreover, each tree in the forest
  is grown using only a (resampled) subset of the original training events.
  The size of the subset as well as the number of variables considered at each
  split can be set using the options~\code{UseNTrainEvents} and \code{UseNVars}.
\end{itemize}

A possible modification of Eq.~(\ref{eq:adaboost}) for the result of
the combined classifier from the forest is to use the training
purity\footnote { The purity of a node is given by the ratio of signal
  events to all events in that node. Hence pure background nodes have
  zero purity.  }  in the leaf node as respective signal or background
{\em weights} rather than relying on the binary decision. This option
is chosen by setting the option \code{UseYesNoLeaf=False}. Such an
approach however should be adopted with care as the purity in the leaf
nodes is sensitive to overtraining and therefore typically
overestimated. Tests performed so far with this option did not show
significant performance increase. Further studies together with tree
pruning are needed to better understand the behaviour of the
purity-weighted BDTs.

\subsubsection*{Training (Building) a decision tree}
\label{sec:treebuilding}

The training, building or {\em growing} of a decision tree is the
process that defines the splitting criteria for each node. The
training starts with the root node, where an initial splitting
criterion for the full training sample is determined. The split
results in two subsets of training events that each go through the
same algorithm of determining the next splitting iteration. This
procedure is repeated until the whole tree is built. At each node, the
split is determined by finding the variable and corresponding cut
value that provides the best separation between signal and background.
The node splitting stops once it has reached the minimum number
of events which is specified in the BDT configuration (option
\code{nEventsMin}).  The leaf nodes are classified as signal or
background according to the class the majority of events belongs to.
If the option \code{UseYesNoLeaf} is set the end-nodes are classified
in the same way. If \code{UseYesNoLeaf} is set to false the end-nodes
are classified according to their purity.
 
A variety of separation criteria can be configured (option
\code{SeparationType} see Option Table~\ref{opt:mva::bdt_2}) 
to assess the performance of a variable and a specific cut requirement. Because a
cut that selects predominantly background is as valuable as one that
selects signal, the criteria are symmetric with respect to the event
classes. All separation criteria have a maximum where the samples are
fully mixed, \ie, at purity $p = 0.5$, and fall off to zero when the
sample consists of one event class only.  Tests have revealed no
significant performance disparity between the following separation
criteria:
\begin{itemize}

\item {\em Gini Index} [default], defined by $p\cdot ( 1 - p )$; \index{Gini Index}

\item {\em Cross entropy}, defined by $-p \cdot \ln (p) - (1-p)\cdot \ln(1-p)$; \index{Cross Entropy}

\item {\em Misclassification error}, defined by $1-{\rm max}(p, 1-p)$; \index{Misclassification error}

\item {\em Statistical significance}, defined by $S/\sqrt{S+B}$;

\item {\em Average squared error}, defined by $1/N\cdot \sum^N (y - {\hat y})^2$ 
  for regression trees where y is the regression target of each event in the node
  and $\hat y$ is its mean value over all events in the node (which would be the
  estimate of y that is given by the node).   
\end{itemize}
Since the splitting criterion is always a cut on a single variable,
the training procedure selects {\em the} variable and cut value that
optimises the {\em increase} in the separation index between the
parent node and the sum of the indices of the two daughter nodes,
weighted by their relative fraction of events. The cut values are
optimised by scanning over the variable range with a granularity that
is set via the option \code{nCuts}. The default value of
\code{nCuts=20} proved to be a good compromise between computing time
and step size. Finer stepping values did not increase noticeably the
performance of the BDTs. However, a truly optimal cut, given the training
sample, is determined by setting \code{nCuts=-1}. This invokes an algorithm
that tests all possible cuts on the training sample and finds the best one.
The latter is of course ``slightly'' slower than the coarse grid.

In principle, the splitting could continue until each leaf node
contains only signal or only background events, which could suggest
that perfect discrimination is achievable.  However, such a decision
tree would be strongly overtrained.  To avoid overtraining a decision
tree must be {\em pruned}.

\subsubsection*{Pruning a decision tree}

Pruning\index{Pruning} is the process of cutting back a tree from the
bottom up after it has been built to its maximum size. Its purpose is
to remove statistically insignificant nodes and thus reduce the
overtraining of the tree. It has been found to be beneficial to first
grow the tree to its maximum size and then cut back, rather than
interrupting the node splitting at an earlier stage. This is because
apparently insignificant splits can nevertheless lead to good splits
further down the tree.  TMVA currently implements two tree pruning
algorithms, which are set by option \code{PruneMethod}.
\begin{itemize}

\item Option \code{PruneMethod=ExpectedError}. For the {\em expected error pruning}~\cite{Quinlan} all leaf nodes for which
      the statistical error estimates of the parent nodes are smaller than
      the combined statistical error estimates of their daughter nodes are
      recursively deleted.  The statistical error
      estimate of each node is calculated using the binomial error
      $\sqrt{p\cdot(1-p)/N}$, where $N$ is the number of training events
      in the node and $p$ its purity.  The amount of pruning is controlled by
      multiplying the error estimate by the fudge factor \code{PruneStrength}.
      Expected error pruning is not available for the regression trees.

\item Option \code{PruneMethod=CostComplexity}. {\em Cost complexity pruning}~\cite{Breiman1984} relates the
  number of nodes in a subtree below a node to the gain in terms of
  misclassified training events by the subtree compared the the node
  itself with {\em no} further splitting. The cost estimate $R$ chosen
  for the misclassification of training events is given by the
  misclassification rate $1-{\rm max}(p, 1-p)$ in a node. The cost
  complexity for this node is then defined by \beq \rho =
  \frac{R(\mbox{node})-R(\mbox{subtree below that node})}
       {\#\mbox{nodes}(\mbox{subtree below that node}) -1} \,.  \eeq
       The node with the smallest $\rho$ value in the tree is
       recursively pruned away as long as $\rho < {\tt
         PruneStrength}$.
       While for classification trees, one typically uses just the misclassification
       error in the pruning, but Gini-Index for the node splitting, regression trees use in
       both cases the squared error loss.
\end{itemize}
Note that the pruning is performed {\em after} the boosting so that
the error fraction used by AdaBoost is derived from the unpruned tree.
 
If the \code{PruneStrength} option is set to a negative value, an
algorithm attempts to automatically detect the optimal strength
parameter. The training sample is divided into two subsamples, of
which only one is used for training, while the other one serves for
validation. The tree is pruned sequentially starting from the node which
has the smallest value of the cost-complexity in the tree. After each pruning
step the performance of the tree is assessed using the validation sample. This
process is repeated until the ROOT node would be pruned. As optimal 
prune strength for this tree the value is chose which corresponds to the best
performing tree using the validation sample.

While this type of pruning obviously gives the ``optimally pruned tree'' given
the training data, it is not completely clear yet if this also applies for the
tree in the forest. Currently it looks as if in TMVA, better results for the
whole forest are often achieved when pruning is not applied, but rather the
maximal tree depth is set to a relatively small value (3 or 4) already during
the tree building phase. 

Note that the Gradient boost does not apply a pruning algorithm
and ignores option \code{PruneMethod}. In this case it is recommended
that the user restricts the number of nodes in the tree to values
between 5 to 20 by using option \code{NNodesMax} or the maximal allowed depth of the
tree \code{MaxDepth}.

\subsubsection{Variable ranking}

A ranking of the BDT input variables is derived by counting how often the
variables are used to split decision tree nodes, and by weighting each 
split occurrence by the separation gain-squared it has achieved and by 
the number of events in the node~\cite{Breiman1984}. This measure of the
variable importance can be used for a single decision tree as well as for a 
forest.

\subsubsection{Performance}

Only limited experience has been gained so far with boosted decision
trees in HEP. In the literature decision trees are sometimes referred
to as the best ``out of the box'' classifiers. This is because little
tuning is required in order to obtain reasonably good results. This is
due to the simplicity of the method where each training step (node
splitting) involves only a one-dimensional cut optimisation. Decision
trees are also insensitive to the inclusion of poorly discriminating
input variables. While for artificial neural networks it is typically
more difficult to deal with such additional variables, the decision
tree training algorithm will basically ignore non-discriminating
variables as for each node splitting only the best discriminating
variable is used.  However, the simplicity of decision trees has the
drawback that their theoretically best performance on a
given problem is generally inferior to other techniques like neural
networks. This is seen for example using the academic training samples
included in the TMVA package. For this sample, which has equal RMS but
shifted mean values for signal and background and linear correlations
between the variables only, the Fisher discriminant provides
theoretically optimal discrimination results. While the artificial
neural networks are able to reproduce this optimal selection
performance the BDTs always fall short in doing so. However, in other
academic examples with more complex correlations or real life
examples, the BDTs often outperform the other techniques.  This is
because either there are not enough training events available that
would be needed by the other classifiers, or the optimal configuration
(\ie\ how many hidden layers, which variables) of the neural network
has not been specified. We have only very limited experience at the
time with the regression, hence cannot really comment on the performance
in this case.

%% file: optiontables/MVA__BDT_1.tex
\begin{optiontableAuto}
                   NTrees  &  \mc{1}{c}{--}  &              200  &  \mc{1}{l}{--}  &  Number of trees in the forest \\
                BoostType  &  \mc{1}{c}{--}  &         AdaBoost  &  AdaBoost, Bagging, RegBoost, AdaBoostR2, Grad  &  Boosting type for the trees in the forest \\
           AdaBoostR2Loss  &  \mc{1}{c}{--}  &         Quadratic &  Linear, Quadratic, Exponential  &  Loss type used in AdaBoostR2 \\
            UseBaggedGrad  &  \mc{1}{c}{--}  &            False  &  \mc{1}{l}{--}  &  Use only a random subsample of all events for growing the trees in each iteration. (Only valid for GradBoost)\\
      GradBaggingFraction  &  \mc{1}{c}{--}  &              0.6  &  \mc{1}{l}{--}  &  Defines the fraction of events to be used in each iteration when \mbox{UseBaggedGrad=kTRUE}.\\
                Shrinkage  &  \mc{1}{c}{--}  &                1  &  \mc{1}{l}{--}  &  Learning rate for GradBoost algorithm \\
             AdaBoostBeta  &  \mc{1}{c}{--}  &                1  &  \mc{1}{l}{--}  &  Parameter for AdaBoost algorithm \\
       UseRandomisedTrees  &  \mc{1}{c}{--}  &            False  &  \mc{1}{l}{--}  &  Choose at each node splitting a random set of variables \\
                 UseNvars  &  \mc{1}{c}{--}  &                4  &  \mc{1}{l}{--}  &  Number of variables used if randomised tree option is chosen \\
           UseNTrainEvent  &  \mc{1}{c}{--}  &                N  &  \mc{1}{l}{--}  &  Number of Training events used in each tree building if randomised tree option is chosen \\
         UseWeightedTrees  &  \mc{1}{c}{--}  &             True  &  \mc{1}{l}{--}  &  Use weighted trees or simple average in classification from the forest \\
             UseYesNoLeaf  &  \mc{1}{c}{--}  &             True  &  \mc{1}{l}{--}  &  Use Sig or Bkg categories, or the purity=S/(S+B) as classification of the leaf node \\
          NodePurityLimit  &  \mc{1}{c}{--}  &              0.5  &  \mc{1}{l}{--}  &  In boosting/pruning, nodes with purity $>$ NodePurityLimit are signal; background otherwise.\\ 
          SeparationType  &  \mc{1}{c}{--}  &        GiniIndex  &  CrossEntropy, GiniIndex, GiniIndexWithLaplace, MisClassificationError, SDivSqrtSPlusB, RegressionVariance  &  Separation criterion for node splitting \\
\end{optiontableAuto}

%% file: optiontables/MVA__BDT_2.tex
\begin{optiontableAuto}
               nEventsMin  &  \mc{1}{c}{--}  &               max(20,NEvtsTrain/NVar$^{\tt 2}$/10)  &  \mc{1}{l}{}  &  Minimum number of events required in a leaf node (default uses given formula)  \\
                    nCuts  &  \mc{1}{c}{--}  &               20  &  \mc{1}{l}{--}  &  Number of steps during node cut optimisation \\
            PruneStrength  &  \mc{1}{c}{--}  &               -1  &  \mc{1}{l}{--}  &  Pruning strength \\
              PruneMethod  &  \mc{1}{c}{--}  &   CostComplexity  &  NoPruning, ExpectedError, CostComplexity  &  Method used for pruning (removal) of statistically insignificant branches \\
         PruneBeforeBoost  &  \mc{1}{c}{--}  &            False  &  \mc{1}{l}{--}  &  Flag to prune the tree before applying boosting algorithm \\
       PruningValFraction  &  \mc{1}{c}{--}  &              0.5  &  \mc{1}{l}{--}  &  Fraction of events to use for optimizing automatic pruning. \\
                NNodesMax  &  \mc{1}{c}{--}  &           100000  &  \mc{1}{l}{--}  &  Max number of nodes in tree \\
                 MaxDepth  &  \mc{1}{c}{--}  &           100000  &  \mc{1}{l}{--}  &  Max depth of the decision tree allowed 
\end{optiontableAuto}

%% file: RuleFit.tex
\subsection{Predictive learning via rule ensembles (RuleFit)}
\label{sec:rulefit}

This classifier is a TMVA implementation of Friedman-Popescu's RuleFit\index{RuleFit}
method described in~\cite{RuleFit}. Its idea is to use an 
ensemble\index{Rules!ensemble of} of so-called {\em rules} to create 
a scoring function with good classification power. Each rule $r_i$ is defined by a 
sequence 
of cuts, such as
\begin{align*}
r_1({\bf x}) & = I(x_2<100.0) \cdot I(x_3>35.0)\,,\\
r_2({\bf x}) & = I(0.45<x_4<1.00) \cdot I(x_1>150.0)\,,\\
r_3({\bf x}) & = I(x_3<11.00)\,,
\end{align*}
where the $x_i$ are discriminating input variables, and $I(\dots)$ returns the truth of 
its argument. A rule applied on a given event is non-zero only if all of its cuts are 
satisfied, in which case the rule returns 1.

The easiest way to create an ensemble of rules is to extract it from a forest 
of decision trees (\cf\  Sec.~\ref{sec:bdt}). Every node in a tree (except the root node) 
corresponds to a sequence of cuts required to reach the node from the root node, 
and can be regarded as a rule.  Hence for the tree illustrated in 
Fig.~\ref{fig:decisiontree} on page~\pageref{fig:decisiontree} 
a total of 8 rules can be formed. Linear combinations of the 
rules in the ensemble are created with coefficients (rule weights) calculated using a 
regularised minimisation procedure~\cite{RuleFitMin}. The resulting linear combination 
of all rules defines a {\em score} function (see below) which provides the RuleFit 
response $\yRF({\bf x})$. 

In some cases a very large rule ensemble is required to obtain a competitive 
discrimination between signal and background. A particularly difficult situation 
is when the true (but unknown) scoring function is described by a linear 
combination of the input variables. \index{Rules!linear terms}
In such cases, \eg, a Fisher discriminant
would perform well. To ease the rule optimisation task, a linear combination of the 
input variables is added to the model. The minimisation procedure will then select the 
appropriate coefficients for the rules {\em and} the linear terms. More details are given in 
Sec.~\ref{sec:RuleFitDescript} below.

\subsubsection{Booking options}

The RuleFit classifier is booked via the command:
\begin{codeexample}
\begin{tmvacode}
factory->BookMethod( Types::kRuleFit, "RuleFit", "<options>" );
\end{tmvacode}
\caption[.]{\codeexampleCaptionSize Booking of RuleFit: the first argument is a
            predefined enumerator, the second argument is a 
            user-defined string identifier, and the third argument is the 
            configuration options string. Individual options are separated by a ':'. 
            See Sec.~\ref{sec:usingtmva:booking} for more information on the booking.}
\end{codeexample}
The RuleFit configuration options are given in Option 
Table~\ref{opt:mva::rulefit}.

% ======= input option table ==========================================
\begin{option}[!t]
\input optiontables/MVA__RuleFit.tex
\caption[.]{\optionCaptionSize 
     Configuration options reference for MVA method: {\em RuleFit}.
     Values given are defaults. If predefined categories exist, the default category 
     is marked by a '$\star$'. The options in Option Table~\ref{opt:mva::methodbase} on 
     page~\pageref{opt:mva::methodbase} can also be configured.     
}
\label{opt:mva::rulefit}
\end{option}
% =====================================================================

\subsubsection{Description and implementation}
\label{sec:RuleFitDescript}

As for all TMVA classifiers, the goal of the rule learning is to find a classification 
function $\yRF({\bf x})$ that optimally classifies an event according to the tuple of 
input observations (variables) ${\bf x}$. The classification function is written as
\beq
\label{eq:F}
 \yRF({\bf x}) = a_0 + \sum_{m=1}^{M_R} a_m f_m({\bf x})\,,
\eeq
where the set $\{f_m({\bf x})\}_{M_R}$ forms an ensemble of {\em base learners}
\index{Rules!base learners}
with $M_R$ elements. A base learner may be any discriminating function derived from 
the training data. In our case, they consist of rules and linear terms as described 
in the introduction. The complete model then reads\index{Rules!linear terms}
\beq
\label{eq:FL}
 \yRF({\bf x}) = a_0 + \sum_{m=1}^{M_R} a_m r_m({\bf x}) + \sum_{i=1}^\Nvar b_i x_i\,.
\eeq
To protect against outliers, the variables in the linear terms are modified to
\beq
 x^\prime_i = \min(\delta^{+}_{i},\max(\delta^{-}_{i}))\,,
\eeq
where $\delta^{\pm}_i$ are the lower and upper $\beta$ quantiles\footnote
{
   Quantiles are points taken at regular intervals from the PDF of a random 
   variable. For example, the 0.5 quantile corresponds to the median of the PDF. 
}
of the variable $x_i$. The value of $\beta$ is set by the option \code{LinQuantile}.
If the variables are used ``as is'', they may have an unequal {\em a priori} 
influence relative to the rules. To counter this effect, the variables are normalised
\beq
 x^\prime_i \rightarrow \sigma_{r} \cdot x^\prime_i/\sigma_{i}\,,
\eeq
where $\sigma_{r}$ and $\sigma_{i}$ are the estimated standard deviations 
of an ensemble of rules and the variable $x^\prime_i$, respectively.

\subsubsection*{Rule generation}

The rules are extracted from a forest of\index{Rules!generation of}
decision trees. There are several ways to generate a forest. In the 
current RuleFit implementation, each tree is generated using 
a fraction of the training sample. The fraction depends on which method is
used for generating the forest. Currently two methods are supported (selected
by option \code{ForestType}); \textit{AdaBoost}
and \textit{Random Forest}. The first method is described in Sec.~\ref{sec:bdt_descr}. 
In that case, the whole training set is used for all trees. The diversity is 
obtained through using different event weights for each tree. For a random forest, 
though, the diversity is created by training each tree using random sub-samples. 
If this method is chosen, the fraction is calculated from the training sample size 
$N$ (signal and background) using the empirical formula~\cite{RuleFitWeb}
\beq
\label{eq:sampfrac}
   f =  \min( 0.5, (100.0 +6.0\cdot\sqrt{N})/N)\,.
\eeq
By default, \code{AdaBoost} is used for creation of the forest.
In general it seems to perform better than the random forest.

The topology of each tree is controlled by the
parameters \code{fEventsMin} and \code{fEventsMax}.
They define a range of fractions which are
used to calculate the minimum number of events required in a node for
further splitting. For each tree, a fraction is drawn from a uniform
distribution within the given range. The obtained fraction is then
multiplied with the number of training events used for the tree,
giving the minimum number of events in a node to allow for splitting.
In this way both large trees (small fraction) giving complex rules and
small trees (large fraction) for simple rules are created.  For a
given forest of $N_t$ trees, where each tree has $n_\ell$ leaf nodes,
the maximum number of possible rules is
\beq
   M_{R,\rm max} = \sum_{i=1}^{N_t} 2(n_{\ell,i} - 1)\,.
\eeq
To prune similar rules, a {\em distance} is defined between two
{\em topologically equal} rules. Two rules are topologically equal if 
their cut sequences follow the same variables only differing in their cut values.
The rule distance used in TMVA is then defined by\index{Rules!distance between}
\beq
   \delta_R^2 = \sum_{i} \frac{\delta_{i,L}^2 + \delta_{i,U}^2}{\sigma_i^2}\,,
\label{eq:ruleDist}
\eeq
where $\delta_{i,L(U)}$ is the difference in lower (upper) limit between
the two cuts containing the variable $x_i$, $i=1,\dots,\Nvar$. The difference is 
normalised to the RMS-squared $\sigma_i^2$ of the variable.  Similar rules with 
a distance smaller than \code{RuleMinDist} are removed from the rule
ensemble. The parameter can be tuned to improve speed and to suppress
noise. In principle, this should be achieved in the fitting procedure. However,
pruning the rule ensemble using a distance cut will reduce the fitting time and 
will probably also reduce the number of rules in the final model.
Note that the cut should be used with care since a too large
cut value will deplete the rule ensemble and weaken its classification
performance.

\subsubsection*{Fitting}
\label{sec:rulefitting}

Once the rules are defined, the coefficients in Eq.~(\ref{eq:FL}) are fitted using
\index{Rules!fitting of}
the training data. For details, the fitting method is described in~\cite{RuleFitMin}. 
A brief description is provided below to motivate the corresponding RuleFit options.

A {\em loss function} $L(\yRF({\bf x})|\hat y)$, given by the ``squared-error 
ramp''~\cite{RuleFitMin} 
\beq
   L(\yRF|\hat y) = \left( \hat y - H(\yRF) \right)^2\,,
\eeq
where $H(y) = \max(-1,\min(\yRF,1))$, quantifies the ``cost'' of misclassifying
\index{Rules!loss function of} an event of given true class $\hat y$. The {\em risk} 
$R$ is defined by the expectation value of $L$ with respect to ${\bf x}$ and 
the true class. Since the true distributions are generally not known, the average 
of $N$ training events is used as an estimate\index{Rules!risk of}
\beq
   R = \frac{1}{N}\sum_{i=1}^{N} L(\yRF({\bf x}_i)|\hat y_i)\,.
\label{eq:risk}
\eeq
A line element in the parameter space of the rule weights (given by the vector ${\bf a}$ 
of all coefficients) is then defined by
\index{Rules!fit path}
\beq
\label{eq:rulefitpath}
{\bf a}(\epsilon + \delta \epsilon) = {\bf a}(\epsilon) + \delta \epsilon \cdot {\bf g}(\epsilon)\,,
\eeq
where $\delta \epsilon$ is a positive small increment and ${\bf
  g}(\epsilon)$ is the negative derivative of the estimated risk $R$,
evaluated at ${\bf a}(\epsilon)$.  The estimated risk-gradient is evaluated
using a sub-sample (\code{GDPathEveFrac}) of the training events.

Starting with all weights set to zero, the consecutive application of 
Eq.~(\ref{eq:rulefitpath}) creates a path in the ${\bf a}$ space.
At each step, the procedure selects only the gradients $g_k$
with absolute values greater than a certain fraction ($\tau$) of the largest
gradient. The fraction $\tau$ is an {\em a priori} unknown quantity between 0 and 1.
With $\tau=0$ all gradients will be used at each step, while
only the strongest gradient is selected for $\tau=1$.
A measure of the ``error'' at each step is calculated by evaluating the risk 
(Eq.~\ref{eq:risk}) using the validation sub-sample (\code{GDValidEveFrac}). 
By construction, the risk will always decrease at each step. However, for the 
validation sample the value will increase once the model starts to be overtrained.
Currently, the fitting is crudely stopped when the error measure is larger than 
\code{GDErrScale} times the minimum error found.
The number of steps is controlled by \code{GDNSteps} and the step size ($\delta \epsilon$ in Eq.~\ref{eq:rulefitpath}) by \code{GDStep}.

\begin{figure}[t]
  \begin{center}
          \includegraphics[width=0.55\textwidth]{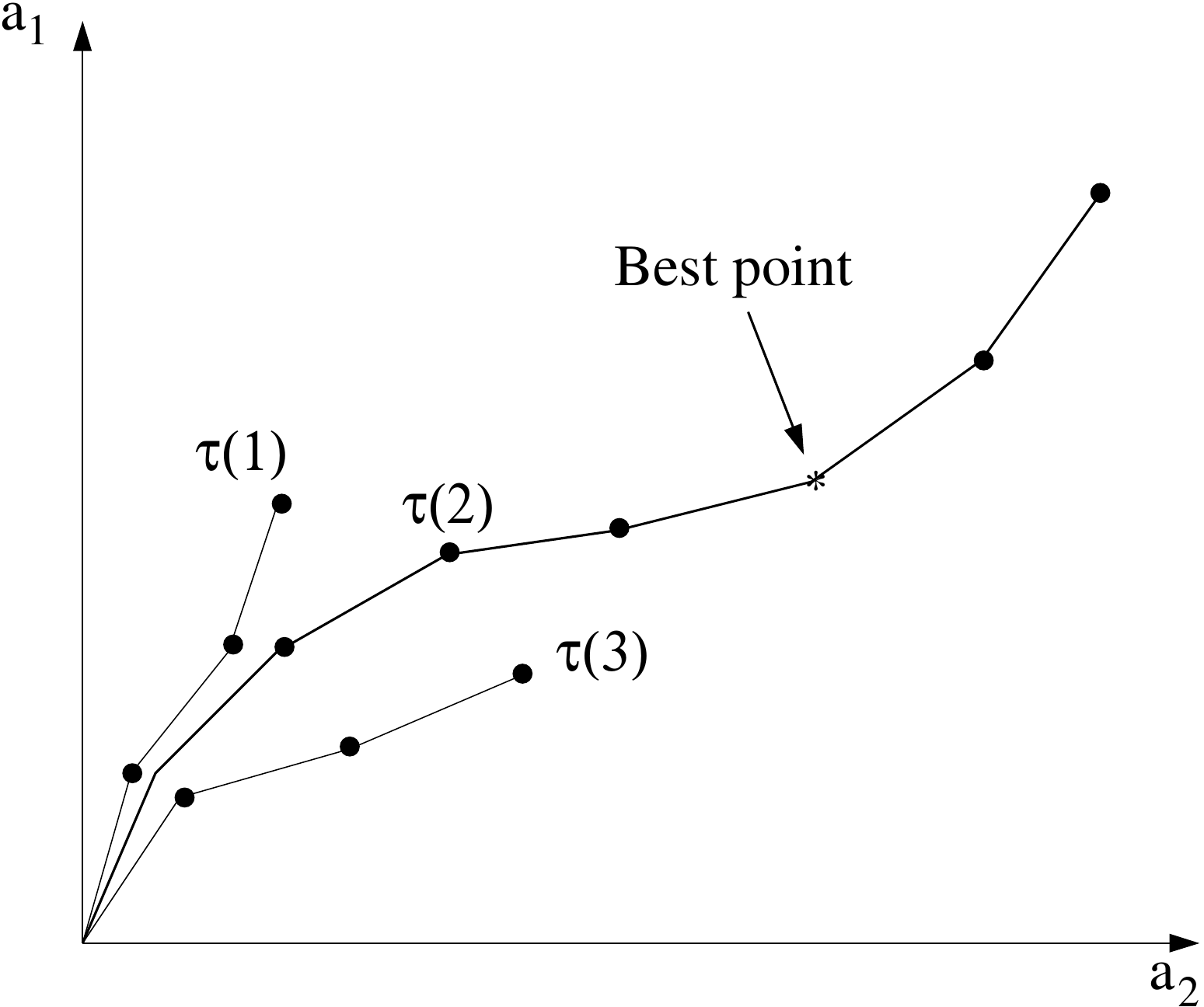}
  \end{center}
  \vspace{-0.3cm}
  \caption[.]{An example of a path scan in two dimensions. Each point represents 
              an $\epsilon$ in Eq.~(\ref{eq:rulefitpath}) and each step is given by 
              $\delta\epsilon$. The direction along the path at each point is given 
              by the vector ${\bf g}$. For the first few points, the paths $\tau(1,2,3)$ 
              are created with different values of $\tau$. After a given number of 
              steps, the best path is chosen and the search is continued. It stops 
              when the best point is found. That is, when the estimated error-rate 
              is minimum.}
\label{fig:rulefitpath}
\end{figure}

If the selected $\tau$ (\code{GDTau}) is a negative number, the best value is 
estimated by means of a scan. In such a case several paths are fitted in parallel, 
each with a different value of $\tau$. The number of paths created depend on the 
required precision on $\tau$ given by \code{GDTauPrec}. By only selecting the 
paths being ``close enough`` to the minimum at each step, the speed for 
the scan is kept down.
The path leading to the lowest estimated error is then selected.
Once the best $\tau$ is found, the fitting proceeds until a minimum is found.
A simple example with a few scan points is illustrated 
in Fig.~\ref{fig:rulefitpath}.

\subsubsection{Variable ranking}

Since the input variables are normalised, the ranking of variables follows naturally 
from the coefficients of the model. To each rule $m$ ($m=1,\dots,M_R$) can be assigned 
an importance defined by
\beq
\label{eq:rulefit:importance}
 I_m = |a_m| \sqrt{s_m (1.0-s_m)}\,,
\eeq
where $s_m$ is the {\em support} of the rule with the following definition
\beq
 s_m = \frac{1}{N} \sum_{n=1}^N r_m( {\bf x}_n )\,.
\eeq
The support is thus the average response for a given rule on the data sample.
A large support implies that many events pass the cuts of the rule. Hence, such 
rules cannot have strong discriminating power. On the other hand, rules with small 
support only accept few events. They may be important for these few events they accept, 
but they are not in the overall picture. The definition~(\ref{eq:rulefit:importance}) 
for the rule importance suppresses rules with both large and small support.

For the linear terms, the definition of importance is
\beq
 I_i = |b_i|\cdot \sigma_{i}\,,
\eeq
so that variables with small overall variation will be assigned a 
small importance.\index{Rules!variable importance}

A measure of the variable importance may then be defined by
\beq
 J_i = I_i + \sum_{m|x_i\in r_m} I_m/q_m\,,
\eeq
where the sum is over all rules containing the variable $x_i$, and $q_m$ is the number of 
variables used in the rule $r_m$. This is introduced in order to share the importance 
equally between all variables in rules with more than one variable.

\subsubsection{Friedman's module}
\label{sec:friedman}
By setting \code{RuleFitModule} to \code{RFFriedman}, the interface to Friedman's RuleFit 
is selected. To use this module, a separate setup is required. If the module is selected 
in a run prior to setting up the environment, TMVA will stop and give instructions on how 
to proceed. A command sequence to setup Friedman's RuleFit in a UNIX environment is:
\begin{codeexample}
\begin{tmvacode}
~> mkdir rulefit
~> cd rulefit
~> wget http://www-stat.stanford.edu/~jhf/r-rulefit/linux/rf_go.exe
~> chmod +x rf_go.exe
\end{tmvacode}
\caption[.]{\codeexampleCaptionSize The first line creates a working directory 
            for Friedman's module. In the third line, the binary executable is fetched 
            from the official web-site. Finally, it is made sure that the module
            is executable.}
\end{codeexample}
As of this writing, binaries exists only for Linux and Windows. Check J.~Friedman's 
home page at \url{http://www-stat.stanford.edu/~jhf} for updated information.
When running this module from TMVA, make sure that the option \code{RFWorkDir}
is set to the proper working directory (default is \code{./rulefit}).
Also note that only the following options are used:
\code{Model}, \code{RFWorkDir}, \code{RFNrules}, \code{RFNendnodes}, 
\code{GDNSteps}, \code{GDStep} and \code{GDErrScale}. The options \code{RFNrules} and \code{RFNendnodes} correspond in the package by Friedman-Popescu to the options \code{max.rules} and \code{tree.size}, respectively. For more details, the reader is referred to Friedman's RuleFit manual~\cite{RuleFitWeb}. 

\subsubsection*{Technical note}
The module \code{rf_go.exe} communicates with the user by means of both ASCII and 
binary files. This makes the input/output from the module machine dependant. TMVA 
reads the output from \code{rf_go.exe} and produces the normal machine independent 
weight (or class) file. This can then be used in other applications and environments.

\subsubsection{Performance}

Rule ensemble based learning machines are not yet well known within the
HEP community, although they start to receive some
attention~\cite{RuleSusy}. Apart from RuleFit~\cite{RuleFit}
other rule ensemble learners exists, such as SLIPPER~\cite{SLIPPER}.

The TMVA implementation of RuleFit follows closely the original design
described in Ref.~\cite{RuleFit}. Currently the performance is however slightly
less robust than the one of the Friedman-Popescu package. Also, the experience 
using the method is still scarce at the time of this writing.

To optimise the performance of RuleFit several
strategies can be employed.  The training consists of two steps, rule
generation and rule ensemble fitting. One approach is to modify the
complexity of the generated rule ensemble by changing either the
number of trees in the forest, or the complexity of each tree. In general, 
large tree ensembles with varying trees sizes perform better than short non-complex
ones. The drawback is of course that fitting becomes slow.  However, if the
fitting performs well, it is likely that a large amount of rules will
have small or zero coefficients. These can be removed, thus
simplifying the ensemble. The fitting performance can be improved by 
increasing the number of steps along with using a smaller step size. Again, 
this will be at the cost of speed performance although only at the training stage.
The setting for the parameter $\tau$ may greatly affect the result. Currently 
an automatic scan is performed by default. In general, it should find the 
optimum $\tau$. If in doubt , the user may set the value explicitly.
In any case, the user is initially advised to use the automatic scan option to 
derive the best path.

%% file: optiontables/MVA__RuleFit.tex
\begin{optiontableAuto}
                    GDTau  &  \mc{1}{c}{--}  &               -1  &  \mc{1}{l}{--}  &  Gradient-directed (GD) path: default fit cut-off \\
                GDTauPrec  &  \mc{1}{c}{--}  &             0.01  &  \mc{1}{l}{--}  &  GD path: precision of tau \\
                   GDStep  &  \mc{1}{c}{--}  &             0.01  &  \mc{1}{l}{--}  &  GD path: step size \\
                 GDNSteps  &  \mc{1}{c}{--}  &            10000  &  \mc{1}{l}{--}  &  GD path: number of steps \\
               GDErrScale  &  \mc{1}{c}{--}  &              1.1  &  \mc{1}{l}{--}  &  Stop scan when error $>$ scale*errmin \\
              LinQuantile  &  \mc{1}{c}{--}  &            0.025  &  \mc{1}{l}{--}  &  Quantile of linear terms (removes outliers) \\
            GDPathEveFrac  &  \mc{1}{c}{--}  &              0.5  &  \mc{1}{l}{--}  &  Fraction of events used for the path search \\
           GDValidEveFrac  &  \mc{1}{c}{--}  &              0.5  &  \mc{1}{l}{--}  &  Fraction of events used for the validation \\
               fEventsMin  &  \mc{1}{c}{--}  &              0.1  &  \mc{1}{l}{--}  &  Minimum fraction of events in a splittable node \\
               fEventsMax  &  \mc{1}{c}{--}  &              0.9  &  \mc{1}{l}{--}  &  Maximum fraction of events in a splittable node \\
                   nTrees  &  \mc{1}{c}{--}  &               20  &  \mc{1}{l}{--}  &  Number of trees in forest. \\
               ForestType  &  \mc{1}{c}{--}  &         AdaBoost  &  AdaBoost, Random  &  Method to use for forest generation \\
              RuleMinDist  &  \mc{1}{c}{--}  &            0.001  &  \mc{1}{l}{--}  &  Minimum distance between rules \\
                   MinImp  &  \mc{1}{c}{--}  &             0.01  &  \mc{1}{l}{--}  &  Minimum rule importance accepted \\
                    Model  &  \mc{1}{c}{--}  &    ModRuleLinear  &  ModRule, ModRuleLinear, ModLinear  &  Model to be used \\
            RuleFitModule  &  \mc{1}{c}{--}  &           RFTMVA  &  RFTMVA, RFFriedman  &  Which RuleFit module to use \\
                RFWorkDir  &  \mc{1}{c}{--}  &        ./rulefit  &  \mc{1}{l}{--}  &  Friedman's RuleFit module (RFF): working dir \\
                 RFNrules  &  \mc{1}{c}{--}  &             2000  &  \mc{1}{l}{--}  &  RFF: Mximum number of rules \\
              RFNendnodes  &  \mc{1}{c}{--}  &                4  &  \mc{1}{l}{--}  &  RFF: Average number of end nodes 
\end{optiontableAuto}

%% file: Combining.tex
\section{Combining MVA Methods}\index{Combining}
\label{sec:combine}

In intricate classification or regression problems with a high demand for 
optimisation, or when treating variable spaces with strongly varying properties, 
it can be useful to combined MVA methods. There is large room for creativity 
inherent in such combinations. For TMVA we distinguish three classes of 
combinations:
\begin{enumerate}

\item {\em boosting} MVA methods,\index{Boosting method}

\item {\em categorising} MVA methods,\index{Category method}

\item building {\em committees} of MVA methods.\index{Committee method}

\end{enumerate}
A general MVA booster is already implemented in TMVA and is discussed in detail below. 
The other methods are under development. {\em Category methods} allow the user to specify
zones of the variables space, assigned by requirements on input variables and defining
distinct sub-populations of the training sample. In each of these zones, an independent
training is performed using the most appropriate MVA method and set of training variables 
in that zone. The division into categories in presence of distinct sub-populations reduces 
the correlations between the training variables and hence increases the classification
and regression performance. {\em Committee methods} allow one to input MVA methods into 
other MVA methods, a procedure that can be arbitrarily chained. 

All of these combined methods are of course MVA methods themselves, treated just like 
any other methods in TMVA for training, evaluation and application. 

\subsection{Boosted classifiers}\index{Boosted}
\label{sec:boosted}

Since generalised boosting is not yet available for regression in TMVA, we 
restrict the following discussion to classification applications.
A boosted\index{Boosting} classifier is a combination of a
collection of classifiers of the same type trained on the same sample
but with different events weights.\footnote{The Boost method is at the
  moment only applicable to classification problems.} The response of
the final classifier is a weighted response of each individual
classifier in the collection. The boosted classifier is potentially
more powerful and more stable with respect to statistical fluctuations
in the training sample.  The latter is particularly the case for
bagging as ``boost'' algorithm (\cf Sec.~\ref{sec:bagging}, page~\pageref{sec:bagging}).

The following sections do not apply to decision trees. We refer to
Sec.~\ref{sec:bdt} (page~\pageref{sec:bdt}) for a description of boosted 
decision trees. In the current version of TMVA only the AdaBoost and Bagging 
algorithms are implemented for the boost of arbitrary classifiers.  The boost
algorithms are described in detail in Sec.~\ref{sec:boost} on page
\pageref{sec:boost}.

\subsubsection{Booking options}

To book a boosted classifier, one needs to add the booster options to
the regular classifier's option string. The minimal option required is
the number of boost iterations \code{Boost_Num}, which must be
set to a value larger than zero.  Once the Factory detects a
\code{Boost_Num>0} in the option string it books a boosted classifier
and passes all boost options (recognised by the prefix \code{Boost_}) to
the Boost method and the other options to the boosted classifier.
%The alternative and more explicit booking method is to book a
%MethodBoost first, and then to book the specific classifier to it:
\begin{codeexample}
\begin{tmvacode}
factory->BookMethod( TMVA::Types::kLikelihood, "BoostedLikelihood",
       "Boost_Num=10:Boost_Type=Bagging:Spline=2:NSmooth=5:NAvEvtPerBin=50" );
\end{tmvacode}
\caption[.]{\codeexampleCaptionSize Booking of the boosted classifier: 
         the first argument is the predefined enumerator, the 
         second argument is a user-defined string identifier, and the third 
         argument is the configuration options string. All options with the 
         prefix \code{Boost_} (in this example the first two options) are 
         passed on to the boost method, the other options are provided to the 
         regular classifier (which in this case is Likelihood). Individual 
         options are separated by a ':'. See Sec.~\ref{sec:usingtmva:booking} 
         for more information on the booking. 
}
\end{codeexample}

The boost configuration options are given in Option 
Table~\ref{opt:mva::boost}.

\begin{option}[!t]
\input optiontables/MVA__Boost.tex
\caption[.]{\optionCaptionSize Boosting configuration options. These
  options can be simply added to a simple classifier's option string
  or used to form the option string of an explicitly booked boosted
  classifier.}
\label{opt:mva::boost}
\end{option}

The options most relevant for the boost process are the number of boost
iterations, \code{Boost_Num}, and the choice of the boost algorithm,
\code{Boost_Type}.  In case of \code{Boost_Type=AdaBoost}, the option
\code{Boost_Num} describes the maximum number of boosts. The algorithm
is iterated until an error rate of 0.5 is reached or until
\code{Boost_Num} iterations occurred. If the algorithm terminates after
to few iterations, the number might be extended by decreasing the
$\beta$ variable (option \code{Boost_AdaBoostBeta}).  Within the
AdaBoost algorithm a decision must be made how to classify an event, a
task usually done by the user. For some classifiers it is straightforward to
set a cut on the MVA response to define signal-like events. For the others, 
the MVA cut is chosen that the error rate is minimised. The option \code{Boost_RecalculateMVACut} 
determines whether this cut should be recomputed for every boosting iteration.
In case of Bagging as boosting algorithm the number of boosting
iterations always reaches \code{Boost_Num}.

By default boosted classifiers are combined as a weighted average with 
weights computed from the misclassification error (option
\code{Boost_MethodWeightType=ByError}). It is also possible to use
the arithmetic average instead (\code{Boost_MethodWeightType=Average}).

\subsubsection{Boostable classifiers}

The boosting process was originally introduced for  simple classifiers.  
The most commonly boosted classifier is the decision tree (DT -- \cf Sec.~\ref{sec:bdt}, 
page~\pageref{sec:bdt}). Decision trees need to be boosted a few hundred
times to effectively stabilise the BDT response and achieve optimal 
performance. 

Another simple classifier in the TMVA package is the Fisher
discriminant~(\cf Sec.~\ref{sec:fisher}, page~\pageref{sec:fisher} -- which is equivalent
to the linear discriminant described in Sec.~\ref{sec:ld}). Because the output 
of a Fisher discriminant represents a linear combination of the input variables, 
a linear combination of different Fisher discriminants is again a Fisher 
discriminant. Hence linear boosting cannot improve the performance. It is
nevertheless possible to effectively boost a linear discriminant by applying
the linear combination not on the discriminant's output, but on the actual
classification results provided.\footnote
{
   Note that in the TMVA standard example, which uses linearly correlated, 
   Gaussian-distributed input variables for signal and background, a
   single Fisher discriminant already provides the theoretically maximum 
   separation power. Hence on this example, no further gain can be
   expected by boosting, no matter what ``tricks'' are applied.
} 
This corresponds to a ``non-linear'' transformation of the
Fisher discriminant output according to a step function. The Boost 
method in TMVA also features a fully non-linear transformation that is 
directly applied to the classifier response value. Overall, the following
transformations are available:
\begin{itemize}
\item{\em linear:} no transformation is applied to the MVA output,
\item{\em step:}   the output is $-1$ below the step
                   and $+1$ above (default setting),
\item{\em log:}    logarithmic transformation of the output.
\end{itemize}

The macro \code{Boost.C} (residing in the \code{macros} (\code{test}) directory 
for the sourceforge (ROOT) version of TMVA) provides examples for the use of 
these transformations to boost a Fisher discriminant. We point out that the 
performance of a boosted classifier strongly depends on its characteristics
as well as on the nature of the input data. A careful adjustment of options is required
if AdaBoost is applied to an arbitrary classifier, since otherwise it might even 
lead to a worse performance than for the unboosted method.

\subsubsection{Monitoring tools}

The current GUI provides figures to monitor the boosting process. Plotted are 
the boost weights, the classifier weights in the boost ensemble, the classifier 
error rates, and the classifier error rates using unboosted event weights.
In addition, when the option \code{Boost_MonitorMethod=T} is set,
monitoring histograms are created for each classifier in the boost ensemble. 
The histograms generated during the boosting process provide useful insight 
into the behaviour of the boosted classifiers and help to adjust to the optimal 
number of boost iterations. These histograms are saved in a separate folder
in the output file, within the folder of {\tt MethodBoost/<Title>/}.
Besides the specific classifier monitoring histograms, this
folder also contains the MVA response of the classifier for the training
and testing samples.

\subsubsection{Variable ranking}

The present boosted classifier implementation does not provide a ranking of 
the input variables.

%% file: optiontables/MVA__Boost.tex
\begin{optiontableAuto}
                Boost\_Num  &  \mc{1}{c}{--}  &              100  &  \mc{1}{l}{--}  &  Number of times the classifier is boosted \\
      Boost\_MonitorMethod  &  \mc{1}{c}{--}  &             True  &  \mc{1}{l}{--}  &  Whether to write monitoring histogram for each boosted classifier \\
               Boost\_Type  &  \mc{1}{c}{--}  &         AdaBoost  &  AdaBoost, Bagging  &  Boosting type for the classifiers \\
   Boost\_MethodWeightType  &  \mc{1}{c}{--}  &          ByError  &  ByError, Average, LastMethod  &  How to set the final weight of the boosted classifiers \\
  Boost\_RecalculateMVACut  &  \mc{1}{c}{--}  &             True  &  \mc{1}{l}{--}  &  Whether to recalculate the classifier MVA Signallike cut at every boost iteration \\
       Boost\_AdaBoostBeta  &  \mc{1}{c}{--}  &                1  &  \mc{1}{l}{--}  &  The ADA boost parameter that sets the effect of every boost step on the events' weights \\
          Boost\_Transform  &  \mc{1}{c}{--}  &             step  &  step, linear, log  &  Type of transform applied to every boosted method linear, log, step 
\end{optiontableAuto}

%% file: Conclusions.tex
\section{Which MVA method should I use for my problem?}
\label{sec:whatMVAshouldIuse}

There is obviously no general answer to that question. To guide the user, we have 
attempted a coarse assessment of various MVA properties in Table~\ref{tab:classQA}. 
Simplicity is a virtue, but only if it is not at the expense of significant loss of
discrimination power. Robustness with respect to overtraining could become 
an issue when the training sample is scarce. Some methods require more attention 
than others in this regard. For example, boosted decision trees are particularly 
vulnerable to overtraining if used without care.\footnote
{
   However, experience shows that the BDT performance is amazingly robust -- even 
   for strongly overtrained decision trees.
} 
To circumvent overtraining a problem-specific adjustment of the pruning strength 
parameter is required. 

To assess whether a linear discriminant analysis (LDA) could be sufficient 
for a classification (regression) problem, the user is advised to analyse the 
correlations among the discriminating variables (among the variables and regression 
target) by inspecting scatter and profile plots (it is not enough to print the 
correlation coefficients, which by definition are linear only). Using an LDA 
greatly reduces the number of parameters to be adjusted and hence allow smaller
training samples. It usually is robust with respect to generalisation
to larger data samples. For moderately intricate problems, the function discriminant 
analysis (FDA) with some added nonlinearity may be found sufficient. It is always 
useful to cross-check its performance against several of the sophisticated nonlinear
methods to see how much can be gained over the use of the simple and very
transparent FDA.

\begin{table}[t]
{\small
\setlength{\tabcolsep}{0.0pc}
\begin{tabular*}{\textwidth}{@{\extracolsep{\fill}}p{1.3cm}p{2.7cm}p{1cm}p{1cm}p{1cm}p{1cm}p{1cm}p{1cm}p{1cm}p{1cm}p{1cm}p{1cm}} 
\hline
&&&&&&&&&\\[\BD]
  && \mc{9}{c}{MVA M{\footnotesize ETHOD}} \\[\AD]
  \mc{2}{c}{C{\footnotesize RITERIA}} 
                       & Cuts & Likeli-hood & PDE-RS~/ k-NN & PDE-Foam & H-Matrix & Fisher /~LD & MLP & BDT & Rule-Fit & SVM \\
&&&&&&&&&&\\[\BD]
\hline
&&&&&&&&&&\\[\BD]
$~~~~~$ Perfor- 
             & No\;or\;linear ~~~~~~ correlations   
                       & \OK & \Good   & \OK   & \OK   & \OK & \Good & \Good & \OK & \Good & \OK \\
mance        & Nonlinear ~~~~~~~~~~~ correlations
                       & \Bad & \Bad   & \Good & \Good & \Bad & \Bad & \Good & \Good & \Good & \Good \\
&&&&&&&&&&\\[\BD]                                       
\hline                                                 
&&&&&&&&&&\\[\BD]                                       
             & Training                                
                       & \Bad  & \Good & \Good & \Good & \Good   & \Good & \OK & \Bad & \OK & \Bad\\
\rs{Speed}   & Response                                
                       & \Good & \Good & \Bad  & \OK   & \Good    & \Good & \Good & \OK & \Good & \OK \\
&&&&&&&&&&\\[\BD]                                       
\hline                                                 
&&&&&&&&&&\\[\BD]                                       
Robust-                                                
             & Overtraining                            
                       & \Good & \OK   & \OK   & \OK   & \Good     & \Good & \OK & \Bad  & \OK & \Good \\
ness         & Weak variables                          
                       & \Good & \OK   & \Bad  & \Bad  & \Good     & \Good & \OK & \Good & \OK & \OK \\
&&&&&&&&&&\\[\BD]                                       
\hline                                                 
&&&&&&&&&&\\[\BD]                                       
\mc{2}{l}{Curse of dimensionality}                     
                       & \Bad & \Good  & \Bad  & \Bad  & \Good    & \Good & \OK & \OK & \OK \\
&&&&&&&&&&\\[\BD]                                       
\hline                                                 
&&&&&&&&&&\\[\BD]                                       
\mc{2}{l}{Transparency}                                
                       & \Good & \Good & \OK   & \OK   & \Good    & \Good & \Bad & \Bad & \Bad & \Bad \\[\AD]
\hline      
\end{tabular*}
}
\caption[.]{\captionfont Assessment of MVA method properties. The symbols stand for 
        the attributes ``good'' ($\star\star$), ``fair'' ($\star$) and ``bad'' ($\circ$).
        ``Curse of dimensionality'' refers to the ``burden'' of required increase in 
        training statistics and processing time when adding more input variables. See also 
        comments in the text. The FDA method is not listed here since its properties
        depend on the chosen function. }
\label{tab:classQA}
\end{table}
For problems that require a high degree of optimisation
and allow to use a large number of input variables, complex nonlinear methods
like neural networks, the support vector machine, boosted decision trees and/or 
RuleFit are more appropriate. 

Very involved multi-dimensional variable correlations with strong nonlinearities 
are usually best mapped by the multidimensional probability density estimators 
such as PDE-RS, k-NN and PDE-Foam, requiring however a reasonably low number of 
input variables.

For RuleFit classification we emphasise that the TMVA implementation differs 
from Friedman-Popescu's original code~\cite{RuleFit}, with slightly better robustness 
and out-of-the-box performance for the latter version. In particular, the behaviour of the 
original code with respect to nonlinear correlations and the curse of dimensionality would 
have merited two stars.\footnote
{
   An interface to Friedman-Popescu's original code can be requested from the TMVA
   authors. See Sec.~\ref{sec:friedman}.
} 
We also point out that the excellent performance for by majority linearly correlated 
input variables is achieved somewhat artificially by adding a Fisher-like term to the 
RuleFit classifier (this is the case for both implementations, \cf\  Sec.~\ref{sec:rulefit}
on page~\pageref{sec:rulefit}). 

\section{TMVA implementation status summary for classification and regression}
\label{sec:classifierSummary}

All TMVA methods are fully operational for user analysis, requiring training, testing,
evaluating and reading for the application to unknown data samples. Additional 
features are optional and -- despite our attempts to provide a fully transparent analysis
-- not yet uniformly available. A status summary is given in Table~\ref{tab:methodStatus}
and annotated below.

Although since TMVA 4 the framework supports multi-dimensional MVA outputs it has not yet 
been implemented for classification. For regression, only a few methods are fully multi-target 
capable so far (see Table \ref{tab:methodStatus}). 

Individual event-weight support is now commonly realised, only missing (and not 
foreseen to be provided) for the less recommended neural network CFMlpANN. Support of
negative event weights occurring, \eg, in NLO MC requires more scrutiny as discussed in 
Sec.~\ref{sec:NegativeEventWeights} on page~\pageref{sec:NegativeEventWeights}.

Ranking of the input variables cannot be defined in a straightforward manner for all 
MVA methods. Transparent and objective variable ranking through performance comparison 
of the MVA method under successive elimination of one input variable at a time is under 
consideration (so far only realised for the naive-Bayes likelihood classifier).

Standalone C++ response classes (not required when using the Reader application) 
are generated by the majority of the classifiers, but not yet for regression analysis. 
The missing ones for PDE-RS, PDE-Foam, k-NN, Cuts and CFMlpANN will only be considered on explicit request.

The availability of help messages, which assist the user with the performance tuning 
and which are printed on standard output when using the booking option 'H', is 
complete. 

Finally, custom macros are provided for some MVA methods to analyse specific 
properties, such as the fidelity of likelihood reference distributions or the 
neural network architecture, etc. More macros can be added upon user request.
\begin{sidewaystable}
\begin{center}
{\small
\setlength{\tabcolsep}{0.3pc}
\begin{tabular}{lcccccccccccccc}\hline
&&&&&&&\\[\BD]
                     & Classi- & Regress- & \mc{2}{c}{Multi-class/target} & \mc{2}{c}{Treats event weights:}  & Variable & Standalone & Help & Custom \\
\rs{MVA method}      & fication
                            & ion 
                                   & classification
                                          & regression
                                                 &  positive 
                                                        & negative 
                                                               & ranking 
                                                                      & response class  
                                                                             & messages 
                                                                                    & macros\\[\AD]
\hline
&&&&&&&\\[\BD]
Cut optimisation     & \YES & \NO  & \NO  & \NO  & \YES & \NO  & \NO  & \NO  & \YES & \NO  \\[\AD]
Likelihood           & \YES & \NO  & \NO  & \NO  & \YES & \YES & \YES & \YES & \YES & \YES \\
PDE-RS               & \YES & \YES & \NO  & \NO  & \YES & \YES & \NO  & \NO  & \YES & \NO  \\
PDE-Foam             & \YES & \YES & \NO  & \YES  & \YES & \YES & \NO  & \NO  & \YES & \YES \\
k-NN                 & \YES & \YES & \NO  & \YES & \YES & \YES & \NO  & \NO  & \YES & \NO  \\[\AD]
H-Matrix             & \YES & \NO  & \NO  & \NO  & \YES & \NO  & \YES & \YES & \YES & \NO  \\
Fisher               & \YES & \NO  & \NO  & \NO  & \YES & \NO  & \YES & \YES & \YES & \NO  \\
LD                   & \YES & \YES & \NO  & \NO  & \YES & \NO  & \YES & \YES & \YES & \NO  \\
FDA                  & \YES & \YES & \NO  & \NO  & \YES & \NO  & \NO  & \YES & \YES & \NO  \\[\AD]
MLP                  & \YES & \YES & \NO  & \YES & \YES & \NO  & \YES & \YES & \YES & \YES \\
TMlpANN$^{(\star)}$   & \YES & \NO  & \NO  & \NO  & \YES & \NO  & \NO  & \YES & \YES & \NO  \\
CFMlpANN             & \YES & \NO  & \NO  & \NO  & \NO  & \NO  & \NO  & \NO  & \NO  & \NO  \\[\AD]
SVM                  & \YES & \NO  & \NO  & \NO  & \YES & \YES & \NO  & \YES & \YES & \NO  \\[\AD]
BDT                  & \YES & \YES & \NO  & \NO  & \YES & \YES & \YES & \YES & \YES & \YES \\
RuleFit              & \YES & \NO  & \NO  & \NO  & \YES & \NO  & \YES & \YES & \YES & \YES \\[\AD]
\hline      
&&&&&&&\\[\BD]
\end{tabular}
}
\end{center}
\vspace{-0.4cm}
\footnotesize \hspace{0.8cm}$^{(\star)}$Not a generic TMVA method $-$ interface to ROOT class 
               \code{TMultiLayerPerceptron}.
\caption[.]{\captionfont Status of the methods with respect to various TMVA features. See 
            text for comments. Note that the column ``Standalone response class'' only refers
            to classification. It is yet unavailable for regression. }
\label{tab:methodStatus}
\end{sidewaystable}

\section{Conclusions and Plans}
\label{sec:conclusions}

TMVA is a toolkit that unifies highly customisable multivariate (MVA) classification and 
regression algorithms in a single framework thus ensuring convenient use and an objective 
performance assessment. It is designed for machine learning applications in high-energy 
physics, but not restricted to these. Source code and library of TMVA-v.3.5.0 and higher
versions are part of the standard ROOT distribution kit (v5.14 and higher). The 
newest TMVA development version can be downloaded from Sourceforge.net at 
\urlsm{http://tmva.sourceforge.net}.

This Users Guide introduced the main steps of a TMVA analysis allowing a user to optimise 
and perform her/his own multivariate classification or regression. Let us recall the main 
features of the TMVA design and purpose:
\begin{itemize}

\item TMVA works in transparent factory mode to allow an unbiased 
      performance assessment and comparison: all MVA methods
      see the same training and test data, and are evaluated following 
      the same prescription. 

\item A complete TMVA analysis consists of two steps:
      \begin{enumerate}

      \item {\bf Training:}
            the ensemble of available and optimally customised MVA methods are 
            trained and tested on independent signal and background data samples; the 
            methods are evaluated and the most appropriate (performing and concise) 
            ones are selected.

      \item {\bf Application:}
            selected trained MVA methods are used for the classification of data samples with
            unknown signal and background composition, or for the estimate of unknown 
            target values (regression).  

      \end{enumerate}
      
\item A Factory class object created by the user organises the 
      customisation and interaction with the MVA methods for the training, 
      testing and evaluation phases of the TMVA analysis. The training results 
      together with the configuration of the methods are written to
      result (``weight'') files in XML format.

\item Standardised outputs during the Factory running, and dedicated ROOT 
      macros allow a refined assessment of each method's behaviour and 
      performance for classification and regression.
     
\item Once appropriate methods have been chosen by the user, they can be 
      applied to data samples with unknown classification or target values. Here, 
      the interaction with the methods occurs through a Reader class 
      object created by the user. A method is booked by giving the path to its
      weight file resulting from the training stage. Then, inside
      the user's event loop, the MVA response is returned by the Reader for 
      each of the booked MVA method, as a function of the event values of the 
      discriminating variables used as input for the classifiers. Alternatively,
      for classification, the user may request from the Reader the probability 
      that a given event belongs to the signal hypothesis and/or the event's Rarity.

\item In parallel to the XML files, TMVA generates standalone C++ classes after the 
      training, which can be used for classification problems (feature not available 
      yet for regression). Such classes are available for all classifiers except for 
      cut optimisation, PDE-RS, PDE-Foam, k-NN and the old CFMlpANN.

\end{itemize}
We give below a summary of the TMVA methods, outlining the current state of their
implementation, their advantages and shortcomings.
\begin{itemize}

\item {\em Rectangular Cut Optimisation} \\
      The current implementation is mature. 
      It includes speed-optimised range searches using binary 
      trees, and three optimisation algorithms: Monte Carlo sampling,
      a Genetic Algorithm and Simulated Annealing. In spite of these 
      tools, optimising the cuts for a large number of discriminating variables 
      remains challenging. The user is advised to reduce the available 
      dimensions to the most significant variables (\eg, using a principal
      component analysis) prior to optimising the cuts.

\item {\em Likelihood} \\
      Automatic non-parametric probability density function (PDF) estimation through 
      histogram smoothing and interpolation with various spline functions and 
      quasi-unbinned kernel density estimators is implemented. The PDF description
      can be individually tuned for each input variable. 

\item {\em PDE-RS} \\
      The multidimensional probability density estimator (PDE) approach is in an advanced 
      development stage featuring adaptive range search, several kernel estimation 
      methods, and speed optimised range search using event sorting in binary trees.
      It has also been extended to regression.

\item {\em PDE-Foam} \\
      This new multidimensional PDE algorithm uses self-adapting phase-space 
      binning and is a fast realisation of PDE-RS in fixed volumes, which are
      determined and optimised during the training phase. Much work went 
      into the development of PDE-Foam. It has been thoroughly tested, and 
      can be considered a mature method. PDE-Foam performs classification 
      and regression analyses.

\item {\em k-NN} \\
      The k-Nearest Neighbour classifier is also in a mature state, featuring
      both classification and regression.
      The code has been well tested and shows satisfactory results. 
      With scarce training statistics it may slightly underperform in comparison
      with PDE-RS, whereas it is significantly faster in the application to large
      data samples.

\item {\em Fisher and H-Matrix}\\
      Both are mature algorithms, featuring linear discrimination for classification 
      only. Higher-order correlations are taken care of by FDA (see below).

\item {\em Linear Discriminant (LD)}\\
      LD is equivalent to Fisher but providing both classification and linear 
      regression.

\item {\em Function Discriminant Analysis (FDA)} \\
      FDA is a mature algorithm, which has not been extensively used yet. It 
      extends the linear discriminant to moderately non-linear correlations that
      are fit to the training data.

\item {\em Artificial Neural Networks} \\
      Significant work went into the 
      implementation of fast feed-forward multilayer perceptron algorithms
      into TMVA. Two external ANNs have been integrated as fully independent
      methods, and another one has been newly developed for TMVA, with emphasis 
      on flexibility and speed. The performance of the latter ANN (MLP) has been 
      cross checked against the Stuttgart ANN (using as an example $\tau$ 
      identification in ATLAS), and was found to achieve competitive performance.
      The MLP ANN also performs multi-target regression.

\item {\em Support Vector Machine}\\
      SVM is a relatively new multivariate analysis algorithm with a strong 
      statistical background. It performs well for nonlinear discrimination 
      and is insensitive to overtraining. Optimisation is 
      straightforward due to a low number of adjustable parameters (only two 
      in the case of Gaussian kernel). The response speed is slower than 
      for a not-too-exhaustive neural network, but comparable with other 
      nonlinear methods. SVM is being extended to multivariate regression.

\item {\em Boosted Decision Trees}\\
      The BDT implementation has received constant attention over the years of
      its development. The current version includes additional features like 
      bagging or gradient boosting, and manual or automatic pruning of 
      statistically insignificant nodes. It is a highly performing MVA method
      that also applies to regression problems. 

\item {\em RuleFit} \\
      The current version has the possibility to run either the original 
      program written by J.~Friedman~\cite{RuleFit} or an independent TMVA 
      implementation. The TMVA version has been improved both in speed and 
      performance and achieves almost equivalent results with respect to the 
      original one, requiring however somewhat more tuning.

\end{itemize}

The new framework introduced with TMVA~4 provides the flexibility to combine
MVA methods in a general fashion. Exploiting these capabilities for classification
and regression however requires to create so-called committee methods for each 
combination. So far, we provide a generalised Boost method, allowing to boost
any classifier by simply setting the variable \code{Boost_Num} in the configuration
options to a positive number (plus possible adjustment of other configuration
parameters). The result is a potentially powerful committee method unifying
the excellent properties of boosting with MVA methods that already represent 
highly optimised algorithms. 

Boosting is not the only combination the new framework allows us to establish. 
We look forward to the implementation of a categorised committee method
that separates the input variables space into regions in which different MVA method
and different variables are applied. Moreover it is planned to develop a 
committee method that allows to insert the result of MVA methods as input 
to another MVA method. 

\subsubsection*{Acknowledgements}
\addcontentsline{toc}{section}{Acknowledgements}
\label{sec:Acknowledgments}

\begin{details}
The fast growth of TMVA would not have been possible without the 
contribution and feedback from many developers (also co-authors of this Users Guide)
and users to whom we are indebted. 
We thank in particular the CERN Summer students Matt Jachowski (Stanford U.) for the 
implementation of TMVA's MLP neural network, and Yair Mahalalel (Tel Aviv U.) for a 
significant improvement of PDE-RS. The Support Vector Machine has been contributed to 
TMVA by Andrzej Zemla and Marcin Wolter (IFJ PAN Krakow), and the k-NN method has 
been written by Rustem Ospanov (Texas U.). 
We are grateful to
Lucian Ancu,
Doug Applegate, 
Kregg Arms, 
Ren\'e Brun and the ROOT team, 
Andrea Bulgarelli,
Marc Escalier,
Zhiyi Liu, 
Colin Mclean,
Elzbieta Richter-Was, 
Alfio Rizzo,
Lydia Roos,
Vincent Tisserand,
Alexei Volk,
Jiahang Zhong
for helpful feedback and bug reports. Thanks also to Lucian Ancu for 
improving the plotting macros.
\end{details}

%% file: Appendix.tex
\begin{appendix}

\section{More Classifier Booking Examples}
\label{sec:appendix:booking}

The Code Examples~\ref{codeex:factoryBookingAll1} and \ref{codeex:factoryBookingAll2}
give a (non-exhaustive) collection of classifier bookings with appropriate 
default options. They correspond to the example training job \code{TMVAClassification.C}.
\begin{codeexample}
\begin{tmvacode}
// Cut optimisation using Monte Carlo sampling
factory->BookMethod( TMVA::Types::kCuts, "Cuts", 
    "!H:!V:FitMethod=MC:EffSel:SampleSize=200000:VarProp=FSmart" );

// Cut optmisation using Genetic Algorithm
factory->BookMethod( TMVA::Types::kCuts, "CutsGA",
    "H:!V:FitMethod=GA:CutRangeMin=-10:CutRangeMax=10:VarProp[1]=FMax:EffSel:\
     Steps=30:Cycles=3:PopSize=400:SC_steps=10:SC_rate=5:SC_factor=0.95" );

// Cut optmisation using Simulated Annealing algorithm
factory->BookMethod( TMVA::Types::kCuts, "CutsSA",
    "!H:!V:FitMethod=SA:EffSel:MaxCalls=150000:KernelTemp=IncAdaptive:\
     InitialTemp=1e+6:MinTemp=1e-6:Eps=1e-10:UseDefaultScale" );

// Likelihood classification (naive Bayes) with Spline PDF parametrisation
factory->BookMethod( TMVA::Types::kLikelihood, "Likelihood", 
    "H:!V:TransformOutput:PDFInterpol=Spline2:NSmoothSig[0]=20:\
     NSmoothBkg[0]=20:NSmoothBkg[1]=10:NSmooth=1:NAvEvtPerBin=50" ); 

// Likelihood with decorrelation of input variables
factory->BookMethod( TMVA::Types::kLikelihood, "LikelihoodD", 
    "!H:!V:!TransformOutput:PDFInterpol=Spline2:NSmoothSig[0]=20:\
     NSmoothBkg[0]=20:NSmooth=5:NAvEvtPerBin=50:VarTransform=Decorrelate" ); 

// Likelihood with unbinned kernel estimator for PDF parametrisation
factory->BookMethod( TMVA::Types::kLikelihood, "LikelihoodKDE", 
    "!H:!V:!TransformOutput:PDFInterpol=KDE:KDEtype=Gauss:KDEiter=Adaptive:\
     KDEFineFactor=0.3:KDEborder=None:NAvEvtPerBin=50" ); 
\end{tmvacode}
\caption[.]{\codeexampleCaptionSize Examples for booking MVA methods in TMVA for
            application to classification and -- where available -- to regression problems. 
            The first argument is 
            a unique type enumerator (the avaliable types can be looked up in \code{src/Types.h}),
            the second is a user-defined name (must be unique among all booked classifiers),
            and the third a configuration option string that is specific to the 
            classifier. For options that are not set in the 
            string default values are used. The syntax of the options should become
            clear from the above examples. Individual options are separated by a 
            ':'. Boolean variables can be set either explicitly as 
            \code{MyBoolVar=True/False}, or just via \code{MyBoolVar/!MyBoolVar}.
            All concrete option variables are explained in the tools and classifier sections
            of this Users Guide. The list is continued in Code Example~\ref{codeex:factoryBookingAll2}.}
\label{codeex:factoryBookingAll1}
\end{codeexample}
\begin{codeexample}
\begin{tmvacode}
// Probability density estimator range search method (multi-dimensional)
factory->BookMethod( TMVA::Types::kPDERS, "PDERS", 
    "!H:V:NormTree=T:VolumeRangeMode=Adaptive:KernelEstimator=Gauss:\
     GaussSigma=0.3:NEventsMin=400:NEventsMax=600" );

// Multi-dimensional PDE using self-adapting phase-space binning
factory->BookMethod( TMVA::Types::kPDEFoam, "PDEFoam", 
    "H:V:SigBgSeparate=F:TailCut=0.001:VolFrac=0.0333:nActiveCells=500:\
     nSampl=2000:nBin=5:CutNmin=T:Nmin=100:Kernel=None:Compress=T" );

// k-Nearest Neighbour method (similar to PDE-RS)
factory->BookMethod( TMVA::Types::kKNN, "KNN", 
    "H:nkNN=20:ScaleFrac=0.8:SigmaFact=1.0:Kernel=Gaus:UseKernel=F:\
     UseWeight=T:!Trim" );

// H-matrix (chi-squared) method
factory->BookMethod( TMVA::Types::kHMatrix, "HMatrix", "!H:!V" ); 

// Fisher discriminant (also creating Rarity distribution of MVA output)
factory->BookMethod( TMVA::Types::kFisher, "Fisher", 
    "H:!V:Fisher:CreateMVAPdfs:PDFInterpolMVAPdf=Spline2:NbinsMVAPdf=60:\
     NsmoothMVAPdf=10" );

// Fisher discriminant with Gauss-transformed input variables
factory->BookMethod( TMVA::Types::kFisher, "FisherG", "VarTransform=Gauss" );

// Fisher discriminant with principle-value-transformed input variables
factory->BookMethod( TMVA::Types::kFisher, "FisherG", "VarTransform=PCA" );

// Boosted Fisher discriminant
factory->BookMethod( TMVA::Types::kFisher, "BoostedFisher", 
    "Boost_Num=20:Boost_Transform=log:\
     Boost_Type=AdaBoost:Boost_AdaBoostBeta=0.2");

// Linear discriminant (same as Fisher, but also performing regression)
factory->BookMethod( TMVA::Types::kLD, "LD", "H:!V:VarTransform=None" );

// Function discrimination analysis (FDA), fitting user-defined function
factory->BookMethod( TMVA::Types::kFDA, "FDA_MT",
    "H:!V:Formula=(0)+(1)*x0+(2)*x1+(3)*x2+(4)*x3:\
     ParRanges=(-1,1);(-10,10);(-10,10);(-10,10);(-10,10):FitMethod=MINUIT:\
     ErrorLevel=1:PrintLevel=-1:FitStrategy=2:UseImprove:UseMinos:SetBatch" );
\end{tmvacode}
\caption[.]{\codeexampleCaptionSize Continuation from Code Example~\ref{codeex:factoryBookingAll1}.
            Continued in Code Example~\ref{codeex:factoryBookingAll2}.}
\label{codeex:factoryBookingAll2}
\end{codeexample}
\begin{codeexample}
\begin{tmvacode}
// Artificial Neural Network (Multilayer perceptron) - TMVA version
factory->BookMethod( TMVA::Types::kMLP, "MLP", 
    "H:!V:NeuronType=tanh:VarTransform=N:NCycles=600:HiddenLayers=N+5:\
     TestRate=5" );

// NN with BFGS quadratic minimisation
factory->BookMethod( TMVA::Types::kMLP, "MLPBFGS", 
    "H:!V:NeuronType=tanh:VarTransform=N:NCycles=600:HiddenLayers=N+5:\
     TestRate=5:TrainingMethod=BFGS" );

// NN (Multilayer perceptron) - ROOT version
factory->BookMethod( TMVA::Types::kTMlpANN, "TMlpANN", 
    "!H:!V:NCycles=200:HiddenLayers=N+1,N:LearningMethod=BFGS:
     ValidationFraction=0.3"  ); 

// NN (Multilayer perceptron) - ALEPH version (depreciated)
factory->BookMethod( TMVA::Types::kCFMlpANN, "CFMlpANN", 
    "!H:!V:NCycles=2000:HiddenLayers=N+1,N"  ); 

// Support Vector Machine
factory->BookMethod( TMVA::Types::kSVM, "SVM", "Gamma=0.25:Tol=0.001" );

// Boosted Decision Trees with adaptive boosting 
factory->BookMethod( TMVA::Types::kBDT, "BDT", 
    "!H:!V:NTrees=400:nEventsMin=400:MaxDepth=3:BoostType=AdaBoost:\
     SeparationType=GiniIndex:nCuts=20:PruneMethod=NoPruning" );

// Boosted Decision Trees with gradient boosting 
factory->BookMethod( TMVA::Types::kBDT, "BDTG", 
    "!H:!V:NTrees=1000:BoostType=Grad:Shrinkage=0.30:UseBaggedGrad:\
     GradBaggingFraction=0.6:SeparationType=GiniIndex:nCuts=20:\
     PruneMethod=CostComplexity:PruneStrength=50:NNodesMax=5" );

// Boosted Decision Trees with bagging
factory->BookMethod( TMVA::Types::kBDT, "BDTB", 
    "!H:!V:NTrees=400:BoostType=Bagging:SeparationType=GiniIndex:\
     nCuts=20:PruneMethod=NoPruning" );

// Predictive learning via rule ensembles (RuleFit)
factory->BookMethod( TMVA::Types::kRuleFit, "RuleFit",
    "H:!V:RuleFitModule=RFTMVA:Model=ModRuleLinear:MinImp=0.001:\
     RuleMinDist=0.001:NTrees=20:fEventsMin=0.01:fEventsMax=0.5:\
     GDTau=-1.0:GDTauPrec=0.01:GDStep=0.01:GDNSteps=10000:GDErrScale=1.02" );
\end{tmvacode}
\caption[.]{\codeexampleCaptionSize Continuation from Code Example~\ref{codeex:factoryBookingAll2}.}
\label{codeex:factoryBookingAll3}
\end{codeexample}

\end{appendix}

%%% Local Variables: 
%%% mode: latex
%%% TeX-master: "TMVAUsersGuide"
%%% End: 